\documentclass[11pt, a4paper, logo, copyright, nonumbering]{map}

\usepackage{natbib}

\usepackage{CJKutf8}

\usepackage{xargs}  

\usepackage{todonotes}  

\usepackage{multirow}

\usepackage{cleveref}

\usepackage{amsmath}
\usepackage{dsfont}

\usepackage{svg}
\usepackage[breakable]{tcolorbox}
\usepackage{fontawesome}
\usepackage{xcolor}

\usepackage{makecell}

\usepackage[utf8]{inputenc}
\usepackage{booktabs}
\usepackage{graphicx}
\usepackage{array}
\usepackage{tabularx}
\usepackage{xcolor,colortbl}  
\definecolor{boxcolor}{HTML}{d92523} 
\definecolor{bulbcolor}{HTML}{e3b87f} 
\usepackage{tikz}
\usepackage{arydshln}
\usepackage{url}
\usepackage{hyperref}

\usepackage{subfigure}
\usepackage{algorithm}
\usepackage{algorithmic}

\hypersetup{
    colorlinks=true,            
    linkcolor=blue,             
    filecolor=magenta,          
    urlcolor=cyan,              
    citecolor=purple,             
    pdftitle={Overleaf Example},
    pdfpagemode=FullScreen,
    }
 
\renewcommand{\today}{}
\newcommand{\firstcolor}[1]{%
  \tikz[baseline,yshift=0.75ex]{\node[fill=green!15, draw=white, rounded corners=5pt, inner sep=8pt] {\makebox[1cm][c]{#1}}}%
}
\newcommand{\secondcolor}[1]{%
  \tikz[baseline,yshift=0.75ex]{\node[fill=cyan!15, draw=white, rounded corners=5pt, inner sep=8pt] {\makebox[1cm][c]{#1}}}%
}
\newcommand{\thirdcolor}[1]{%
  \tikz[baseline,yshift=0.75ex]{\node[fill=orange!15, draw=white, rounded corners=5pt, inner sep=8pt] {\makebox[1cm][c]{#1}}}%
}

\newcommand{\ourmethod}{CryptoX}
\newcommand{\benchmark}{CryptoBench}

\title{\centering CryptoX : Compositional Reasoning Evaluation of Large Language Models}

\makeatletter
\renewcommand{\@makefnmark}{} 
\footnotetext{*$\ $ Equal Technical Contributions.}
\footnotetext{\textsuperscript{\dag} $\ $Corresponding Authors.}
\AtBeginDocument{%
  \footnotetext{\protect\url{https://github.com/multimodal-art-projection/CryptoX}}%
}
\makeatother

\author{
\textbf{Jiajun Shi\textsuperscript{\rm 1 2 *}, Chaoren Wei\textsuperscript{\rm 1 2 *}, Liqun Yang\textsuperscript{\rm 2 *}, Zekun Moore Wang\textsuperscript{\rm 1 2}, Chenghao Yang\textsuperscript{\rm 3 4}, \\ Ge Zhang\textsuperscript{\rm 1 3},  Stephen Huang\textsuperscript{\rm 1 3}, Tao Peng\textsuperscript{\rm 3}, Jian Yang\textsuperscript{\rm 2 \dag}, Zhoufutu Wen\textsuperscript{\rm 1 3 \dag}} \\
 \textsuperscript{\rm 1}M-A-P; \textsuperscript{\rm 2}Beihang University; \textsuperscript{\rm 3}ByteDance.Inc;\\\textsuperscript{\rm 4}University of Science and Technology of China;
}

\begin{abstract}
The compositional reasoning ability has long been regarded as critical to the generalization and intelligence emergence~\citep{CompositionalReasoning,wang2024grokked} of large language models (\textbf{LLMs}). 
However, despite numerous reasoning-related benchmarks~\citep{gui2024logicgame,zebralogicbench2024,ma2024kor}, the compositional reasoning capacity of LLMs is rarely studied or quantified in the existing benchmarks.
In this paper, we introduce \textbf{\ourmethod{}}, an evaluation framework that, for the first time, combines existing benchmarks and cryptographic, to quantify the compositional reasoning capacity of LLMs.
Building upon \ourmethod{}, we construct \textbf{\benchmark{}}, which integrates these principles into several benchmarks for systematic evaluation.
We conduct detailed experiments on widely used open-source and closed-source LLMs using \benchmark{}, revealing a huge gap between open-source and closed-source LLMs.
We further conduct thorough mechanical interpretability experiments to reveal the inner mechanism of LLMs' compositional reasoning, involving subproblem decomposition, subproblem inference, and summarizing subproblem conclusions. 
Through analysis based on \benchmark{}, we highlight the value of independently studying compositional reasoning and emphasize the need to enhance the compositional reasoning abilities of LLMs.
\end{abstract}

\begin{document}
\begin{CJK*}{UTF8}{gbsn}

\maketitle

\newpage

\tableofcontents

\newpage

\section{Introduction}
Compositional reasoning (\textbf{CR}) refers to the ability to break down complex problems into simpler components and then use those components to form new ideas~\citep{hou-etal-2023-towards}.
Existing research works~\citep{wang2024grokked, hou-etal-2023-towards, cheng2024understanding} have pointed out that compositional reasoning plays a key role in the generalization and intelligence emergence of LLMs.
Quantifying the compositional reasoning ability and behavior of LLMs helps reveal how they transfer knowledge and skills from pretraining and alignment data to solve new problems, and uncover patterns of emergent generalization~\citep{Measuring_Massive_Multitask} as the size of LLMs increases.

\begin{figure}[t]
    \centering
    \vskip 0.2in
    \subfigure[Concept of compositional reasoning]{
        \includegraphics[width=0.8\textwidth]{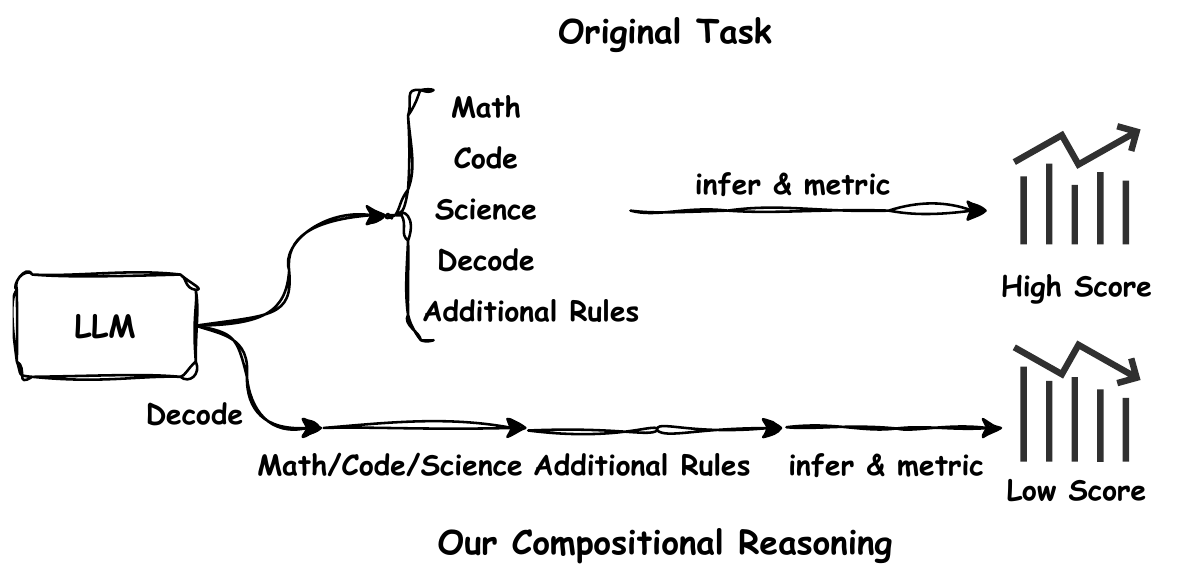}
    }
    \subfigure[Performance gap between original benchmarks and CryptoBench]{
        \includegraphics[width=0.8\textwidth]{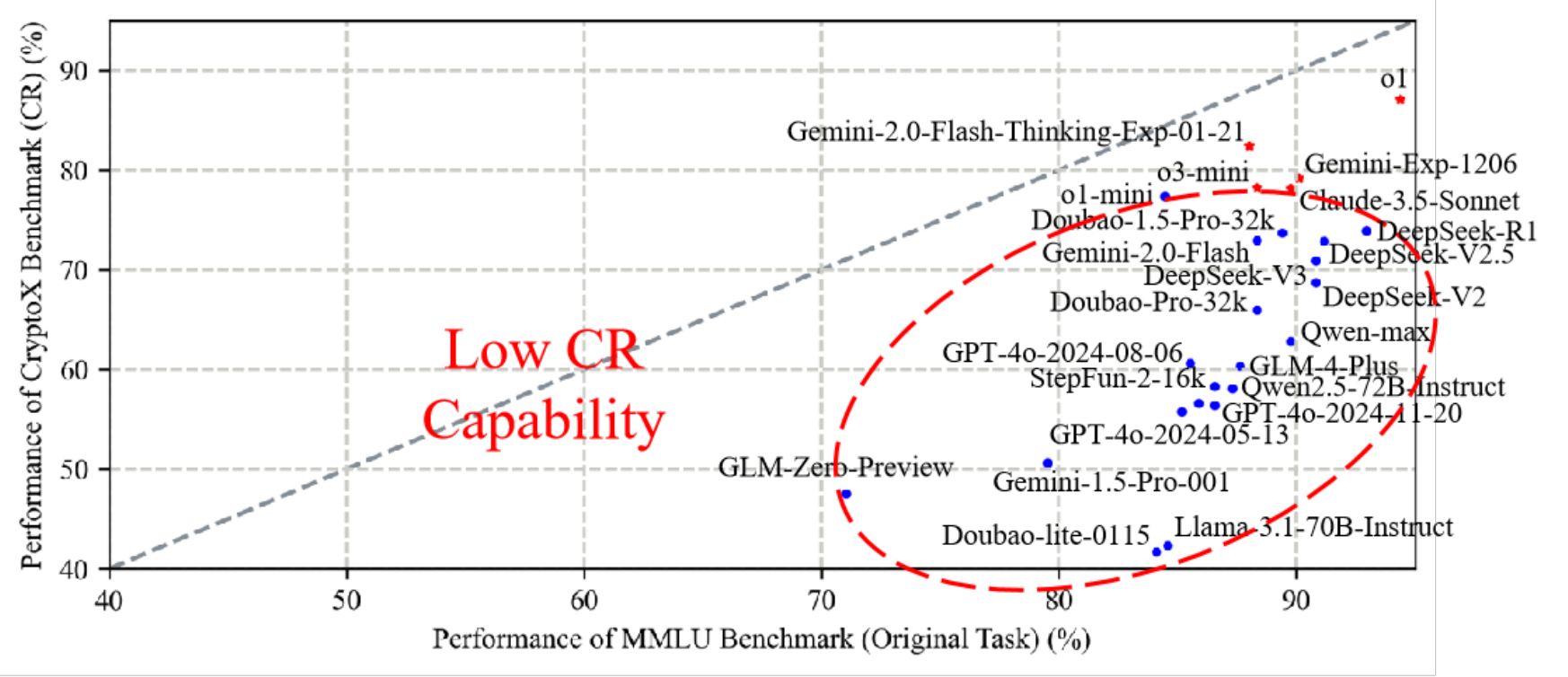}
    }
    \caption{(a) shows the concept of compositional reasoning (CR) which involves combining different abilities in one single model run (e.g., A+B) instead of reasoning via individual ability (e.g., A \textbf{or} B). (b) shows the evaluation result of CryptoX. Some LLMs with strong reasoning abilities on the original benchmark have the low-CR abilities on \benchmark{}.}
    \label{fig:concept}
    \vskip -0.2in
\end{figure}

However, existing reasoning-related benchmarks are either tightly coupled with specific domains~\citep{cobbe2021training,hendrycks2021measuring,han2022folio} or pursuing orthogonality of pretraining knowledge~\citep{gui2024logicgame,ma2024kor}.
As a result, despite numerous existing reasoning-related benchmarks, the CR capabilities of LLMs have not been well studied or quantified~\citep{hou-etal-2023-towards}.
Previous research work~\citep{wang2024grokked} emphasizes the importance of CR capabilities by showing how pre-trained models generalize to unseen data processing under toy experiment settings. But these analysis based on toy data suffer from generalizing to LLMs, as the nature of toy data differs significantly from text pretraining data~\citep{ma2024kor}, which exhibit more diverse reasoning patterns.

To address the gap in evaluating CR capabilities, we propose \textbf{\ourmethod{}}\footnote{\url{https://github.com/multimodal-art-projection/CryptoX}}, inspired by cryptographic techniques~\citep{crypto}. 
\ourmethod{} flexibly transforms existing benchmarks into \benchmark{} using \textbf{instruction encryption} and \textbf{instruction transformation}.  
Instruction encryption randomly encodes part of each instruction in the benchmarks using a given codebook.
Instruction transformation defines additional projection rules from the original answer to the \ourmethod{} answer, e.g. the original correct choice answers in MMLU~\citep{MMLU} require an additional Numeric Transformation operation($A \rightarrow 1,B \rightarrow 2
 ,\dots $) to be viewed correct in Crypto-MMLU.
All the additional rules for instruction encryption and transformation are clearly stated in the given concatenated instructions.
By incorporating instruction encryption and instruction transformation, \benchmark{} benchmarks aim to assess LLM's CR capabilities and reveal LLM's inner mechanism of CR in a flexible manner.
We further conduct Mechanistic Interpretability experiments on analysis of LLMs' neuron activation and inner workings via logit lens~\citep{logitlens} to provide more insights about CR behaviour.

Given the results of \benchmark{} and related mechanical interpretability experiments, we share several key insights about LLMs' compositional reasoning: (1) Most existing LLMs have weak CR abilities, and the proposed \benchmark{} can measure the CR ability gap between different LLMs. (2) The CR ability of the model is influenced by various factors, such as model size, architecture, and other relevant factors.
(3) The probing experiments indicate that the LLMs summarize the reasoning results of the subtasks to obtain the answer to the CR problem, emphasizing the importance of the \benchmark{} for evaluating the CR reasoning abilities. (4) The layers of LLMs exhibit a clear hierarchical pattern of executing different subtasks in different layers and then aggregating for compositional reasoning. 

The contributions of this paper are as follows:
\vspace{-10pt}
\begin{itemize}
\item We propose CryptoX, a flexible evaluation method for LLMs' compositional reasoning.
\item We build \benchmark{} and evaluate LLMs' compositional reasoning ability based on it.
\item We reveal the internal mechanism of LLMs' compositional reasoning by analyzing LLMs' neuron activation and inner workings via logit lens.
\end{itemize}

\section{\ourmethod{}: Compositional Reasoning Evaluation Method}
\label{sec:methods}

\subsection{Overview}
\begin{figure*}[ht]
\centering
\vskip 0.2in
\includegraphics[width=\textwidth]{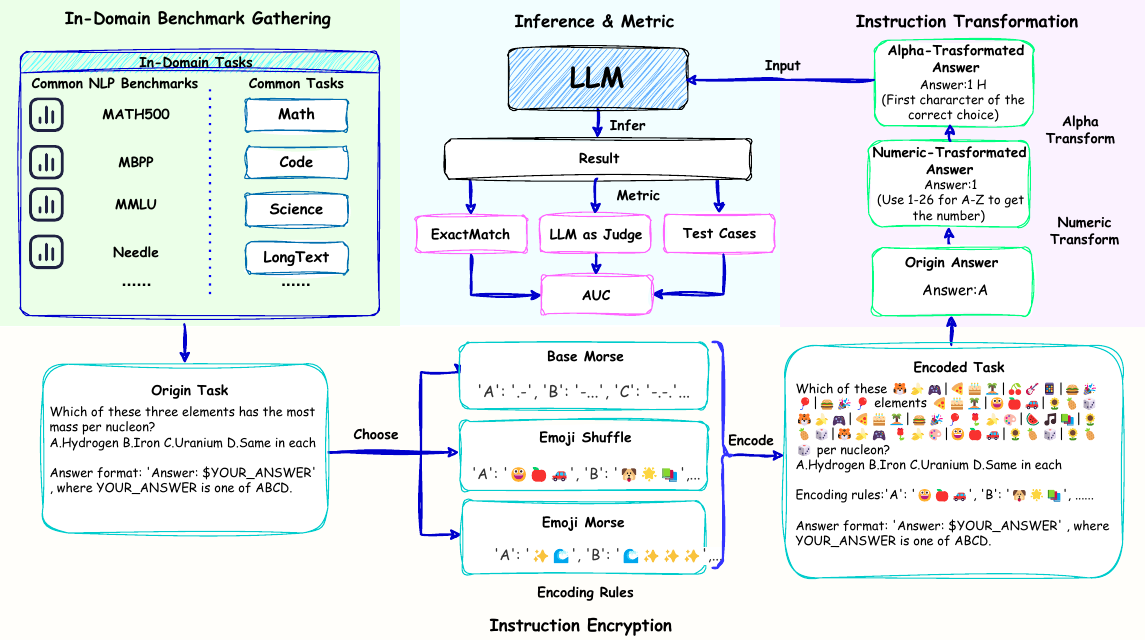}
\caption{Overview of the \benchmark{} Construction Process. We apply instruction encryption and transformation to the tasks from common NLP benchmarks and combine them to construct our \benchmark{} Task. Then we use Exact Match, LLM as judge, UnitTest and AUC as our Evaluation Metrics to judge LLM's performance. }
\label{fig:overview}
\vskip -0.2in
\end{figure*}

\ourmethod{} presents a novel approach to evaluate LLMs’ compositional reasoning ability by integrating existing benchmarks with cipher-based transformations.
The method first encodes specific words in prompts into new ciphered characters and explicitly includes encoding rules within instructions.
To further increase difficulty, some instructions require multi-step reasoning, guiding models to follow structured solution steps.
LLMs must first decode ciphered characters before answering original questions, ensuring a rigorous assessment of compositional reasoning.
Building upon this framework, we construct \benchmark{}, a benchmark set that systematically applies these principles for evaluation.

\subsection{Task Definition and Methodology}
\begin{algorithm}[h!]
\caption{Instruction Encryption}
\label{alg:encode_question}
\begin{algorithmic}
\STATE \textbf{Input:} Original question $Q=(w_1,\dots,w_n)$, where \( w_i \) are words split by spaces, encoding mapping table $\mathcal{M}$, and the number of encoded words $m$
\STATE \textbf{Output:} Encoded question $Q_{e}$
\STATE $selected\_words$ $\gets$ $[]$
\FOR{1 to $m$}
    \STATE $w_{r} \gets \text{RandomSelect}(Q)$
    \IF{$|w_{r}| > 1 \land w_{r} \notin selected\_words$}
        \STATE $w_{r} \gets \mathcal{M}(w_{r}) $
        \STATE $selected\_words.\text{append}(w_{r})$
    \ENDIF
\ENDFOR
\end{algorithmic}
\end{algorithm}
Given a benchmark \( B = \{x_1, x_2, \dots, x_N\} \), where \( x_i \) is a prompt consisting of an instruction and a question \( Q \), \( B \) contains \( N \) data points.
To transform the existing prompt into a compositional reasoning prompt, we define a set of transformation rules, where each data point undergoes a series of transformations. 
Specifically, the transformation process can be described as:  

\begin{align}
x' = r_m \circ r_{m-1} \circ \dots \circ r_1(x),
\label{eq:composition}
\end{align}

where \( r_i \) denotes the \( i^{\text{th}} \) transformation rule, and \( \circ \) represents the composition of transformations.  
Based on our formal definitions, we explore two implementation approaches of  \( r \) : \textbf{instruction encryption}, which encodes parts of the prompt, and \textbf{instruction transformation}, which restructures it. 
These transformations enable the creation of diverse compositional reasoning tasks.

\paragraph{Instruction Encryption}

Specifically, as shown in Figure \ref{fig:overview}, Instruction Encryption applies three encoding rules to the prompt, with its detailed process described in Algorithm \ref{alg:encode_question}.  
It allows encoding any number of words in the prompt, ensuring that the encoded prompts, generated by flexible self-defined rules, are unlikely to overlap with the pre-training corpus. For simplicity, we use emoji shuffle as an exemplifying encoding rule in the following sections and detailed encoding rules are explained in Appendix \ref{appendix: encoding rule}.
\nocite{langley00}
\paragraph{Instruction Transformation}

The vanilla approaches involve prompting the LLM to provide an answer in the format of ``Answer: A'' for the multi-choice question tasks. 
To further evaluate the compositional reasoning capabilities of LLMs in handling OOD scenarios, as shown in Figure \ref{fig:overview}, we establish instruction transformation to further increase the number of reasoning hops.
(1) \textbf{Numeric Transformation}: Based on $Q_{e}$, we perform numeric transformation. For example, mapping ``$A \rightarrow 1, B \rightarrow 2, C \rightarrow 3, D \rightarrow 4$'' will require the LLM to answer ``Answer: 1'' if the answer is ``Answer: A'', which forces the LLM to perform further reasoning after obtaining the original answer.
(2) \textbf{Alpha Transformation}: Additionally, the task can be made more complex by requiring the LLM to provide both the numerical answer and the first alphanumeric character of the corresponding answer content. For example, if the original answer is ``Answer: A'' and the answer content is ``Happiness'', the LLM would output ``Answer: 1 H''.

\section{Experiment}
\label{sec:experiment}

\begin{table}[h]
    \centering
    \caption{\textbf{Statistics of our Benchmarks}: Avg. Len refers to the total number of characters in each question with 0-, 5-, or 10-word encoding numbers, respectively. Answer Format includes the following types: ME (Mathematical Expression), SC (Single Choice), CB (Code Blocks), TE (Textual Expression), and MC (Multiple Choices).}
    \label{tab:benchmark_statistics}
    \vskip 0.15in
    \resizebox{0.9\textwidth}{!}{
    \begin{tabular}{@{}lcccc@{}}
        \toprule
        \textbf{Category}   & \textbf{Total Nums} & \textbf{Avg. Len} & \textbf{Ans. Fmt}\\ 
        \midrule
        Crypto-Math    & 500 & 441.89 / 1410.86 / 1471.17 &ME\\ 
        Crypto-MMLU  & 285 & 627.97 / 1274.82 / 1333.6 &SC\\ 
        Crypto-MMLU-Num  & 285 & 699.97 / 1346.82 / 1405.6 &SC\\ 
        Crypto-MMLU-Alpha  & 285 & 922.97 / 1569.82 / 1628.6 &SC\\ 
        Crypto-MBPP      & 427 & 621.17 / 1268.54 / 1327.14 &CB\\ 
        Crypto-BBH        & 405 & 1585.3 / 3464.98 / 3517.24 &TE\&MC\\  
        Crypto-Needle-30K     & 100 & 64811.41 / 63535.62 / 62111.8&TE\\
        Crypto-HighResolution     & 497 & 1177.49 / 2360.61 / 2426.62 &ME\&SC\&TE\&MC\&CB\\
        \bottomrule
    \end{tabular}
    }
    \vskip -0.1in
\end{table}

\paragraph{LLMs}
We evaluate both open-source and closed-source models on \benchmark{}. For closed-source LLMs, we evaluate GPT series~\citep{gpt4} (GPT-4o), Claude series~\citep{claude35addendum}, and o1 series~\citep{openaiO1}, and the like. 
For open-source LLMs, we evaluate Qwen2.5 series~\citep{qwen2.5}, Llama-3.1 series~\citep{llama3}, Codestral~\citep{mistral} and Jamba-1.5-mini~\citep{Jamba}, and the like.
For all LLMs, we use temperature $T=0.7 \text{ to } 1.0$ and $\text{top}\ p=0.75 \text{ to } 1.0$.

\paragraph{Vanilla Benchmark}  
As shown in Table \ref{tab:benchmark_statistics}, we apply our method on five benchmarks: MATH~\citep{hendrycks2021measuring}, MMLU~\citep{MMLU}, BBH~\citep{suzgun2022challenging}, MBPP~\citep{austin2021program}, and Needle~\citep{Needle}.  
For MATH, we use the MATH 500 subset; for MMLU, we adopt MMLU-dev; for MBPP, we adopt MBPP-sanitized; and for Needle, we use the three-needle setting.  
Each benchmark is tested under three levels of Instruction Encryption: 0, 5, and 10, where 0 represents the original task without modification.  
Additionally, we construct a high-resolution (Applying Instruction Encryption with values ranging from 0 to 10) subset Crypto-HighResolution by sampling data from these benchmarks.  

\paragraph{Prompt Setting}
To systematically evaluate the models under various experimental settings, we design the following prompt templates (c.f., Appendix \ref{appendix: encoding rule}): (1) \textbf{Zero-shot Prompts} are applied on subsets including Crypto-Needle-30K, Crypto-Math, Crypto-MBPP, and Crypto-MMLU. (2) \textbf{Few-shot Prompts} are applied on the Crypto-BBH subset.

\paragraph{Evaluation Metrics}
(1) \textbf{Exact Match (EM)}: For simple answers (e.g., multiple-choice questions), we use regular expressions to extract the answer from the LLM's response and standardize its format. The standardized answer is then compared directly with the correct answer.  
(2) \textbf{LLM as judge}: For more complex answers (e.g., mathematical expressions), we employ the LLM-as-judge (Doubao-Pro-256K) to judge the answer against the correct answer. 
(3) \textbf{UnitTest}: For code-related questions, we evaluate responses using a test-case-based approach. 
The number of passed test cases is then counted, and the final score is calculated as:
score = (number of passed test cases) / (total number of test cases).
\paragraph{AUC of Compositional Reasoning}  
To better assess a model’s overall performance across different difficulty levels, we compute an area under the curve (AUC) score.  
After obtaining the evaluation metrics, we plot the number of encoded words \( k \) as the x-axis and the corresponding model performance as the y-axis.  
The AUC is then calculated using the trapezoidal rule:  

\begin{equation}
    \text{AUC} = \int_{k_{\min}}^{k_{\max}} f(k) \,dk \approx \sum_{i=1}^{N-1} (k_{i+1} - k_i) \frac{y_i + y_{i+1}}{2}
\end{equation}

where \( k_i \) represents the number of encoded words (Instruction Encryption level), and \( y_i \) is the corresponding model performance.  
A higher AUC indicates better compositional reasoning ability under varying levels of instruction encryption.

\subsection{Main Results}
Table \ref{tab:common_test} presents the performance of diverse models on our \benchmark{} across various domains, revealing several key insights below. 

\paragraph{Current Models Demonstrate Limited CR Abilities} We find that as the number of encoded words increases, the reasoning difficulty also increases, leading to a decline in evaluation metrics across all models. 
As mentioned earlier, \benchmark{} forces models to conduct compositional reasoning, requiring them to decode the question before answering it. 
The increase in the number of encoded words corresponds to a higher likelihood of the models conducting compositional reasoning. 
The experimental results indicate that while most models perform poorly in compositional reasoning tasks (we analyze specific cases in the Appendix \ref{casestudy}), o1, o3-mini, and Gemini-2.0-Flash-Thinking-Exp-01-21 still demonstrate relatively strong performance. Open-source models are generally weaker than closed-source models, where open-source models exhibit a significant accuracy gap when answering questions compared to closed-source models.


\begin{table*}[ht]
\renewcommand{\arraystretch}{1.1}
\setlength{\extrarowheight}{3pt} 
\centering
\caption{Performance of different models on \benchmark{}. 0 / 5 / 10 represents the number of words encoded. For Gemini-2.0-Flash-Thinking, we use a version of Exp-01-21. \colorbox{green!15}{\textcolor{black}{green}} represents the number with the highest accuracy, \colorbox{cyan!15}{\textcolor{black}{blue}} represents the number with the second highest accuracy, and \colorbox{orange!15}{\textcolor{black}{orange}} represents the value with the third highest accuracy.}
\label{tab:common_test}
\vskip 0.15in
\resizebox{\textwidth}{!}{
\LARGE
\begin{tabular}{lccccc ccccc ccccc ccccc ccc}
\toprule
\multicolumn{1}{c}{\multirow{2}*{\textbf{Model}}} & \multicolumn{1}{c}{\multirow{2}*{\textbf{AUC}}} & \multicolumn{1}{c}{\multirow{2}*{\textbf{Avg}}} & \multicolumn{3}{c}{\textbf{Crypto-Math}} & \multicolumn{3}{c}{\textbf{Crypto-MBPP}} & \multicolumn{3}{c}{\textbf{Crypto-BBH}} & \multicolumn{3}{c}{\textbf{Crypto-MMLU}} & \multicolumn{3}{c}{\textbf{Crypto-MMLU-Num}} & \multicolumn{3}{c}{\textbf{Crypto-MMLU-Alpha}}  & \multicolumn{3}{c}{\textbf{Crypto-Needle-30K}} \\
\cmidrule{4-24}
\multicolumn{1}{c}{~} & \multicolumn{1}{c}{~} & \multicolumn{1}{c}{~} & \textbf{0} & \textbf{5} & \textbf{10} & \textbf{0} & \textbf{5} & \textbf{10} & \textbf{0} & \textbf{5} & \textbf{10} & \textbf{0} & \textbf{5} & \textbf{10} & \textbf{0} & \textbf{5} & \textbf{10} & \textbf{0} & \textbf{5} & \textbf{10} & \textbf{0} & \textbf{5} & \textbf{10}\\
\midrule
\multicolumn{24}{c}{\textit{Closed-source LLMs}} \\
\midrule
\multicolumn{1}{l}{o1} & \multicolumn{1}{c}{\firstcolor{4.05}} & \multicolumn{1}{c}{\firstcolor{83.69}} & \multicolumn{1}{c}{\firstcolor{96.99}} & \multicolumn{1}{c}{\firstcolor{89.66}} & \multicolumn{1}{c}{\firstcolor{84.48}} & \multicolumn{1}{c}{64.93} & \multicolumn{1}{c}{\secondcolor{68.04}} & \multicolumn{1}{c}{63.9} & \multicolumn{1}{c}{84.08} & \multicolumn{1}{c}{\secondcolor{83.66}} & \multicolumn{1}{c}{\secondcolor{82.13}} & \multicolumn{1}{c}{\firstcolor{94.35}} & \multicolumn{1}{c}{\firstcolor{92.25}} & \multicolumn{1}{c}{\secondcolor{90.53}} & \multicolumn{1}{c}{\firstcolor{92.5}} & \multicolumn{1}{c}{\firstcolor{91.43}} & \multicolumn{1}{c}{\firstcolor{90.0}} & \multicolumn{1}{c}{\firstcolor{90.07}} & \multicolumn{1}{c}{\firstcolor{87.99}} & \multicolumn{1}{c}{\firstcolor{87.32}} & \multicolumn{1}{c}{\firstcolor{99.66}} & \multicolumn{1}{c}{\firstcolor{80.33}} & \multicolumn{1}{c}{\firstcolor{43.2}} \\

\multicolumn{1}{l}{o3-mini} & \multicolumn{1}{c}{\secondcolor{3.67}} & \multicolumn{1}{c}{\secondcolor{76.38}} & \multicolumn{1}{c}{\secondcolor{95.99}} & \multicolumn{1}{c}{\secondcolor{85.4}} & \multicolumn{1}{c}{\secondcolor{77.62}} & \multicolumn{1}{c}{46.96} & \multicolumn{1}{c}{60.39} & \multicolumn{1}{c}{57.49} & \multicolumn{1}{c}{80.6} & \multicolumn{1}{c}{72.03} & \multicolumn{1}{c}{72.7} & \multicolumn{1}{c}{88.34} & \multicolumn{1}{c}{86.97} & \multicolumn{1}{c}{\thirdcolor{84.91}} & \multicolumn{1}{c}{88.93} & \multicolumn{1}{c}{\thirdcolor{88.57}} & \multicolumn{1}{c}{\secondcolor{86.07}} & \multicolumn{1}{c}{\secondcolor{87.23}} & \multicolumn{1}{c}{\secondcolor{85.87}} & \multicolumn{1}{c}{\secondcolor{83.1}} & \multicolumn{1}{c}{88.55} & \multicolumn{1}{c}{52.0} & \multicolumn{1}{c}{34.35} \\

\multicolumn{1}{l}{Gemini-2.0-Flash-Thinking} & \multicolumn{1}{c}{\thirdcolor{3.58}} & \multicolumn{1}{c}{\thirdcolor{76.06}} & \multicolumn{1}{c}{87.98} & \multicolumn{1}{c}{78.3} & \multicolumn{1}{c}{\thirdcolor{73.79}} & \multicolumn{1}{c}{63.31} & \multicolumn{1}{c}{51.05} & \multicolumn{1}{c}{47.05} & \multicolumn{1}{c}{82.34} & \multicolumn{1}{c}{\firstcolor{84.65}} & \multicolumn{1}{c}{\firstcolor{82.63}} & \multicolumn{1}{c}{87.99} & \multicolumn{1}{c}{\thirdcolor{88.03}} & \multicolumn{1}{c}{83.86} & \multicolumn{1}{c}{\thirdcolor{91.07}} & \multicolumn{1}{c}{\secondcolor{89.29}} & \multicolumn{1}{c}{\thirdcolor{85.36}} & \multicolumn{1}{c}{84.75} & \multicolumn{1}{c}{\thirdcolor{83.04}} & \multicolumn{1}{c}{\thirdcolor{81.34}} & \multicolumn{1}{c}{98.32} & \multicolumn{1}{c}{51.0} & \multicolumn{1}{c}{22.11} \\

\multicolumn{1}{l}{Gemini-Exp-1206} & \multicolumn{1}{c}{3.52} & \multicolumn{1}{c}{74.72} & \multicolumn{1}{c}{85.57} & \multicolumn{1}{c}{73.22} & \multicolumn{1}{c}{68.14} & \multicolumn{1}{c}{53.3} & \multicolumn{1}{c}{56.15} & \multicolumn{1}{c}{53.99} & \multicolumn{1}{c}{79.85} & \multicolumn{1}{c}{80.45} & \multicolumn{1}{c}{76.18} & \multicolumn{1}{c}{90.11} & \multicolumn{1}{c}{84.51} & \multicolumn{1}{c}{77.9} & \multicolumn{1}{c}{\secondcolor{92.14}} & \multicolumn{1}{c}{83.93} & \multicolumn{1}{c}{80.0} & \multicolumn{1}{c}{\thirdcolor{85.82}} & \multicolumn{1}{c}{79.15} & \multicolumn{1}{c}{73.94} & \multicolumn{1}{c}{97.64} & \multicolumn{1}{c}{\secondcolor{60.67}} & \multicolumn{1}{c}{\secondcolor{36.39}} \\

\multicolumn{1}{l}{Claude-3.5-Sonnet} & \multicolumn{1}{c}{3.45} & \multicolumn{1}{c}{74.07} & \multicolumn{1}{c}{74.75} & \multicolumn{1}{c}{66.33} & \multicolumn{1}{c}{60.69} & \multicolumn{1}{c}{\thirdcolor{69.18}} & \multicolumn{1}{c}{\thirdcolor{66.93}} & \multicolumn{1}{c}{\thirdcolor{67.54}} & \multicolumn{1}{c}{\thirdcolor{84.58}} & \multicolumn{1}{c}{81.68} & \multicolumn{1}{c}{77.42} & \multicolumn{1}{c}{89.75} & \multicolumn{1}{c}{86.97} & \multicolumn{1}{c}{83.51} & \multicolumn{1}{c}{90.36} & \multicolumn{1}{c}{82.5} & \multicolumn{1}{c}{81.07} & \multicolumn{1}{c}{82.98} & \multicolumn{1}{c}{77.74} & \multicolumn{1}{c}{78.17} & \multicolumn{1}{c}{\thirdcolor{98.99}} & \multicolumn{1}{c}{38.0} & \multicolumn{1}{c}{16.33} \\

\multicolumn{1}{l}{o1-mini} & \multicolumn{1}{c}{3.43} & \multicolumn{1}{c}{73.98} & \multicolumn{1}{c}{\thirdcolor{90.78}} & \multicolumn{1}{c}{78.3} & \multicolumn{1}{c}{66.33} & \multicolumn{1}{c}{\secondcolor{72.69}} & \multicolumn{1}{c}{65.37} & \multicolumn{1}{c}{\secondcolor{67.88}} & \multicolumn{1}{c}{83.33} & \multicolumn{1}{c}{77.23} & \multicolumn{1}{c}{75.93} & \multicolumn{1}{c}{84.45} & \multicolumn{1}{c}{80.99} & \multicolumn{1}{c}{79.3} & \multicolumn{1}{c}{85.36} & \multicolumn{1}{c}{77.86} & \multicolumn{1}{c}{80.36} & \multicolumn{1}{c}{80.5} & \multicolumn{1}{c}{73.5} & \multicolumn{1}{c}{74.3} & \multicolumn{1}{c}{95.96} & \multicolumn{1}{c}{35.0} & \multicolumn{1}{c}{28.23} \\

\multicolumn{1}{l}{DeepSeek-R1} & \multicolumn{1}{c}{3.2} & \multicolumn{1}{c}{68.1} & \multicolumn{1}{c}{90.38} & \multicolumn{1}{c}{\thirdcolor{78.91}} & \multicolumn{1}{c}{70.36} & \multicolumn{1}{c}{68.53} & \multicolumn{1}{c}{\firstcolor{70.61}} & \multicolumn{1}{c}{\firstcolor{69.35}} & \multicolumn{1}{c}{\secondcolor{86.07}} & \multicolumn{1}{c}{80.45} & \multicolumn{1}{c}{\thirdcolor{78.66}} & \multicolumn{1}{c}{\secondcolor{92.93}} & \multicolumn{1}{c}{\secondcolor{90.14}} & \multicolumn{1}{c}{\firstcolor{90.88}} & \multicolumn{1}{c}{61.43} & \multicolumn{1}{c}{42.86} & \multicolumn{1}{c}{37.5} & \multicolumn{1}{c}{46.81} & \multicolumn{1}{c}{45.94} & \multicolumn{1}{c}{45.77} & \multicolumn{1}{c}{87.88} & \multicolumn{1}{c}{\thirdcolor{59.67}} & \multicolumn{1}{c}{\thirdcolor{35.03}} \\

\multicolumn{1}{l}{Gemini-2.0-Flash} & \multicolumn{1}{c}{3.14} & \multicolumn{1}{c}{68.62} & \multicolumn{1}{c}{89.18} & \multicolumn{1}{c}{65.11} & \multicolumn{1}{c}{58.87} & \multicolumn{1}{c}{49.65} & \multicolumn{1}{c}{50.86} & \multicolumn{1}{c}{46.37} & \multicolumn{1}{c}{79.1} & \multicolumn{1}{c}{72.28} & \multicolumn{1}{c}{68.73} & \multicolumn{1}{c}{88.34} & \multicolumn{1}{c}{79.93} & \multicolumn{1}{c}{77.54} & \multicolumn{1}{c}{87.14} & \multicolumn{1}{c}{79.29} & \multicolumn{1}{c}{72.14} & \multicolumn{1}{c}{73.05} & \multicolumn{1}{c}{74.56} & \multicolumn{1}{c}{66.55} & \multicolumn{1}{c}{96.63} & \multicolumn{1}{c}{44.0} & \multicolumn{1}{c}{21.77} \\

\multicolumn{1}{l}{DeepSeek-V2} & \multicolumn{1}{c}{3.08} & \multicolumn{1}{c}{68.29} & \multicolumn{1}{c}{85.37} & \multicolumn{1}{c}{64.5} & \multicolumn{1}{c}{54.23} & \multicolumn{1}{c}{60.83} & \multicolumn{1}{c}{57.86} & \multicolumn{1}{c}{57.28} & \multicolumn{1}{c}{84.08} & \multicolumn{1}{c}{77.23} & \multicolumn{1}{c}{67.49} & \multicolumn{1}{c}{90.81} & \multicolumn{1}{c}{80.99} & \multicolumn{1}{c}{75.79} & \multicolumn{1}{c}{90.36} & \multicolumn{1}{c}{80.71} & \multicolumn{1}{c}{73.93} & \multicolumn{1}{c}{71.99} & \multicolumn{1}{c}{64.31} & \multicolumn{1}{c}{58.8} & \multicolumn{1}{c}{87.54} & \multicolumn{1}{c}{31.33} & \multicolumn{1}{c}{18.71} \\

\multicolumn{1}{l}{DeepSeek-V2.5} & \multicolumn{1}{c}{3.08} & \multicolumn{1}{c}{68.23} & \multicolumn{1}{c}{85.37} & \multicolumn{1}{c}{65.31} & \multicolumn{1}{c}{56.65} & \multicolumn{1}{c}{61.45} & \multicolumn{1}{c}{61.66} & \multicolumn{1}{c}{57.93} & \multicolumn{1}{c}{82.84} & \multicolumn{1}{c}{74.5} & \multicolumn{1}{c}{68.73} & \multicolumn{1}{c}{\thirdcolor{91.17}} & \multicolumn{1}{c}{79.22} & \multicolumn{1}{c}{74.74} & \multicolumn{1}{c}{88.93} & \multicolumn{1}{c}{77.86} & \multicolumn{1}{c}{75.0} & \multicolumn{1}{c}{75.89} & \multicolumn{1}{c}{63.96} & \multicolumn{1}{c}{57.75} & \multicolumn{1}{c}{86.19} & \multicolumn{1}{c}{28.67} & \multicolumn{1}{c}{19.05} \\

\multicolumn{1}{l}{DeepSeek-V3} & \multicolumn{1}{c}{3.07} & \multicolumn{1}{c}{68.35} & \multicolumn{1}{c}{85.97} & \multicolumn{1}{c}{66.94} & \multicolumn{1}{c}{55.44} & \multicolumn{1}{c}{61.95} & \multicolumn{1}{c}{57.34} & \multicolumn{1}{c}{58.01} & \multicolumn{1}{c}{83.08} & \multicolumn{1}{c}{77.72} & \multicolumn{1}{c}{67.0} & \multicolumn{1}{c}{90.81} & \multicolumn{1}{c}{80.99} & \multicolumn{1}{c}{77.9} & \multicolumn{1}{c}{90.36} & \multicolumn{1}{c}{79.29} & \multicolumn{1}{c}{72.86} & \multicolumn{1}{c}{75.89} & \multicolumn{1}{c}{60.42} & \multicolumn{1}{c}{57.75} & \multicolumn{1}{c}{87.54} & \multicolumn{1}{c}{27.0} & \multicolumn{1}{c}{21.09} \\

\multicolumn{1}{l}{Doubao-1.5-Pro-32k} & \multicolumn{1}{c}{2.94} & \multicolumn{1}{c}{66.22} & \multicolumn{1}{c}{81.76} & \multicolumn{1}{c}{61.46} & \multicolumn{1}{c}{49.4} & \multicolumn{1}{c}{\firstcolor{72.99}} & \multicolumn{1}{c}{64.13} & \multicolumn{1}{c}{59.41} & \multicolumn{1}{c}{84.08} & \multicolumn{1}{c}{75.25} & \multicolumn{1}{c}{71.71} & \multicolumn{1}{c}{89.4} & \multicolumn{1}{c}{80.63} & \multicolumn{1}{c}{75.79} & \multicolumn{1}{c}{90.36} & \multicolumn{1}{c}{78.21} & \multicolumn{1}{c}{68.57} & \multicolumn{1}{c}{53.19} & \multicolumn{1}{c}{56.54} & \multicolumn{1}{c}{50.7} & \multicolumn{1}{c}{95.62} & \multicolumn{1}{c}{28.0} & \multicolumn{1}{c}{3.4} \\

\multicolumn{1}{l}{GPT-4o-2024-08-06} & \multicolumn{1}{c}{2.73} & \multicolumn{1}{c}{61.9} & \multicolumn{1}{c}{75.95} & \multicolumn{1}{c}{47.67} & \multicolumn{1}{c}{36.49} & \multicolumn{1}{c}{48.03} & \multicolumn{1}{c}{52.1} & \multicolumn{1}{c}{47.47} & \multicolumn{1}{c}{81.09} & \multicolumn{1}{c}{70.3} & \multicolumn{1}{c}{64.52} & \multicolumn{1}{c}{85.51} & \multicolumn{1}{c}{78.17} & \multicolumn{1}{c}{75.09} & \multicolumn{1}{c}{83.93} & \multicolumn{1}{c}{71.07} & \multicolumn{1}{c}{65.0} & \multicolumn{1}{c}{63.12} & \multicolumn{1}{c}{51.24} & \multicolumn{1}{c}{37.68} & \multicolumn{1}{c}{97.98} & \multicolumn{1}{c}{42.0} & \multicolumn{1}{c}{25.51} \\

\multicolumn{1}{l}{Doubao-Pro-32k} & \multicolumn{1}{c}{2.64} & \multicolumn{1}{c}{62.32} & \multicolumn{1}{c}{87.58} & \multicolumn{1}{c}{52.94} & \multicolumn{1}{c}{36.49} & \multicolumn{1}{c}{64.64} & \multicolumn{1}{c}{65.7} & \multicolumn{1}{c}{60.05} & \multicolumn{1}{c}{81.84} & \multicolumn{1}{c}{74.26} & \multicolumn{1}{c}{62.78} & \multicolumn{1}{c}{88.34} & \multicolumn{1}{c}{72.18} & \multicolumn{1}{c}{64.21} & \multicolumn{1}{c}{87.14} & \multicolumn{1}{c}{75.0} & \multicolumn{1}{c}{64.29} & \multicolumn{1}{c}{60.64} & \multicolumn{1}{c}{53.71} & \multicolumn{1}{c}{47.89} & \multicolumn{1}{c}{97.98} & \multicolumn{1}{c}{11.0} & \multicolumn{1}{c}{0.0} \\

\multicolumn{1}{l}{GPT-4o-2024-05-13} & \multicolumn{1}{c}{2.6} & \multicolumn{1}{c}{60.4} & \multicolumn{1}{c}{73.95} & \multicolumn{1}{c}{44.42} & \multicolumn{1}{c}{29.23} & \multicolumn{1}{c}{49.33} & \multicolumn{1}{c}{58.53} & \multicolumn{1}{c}{50.33} & \multicolumn{1}{c}{80.6} & \multicolumn{1}{c}{69.06} & \multicolumn{1}{c}{57.82} & \multicolumn{1}{c}{85.16} & \multicolumn{1}{c}{76.06} & \multicolumn{1}{c}{70.17} & \multicolumn{1}{c}{84.29} & \multicolumn{1}{c}{64.64} & \multicolumn{1}{c}{54.64} & \multicolumn{1}{c}{68.79} & \multicolumn{1}{c}{52.3} & \multicolumn{1}{c}{36.97} & \multicolumn{1}{c}{97.31} & \multicolumn{1}{c}{38.33} & \multicolumn{1}{c}{26.53} \\

\multicolumn{1}{l}{Qwen-max} & \multicolumn{1}{c}{2.57} & \multicolumn{1}{c}{61.48} & \multicolumn{1}{c}{81.16} & \multicolumn{1}{c}{45.03} & \multicolumn{1}{c}{27.82} & \multicolumn{1}{c}{59.48} & \multicolumn{1}{c}{55.27} & \multicolumn{1}{c}{46.67} & \multicolumn{1}{c}{81.84} & \multicolumn{1}{c}{71.54} & \multicolumn{1}{c}{58.56} & \multicolumn{1}{c}{89.75} & \multicolumn{1}{c}{73.94} & \multicolumn{1}{c}{64.91} & \multicolumn{1}{c}{89.29} & \multicolumn{1}{c}{73.57} & \multicolumn{1}{c}{64.64} & \multicolumn{1}{c}{81.92} & \multicolumn{1}{c}{62.19} & \multicolumn{1}{c}{52.46} & \multicolumn{1}{c}{88.55} & \multicolumn{1}{c}{16.0} & \multicolumn{1}{c}{6.46} \\

\multicolumn{1}{l}{GPT-4o-2024-11-20} & \multicolumn{1}{c}{2.55} & \multicolumn{1}{c}{57.91} & \multicolumn{1}{c}{68.94} & \multicolumn{1}{c}{47.87} & \multicolumn{1}{c}{35.08} & \multicolumn{1}{c}{41.27} & \multicolumn{1}{c}{25.88} & \multicolumn{1}{c}{37.14} & \multicolumn{1}{c}{81.34} & \multicolumn{1}{c}{70.55} & \multicolumn{1}{c}{63.52} & \multicolumn{1}{c}{86.57} & \multicolumn{1}{c}{77.82} & \multicolumn{1}{c}{71.23} & \multicolumn{1}{c}{73.57} & \multicolumn{1}{c}{68.93} & \multicolumn{1}{c}{63.93} & \multicolumn{1}{c}{51.77} & \multicolumn{1}{c}{44.88} & \multicolumn{1}{c}{38.38} & \multicolumn{1}{c}{\secondcolor{99.33}} & \multicolumn{1}{c}{43.33} & \multicolumn{1}{c}{24.83} \\

\multicolumn{1}{l}{GLM-4-Plus} & \multicolumn{1}{c}{2.53} & \multicolumn{1}{c}{59.29} & \multicolumn{1}{c}{71.34} & \multicolumn{1}{c}{44.83} & \multicolumn{1}{c}{29.64} & \multicolumn{1}{c}{45.09} & \multicolumn{1}{c}{57.79} & \multicolumn{1}{c}{51.07} & \multicolumn{1}{c}{81.34} & \multicolumn{1}{c}{69.55} & \multicolumn{1}{c}{58.56} & \multicolumn{1}{c}{87.63} & \multicolumn{1}{c}{72.54} & \multicolumn{1}{c}{60.0} & \multicolumn{1}{c}{80.0} & \multicolumn{1}{c}{71.43} & \multicolumn{1}{c}{62.14} & \multicolumn{1}{c}{74.47} & \multicolumn{1}{c}{66.08} & \multicolumn{1}{c}{52.82} & \multicolumn{1}{c}{97.64} & \multicolumn{1}{c}{8.0} & \multicolumn{1}{c}{3.06} \\

\multicolumn{1}{l}{StepFun-2-16k} & \multicolumn{1}{c}{2.35} & \multicolumn{1}{c}{52.81} & \multicolumn{1}{c}{75.75} & \multicolumn{1}{c}{48.48} & \multicolumn{1}{c}{32.46} & \multicolumn{1}{c}{58.55} & \multicolumn{1}{c}{57.88} & \multicolumn{1}{c}{52.8} & \multicolumn{1}{c}{\firstcolor{88.81}} & \multicolumn{1}{c}{\thirdcolor{82.18}} & \multicolumn{1}{c}{65.51} & \multicolumn{1}{c}{86.57} & \multicolumn{1}{c}{70.42} & \multicolumn{1}{c}{60.7} & \multicolumn{1}{c}{83.93} & \multicolumn{1}{c}{63.57} & \multicolumn{1}{c}{57.5} & \multicolumn{1}{c}{37.59} & \multicolumn{1}{c}{31.45} & \multicolumn{1}{c}{25.7} & \multicolumn{1}{c}{20.2} & \multicolumn{1}{c}{8.33} & \multicolumn{1}{c}{0.68} \\

\multicolumn{1}{l}{Gemini-1.5-Pro-001} & \multicolumn{1}{c}{2.09} & \multicolumn{1}{c}{49.06} & \multicolumn{1}{c}{64.53} & \multicolumn{1}{c}{39.96} & \multicolumn{1}{c}{27.02} & \multicolumn{1}{c}{42.07} & \multicolumn{1}{c}{50.83} & \multicolumn{1}{c}{43.84} & \multicolumn{1}{c}{60.95} & \multicolumn{1}{c}{51.98} & \multicolumn{1}{c}{47.39} & \multicolumn{1}{c}{79.51} & \multicolumn{1}{c}{60.91} & \multicolumn{1}{c}{58.6} & \multicolumn{1}{c}{73.93} & \multicolumn{1}{c}{69.29} & \multicolumn{1}{c}{57.86} & \multicolumn{1}{c}{58.51} & \multicolumn{1}{c}{41.34} & \multicolumn{1}{c}{31.34} & \multicolumn{1}{c}{66.67} & \multicolumn{1}{c}{3.67} & \multicolumn{1}{c}{0.0} \\

\multicolumn{1}{l}{GPT-4-Turbo} & \multicolumn{1}{c}{2.07} & \multicolumn{1}{c}{52.48} & \multicolumn{1}{c}{70.34} & \multicolumn{1}{c}{38.34} & \multicolumn{1}{c}{21.57} & \multicolumn{1}{c}{43.76} & \multicolumn{1}{c}{36.53} & \multicolumn{1}{c}{21.13} & \multicolumn{1}{c}{81.09} & \multicolumn{1}{c}{66.83} & \multicolumn{1}{c}{55.34} & \multicolumn{1}{c}{85.87} & \multicolumn{1}{c}{69.01} & \multicolumn{1}{c}{56.14} & \multicolumn{1}{c}{80.71} & \multicolumn{1}{c}{57.86} & \multicolumn{1}{c}{47.14} & \multicolumn{1}{c}{62.41} & \multicolumn{1}{c}{50.18} & \multicolumn{1}{c}{45.07} & \multicolumn{1}{c}{98.65} & \multicolumn{1}{c}{9.33} & \multicolumn{1}{c}{4.76} \\

\multicolumn{1}{l}{GLM-Zero-Preview} & \multicolumn{1}{c}{2.06} & \multicolumn{1}{c}{48.89} & \multicolumn{1}{c}{69.34} & \multicolumn{1}{c}{31.03} & \multicolumn{1}{c}{18.55} & \multicolumn{1}{c}{45.65} & \multicolumn{1}{c}{39.84} & \multicolumn{1}{c}{37.7} & \multicolumn{1}{c}{73.88} & \multicolumn{1}{c}{62.87} & \multicolumn{1}{c}{48.39} & \multicolumn{1}{c}{71.03} & \multicolumn{1}{c}{63.38} & \multicolumn{1}{c}{54.74} & \multicolumn{1}{c}{74.29} & \multicolumn{1}{c}{66.79} & \multicolumn{1}{c}{50.71} & \multicolumn{1}{c}{57.8} & \multicolumn{1}{c}{55.48} & \multicolumn{1}{c}{43.66} & \multicolumn{1}{c}{57.91} & \multicolumn{1}{c}{2.33} & \multicolumn{1}{c}{1.36} \\

\multicolumn{1}{l}{Doubao-lite-0115} & \multicolumn{1}{c}{1.55} & \multicolumn{1}{c}{41.34} & \multicolumn{1}{c}{74.35} & \multicolumn{1}{c}{32.66} & \multicolumn{1}{c}{16.13} & \multicolumn{1}{c}{43.4} & \multicolumn{1}{c}{39.37} & \multicolumn{1}{c}{29.57} & \multicolumn{1}{c}{72.64} & \multicolumn{1}{c}{56.68} & \multicolumn{1}{c}{46.15} & \multicolumn{1}{c}{84.1} & \multicolumn{1}{c}{63.73} & \multicolumn{1}{c}{52.98} & \multicolumn{1}{c}{71.07} & \multicolumn{1}{c}{42.14} & \multicolumn{1}{c}{37.5} & \multicolumn{1}{c}{4.25} & \multicolumn{1}{c}{3.53} & \multicolumn{1}{c}{4.58} & \multicolumn{1}{c}{84.85} & \multicolumn{1}{c}{5.67} & \multicolumn{1}{c}{2.72} \\

\midrule
\multicolumn{24}{c}{\textit{Open-source LLMs}} \\
\midrule

\multicolumn{1}{l}{Qwen2.5-72B-Instruct} & \multicolumn{1}{c}{2.44} & \multicolumn{1}{c}{59.99} & \multicolumn{1}{c}{82.6} & \multicolumn{1}{c}{41.0} & \multicolumn{1}{c}{23.8} & \multicolumn{1}{c}{66.2} & \multicolumn{1}{c}{54.58} & \multicolumn{1}{c}{43.35} & \multicolumn{1}{c}{81.48} & \multicolumn{1}{c}{68.15} & \multicolumn{1}{c}{55.56} & \multicolumn{1}{c}{88.07} & \multicolumn{1}{c}{73.68} & \multicolumn{1}{c}{60.7} & \multicolumn{1}{c}{90.18} & \multicolumn{1}{c}{75.09} & \multicolumn{1}{c}{62.81} & \multicolumn{1}{c}{70.88} & \multicolumn{1}{c}{56.14} & \multicolumn{1}{c}{47.37} & \multicolumn{1}{c}{97.78} & \multicolumn{1}{c}{12.59} & \multicolumn{1}{c}{7.78} \\

\multicolumn{1}{l}{Qwen2.5-72B} & \multicolumn{1}{c}{2.18} & \multicolumn{1}{c}{48.83} & \multicolumn{1}{c}{52.0} & \multicolumn{1}{c}{22.6} & \multicolumn{1}{c}{9.6} & \multicolumn{1}{c}{66.51} & \multicolumn{1}{c}{59.04} & \multicolumn{1}{c}{52.91} & \multicolumn{1}{c}{55.56} & \multicolumn{1}{c}{42.96} & \multicolumn{1}{c}{35.31} & \multicolumn{1}{c}{59.65} & \multicolumn{1}{c}{57.89} & \multicolumn{1}{c}{43.86} & \multicolumn{1}{c}{65.61} & \multicolumn{1}{c}{58.25} & \multicolumn{1}{c}{44.91} & \multicolumn{1}{c}{56.49} & \multicolumn{1}{c}{54.03} & \multicolumn{1}{c}{41.75} & \multicolumn{1}{c}{/} & \multicolumn{1}{c}{/} & \multicolumn{1}{c}{/} \\

\multicolumn{1}{l}{Llama-3.1-70B-Instruct} & \multicolumn{1}{c}{1.74} & \multicolumn{1}{c}{44.65} & \multicolumn{1}{c}{56.2} & \multicolumn{1}{c}{27.6} & \multicolumn{1}{c}{15.2} & \multicolumn{1}{c}{50.27} & \multicolumn{1}{c}{48.59} & \multicolumn{1}{c}{44.84} & \multicolumn{1}{c}{68.39} & \multicolumn{1}{c}{59.01} & \multicolumn{1}{c}{52.59} & \multicolumn{1}{c}{84.56} & \multicolumn{1}{c}{70.53} & \multicolumn{1}{c}{55.44} & \multicolumn{1}{c}{54.03} & \multicolumn{1}{c}{36.49} & \multicolumn{1}{c}{31.23} & \multicolumn{1}{c}{43.51} & \multicolumn{1}{c}{23.51} & \multicolumn{1}{c}{17.54} & \multicolumn{1}{c}{94.44} & \multicolumn{1}{c}{3.7} & \multicolumn{1}{c}{0.0} \\

\multicolumn{1}{l}{Qwen2.5-7B-Instruct} & \multicolumn{1}{c}{1.24} & \multicolumn{1}{c}{34.56} & \multicolumn{1}{c}{48.2} & \multicolumn{1}{c}{14.6} & \multicolumn{1}{c}{4.8} & \multicolumn{1}{c}{50.06} & \multicolumn{1}{c}{14.29} & \multicolumn{1}{c}{7.21} & \multicolumn{1}{c}{60.99} & \multicolumn{1}{c}{44.2} & \multicolumn{1}{c}{33.09} & \multicolumn{1}{c}{78.25} & \multicolumn{1}{c}{51.58} & \multicolumn{1}{c}{40.7} & \multicolumn{1}{c}{70.17} & \multicolumn{1}{c}{40.7} & \multicolumn{1}{c}{34.03} & \multicolumn{1}{c}{17.89} & \multicolumn{1}{c}{7.37} & \multicolumn{1}{c}{3.86} & \multicolumn{1}{c}{/} & \multicolumn{1}{c}{/} & \multicolumn{1}{c}{/} \\

\multicolumn{1}{l}{Codestral-22B-V0.1} & \multicolumn{1}{c}{0.83} & \multicolumn{1}{c}{22.09} & \multicolumn{1}{c}{32.2} & \multicolumn{1}{c}{9.4} & \multicolumn{1}{c}{2.2} & \multicolumn{1}{c}{60.34} & \multicolumn{1}{c}{37.47} & \multicolumn{1}{c}{22.13} & \multicolumn{1}{c}{40.99} & \multicolumn{1}{c}{26.91} & \multicolumn{1}{c}{22.47} & \multicolumn{1}{c}{32.63} & \multicolumn{1}{c}{26.67} & \multicolumn{1}{c}{21.4} & \multicolumn{1}{c}{19.3} & \multicolumn{1}{c}{15.79} & \multicolumn{1}{c}{7.72} & \multicolumn{1}{c}{12.3} & \multicolumn{1}{c}{4.56} & \multicolumn{1}{c}{3.16} & \multicolumn{1}{c}{/} & \multicolumn{1}{c}{/} & \multicolumn{1}{c}{/} \\

\multicolumn{1}{l}{Qwen2.5-7B} & \multicolumn{1}{c}{0.8} & \multicolumn{1}{c}{20.03} & \multicolumn{1}{c}{22.8} & \multicolumn{1}{c}{8.4} & \multicolumn{1}{c}{3.8} & \multicolumn{1}{c}{44.07} & \multicolumn{1}{c}{30.29} & \multicolumn{1}{c}{27.84} & \multicolumn{1}{c}{25.43} & \multicolumn{1}{c}{17.28} & \multicolumn{1}{c}{13.33} & \multicolumn{1}{c}{36.49} & \multicolumn{1}{c}{20.7} & \multicolumn{1}{c}{22.46} & \multicolumn{1}{c}{29.47} & \multicolumn{1}{c}{18.95} & \multicolumn{1}{c}{18.6} & \multicolumn{1}{c}{9.47} & \multicolumn{1}{c}{8.07} & \multicolumn{1}{c}{3.16} & \multicolumn{1}{c}{/} & \multicolumn{1}{c}{/} & \multicolumn{1}{c}{/} \\

\multicolumn{1}{l}{Llama-3.1-8B-Instruct} & \multicolumn{1}{c}{0.73} & \multicolumn{1}{c}{22.86} & \multicolumn{1}{c}{34.0} & \multicolumn{1}{c}{7.0} & \multicolumn{1}{c}{3.0} & \multicolumn{1}{c}{33.83} & \multicolumn{1}{c}{11.78} & \multicolumn{1}{c}{8.88} & \multicolumn{1}{c}{37.78} & \multicolumn{1}{c}{20.74} & \multicolumn{1}{c}{18.02} & \multicolumn{1}{c}{65.96} & \multicolumn{1}{c}{31.93} & \multicolumn{1}{c}{26.67} & \multicolumn{1}{c}{55.09} & \multicolumn{1}{c}{25.26} & \multicolumn{1}{c}{19.3} & \multicolumn{1}{c}{9.83} & \multicolumn{1}{c}{1.05} & \multicolumn{1}{c}{1.4} & \multicolumn{1}{c}{/} & \multicolumn{1}{c}{/} & \multicolumn{1}{c}{/} \\

\multicolumn{1}{l}{AI21-Jamba-1.5-mini} & \multicolumn{1}{c}{0.57} & \multicolumn{1}{c}{16.28} & \multicolumn{1}{c}{29.4} & \multicolumn{1}{c}{3.2} & \multicolumn{1}{c}{2.6} & \multicolumn{1}{c}{52.7} & \multicolumn{1}{c}{18.64} & \multicolumn{1}{c}{5.97} & \multicolumn{1}{c}{28.89} & \multicolumn{1}{c}{21.98} & \multicolumn{1}{c}{22.22} & \multicolumn{1}{c}{39.65} & \multicolumn{1}{c}{24.91} & \multicolumn{1}{c}{29.12} & \multicolumn{1}{c}{6.7} & \multicolumn{1}{c}{3.86} & \multicolumn{1}{c}{3.16} & \multicolumn{1}{c}{0.0} & \multicolumn{1}{c}{0.0} & \multicolumn{1}{c}{0.0} & \multicolumn{1}{c}{/} & \multicolumn{1}{c}{/} & \multicolumn{1}{c}{/} \\

\multicolumn{1}{l}{Qwen2.5-1.5B-Instruct} & \multicolumn{1}{c}{0.39} & \multicolumn{1}{c}{14.45} & \multicolumn{1}{c}{31.6} & \multicolumn{1}{c}{4.4} & \multicolumn{1}{c}{1.4} & \multicolumn{1}{c}{41.88} & \multicolumn{1}{c}{5.7} & \multicolumn{1}{c}{1.62} & \multicolumn{1}{c}{26.17} & \multicolumn{1}{c}{20.74} & \multicolumn{1}{c}{17.78} & \multicolumn{1}{c}{52.63} & \multicolumn{1}{c}{17.19} & \multicolumn{1}{c}{16.14} & \multicolumn{1}{c}{13.3} & \multicolumn{1}{c}{4.91} & \multicolumn{1}{c}{3.51} & \multicolumn{1}{c}{0.7} & \multicolumn{1}{c}{0.0} & \multicolumn{1}{c}{0.35} & \multicolumn{1}{c}{/} & \multicolumn{1}{c}{/} & \multicolumn{1}{c}{/} \\

\multicolumn{1}{l}{Qwen2.5-1.5B} & \multicolumn{1}{c}{0.07} & \multicolumn{1}{c}{2.46} & \multicolumn{1}{c}{4.4} & \multicolumn{1}{c}{0.2} & \multicolumn{1}{c}{0.2} & \multicolumn{1}{c}{7.44} & \multicolumn{1}{c}{0.55} & \multicolumn{1}{c}{0.47} & \multicolumn{1}{c}{4.69} & \multicolumn{1}{c}{5.43} & \multicolumn{1}{c}{2.72} & \multicolumn{1}{c}{7.37} & \multicolumn{1}{c}{2.1} & \multicolumn{1}{c}{1.05} & \multicolumn{1}{c}{3.86} & \multicolumn{1}{c}{1.75} & \multicolumn{1}{c}{1.75} & \multicolumn{1}{c}{0.35} & \multicolumn{1}{c}{0.0} & \multicolumn{1}{c}{0.0} & \multicolumn{1}{c}{/} & \multicolumn{1}{c}{/} & \multicolumn{1}{c}{/} \\

\bottomrule
\end{tabular}
}
\vskip -0.1in
\end{table*}

\begin{figure}[t!]
    \centering
    \vskip 0.2in
    \includegraphics[width=0.8\linewidth]{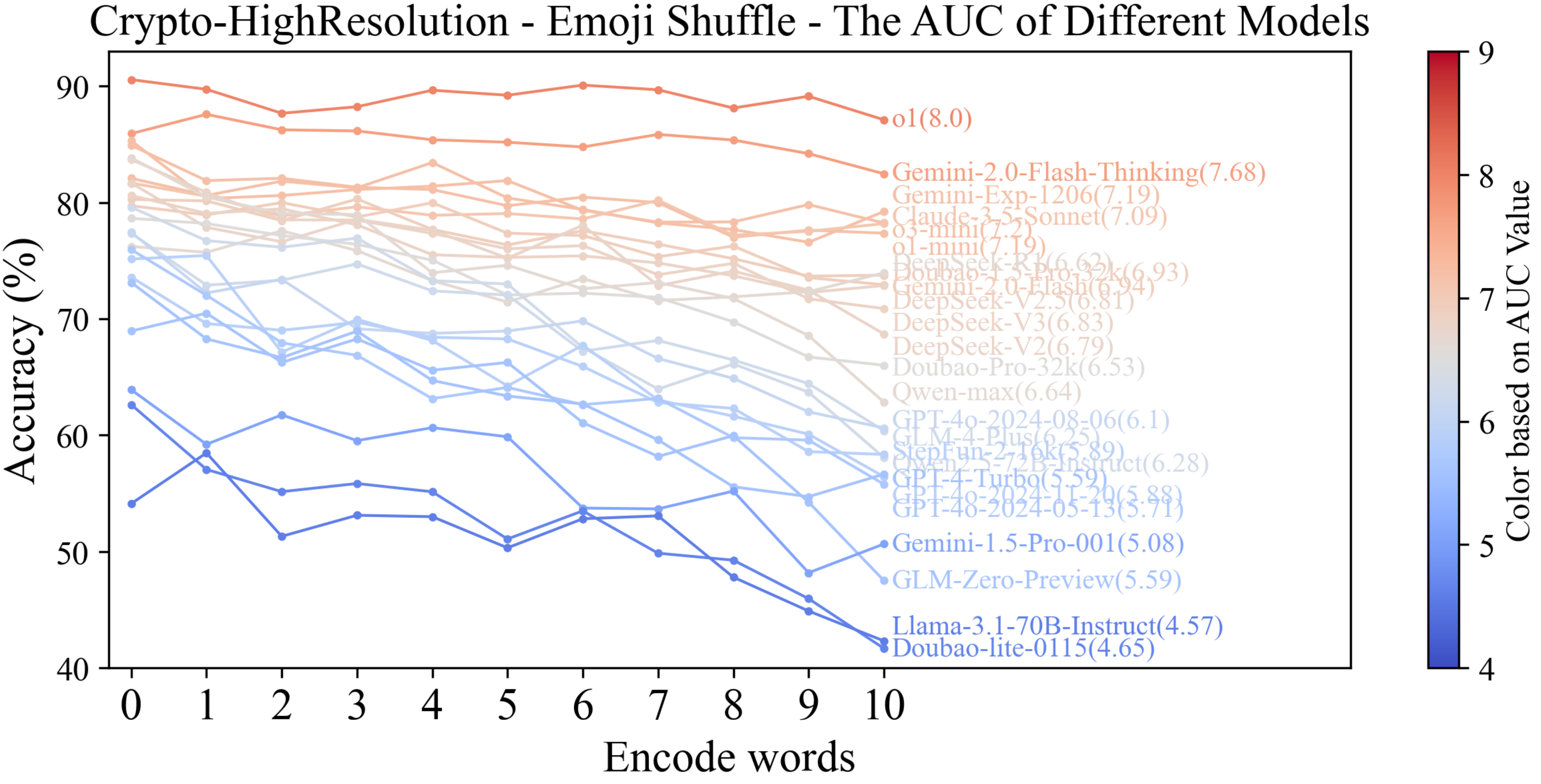}
    \caption{The performance of different models on \textit{Crypto-HighResolution} with 10 varying degrees of encoding.}
    \label{fig:subset}
    \vskip -0.2in
\end{figure}

\paragraph{CR Ability can be Better Evaluated Using AUC}
Currently, many popular benchmarks, such as MMLU, are no longer effective at capturing the performance gaps between different models. The accuracy gaps between different models are quite limited recently, which fails to fully reflect their actual abilities. 
To help LLM practitioners more accurately assess the CR ability of models, we propose using the AUC to evaluate the model's CR ability. 
The AUC provides a more effective measure of the model’s CR ability. In Figure \ref{fig:subset}, the redder the lines, the higher the corresponding AUC values, which also indicate stronger CR abilities of the model. It can be observed that o1, o3-mini, and Gemini-2.0-Flash-Thinking-Exp-01-21 exhibit relatively strong CR abilities. 
Furthermore, the variance of the models' AUC when evaluated in our \ourmethod{} settings is higher than that when evaluated in the vanilla settings, as illustrated in Appendix \ref{Variance of AUC}.

\paragraph{CR Ability in Models is Influenced by Various Factors}
We design three subtasks: question decoding, multi-hop reasoning, and answering the original question. 
We observe that as the number of subtasks or the number of encoded words increases, the model's accuracy in answering questions decreases. This indicates that the evaluation of the model's CR ability is influenced by the complexity of the question composition and the difficulty of the subtasks. 
We test the performance of Qwen2.5 models with different sizes on the hardest Crypto-MMLU-Alpha question, and the results show that when the number of encoded words reached 10, Qwen2.5-1.5B-Instruct has near-zero accuracy, while Qwen2.5-72B-Instruct demonstrates some CR ability. The same performance is observed when the number of encoded words is zero. 
To validate our findings, we also compare base models with instruct models and examine the differences across various model architectures in Section \ref{section: ablation experiments}.


\begin{figure}[h!]
    \centering
    \vskip 0.2in
    \includegraphics[width=0.8\linewidth]{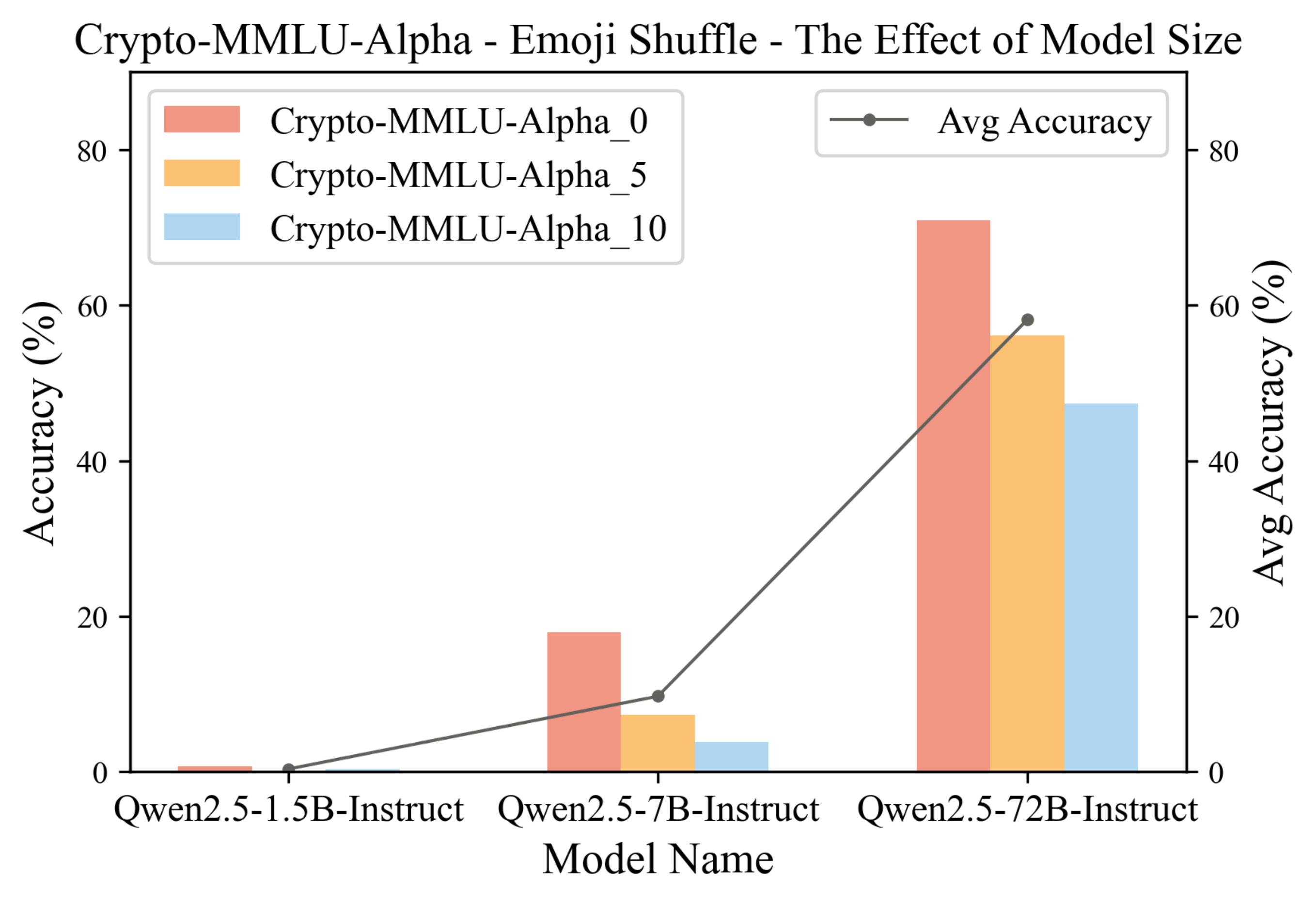}
    \caption{The performance of models with different model size on Crypto-MMLU-Alpha. In \textit{Crypto-MMLU-Alpha\_Words}, \textit{Words} denotes the number of words encoded in the given question.}
    \label{fig:model_size}
    \vskip -0.2in
\end{figure}

\subsection{Ablation Experiments}
\label{section: ablation experiments}
To validate our conclusion, we consider various factors affecting the model's CR abilities (e.g., multi-turn dialogue vs single-turn dialogue, base model vs instruct model, decoding capacity, different architectures, Doubao-Moe vs Doubao-Dense). Complete experimental results and other ablation experiments can be seen in Appendix \ref{appendix: Ablation Experiments}.

\paragraph{Multi-Turn Decomposition Reasoning is Simpler than Single-Turn Composition Reasoning}
We decompose the CR task into multiple turns, where multi-turn focuses more on individual questions, while single-turn focuses on the compositional question (The detailed design is provided in Appendix \ref{appendix: multi-turn setup}).
In Figure \ref{fig:multi-turns}, models with higher accuracy in answering questions generally perform better in multi-turn dialogues than in single-turn dialogues, indicating that the model has certain limitations in composition reasoning compared to solving individual questions.

\begin{figure}[ht!]
    \centering
    \vskip 0.2in
    \includegraphics[width=0.8\linewidth]{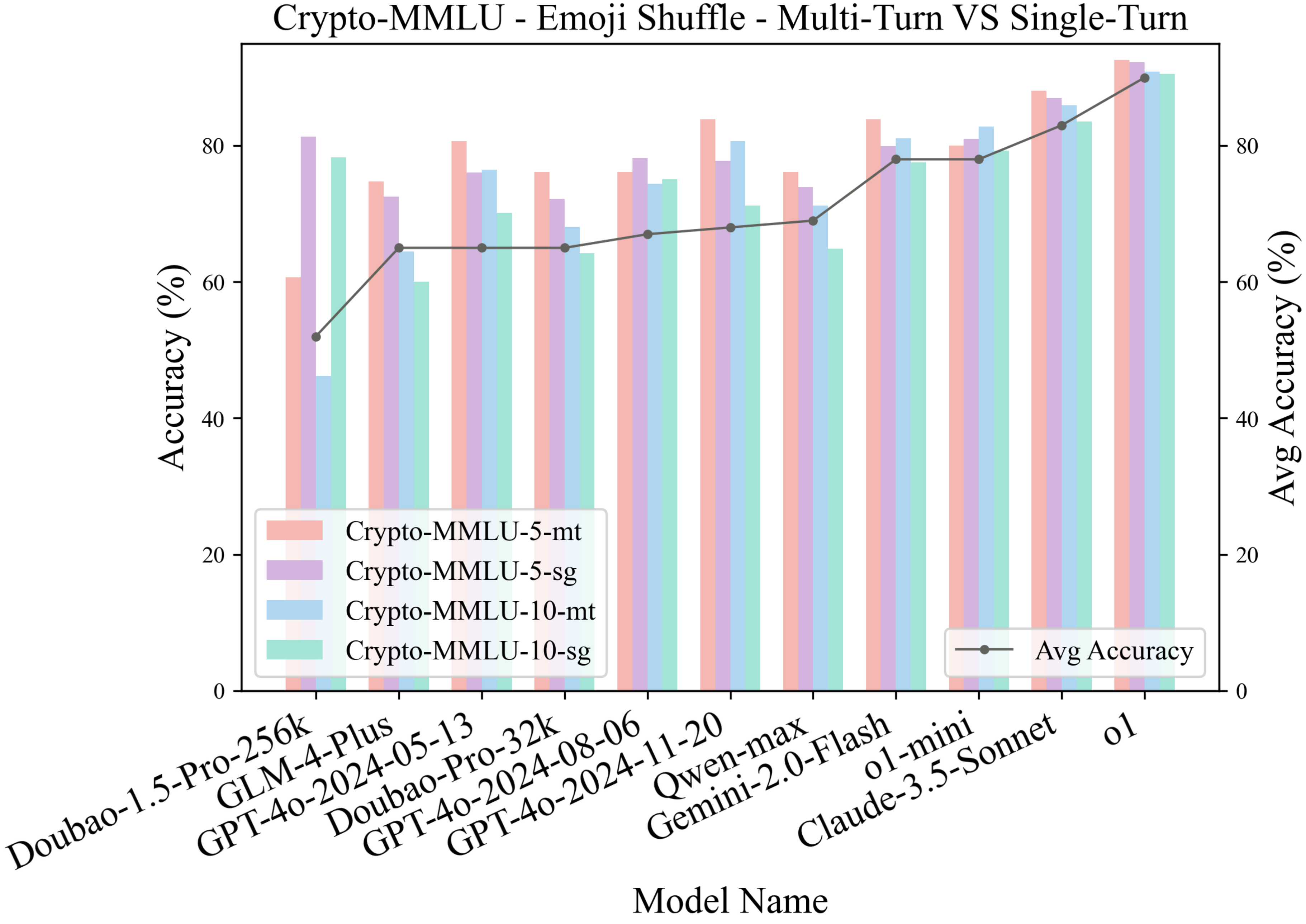}
    \caption{The comparison of the performance of multi-turn and single-turn dialogues. In \textit{Crypto-MMLU-Words-Type}, \textit{Words} denotes the number of words encoded in the given question. \textit{Type} denotes multi-turn(mt) or single-turn(sg).}
    \label{fig:multi-turns}
    \vskip -0.2in
\end{figure}
\begin{figure}[h!]
    \centering
    \vskip 0.2in
    \includegraphics[width=0.8\linewidth]{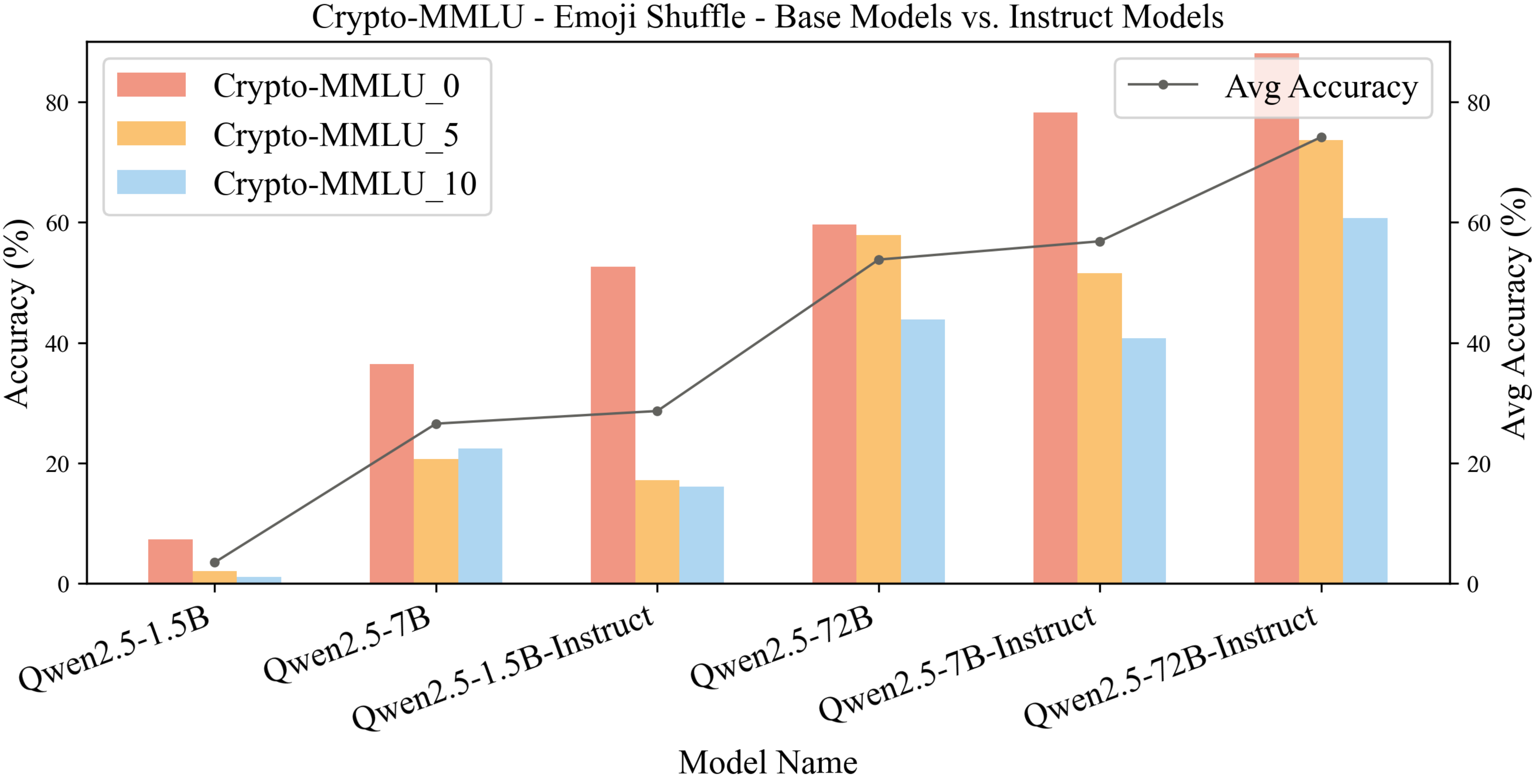}
    \caption{The performance of the Base models and the Instrcut models on Crypto-MMLU. In \textit{Crypto-MMLU\_Words}, \textit{Words} denotes the number of words encoded in the given question.}
    \label{fig:base}
    \vskip -0.2in
\end{figure}

\vspace{-10pt} 
\paragraph{Instruct Fine-tuning Enhances the CR Ability of the Model}
With increasing encoding complexity and multi-hops, base models demonstrate degraded performance in handling complex compositional tasks, showing significant accuracy decline. 
Figure \ref{fig:base} reveals that base models exhibit markedly inferior CR abilities compared to instruct models. The average accuracy of question answering in base models decreased by no less than 20\%.
There is even a case where the larger model, Qwen2.5-72B, performs worse than the smaller model, Qwen2.5-7B-Instruct.

\definecolor{mygreen}{rgb}{0.173, 0.627, 0.173} 
\definecolor{myorange}{rgb}{1.0, 0.498, 0.055}
\definecolor{myblue}{rgb}{0.122, 0.467, 0.706}
\definecolor{myred}{rgb}{0.839, 0.153, 0.157}
\definecolor{mypurple}{rgb}{0.580, 0.404, 0.741}

\section{Analysis}
\label{sec:analysis}
To further analyze the internal mechanism of the CR of LLMs instead of simply measuring the CR performance of LLMs,
We try to explore the reasons why most LLMs perform well on original tasks, but only a few models excel at CR tasks. Therefore, we further conduct analytic experiments, including the logit lens~\citep{logitlens}, the neuron activation of LLMs, and analysis on compositional reasoning stages, to explore the reasoning from a more mechanistic and interpretable perspective.

\begin{figure}[h!]
    \centering
    \vskip 0.2in
    \subfigure[Result of Qwen2.5-3B-Instruct]{
        \includegraphics[width=0.3\linewidth]{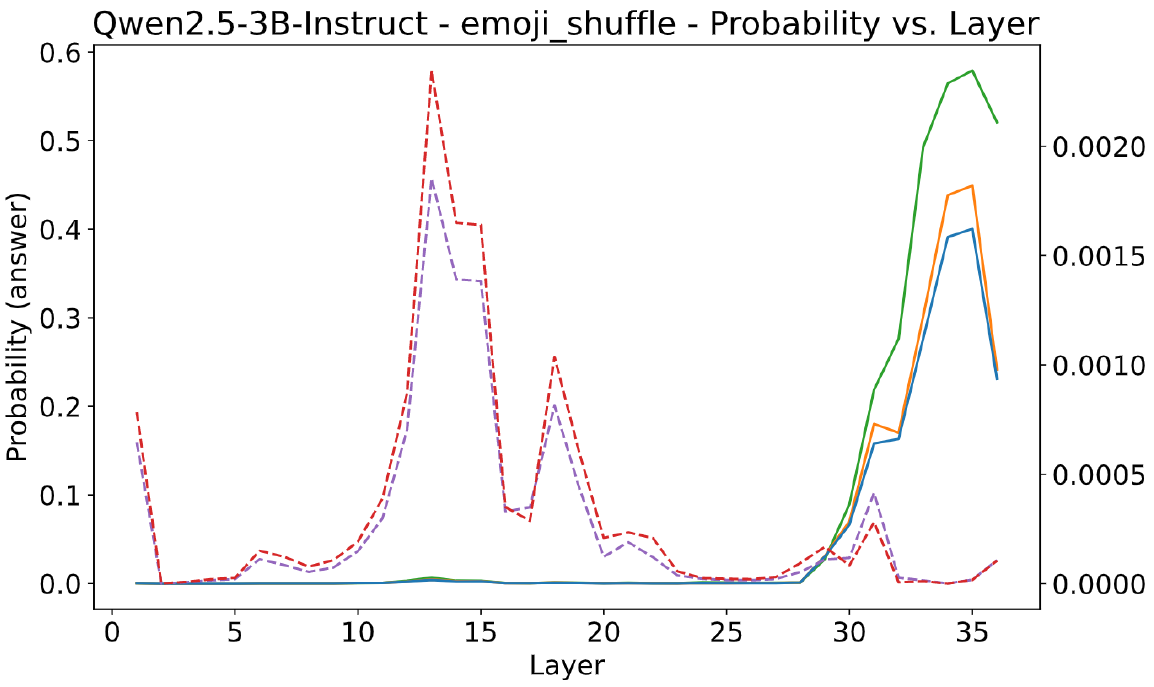}
    }
    \subfigure[Result of Qwen2.5-7B-Instruct]{
        \includegraphics[width=0.3\linewidth]{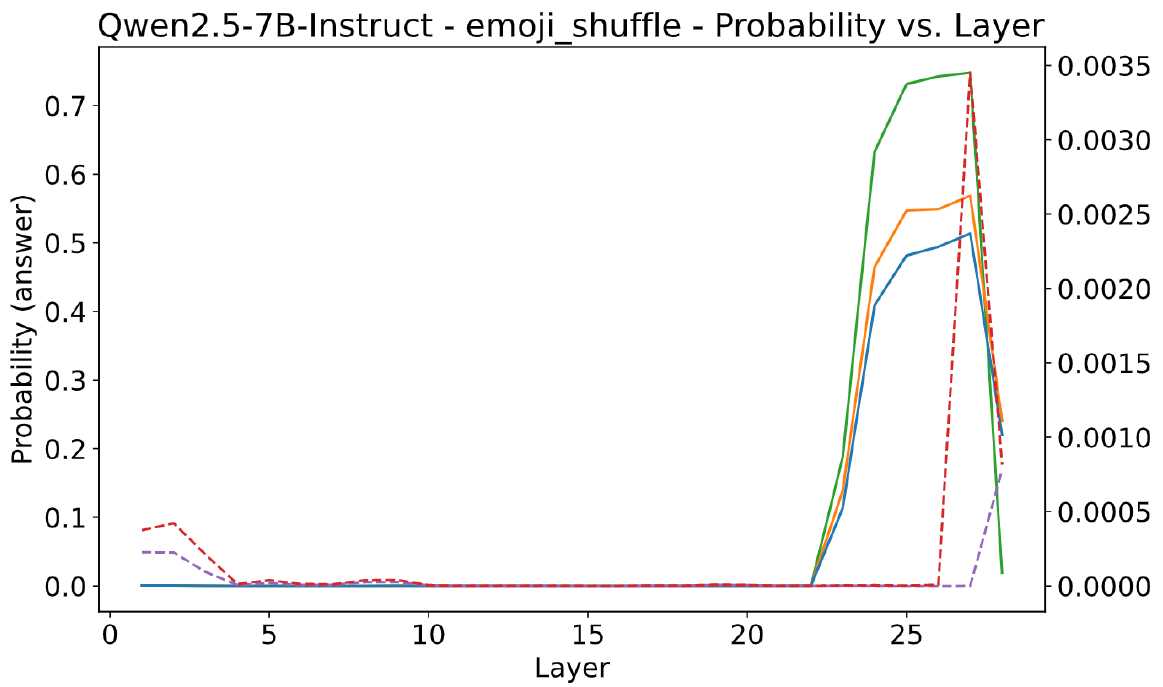}
    }
    \subfigure[Result of Llama-3.1-8B-Instruct]{
        \includegraphics[width=0.3\linewidth]{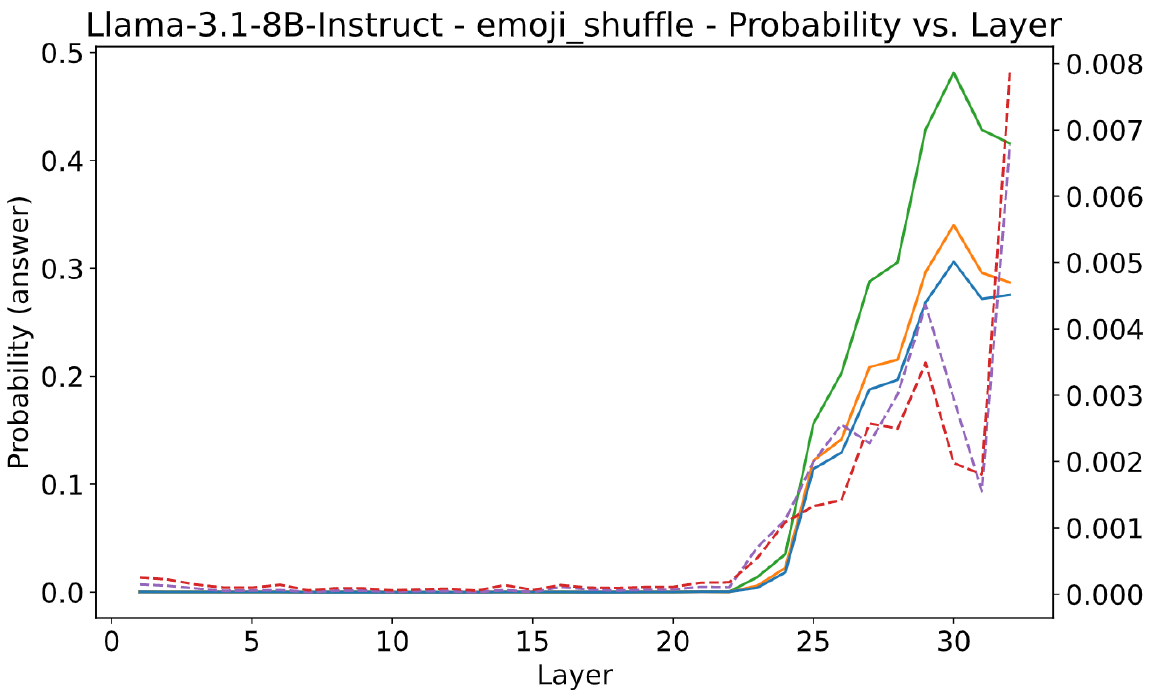}
    }
    \caption{The result of logit lens experiments.
The green valid line ``\protect\tikz[baseline] \protect\draw[line width=0.5mm,color=mygreen,yshift=1.2mm] (0,0) -- (0.4,0);'' corresponds to the answer set when encoding 0 words; 
The orange valid line ``\protect\tikz[baseline] \protect\draw[ line width=0.5mm,color=myorange,yshift=1.2mm] (0,0) -- (0.4,0);'' corresponds to the answer set when encoding 3 words;
The blue valid line ``\protect\tikz[baseline] \protect\draw[ line width=0.5mm,color=myblue,yshift=1.2mm] (0,0) -- (0.4,0);'' corresponds to the answer set when encoding 5 words;
The purple dashed line ``\protect\tikz[baseline] \protect\draw[dashed, line width=0.5mm,color=mypurple,yshift=1.2mm] (0,0) -- (0.4,0);'' corresponds to the decoded words set when encoding 3 words;
The red dashed line ``\protect\tikz[baseline] \protect\draw[dashed, line width=0.5mm,color=myred,yshift=1.2mm] (0,0) -- (0.4,0);'' corresponds to the decoded words set when encoding 5 words;}
\label{fig:logit_lens_res}
\vskip -0.2in
\end{figure}

\subsection{Logit Lens}

\begin{table}[h!]
\centering
\caption{Scores of models used in logit lens analysis. We tested these models using 0/3/5 words encoded dataset in the same configuration and methods of the Experiment section.}
\label{tab:scores_logit_lens}
\vskip 0.15in
\resizebox{0.5\linewidth}{!}{%
\begin{tabular}{cccc}
\toprule
\multirow{2}{*}{\textbf{Model}} & \multicolumn{3}{c}{\textbf{MMLU}} \\
\cmidrule{2-4}
 &  \textbf{0} & \textbf{3} & \textbf{5} \\
\midrule
Qwen2.5-3B-Instruct  & 0.722 & 0.305 & 0.266 \\
Qwen2.5-7B-Instruct  & 0.771 & 0.592 & 0.512 \\
Llama-3.1-8B-Instruct & 0.663 & 0.382 & 0.301 \\
\bottomrule
\end{tabular}
}
\vskip -0.1in
\end{table}
Logit lens~\citep{logitlens} is a technique used to interpret the intermediate outputs of language models by mapping hidden states directly to the output vocabulary.
In logit lens analysis experiments, we defined two target sets, $T_1$ and $T_2$. 
$T_1$ consists of the options and content of the answers, while $T_2$ consists of the decoded content from the encoded elements in the question. $T_1=\bigcup\limits_{a\in \mathcal{A}} \sigma(a)$ and $T_2=\bigcup\limits_{w\in \mathcal{E}} \sigma(\mathcal{D}(w))$, where $\mathcal{D}(w)$ denotes the correct decoding process using our codetable. $\mathcal{A}$ denotes the answer set and $\mathcal{E}$ denotes the encoded words set. $\sigma(.)$ denotes the function that transforms the original word into a series of candidate prefix tokens with different tokenization. For example, $\sigma(``water'') = [``wa'', \dots, ``water'']$.

When the question and encoding rules are provided as input to the LLM, we extract the hidden state of the last token $h_{e}$ during the reasoning process and then get the word distribution $p_{e}$ through the word projection. Then, we accumulate the probabilities of all candidate tokens from $T_{1}$ or $T_{2}$ to get the total probability: $p=\sum\limits_{v\in T_{1} \ \text{or} \ v\in T_{2}}\texttt{Softmax}(h_iW_{p})[v]$, where $p$ is denoted as logits lens.

The logit lens result is shown in Figure \ref{fig:logit_lens_res} and the performance of corresponding models is shown in Table \ref{tab:scores_logit_lens}. Through observation and analysis, we reached the following conclusions:

(1) \textbf{Task decomposition and sequential reasoning}. The probability on $ T_1 $ and $ T_2 $ both showed peaks, with the peak for $ T_2 $ occurring earlier than $ T_1 $, which indicates that the model decomposes the compositional problem into a translation problem, an original MMLU problem and then solves them in order.

(2) \textbf{Summary of subtask answers}.We find that there are two peaks on $ T_1 $ and the second peak typically aligns closely with the peak of $ T_2 $ in the last layers. We hypothesize that LLMs tend to summarize the answers of the subtasks to derive the answer of the compositonal quesiton, when providing the final answer to a compositional problem.

\subsection{Neuron Activation Analysis}
Neuron Activation Analysis examines the activation patterns of individual neurons within a language model to understand their specific roles and contributions during the reasoning or prediction process. Specifically, in order to facilitate the division of the token set, this experiment selects base morse as the encoding method.
We categorize the tokens into two sets to conduct Neuron Activation Analysis: 
(1)\textbf{Vocab} contains tokens about the encoding rules (e.g. 'A': '$.-$', 'B': '$-...$',\dots).
(2)\textbf{Encoded} contains encoded tokens of the question (e.g. $..-.|..-|-.|-.-.$ \dots).

We extract the activation values of the neurons in the MLP layers of LLM during reasoning and then normalize them to a range of 0 to 10. 
For each neuron, we examine the activation values of the tokens in the Vocab and Encoded set. 
If the activation value of a token in this range exceeds 7, we classify that neuron as highly activated for this set. 
As shown in the Figure \ref{fig:neuron-level}, we record the number of highly activated neurons of each layer and have following conclusions:

\begin{figure}[h!]
\vskip 0.2in
\begin{center}
\centerline{\includegraphics[width=0.95\textwidth]{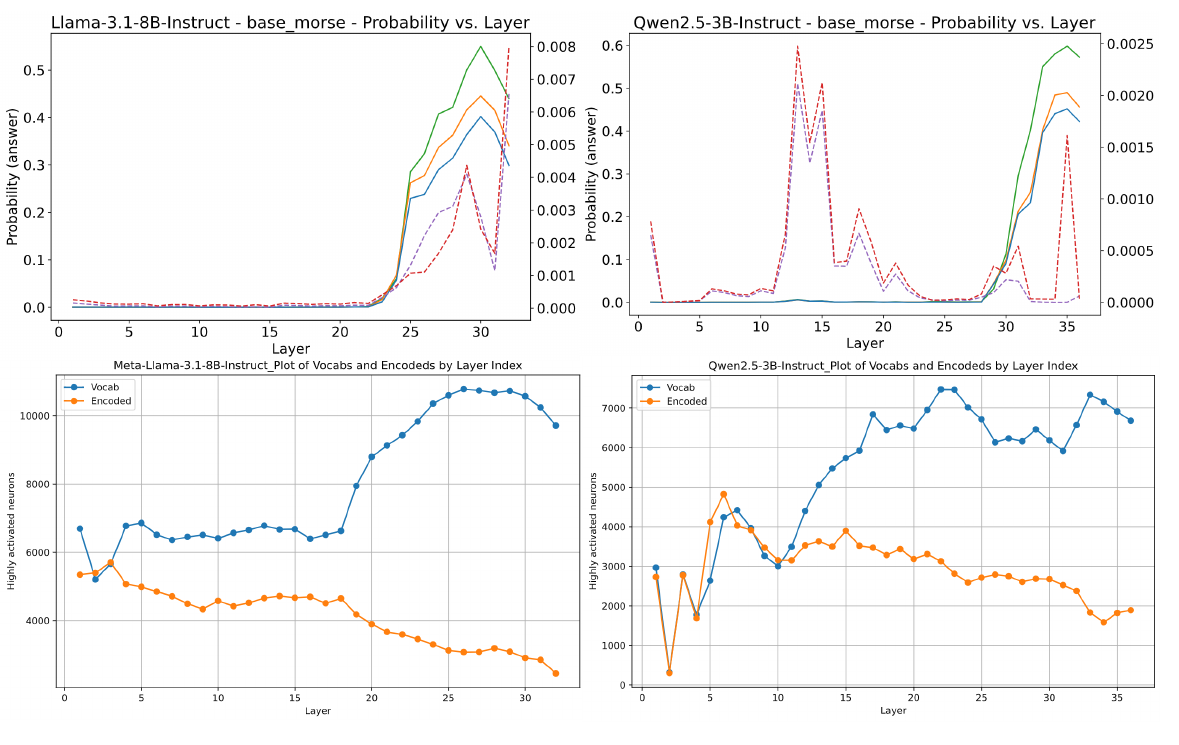}}
\caption{Neuron Activation Analysis:the comparison figure of neuron activation analysis and logit lens on Llama-3.1-8B-Instruct and Qwen2.5-3B-Instruct.}
\label{fig:neuron-level}
\end{center}
\vskip -0.2in
\end{figure}

\begin{table*}[h!]
\centering
\caption{Key tokens, corresponding layers, and their corresponding functions
}
\label{tab:layer_function_table}
\vskip 0.15in
\resizebox{1.0\linewidth}{!}{
\begin{tabular}{ccc}
\toprule
\textbf{Token} & \textbf{Layer} & \textbf{Functions} \\ 
\midrule
createCommand & 6 & Related to the subprocesses in the compositional reasoning process. \\ 
addCriterion & 6 & Related to rules in the reasoning process. \\ 
.GetObject & 2$\sim$5 & Related to capturing key content in the previous layers. \\ 
iteDatabase & 8$\sim$10, 16 & Related to content retrieval, particularly frequent when the original problem is a knowledge-based question. \\ 
Options & 22$\sim$25 & Related to making choices. \\ 
DataExchange & 17$\sim$20 & \makecell{Related to data updates within sub-tasks; these layers are highly sensitive to encoding rules, possibly related to decoding.} \\ 
\bottomrule
\end{tabular}}
\vskip -0.1in
\end{table*}

(1) \textbf{There is a clear nueron activation order in the compositional reasoning}.In the initial layers, both the Vocab and Encoded curves show a brief upward trend, followed by a continuous decline in Encoded curve and a rise and fall in the Vocab curve. This suggests that during the compositional reasoning process, the LLM first focuses on reading the entire question and then shifts its attention to encoding method, decodes the encoded question, and places less focus on the encoding tokens corresponding to already decoded words. Finally, the model concentrates on solving the problem and summarizing the answer. Thereby, When LLMs perform compositional reasoning, neurons exhibit a clear activation sequence corresponding to the order of subtasks.

(2) \textbf{Neuron activation precedes subtasks resolution}. By comparing the logit lens and neuron activation analysis experiments, we observe that in the logit lens, there is a noticeable increase in the decoded words set in the initial layers. Similarly, in the neuron activation analysis, Vocab also shows a significant rise which precedes the increase observed in the Logit Lens. This indicates that, before performing the decoding operation, the LLM first focuses its attention on the encoding rules and uses these rules to decode. Thereby, the activation of neurons helps LLM understand and solve the subtasks during compositional reasoning.

\subsection{Reasoning Stage Analysis}
Reasoning Stage Analysis investigates the sequential steps and intermediate stages a language model goes through during its reasoning process, aiming to identify how the model transitions from input to final output.
To explore the reasoning stages of CR, we extract the top 30 normalized logits and their corresponding tokens for each layer and use \textbf{Explainer LLM} (Doubao-Pro-256K) to explain the functions of each layer. During this process, we divide the reasoning stage according to layer function and identify some key functional tokens that appeared frequently:

\textbf{Layer Functions}.The stages of the compositional reasoning can be summarized as \textbf{Shallow
Reasoning} (layer 1$\sim$11, responsible for task decomposition and information collection), \textbf{Intermediate
Reasoning} (layer 10$\sim$26, responsible for subtask solving)and \textbf{Deep
Reasoning} (layer 18$\sim$31, responsible for Information summarizing and final answer generation). 

\textbf{Functional Tokens}. We find that some tokens appearing in the top 30 logits may be related to the key functions of LLMs during compositional reasoning, which is presented in Table \ref{tab:layer_function_table}. These functional tokens demonstrate that LLMs trigger specific thinking mechanisms when handling complex compositional reasoning tasks, which assist the model in decomposing, processing and summarizing CR tasks and are clearly distributed in specific reasoning stages.

\section{Related Work}
\subsection{Compostional Reasoning of LLMs}
To systematically evaluate the reasoning capabilities of LLMs~\cite{map_neo,llama3,qwen25,execrepobench,codearena,mdeval}, researchers have developed benchmarks encompassing multiple cognitive dimensions, including commonsense reasoning~\citep{ponti-etal-2020-xcopa,singh2021com2sense,onoe2021creak}, multi-hop reasoning~\citep{ho2020constructing, shi2022stepgame, trivedi2022musique}, and logical reasoning~\citep{saparov2023language,han2022folio,masry2022chartqa}. However, LLMs still face severe challenges when faced with CR tasks that require cross-domain integration~\citep{dziri2024faith, press2022measuring}. Early research in CR primarily employed CoT with compositional queries to guide model reasoning~\citep{zhou2022least, drozdov2022compositional}. Existing studies on CR largely utilize multi-hop reasoning benchmarks to investigate models' internal reasoning mechanisms~\citep{sakarvadia2023memory, ghandeharioun2024patchscope, li2024understanding}. This study proposes a novel framework aimed at enabling an effective combination of diverse reasoning tasks, thereby comprehensively assessing models' CR capabilities.

\subsection{Mechanistic Analysis Methods}
To further study internal reasoning mechanisms of LLM, many researchers focus on the changes in hidden states during the reasoning process. Logit lens~\citep{logitlens} is a typical analytical tool, which extracts the hidden states between layers to obtain probability distribution of each layer. Researchers use logit lens to analyze the performance of LLMs in multilingual settings~\citep{LanguageMoel} and the analysis of compositional reasoning~\citep{CompositionalReasoning}, aiming to study the implicit results in reasoning process. The research about Retrieval Head~\citep{Wu2024RetrievalHM} proposes an attention-based interpretability analysis method based on the Retrieval Head in the context, providing a visual analysis of the attention distributions during reasoning process. Meanwhile, the OpenAI team introduced a method using LLMs to generate behavior explanations through neuron activation~\citep{bills2023language}, allowing the behavior of pivotal neurons to be more intuitively presented to researchers.

\section{Conclusion}
In this paper, we propose a bench-free compositional reasoning task generation framework \ourmethod{} and develop a CR evaluation benchmark \benchmark{}, which includes 21 sub-datasets and nearly 7K cases collected from multiple domains. We conduct several experiments on 20+ open-source and closed-source LLMs to demonstrate that there is still a huge gap between open-source models and closed-source LLMs, which shows significant performance in CR ability for o1, with an accuracy rate of 83.7\% on \benchmark{}. Furthermore, we conduct logit lens, neuron activation analysis and reasoning stage analysis to describe the inner mechanism of CR. In the future, we plan to continuously update and expand the dataset to support a wider variety of problem types. We also aim to incorporate dynamic encoding methods (e.g., Huffman coding, RSA encryption) and extend \benchmark{} to multi-modal datasets, driving further advancements in this field.

\section{Acknowledgements}
Multimodal Art Projection (M-A-P) is an open-source AI research community.
The community members are working on research topics in a wide range of spectrum, including but not limited to pre-training paradigm of foundation models, large-scale data collection and processing, and the derived applciations on coding, reasoning and music generation.

\newpage

\bibliography{main.bib}

\newpage
\appendix
\section{Encoding Rule}
\label{appendix: encoding rule}
Since sections \ref{sec:methods} and \ref{sec:experiment} primarily introduce and utilize the emoji shuffle encoding method, this section will present and supplement the other two encoding methods. Content \ref{appendix: encoding rule} below shows the encoding rules.
\begin{figure}[H]
    \centering
    \vskip 0.2in
    \subfigure[Morse Base]{
        \includegraphics[width=0.36\textwidth]{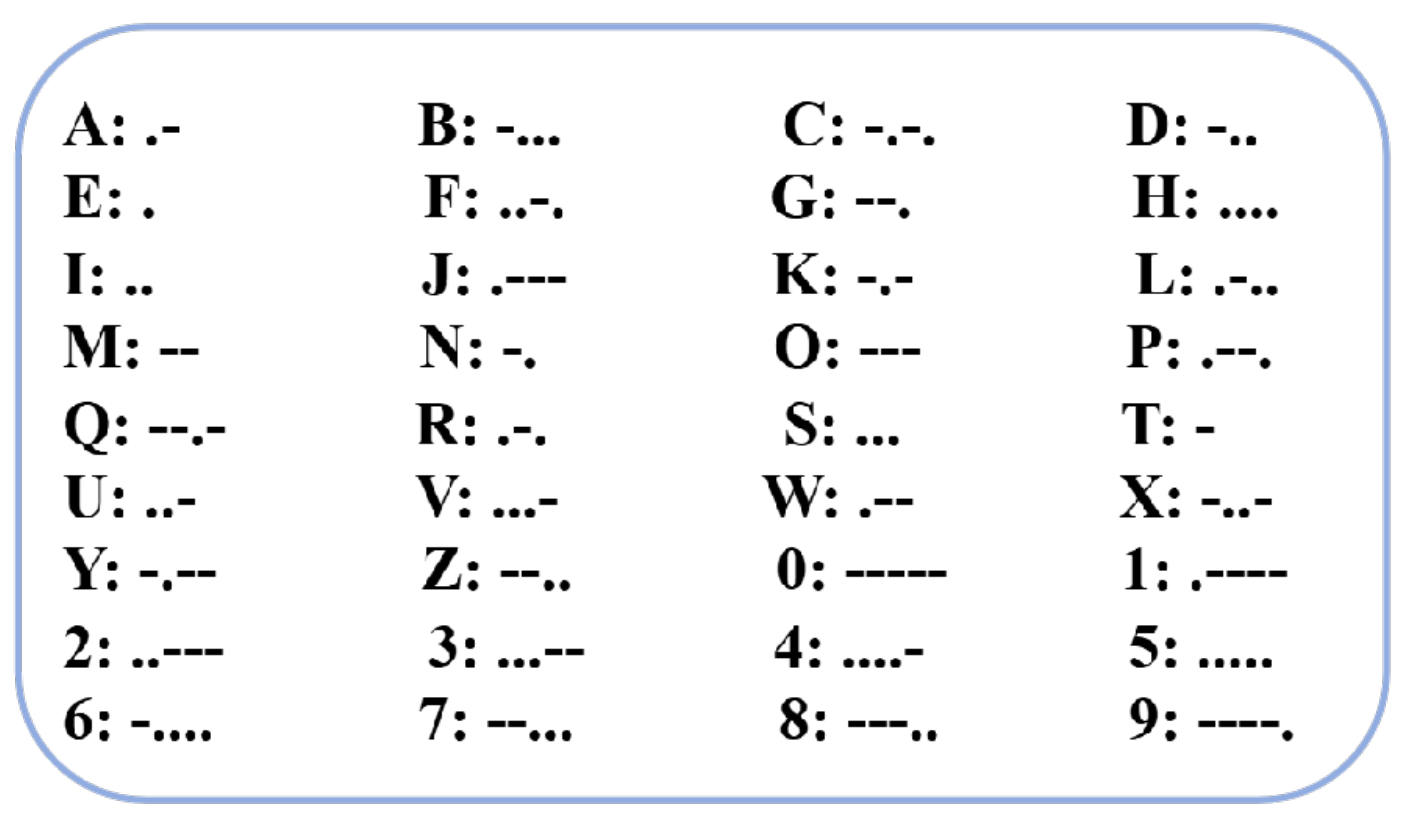}
    }
    \subfigure[Emoji Base]{
        \includegraphics[width=0.54\textwidth]{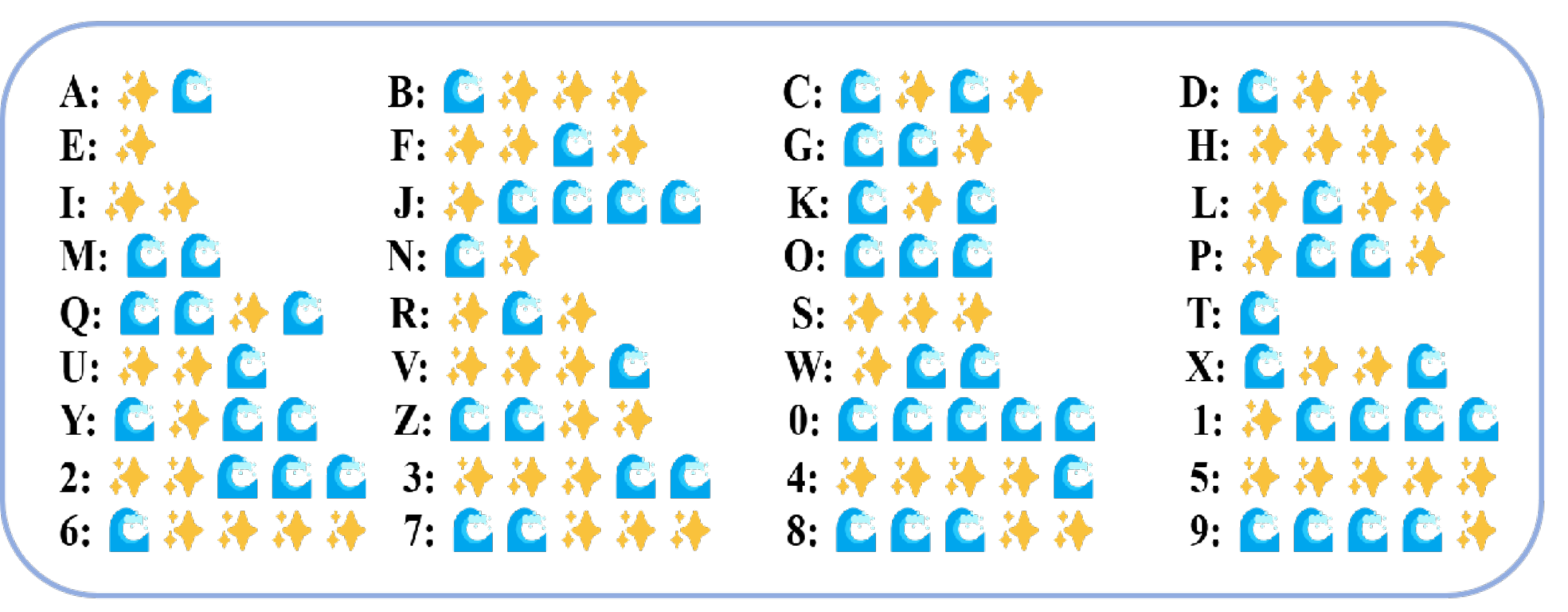}
    }
    \subfigure[Emoji Shuffle]{
        \includegraphics[width=0.54\textwidth]{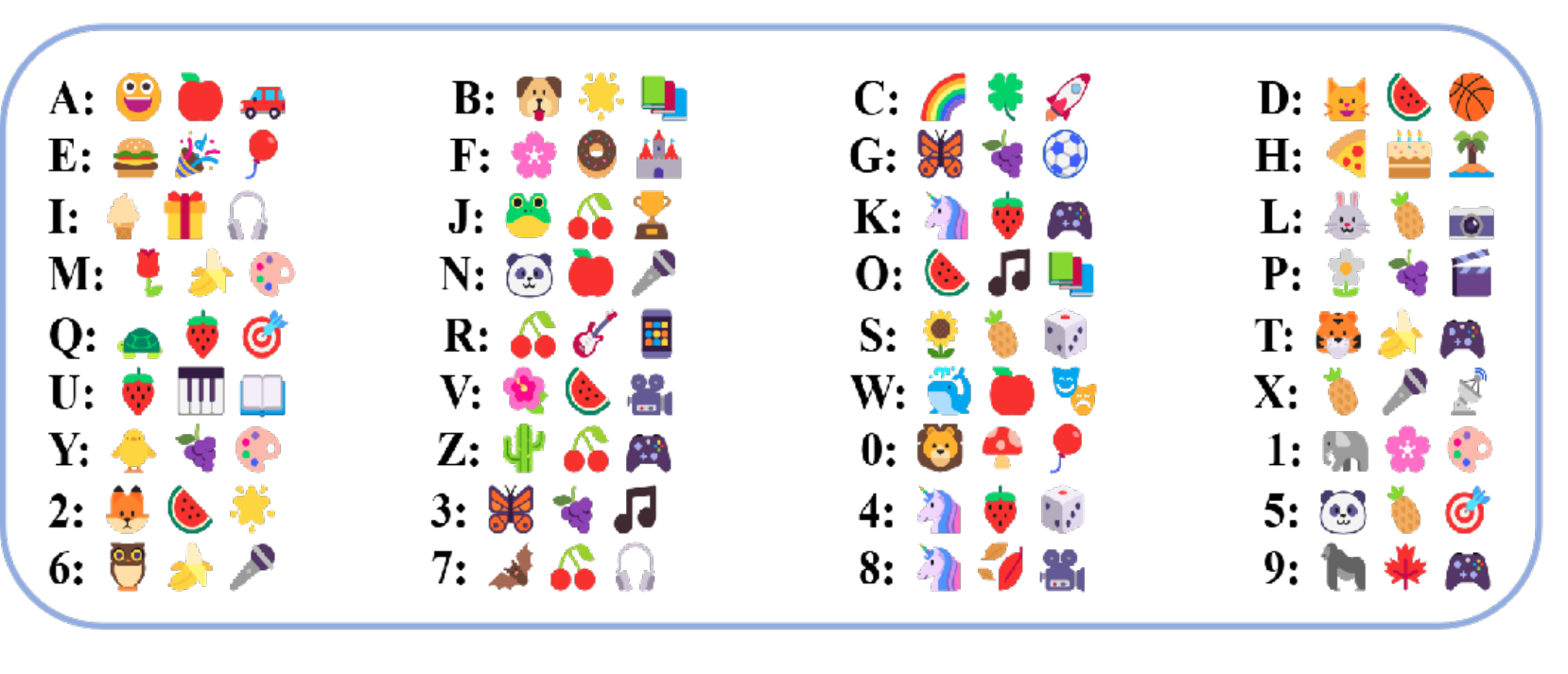}
    }
    \subfigure[Morse Base Math]{
        \includegraphics[width=0.36\textwidth]{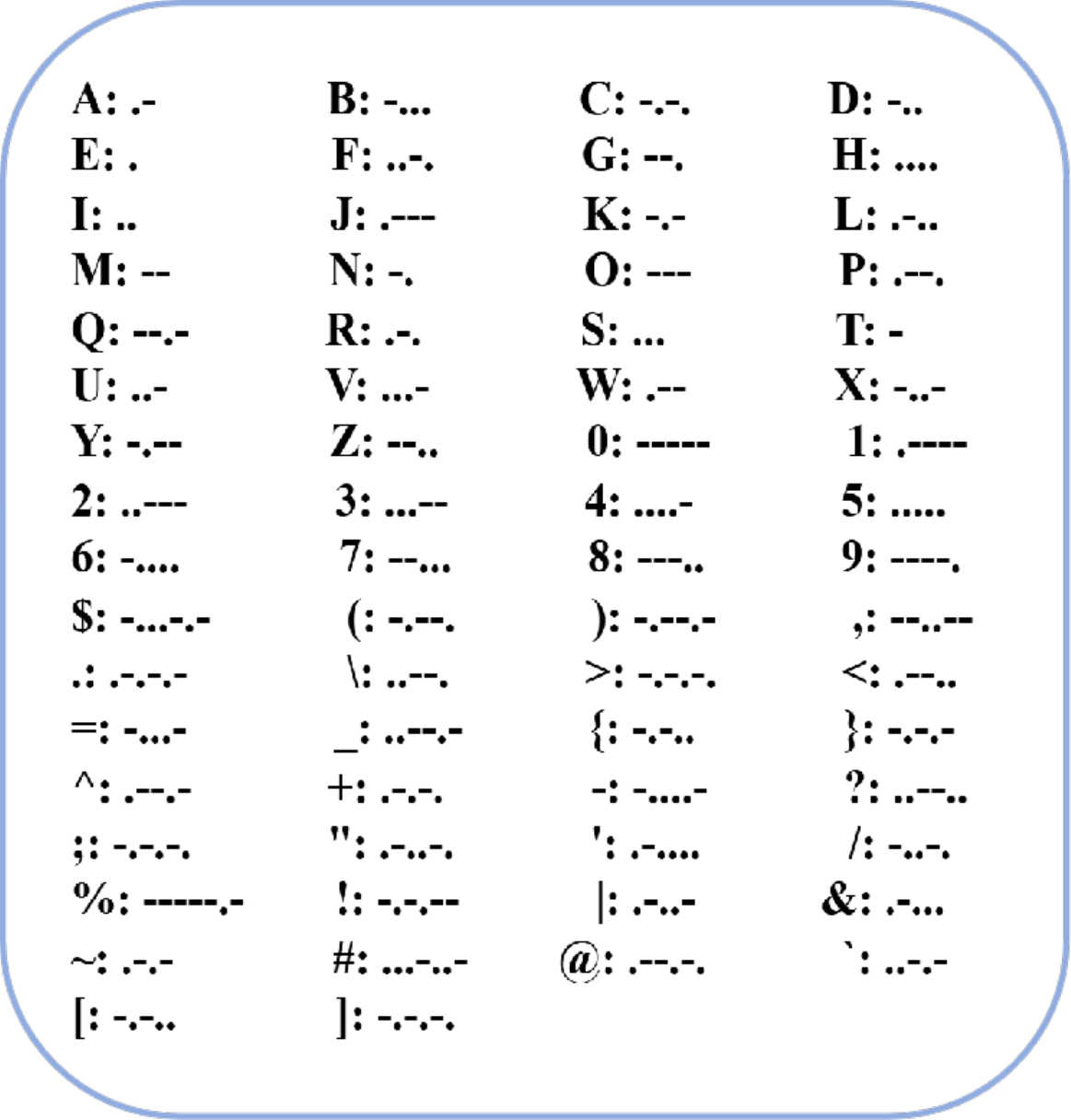}
    }
    \subfigure[Emoji Base Math]{
        \includegraphics[width=0.49\textwidth]{pics/encoding_rule/emoji_base.pdf}
    }
    \subfigure[Emoji Shuffle Math]{
        \includegraphics[width=0.41\textwidth]{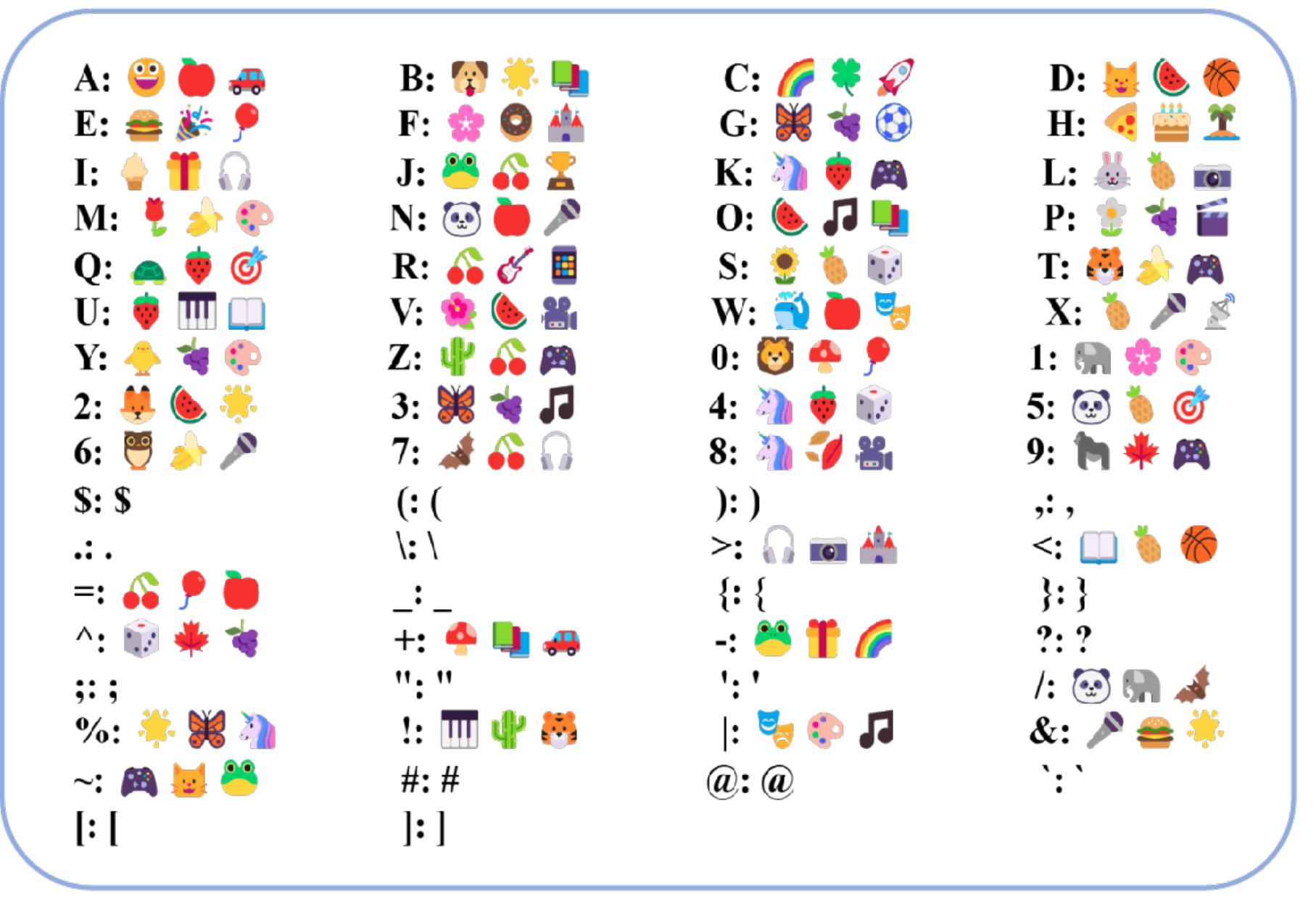}
    }
    \caption{Encoding rule for our experiments.}
    \label{fig:encoding rule}
    \vskip -0.2in
\end{figure}
\vfill

\newpage
\section{Prompt Templates}
\label{appendix: prompt Template}
Content \ref{appendix: prompt Template} below shows the prompt templates used in our \benchmark{}.
\subsection{Prompt for Crypto-Math, Crypto-MBPP and Crypto-BBH}
\subsubsection{Zero-Shot Prompt}
\begin{figure}[H]
    \centering
    \vskip 0.2in
    \subfigure[Zero-shot prompt used for non-encoded question]{
        \includegraphics[width=0.47\textwidth]{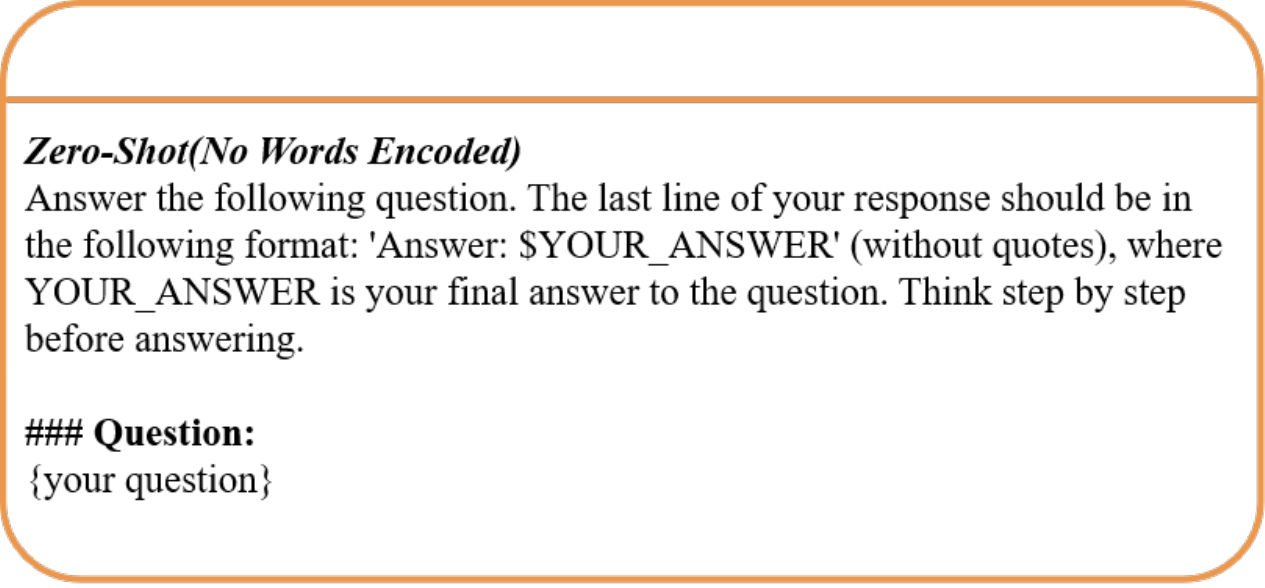}
    }
    \subfigure[Zero-shot prompt used for encoded question]{
        \includegraphics[width=0.42\textwidth]{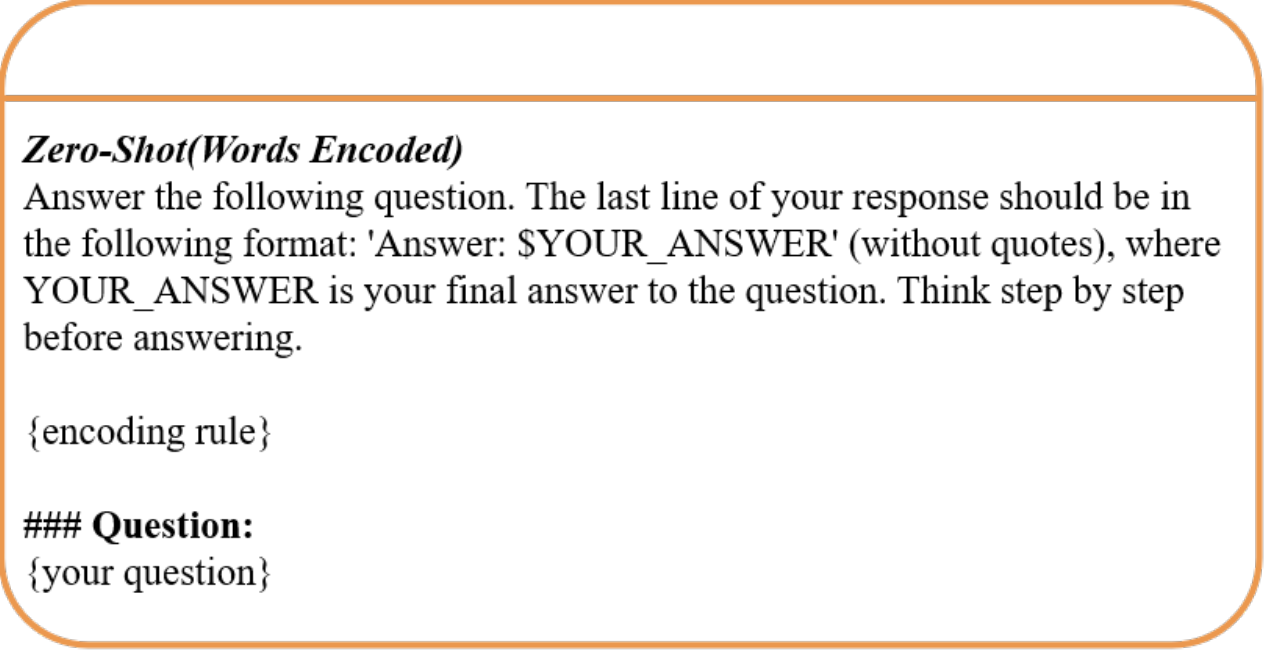}
    }
    \caption{Zero-shot prompt for Crypto-Math and Crypto-MMBP}
    \label{fig:0shotall}
    \vskip -0.2in
\end{figure}

\subsubsection{Three-Shot Prompt}
\begin{figure}[H]
    \centering
    \vskip 0.2in
    \subfigure[Three-shot prompt used for non-encoded question]{
        \includegraphics[width=0.45\textwidth]{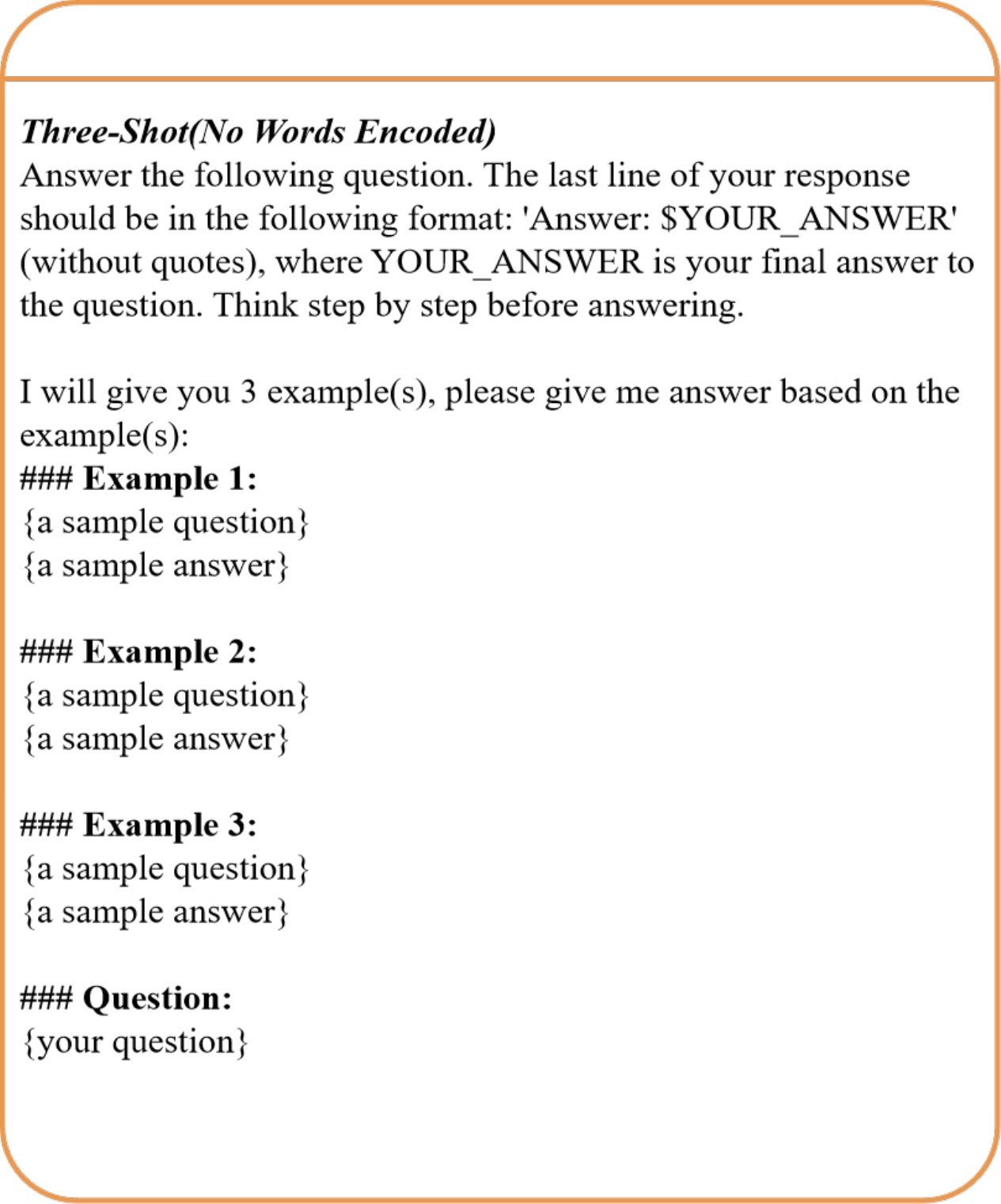}
    }
    \subfigure[Three-shot prompt used for encoded question]{
        \includegraphics[width=0.45\textwidth]{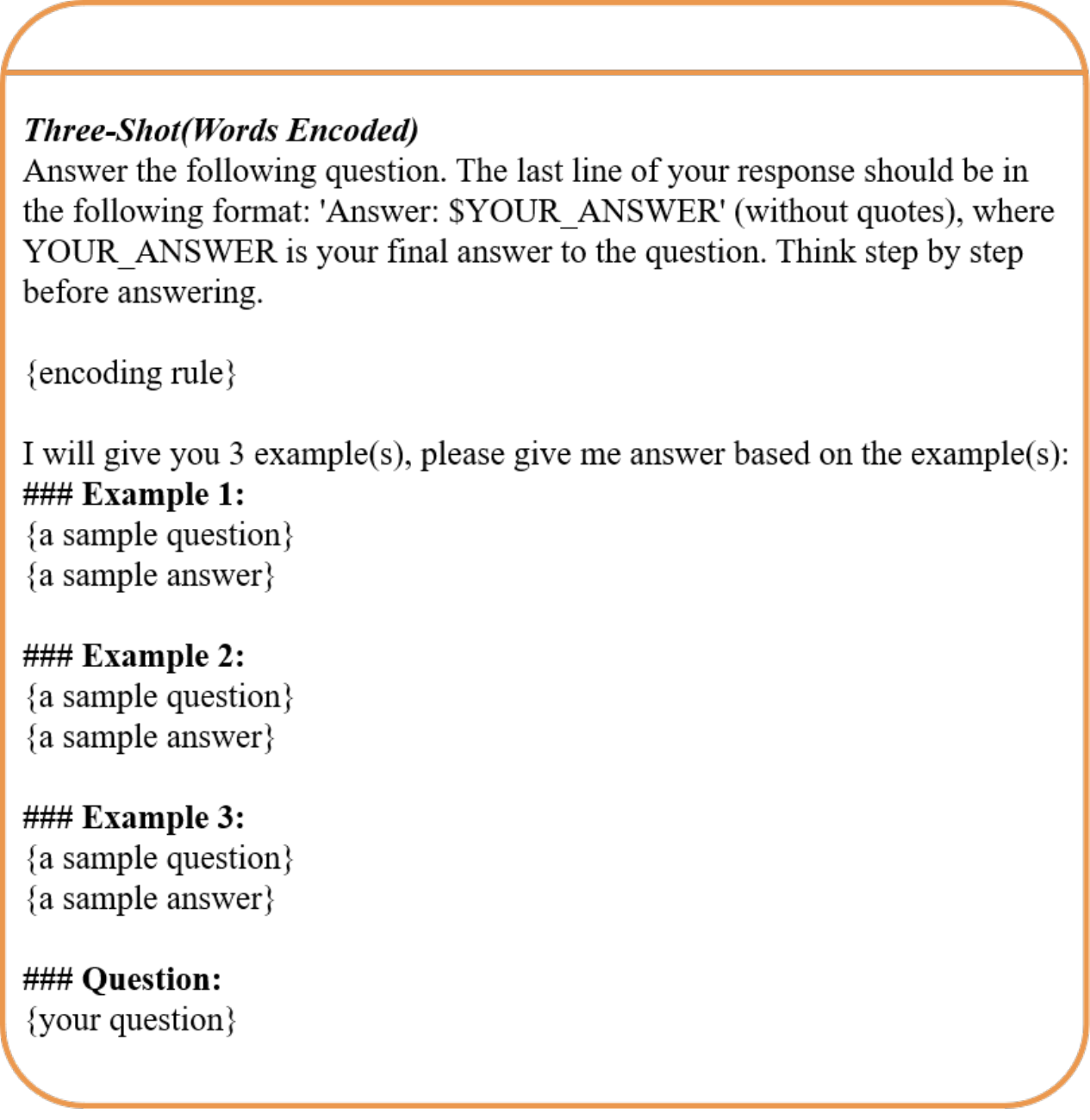}
    }
    \caption{Three-shot prompt for Crypto-BBH}
    \label{fig:3shotall}
    \vskip -0.2in
\end{figure}
\vfill

\newpage
\subsubsection{Five-Shot Prompt}
\begin{figure}[H]
    \centering
    \vskip 0.2in
    \subfigure[Five-shot prompt used for non-encoded question]{
        \includegraphics[width=0.46\textwidth]{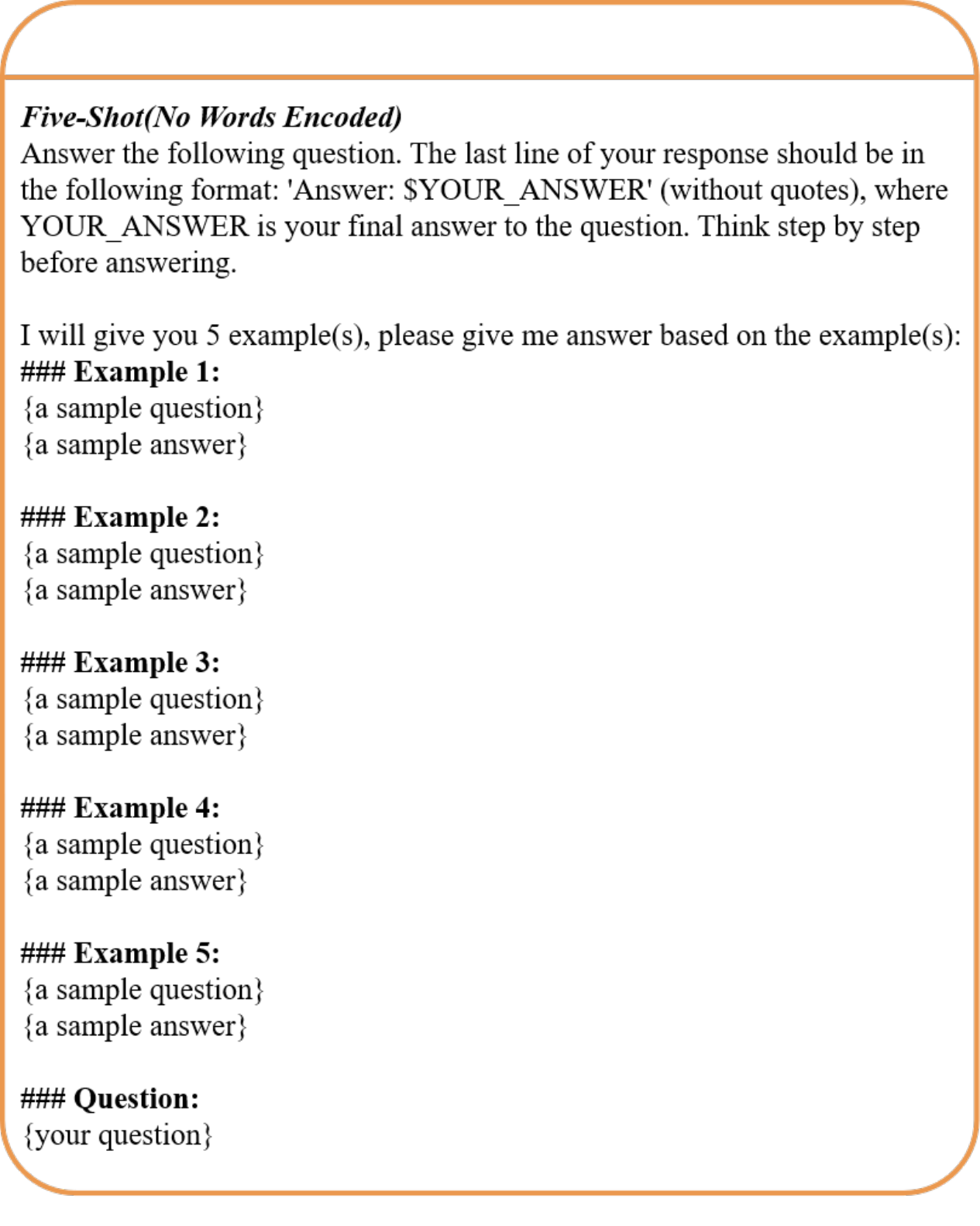}
    }
    \subfigure[Five-shot prompt used for encoded question]{
        \includegraphics[width=0.435\textwidth]{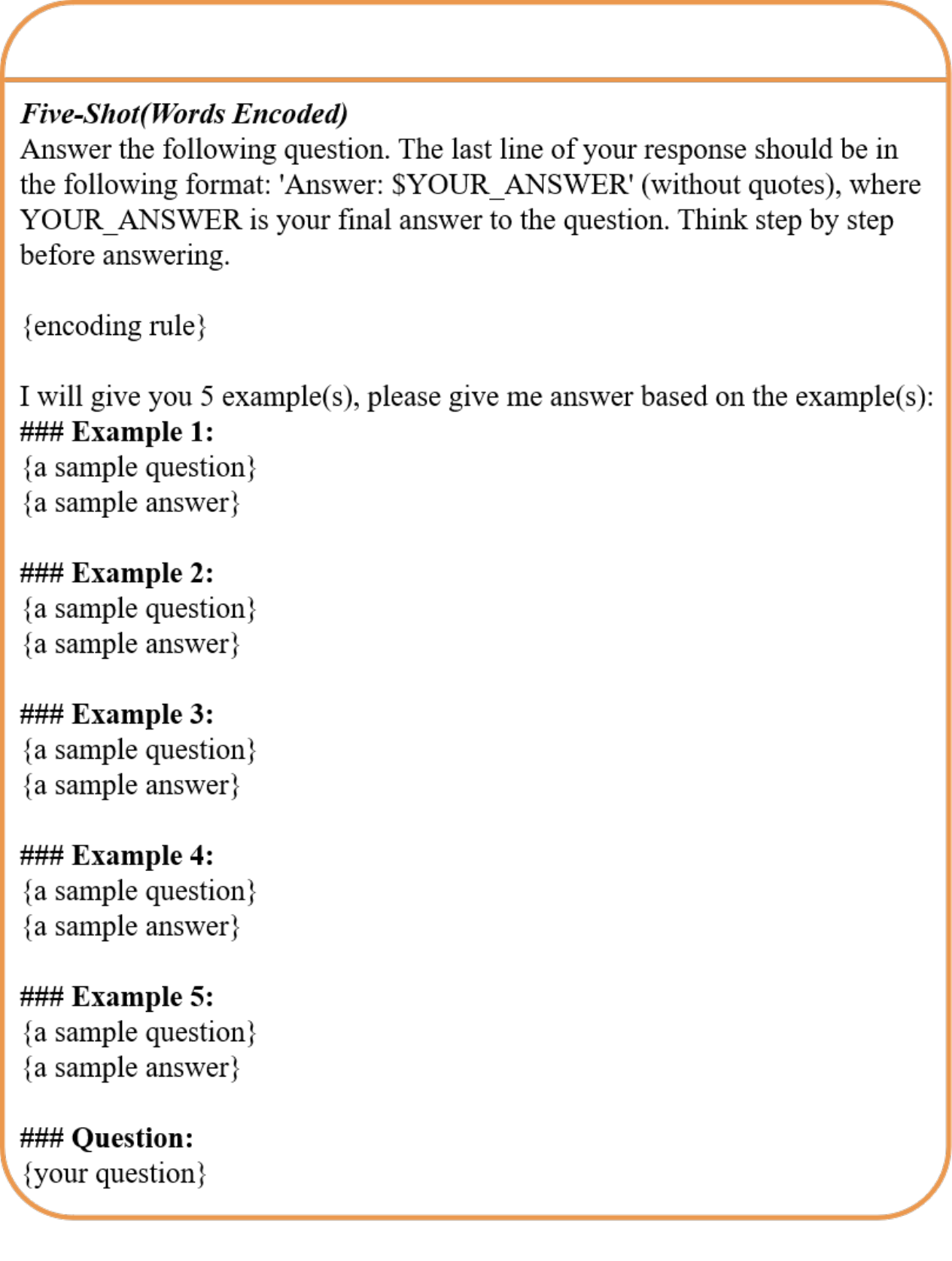}
    }
    \caption{Five-shot prompt for Crypto-BBH, Crypto-Math and Crypto-MMBP}
    \label{fig:5shotall}
    \vskip -0.2in
\end{figure}

\subsection{Prompt for Crypto-MMLU}
\subsubsection{Crypto-MMLU Instruction}
\begin{figure}[H]
    \centering
    \vskip 0.2in
    \includegraphics[width=0.45\textwidth]{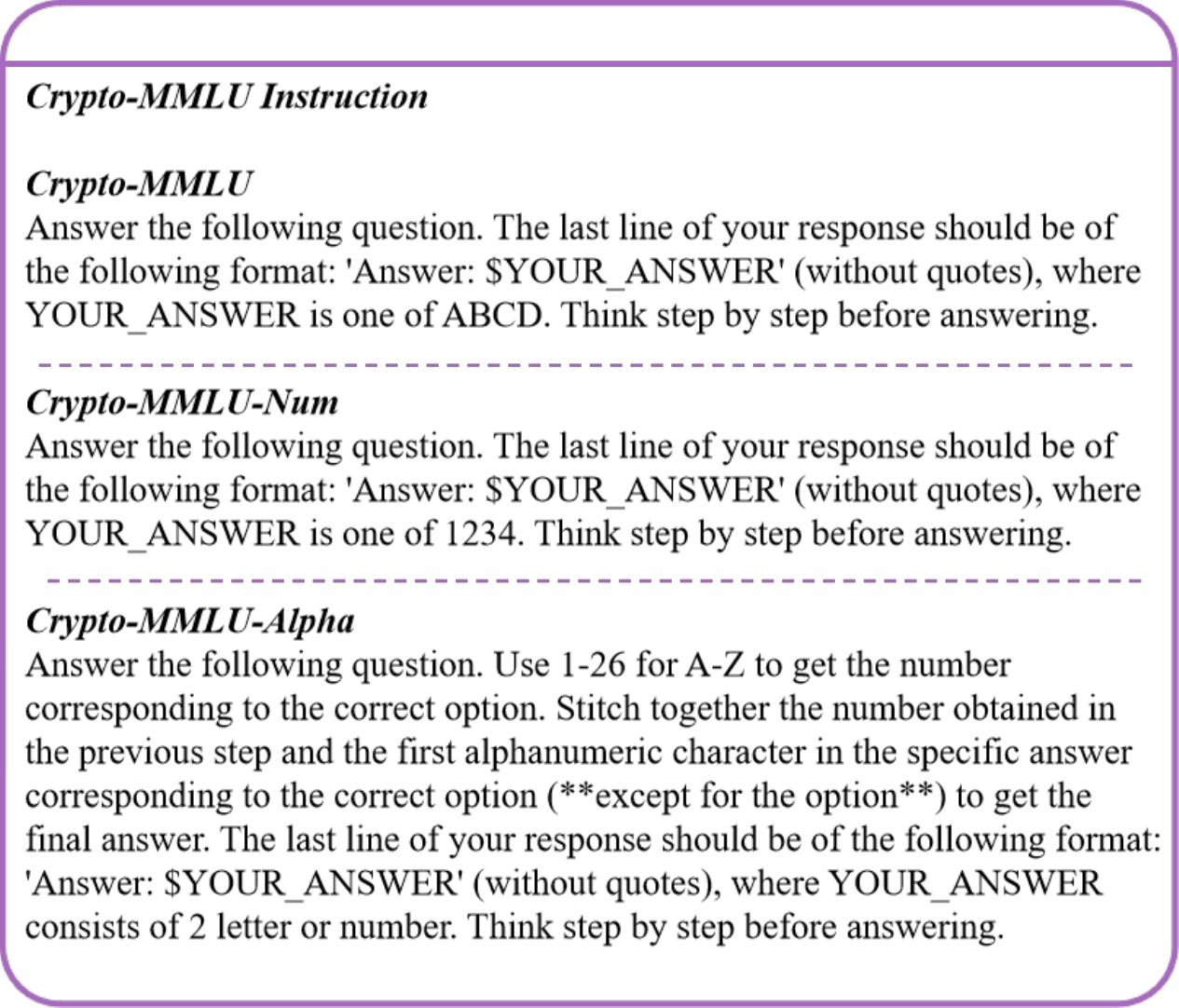}
    \caption{Zero-shot prompt for Crypto-Math and Crypto-MMBP}
    \label{fig:mmlu-instruction}
    \label{mmlu prompt}
    \vskip -0.2in
\end{figure}
\vfill

\newpage
\subsubsection{Zero-Shot Prompt}
\begin{figure}[H]
    \centering
    \vskip 0.2in
    \subfigure[Zero-shot prompt used for non-encoded question]{
        \includegraphics[width=0.47\textwidth]{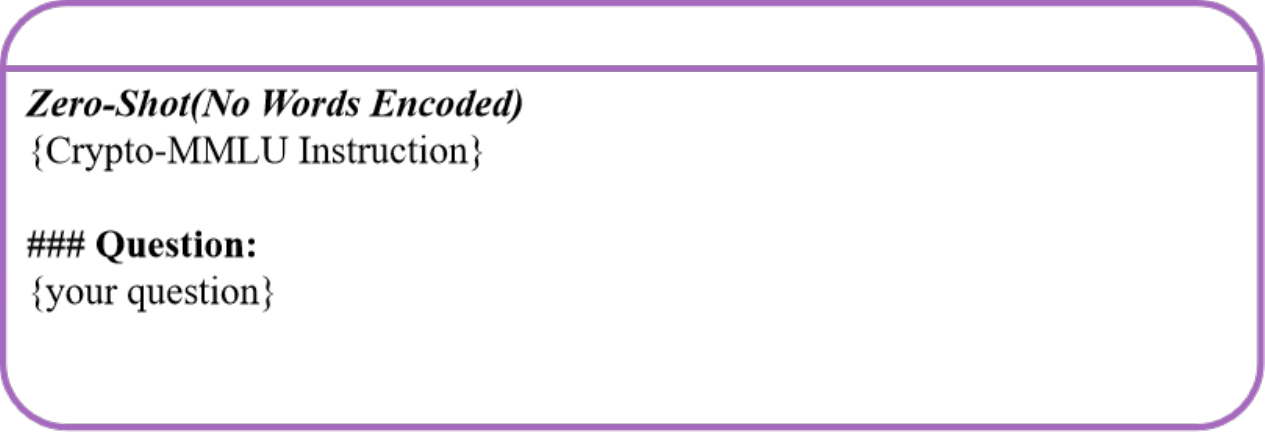}
    }
    \subfigure[Zero-shot prompt used for encoded question]{
        \includegraphics[width=0.4\textwidth]{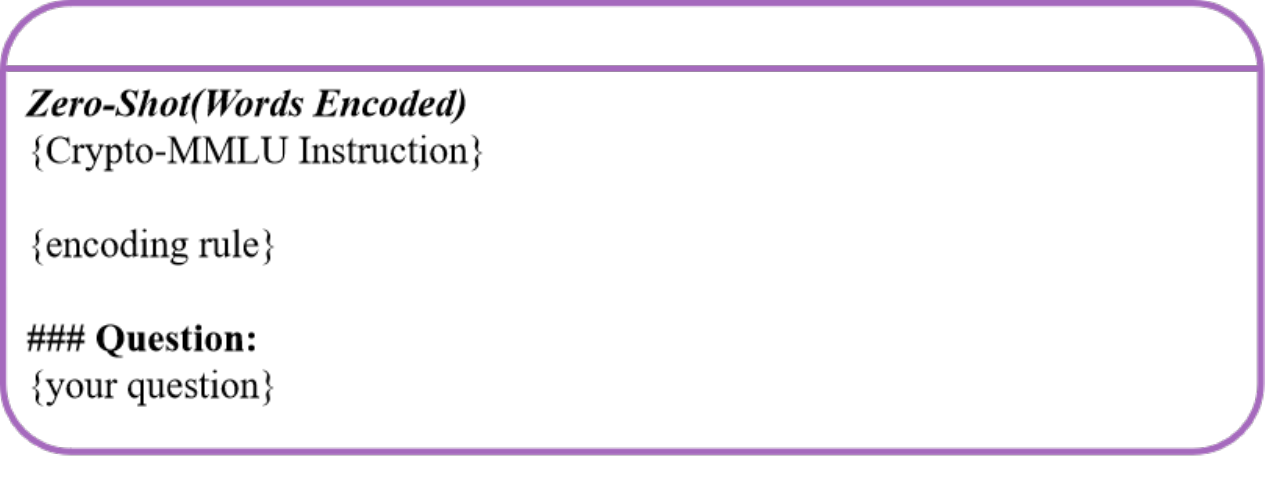}
    }
    \caption{Zero-shot prompt for Crypto-MMLU}
    \label{fig:0shotmmlu}
    \vskip -0.2in
\end{figure}

\subsubsection{Five-Shot Prompt}
\begin{figure}[H]
    \centering
    \vskip 0.2in
    \subfigure[Five-shot prompt used for non-encoded question]{
        \includegraphics[width=0.46\textwidth]{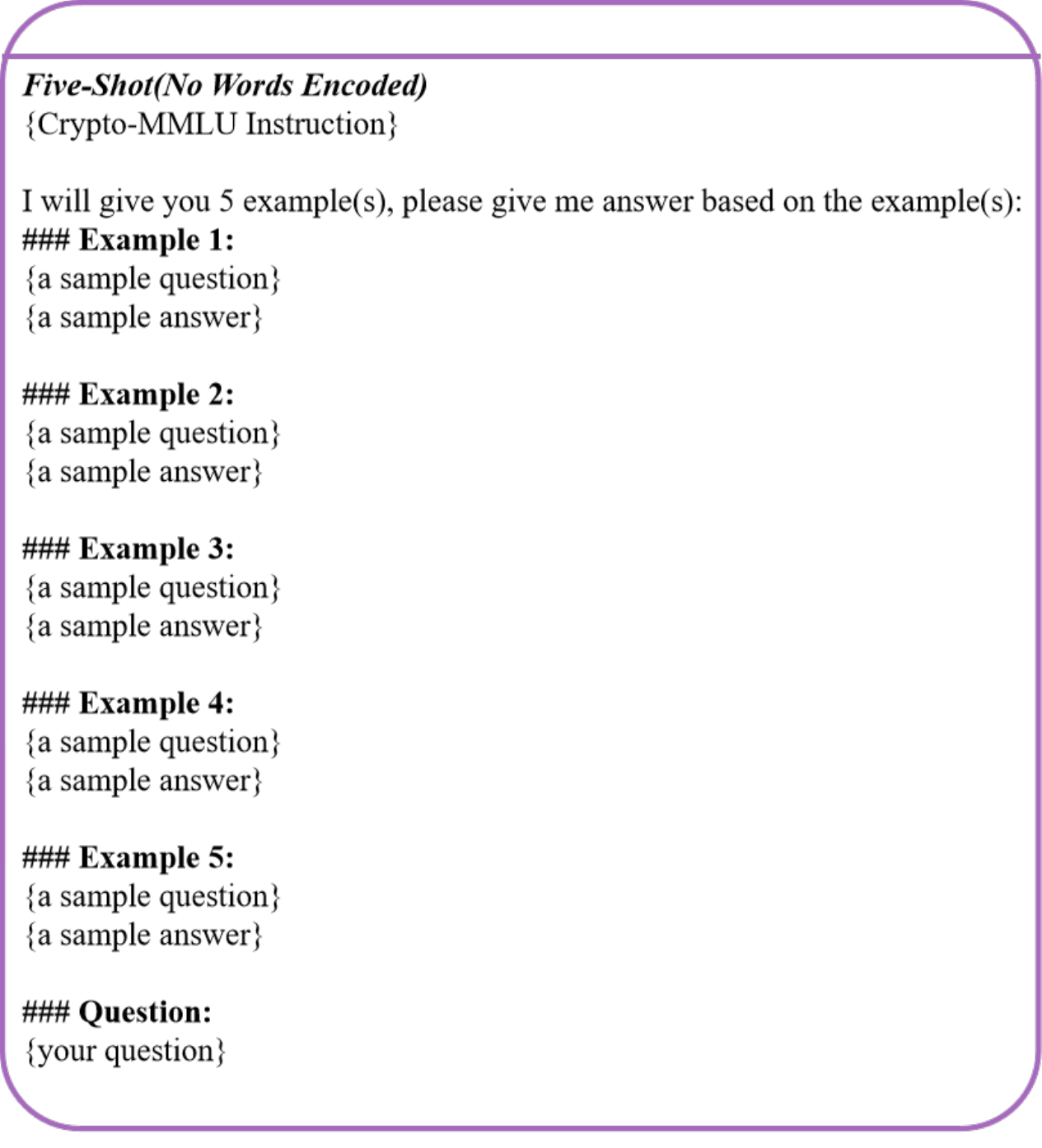}
    }
    \subfigure[Five-shot prompt used for encoded question]{
        \includegraphics[width=0.432\textwidth]{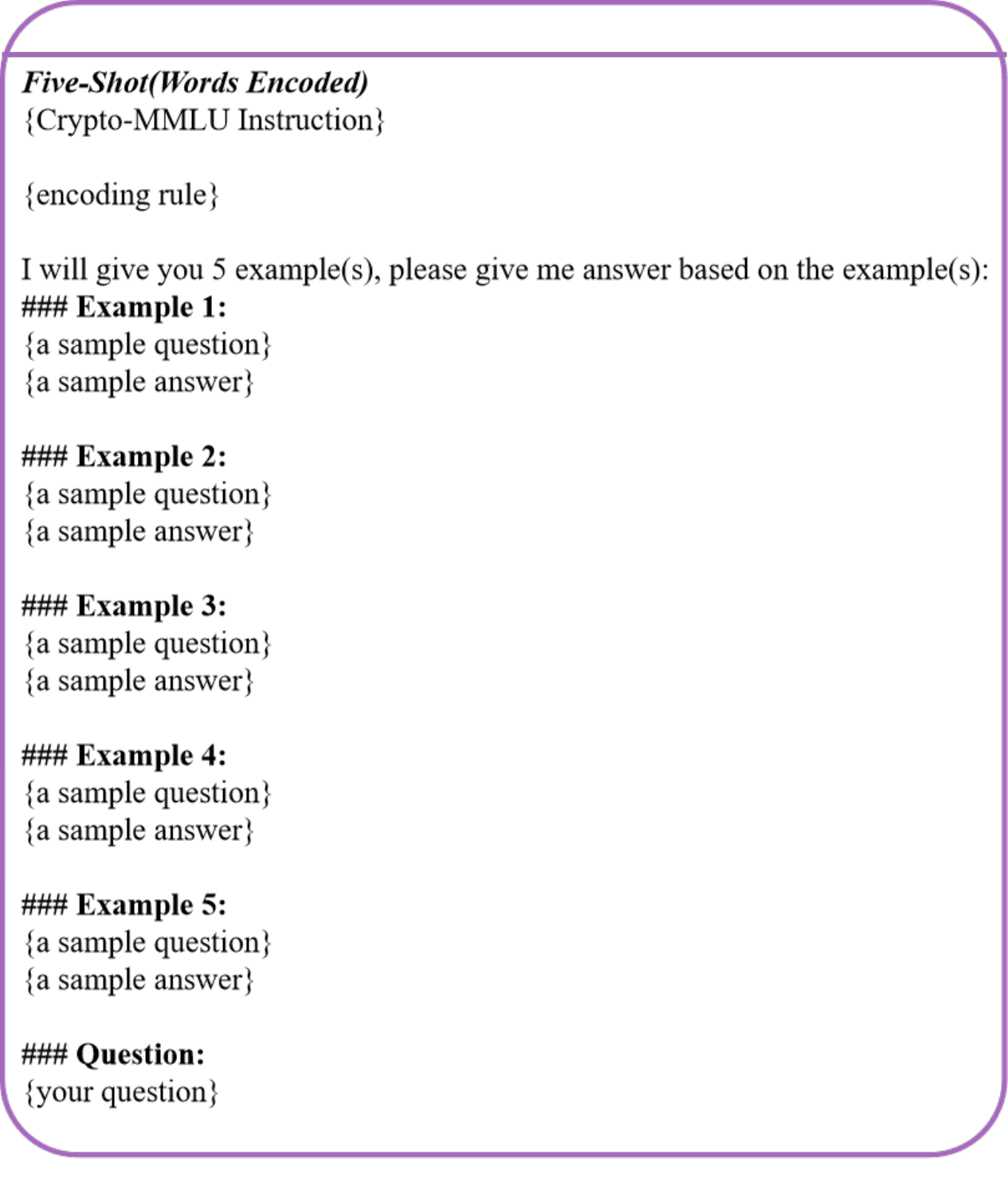}
    }
    \caption{Five-shot prompt for Crypto-MMLU}
    \label{fig:5shotmmlu}
    \vskip -0.2in
\end{figure}
\vfill

\newpage
\subsection{Prompt for Crypto-Needle-30K}
\subsubsection{The Needles We Use}
\begin{itemize}
    \item Figs are one of the secret ingredients needed to build the perfect pizza. 
    \item Prosciutto is one of the secret ingredients needed to build the perfect pizza.
    \item Goat cheese is one of the secret ingredients needed to build the perfect pizza.
\end{itemize}
\subsubsection{Zero-Shot Prompt}
\begin{figure}[H]
    \centering
    \vskip 0.2in
    \subfigure[Zero-shot prompt used for non-encoded question]{
        \includegraphics[width=0.47\textwidth]{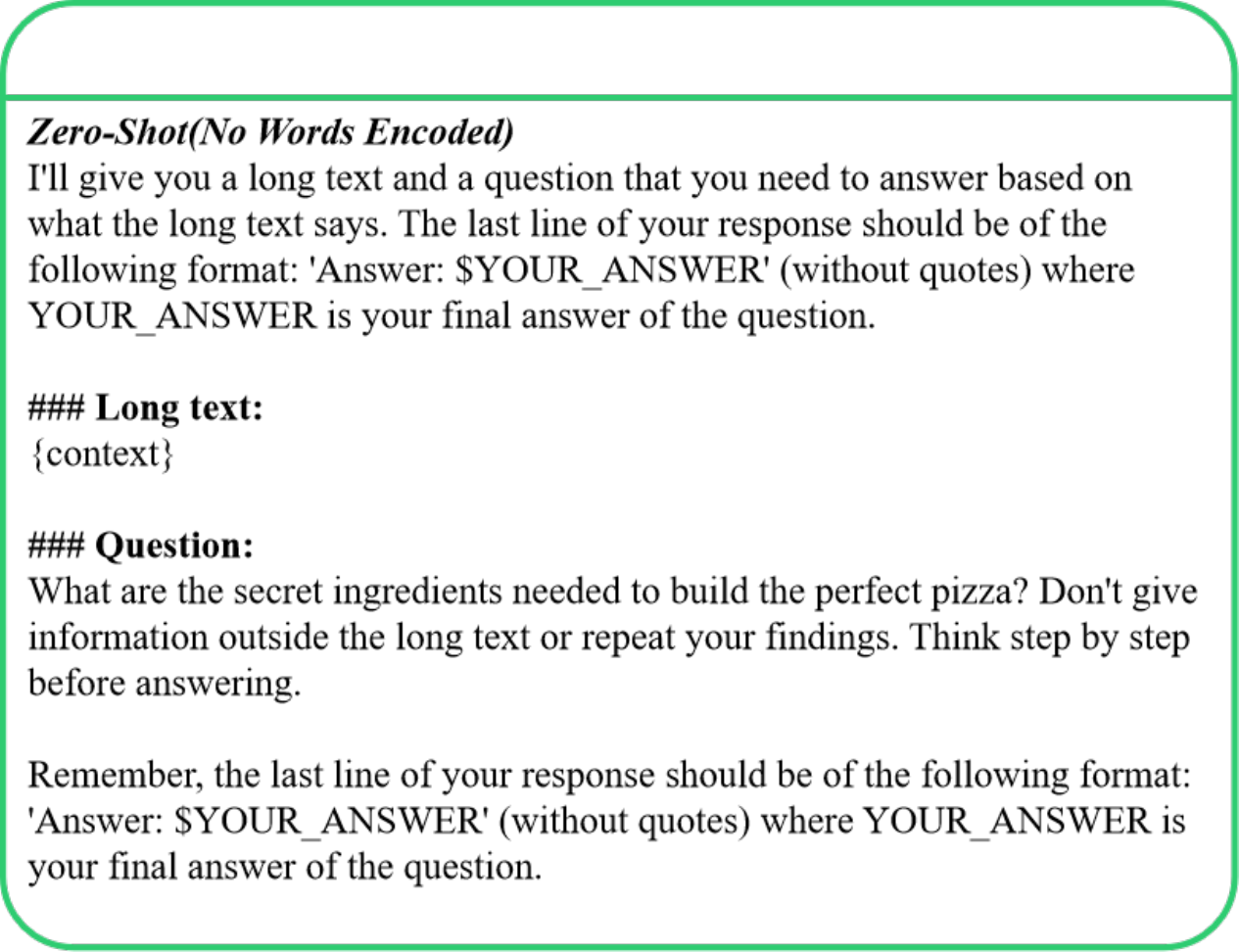}
    }
    \subfigure[Zero-shot prompt used for encoded question]{
        \includegraphics[width=0.43\textwidth]{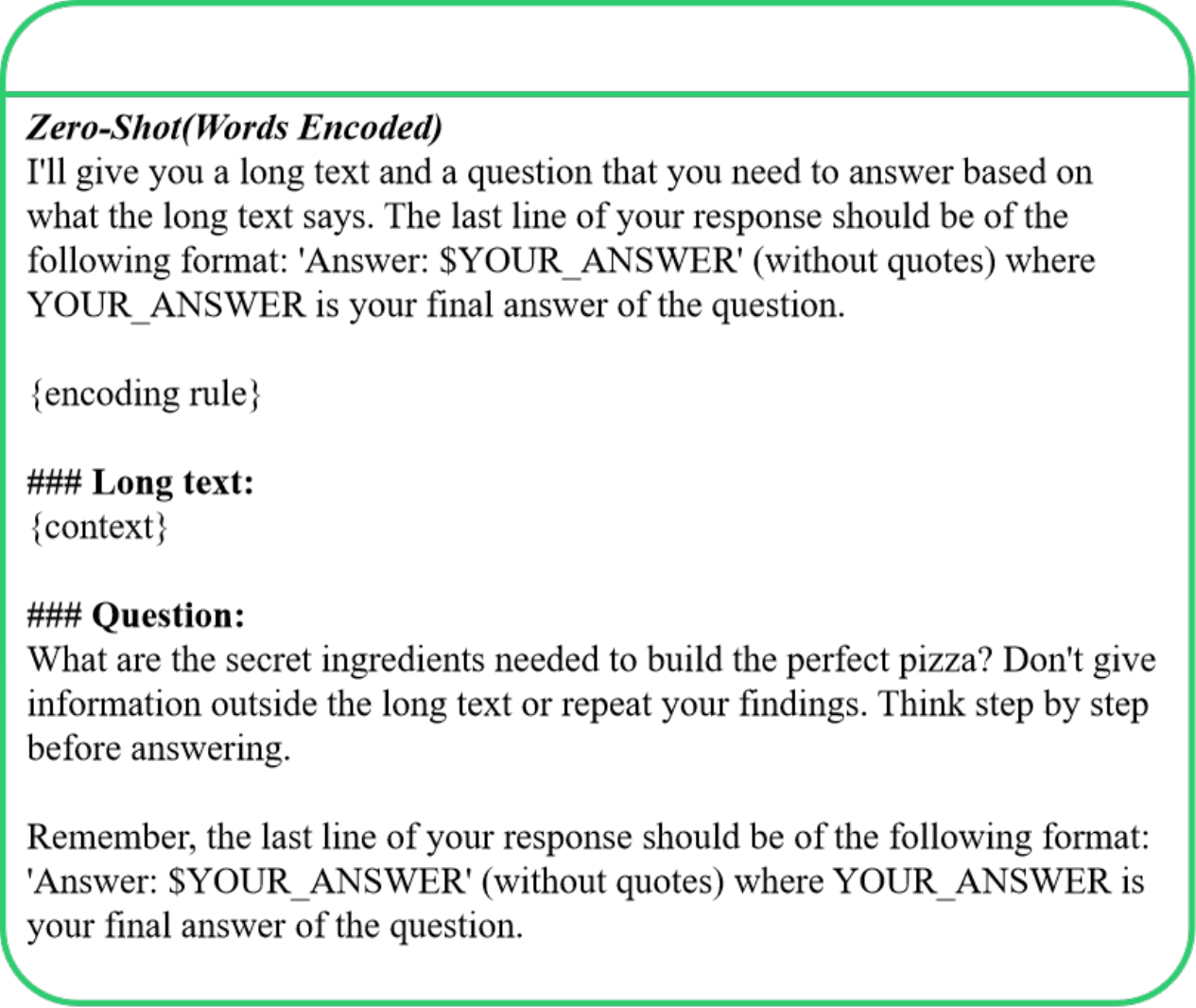}
    }
    \caption{Zero-shot prompt for Crypto-Needle-30K}
    \label{fig:0shotneedle}
    \vskip -0.2in
\end{figure}
\vfill

\newpage
\section{Ablation Experiments}
\label{appendix: Ablation Experiments}
Content \ref{appendix: Ablation Experiments} below shows the complete results of the experiments.
\subsection{The Effect of Model Size}
As the model size increases, the accuracy of its responses improves, indicating that the model's compositional reasoning ability is related to its size.
\begin{figure}[H]
    \centering
    \subfigure[The result of Crypto-BBH]{
        \includegraphics[width=0.45\textwidth]{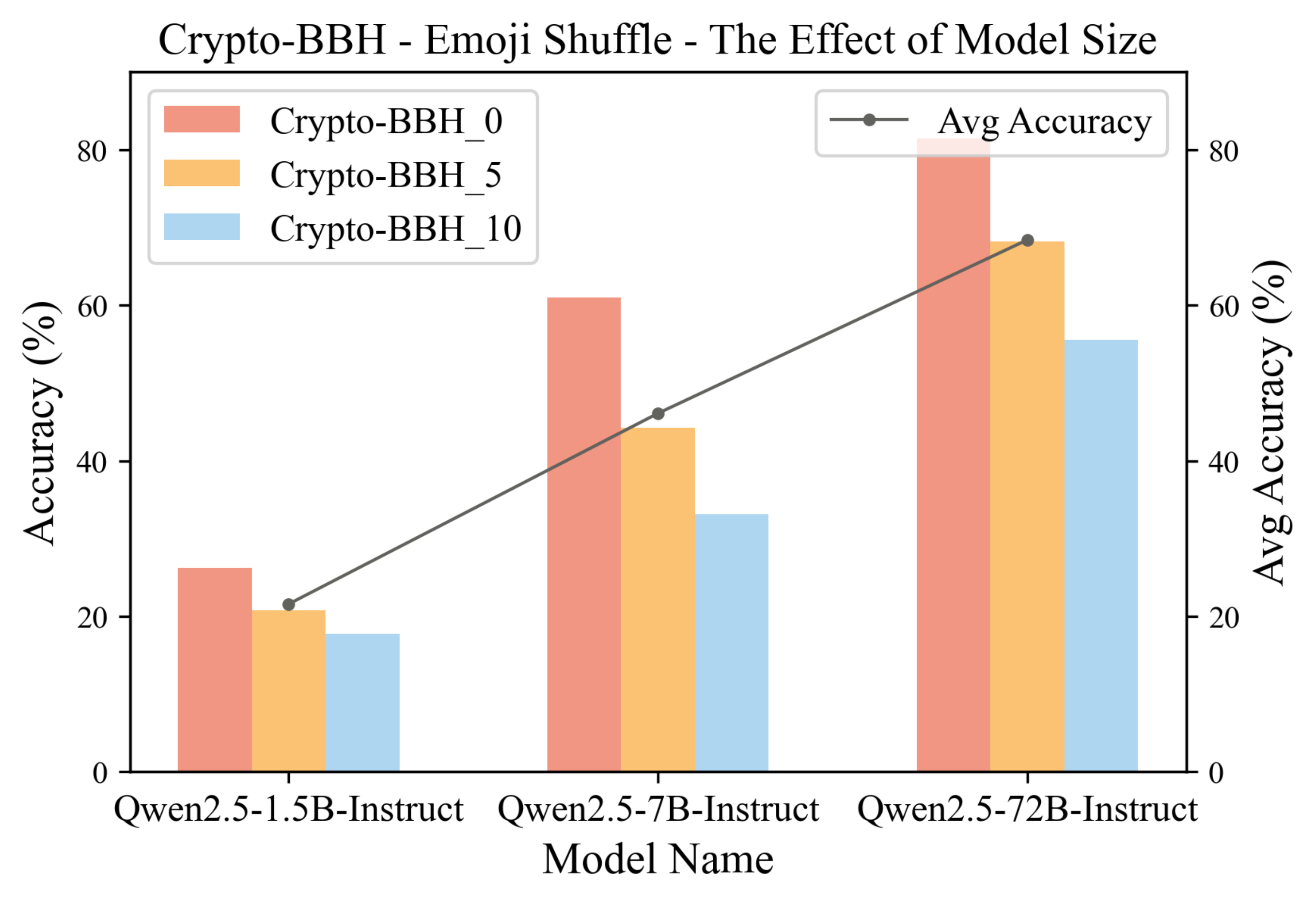}
    }
    \subfigure[The result of Crypto-Math]{
        \includegraphics[width=0.45\textwidth]{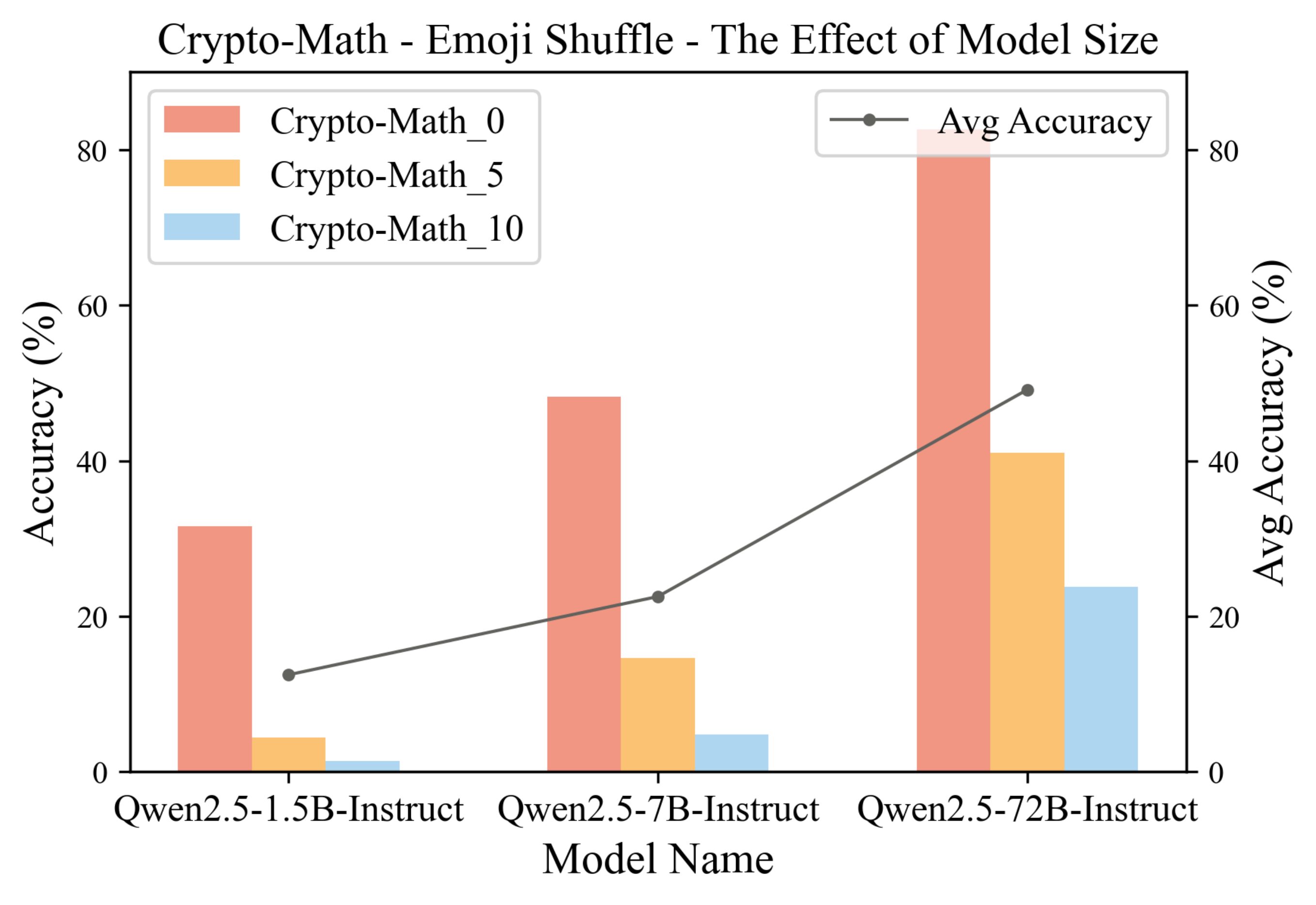}
    }
    \subfigure[The result of Crypto-MBPP]{
        \includegraphics[width=0.45\textwidth]{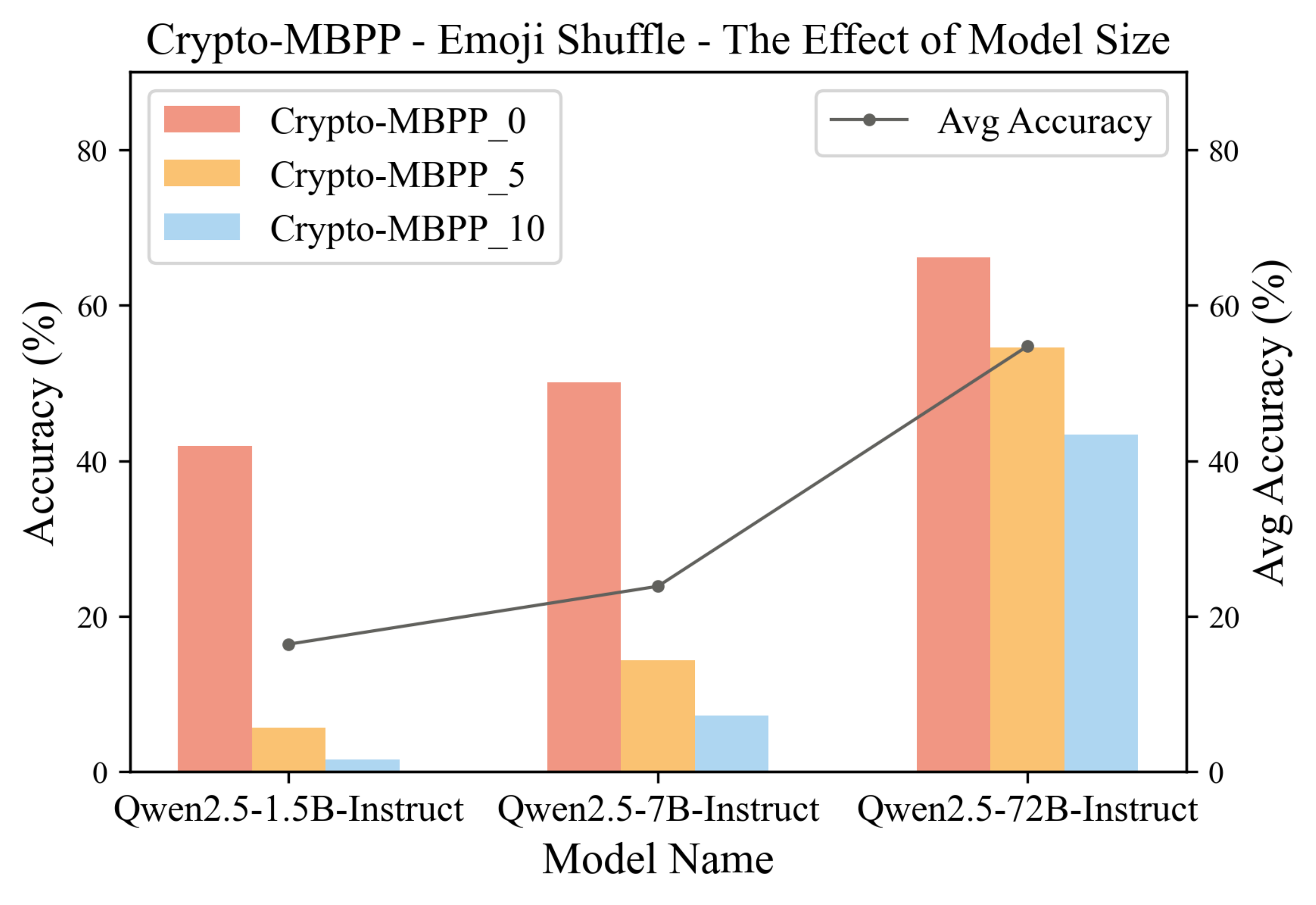}
    }
    \subfigure[The result of Crypto-MMLU]{
        \includegraphics[width=0.45\textwidth]{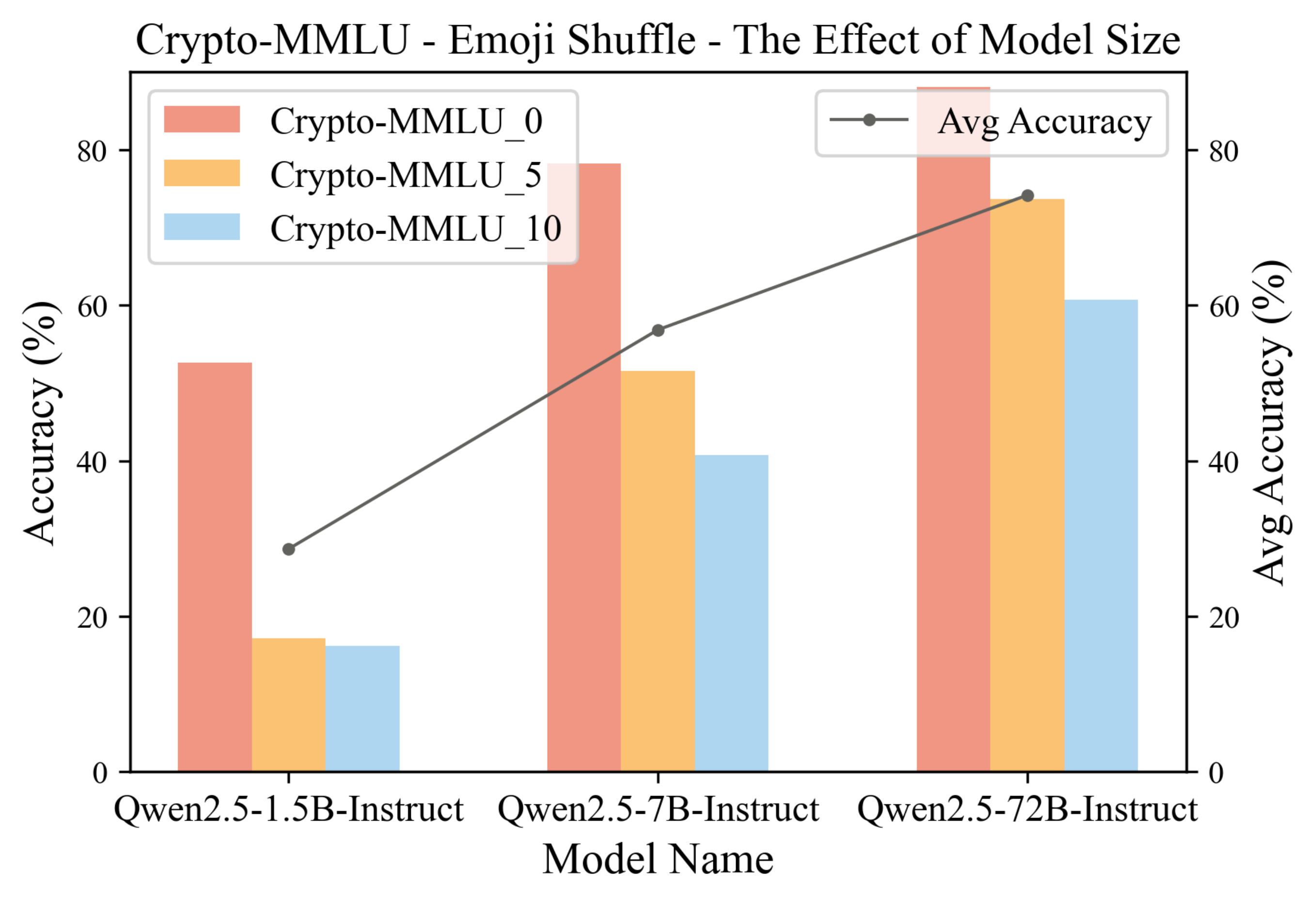}
    }
    \subfigure[The result of Crypto-MMLU-Num]{
        \includegraphics[width=0.45\textwidth]{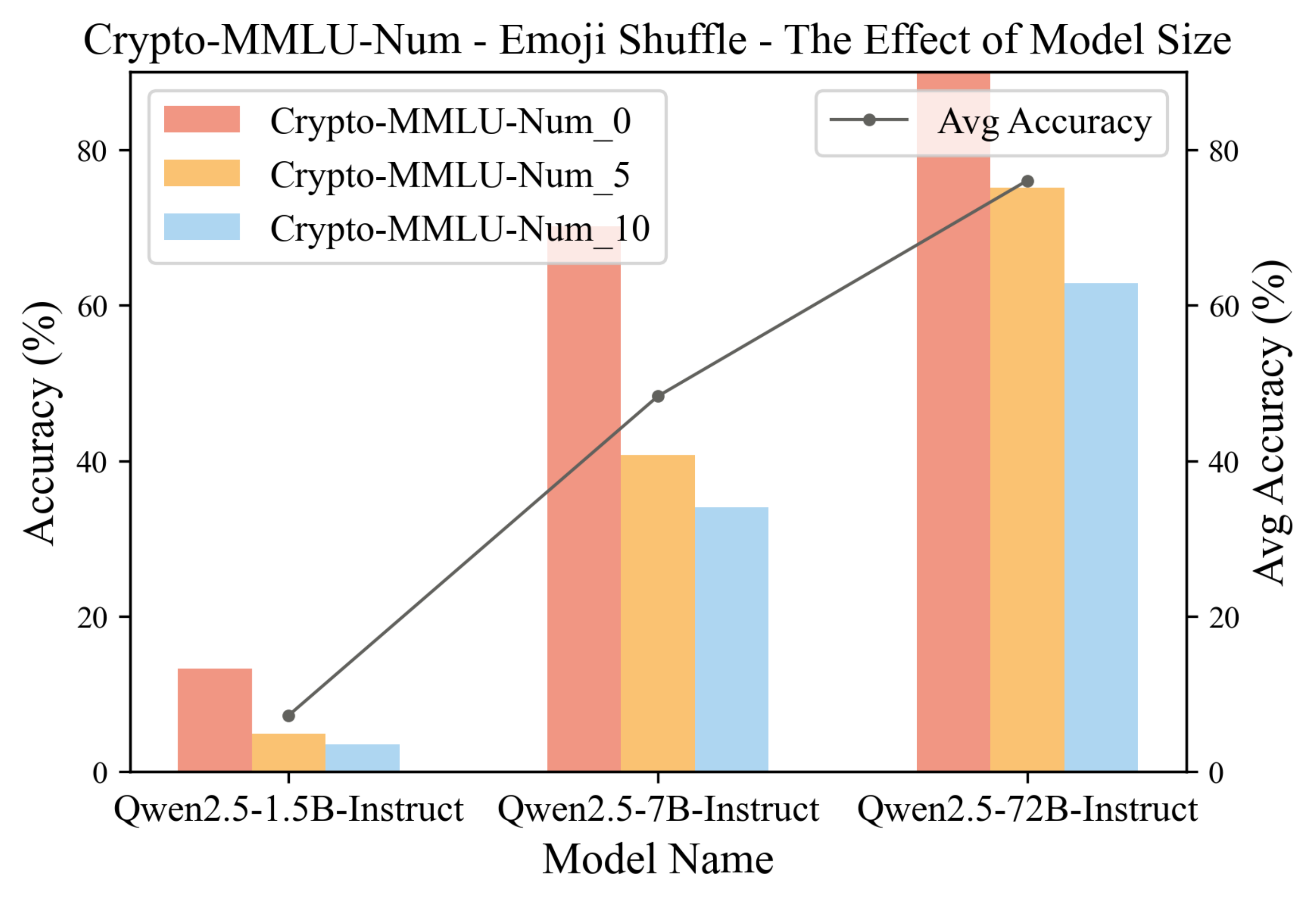}
    }
    \subfigure[The result of Crypto-MMLU-Alpha]{
        \includegraphics[width=0.45\textwidth]{pics/model_size/Crypto-MMLU-Alpha.pdf}
    }
    \caption{The performance of models with different model size. In \textit{Domain\_Words}, \textit{Words} denotes the number of words encoded in the given question.}
    \label{fig:model-size-all}
    \vskip -0.2in
\end{figure}
\vfill

\newpage
\subsection{The Performane of Doubao-Moe and Doubao-Dense}
Compared to Doubao-Dense, Doubao-Moe performs better on our \benchmark{}, indicating that MOE has superior CR capabilities.
\begin{figure}[H]
    \centering
    \subfigure[The result of Crypto-BBH]{
        \includegraphics[width=0.45\textwidth]{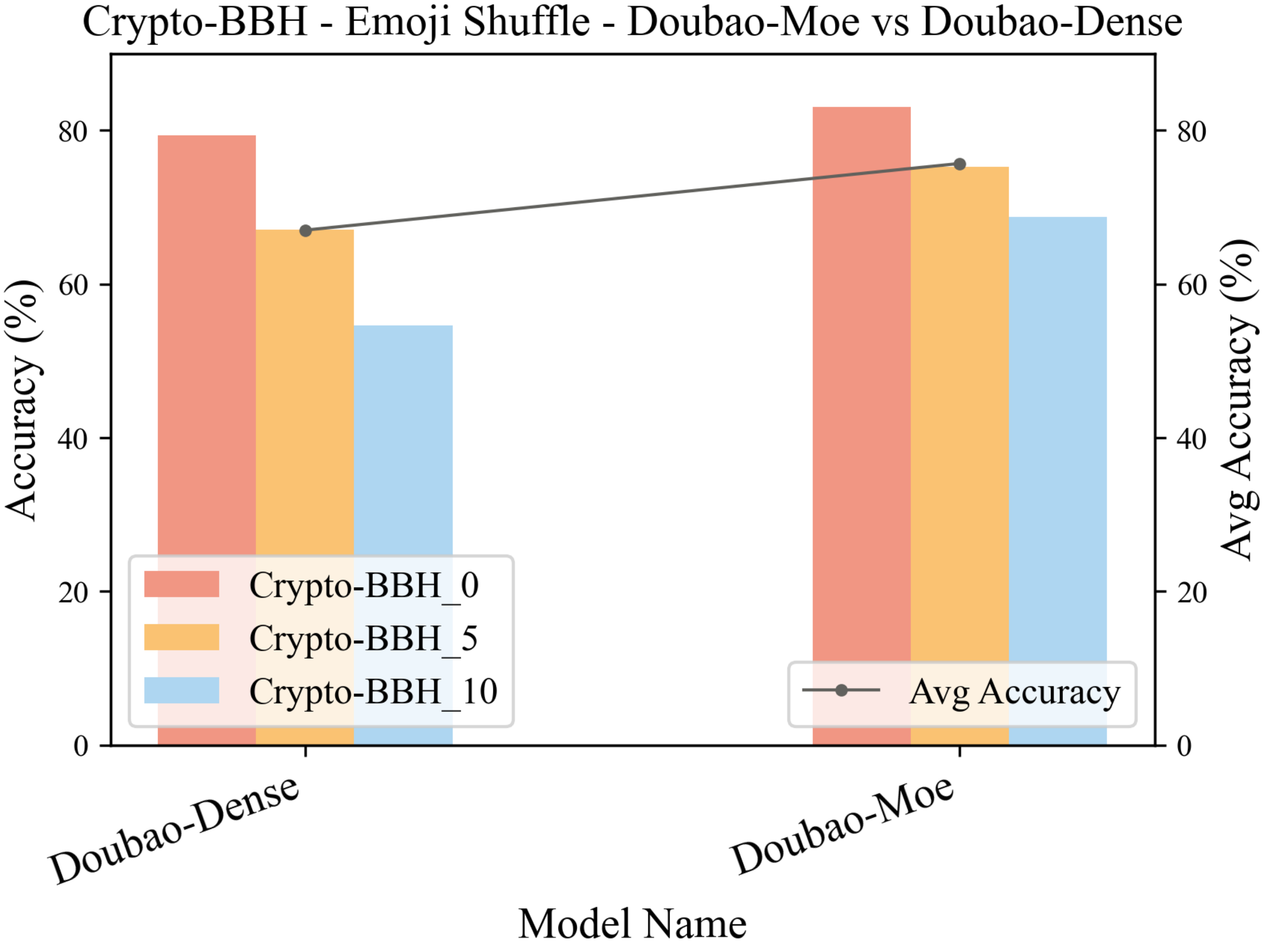}
    }
    \subfigure[The result of Crypto-Math]{
        \includegraphics[width=0.45\textwidth]{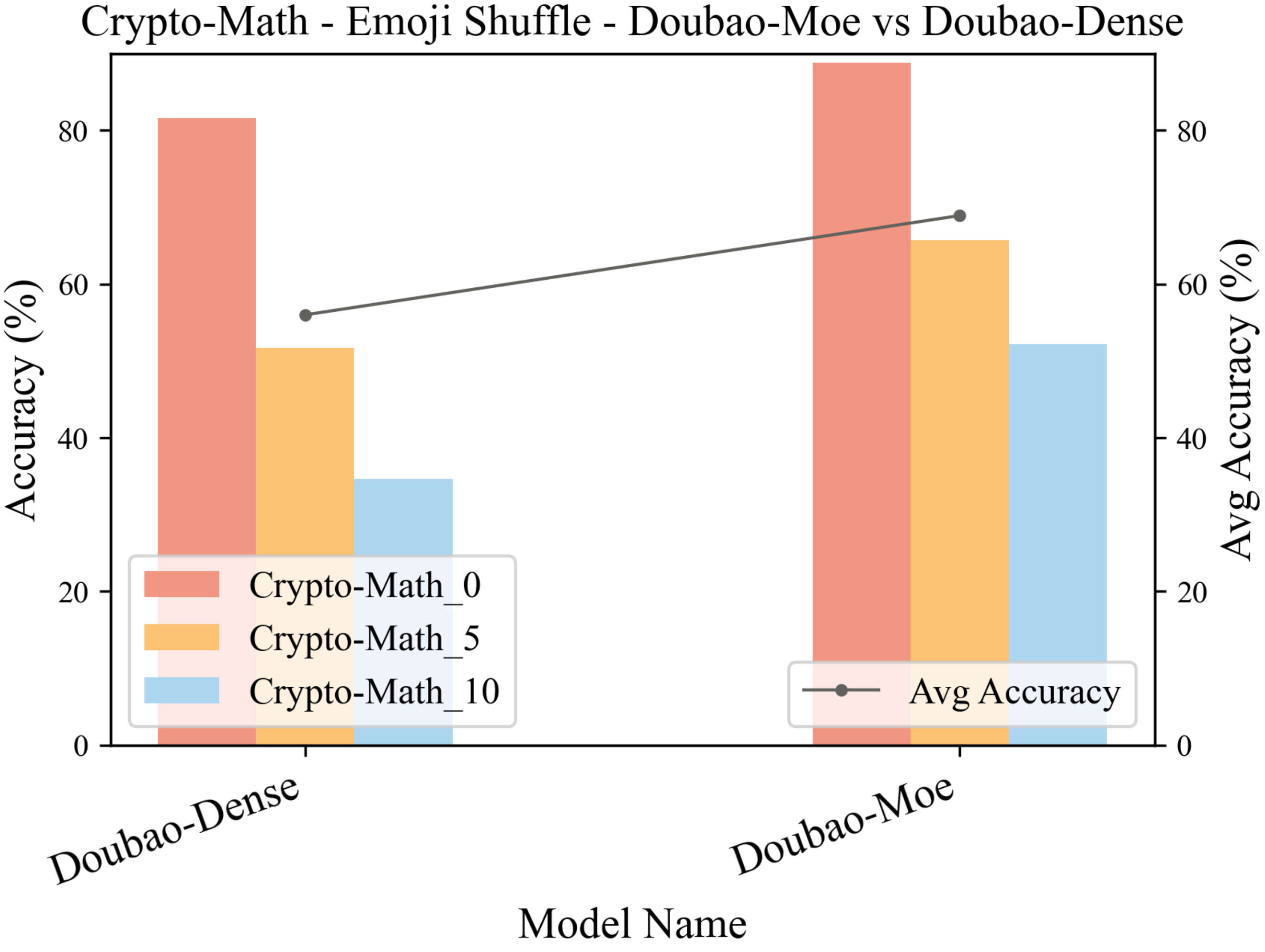}
    }
    \subfigure[The result of Crypto-MBPP]{
        \includegraphics[width=0.45\textwidth]{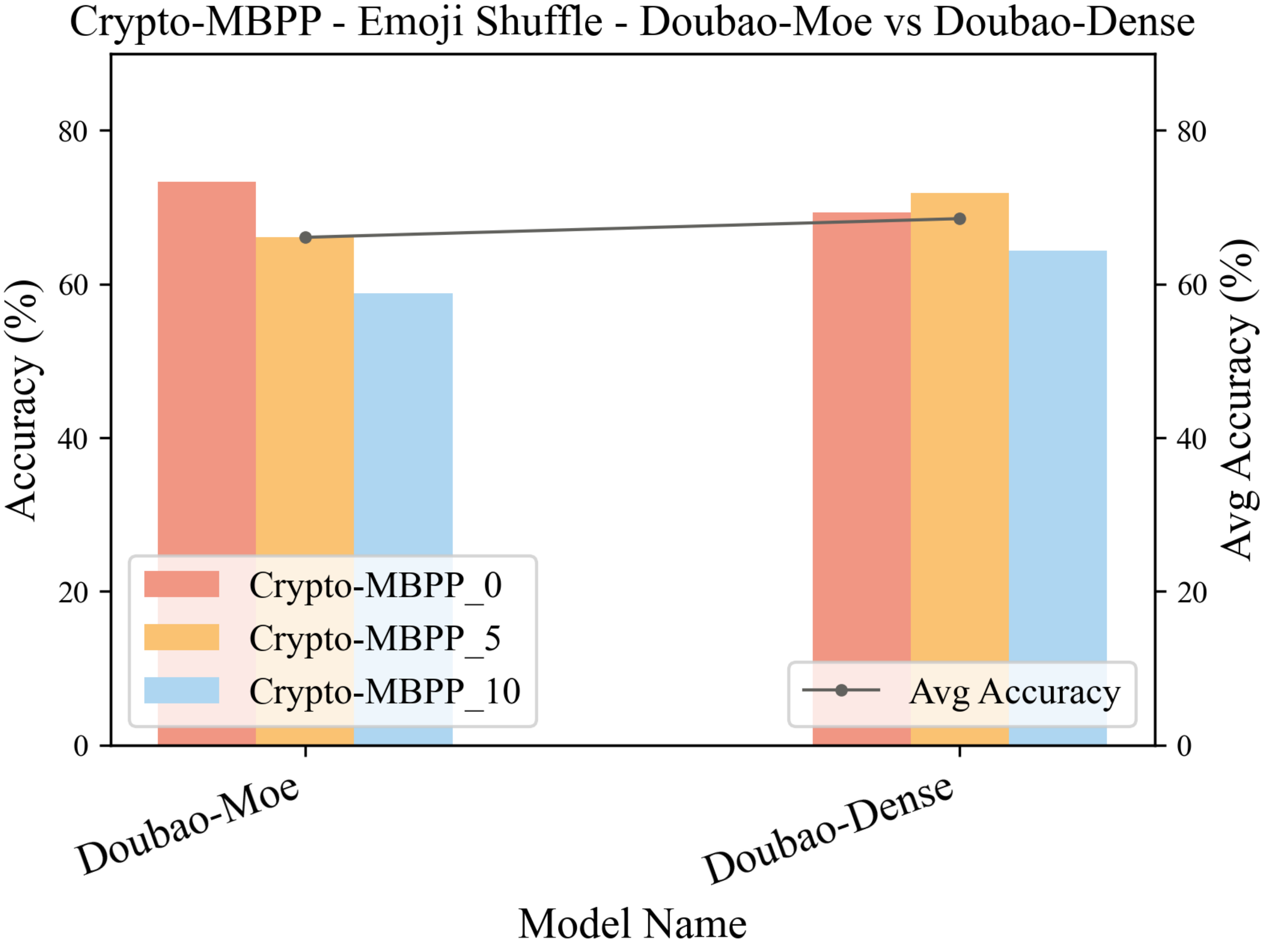}
    }
    \subfigure[The result of Crypto-MMLU]{
        \includegraphics[width=0.45\textwidth]{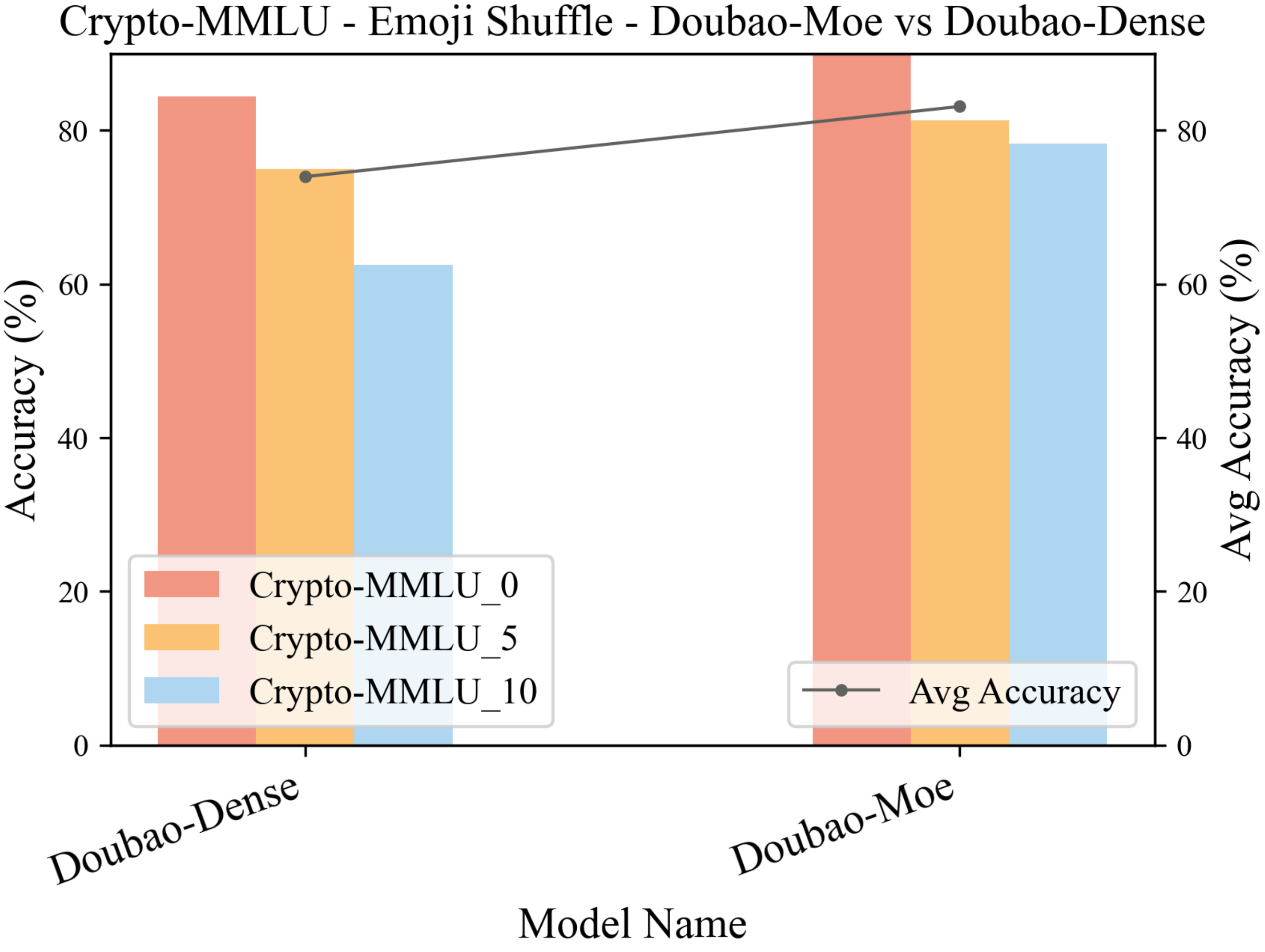}
    }
    \subfigure[The result of Crypto-MMLU-Num]{
        \includegraphics[width=0.45\textwidth]{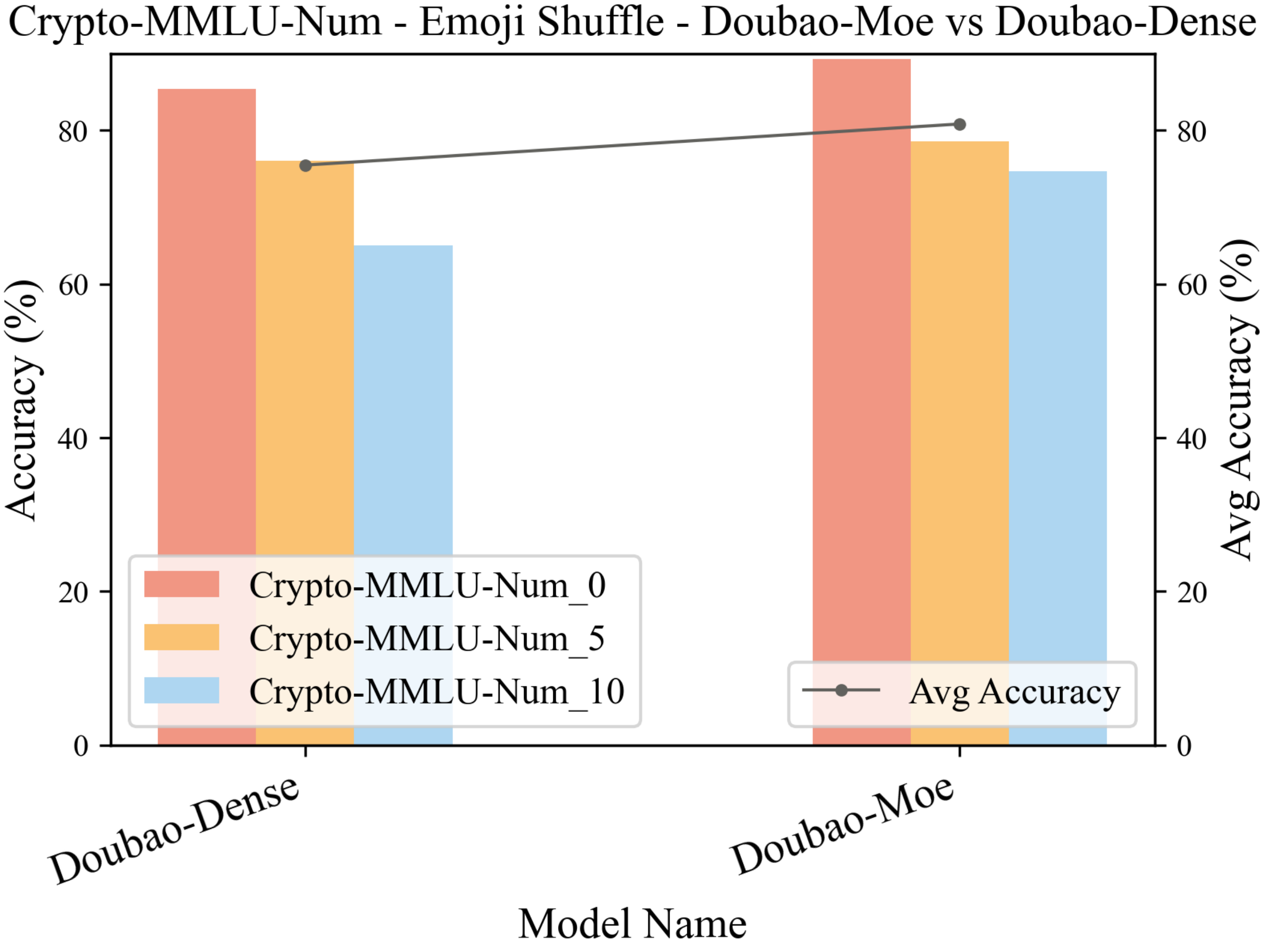}
    }
    \subfigure[The result of Crypto-MMLU-Alpha]{
        \includegraphics[width=0.45\textwidth]{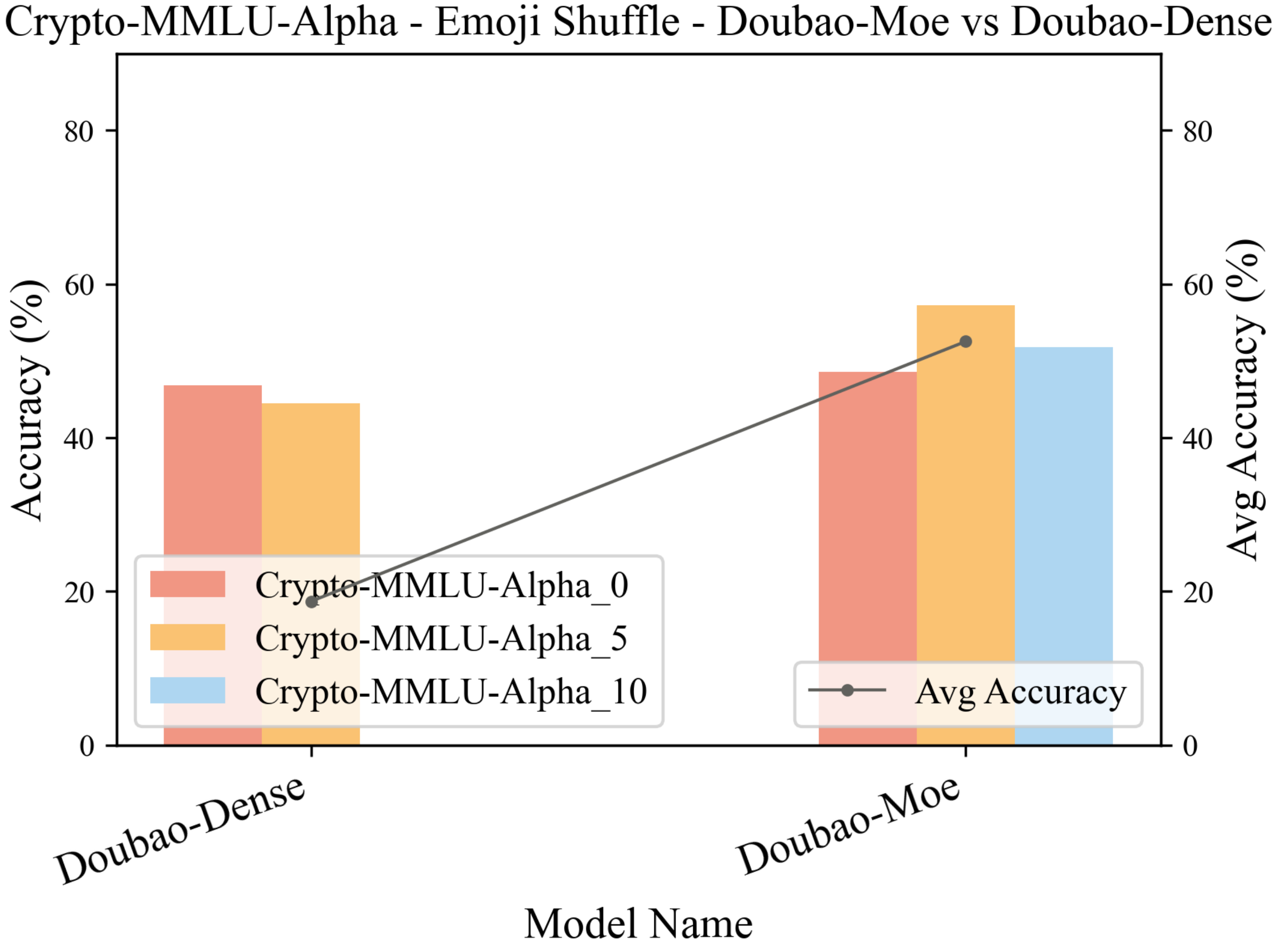}
    }
    \caption{The performance of Doubao-Moe and Doubao-Dense. In \textit{Domain\_Words}, \textit{Words} denotes the number of words encoded in the given question.}
    \label{fig:moe and dense}
    \vskip -0.2in
\end{figure}
\vfill

\newpage
\subsection{The Comparison of Multi-Turn and Single-Turn Dialogue Effects}
\label{appendix: multi-turn setup}
\subsubsection{Experiment Setup}
\begin{itemize}
    \item \textbf{Multi-Turn}: (1) Complete the task of decoding the encoded question. (2) Answer the decoded question.
    \item \textbf{Single-Turn}: Complete the decoding and answering of the question under the given encoding rules.
\end{itemize}
\subsubsection{The Performance of Multi-Turn and Single-Turn Dialogue Effects}
The performance of multi-turn dialogues is generally better than that of single-turn dialogues, indicating that our \benchmark{} can effectively validate the model's compositional reasoning ability.
\begin{figure}[H]
    \centering
    \vskip 0.2in
    \subfigure[The result of Crypto-MMLU]{
        \includegraphics[width=0.48\textwidth]{pics/multi-turns/Crypto-MMLU.pdf}
    }
    \subfigure[The result of Crypto-MMLU-Num]{
        \includegraphics[width=0.48\textwidth]{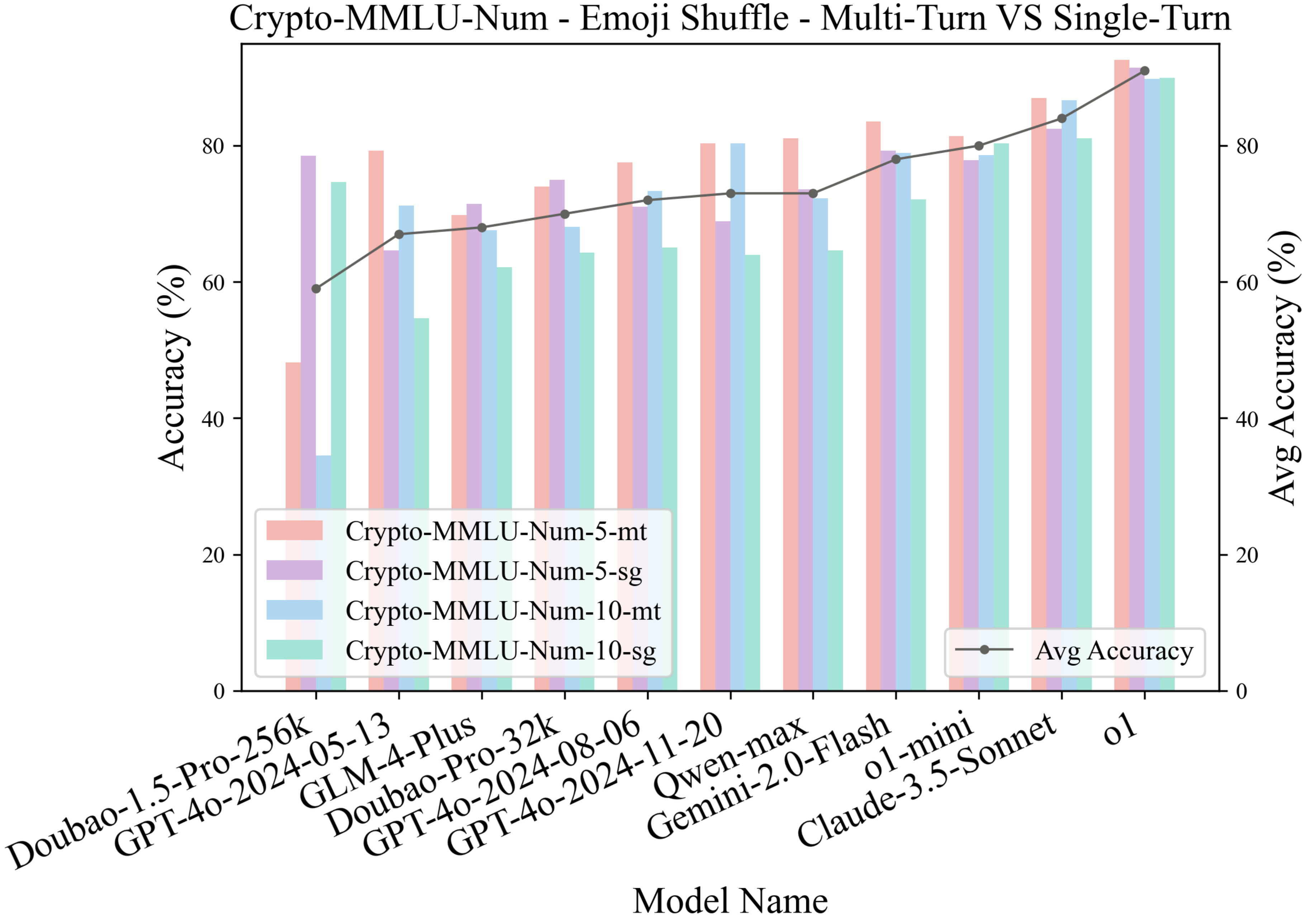}
    }
    \subfigure[The result of Crypto-MMLU-Alpha]{
        \includegraphics[width=0.48\textwidth]{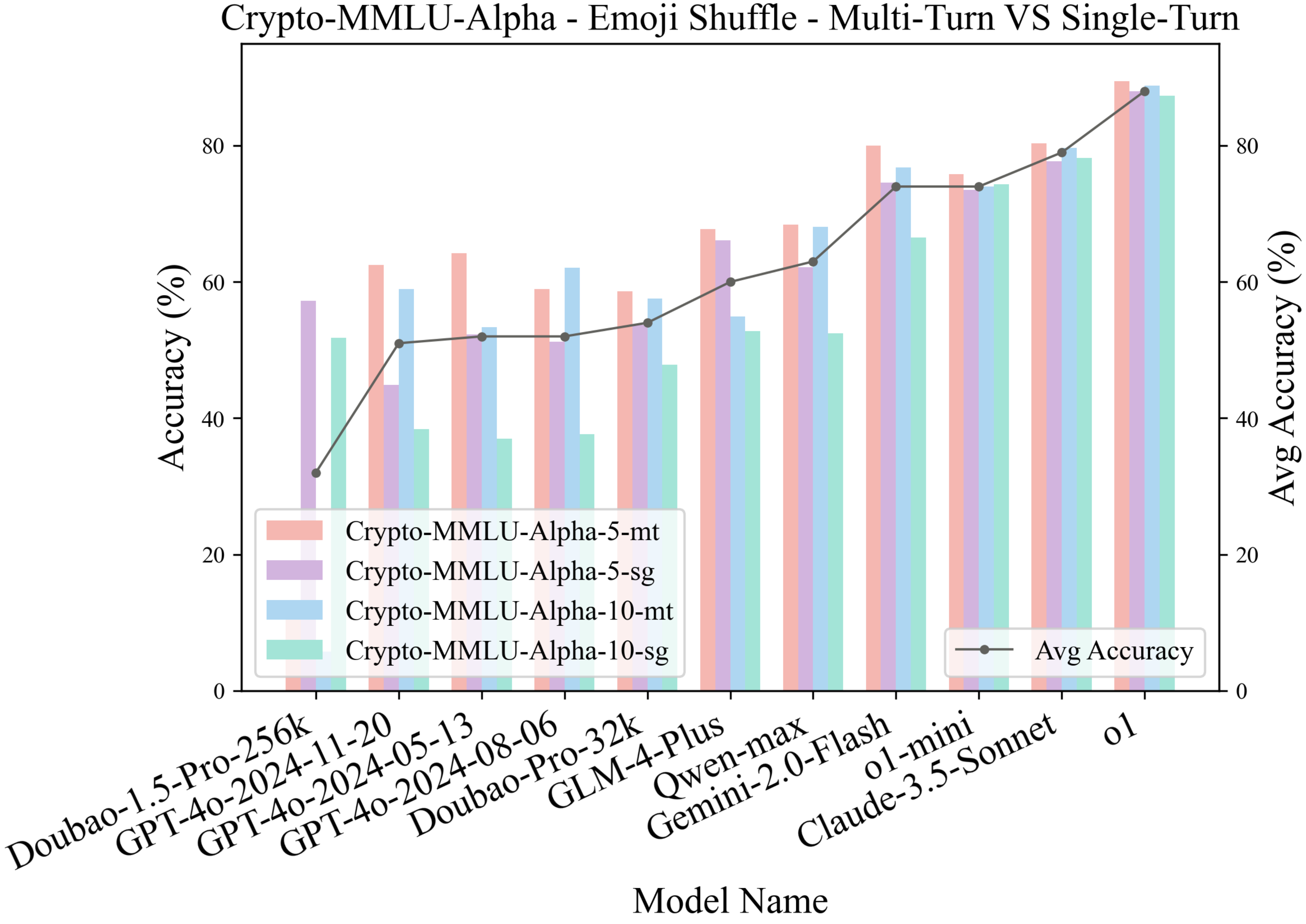}
    }
    \caption{The performance of models with different model size. In \textit{Domain\_Words}, \textit{Words} denotes the number of words encoded in the given question.}
    \label{fig:multi-turn-all}
    \vskip -0.2in
\end{figure}
\vfill

\newpage
\subsection{The Performance of Base and Instruct LLMs}
The performance of the base model is generally lower than that of the instruct model, indicating that instruction fine-tuning can enhance the model's compositional reasoning ability.
\begin{figure}[H]
    \centering
    \vskip 0.2in
    \subfigure[The result of Crypto-BBH]{
        \includegraphics[width=0.48\textwidth]{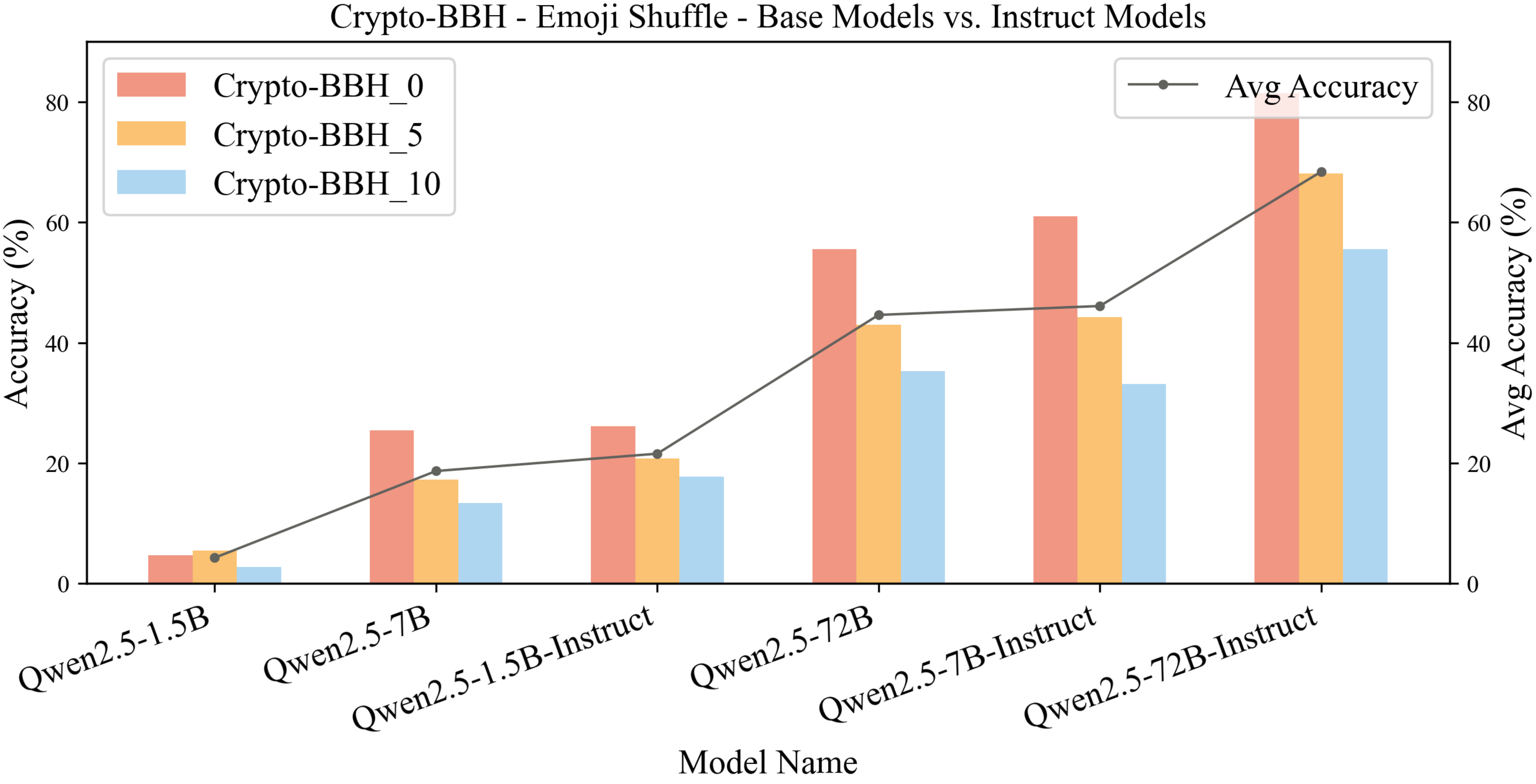}
    }
    \subfigure[The result of Crypto-Math]{
        \includegraphics[width=0.48\textwidth]{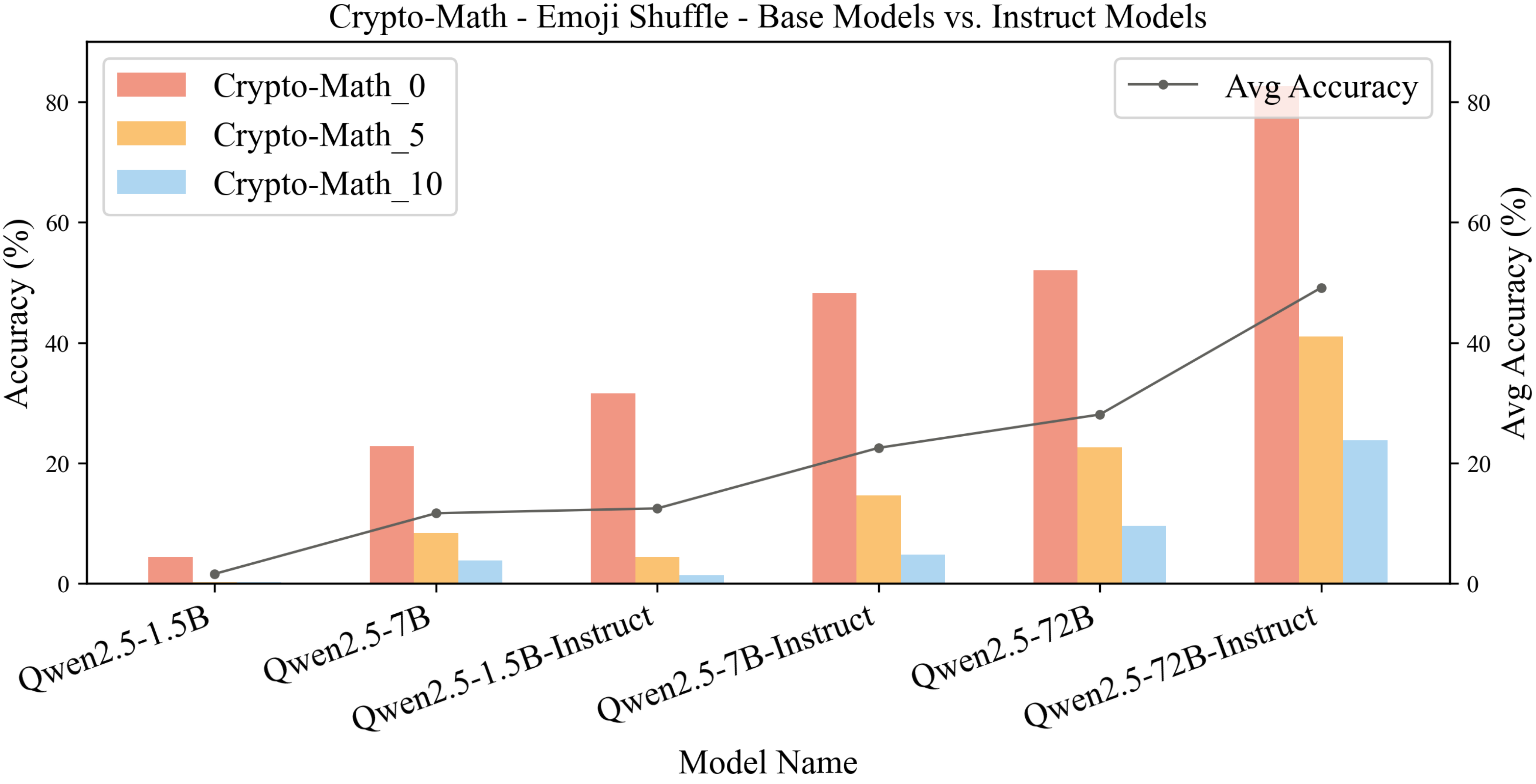}
    }
    \subfigure[The result of Crypto-MBPP]{
        \includegraphics[width=0.48\textwidth]{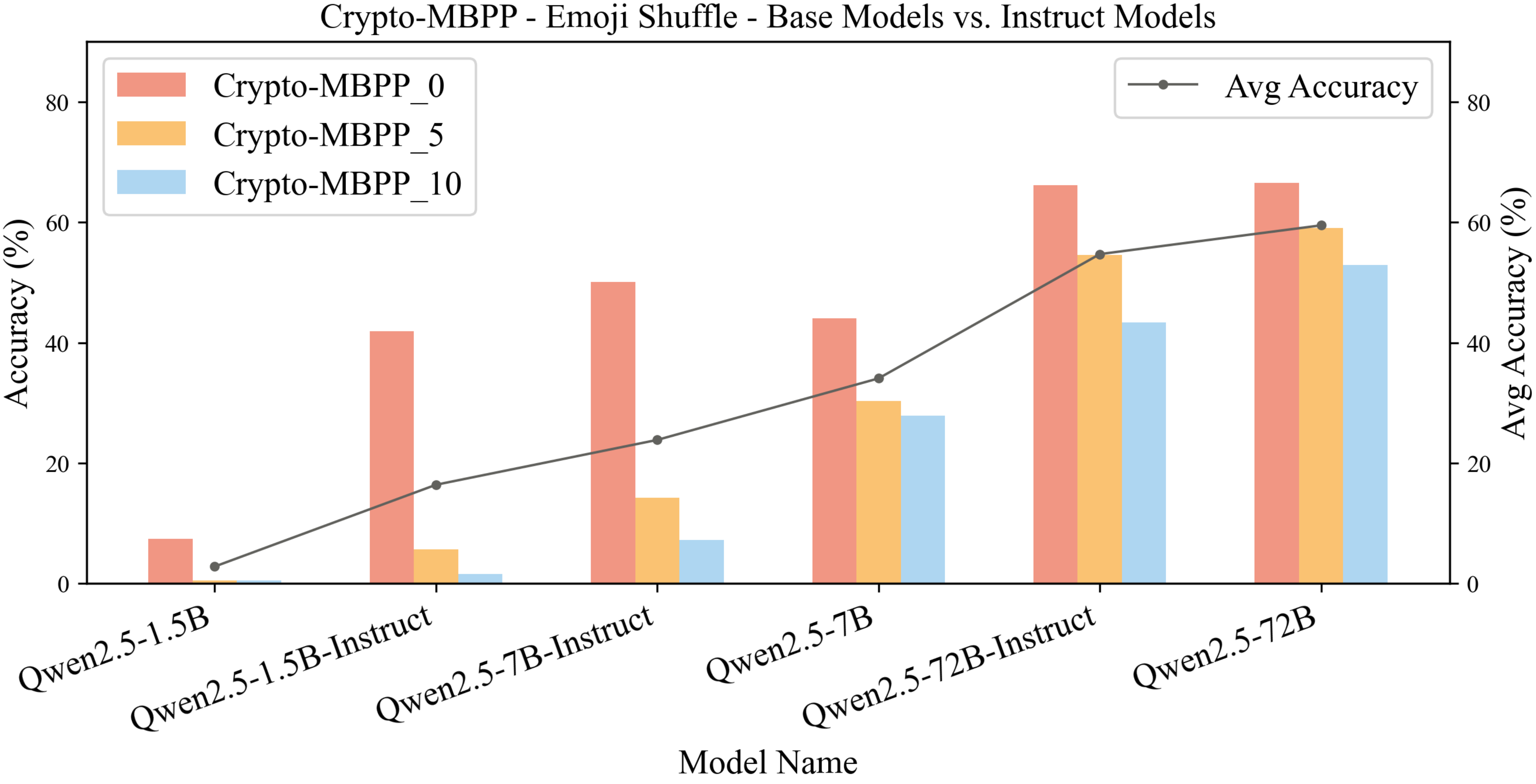}
    }
    \subfigure[The result of Crypto-MMLU]{
        \includegraphics[width=0.48\textwidth]{pics/base_and_instruct/Crypto-MMLU.pdf}
    }
    \subfigure[The result of Crypto-MMLU-Num]{
        \includegraphics[width=0.48\textwidth]{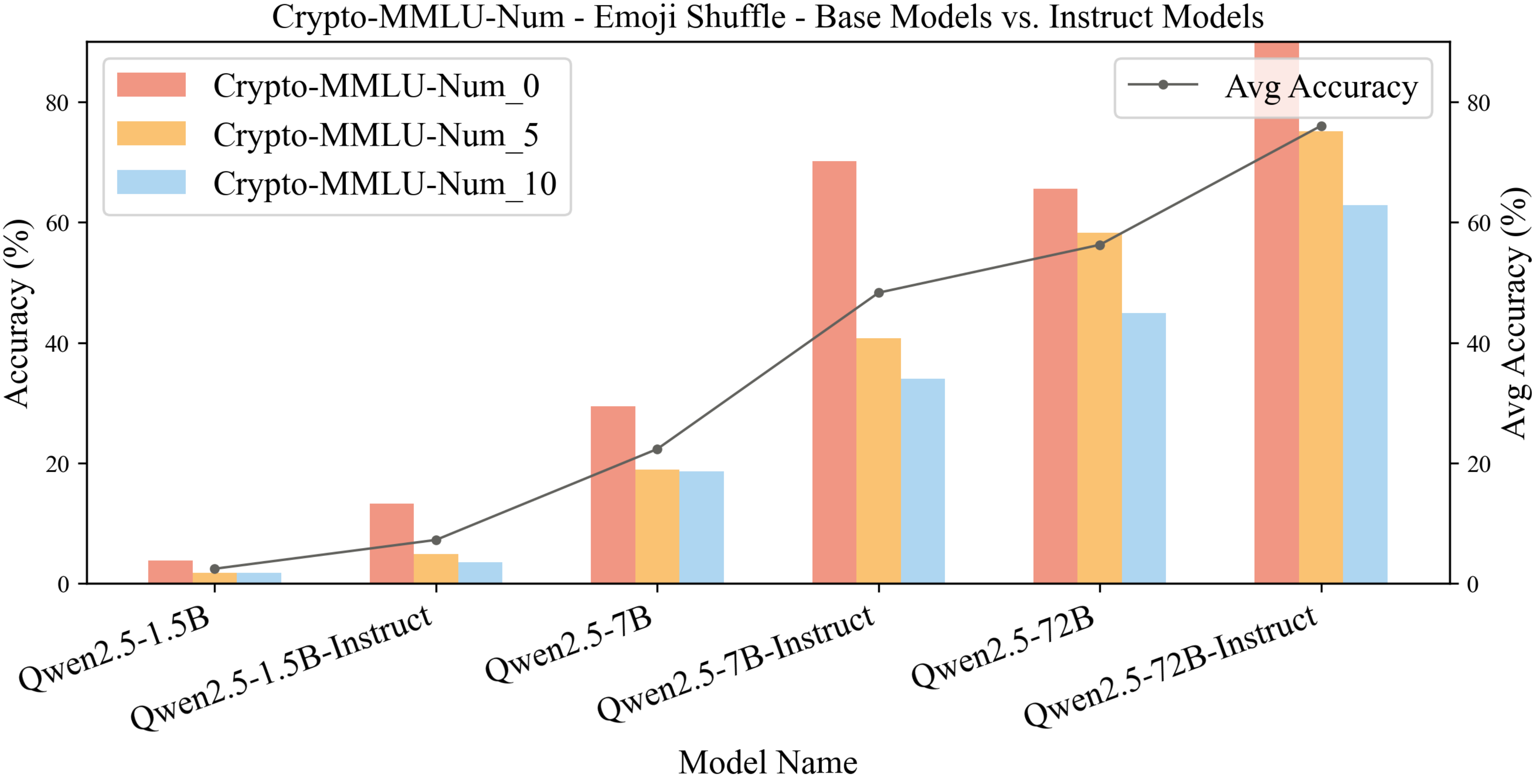}
    }
    \subfigure[The result of Crypto-MMLU-Alpha]{
        \includegraphics[width=0.48\textwidth]{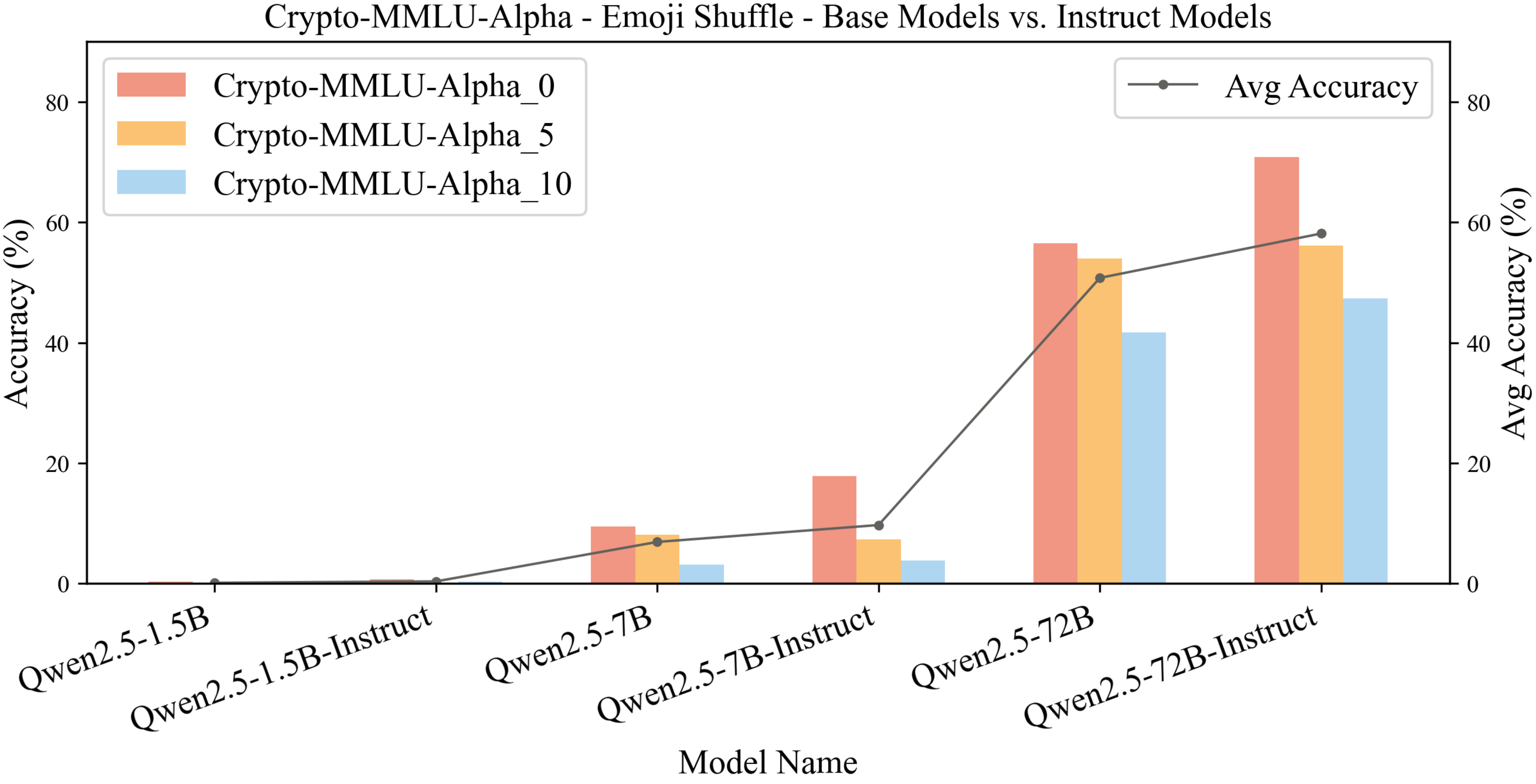}
    }
    \caption{The performance between base models and instruct models. In \textit{Domain\_Words}, \textit{Words} denotes the number of words encoded in the given question.}
    \label{fig:base-instruct-all}
    \vskip -0.2in
\end{figure}
\vfill

\newpage
\subsection{Thr Effect of Different Architectures}
Models with non-mainstream architectures perform even worse than smaller models like Qwen2.5 and Llama-3.1, suggesting that the model architecture is an important factor affecting compositional reasoning ability.
\begin{figure}[H]
    \centering
    \vskip 0.2in
    \subfigure[The result of Crypto-BBH]{
        \includegraphics[width=0.48\textwidth]{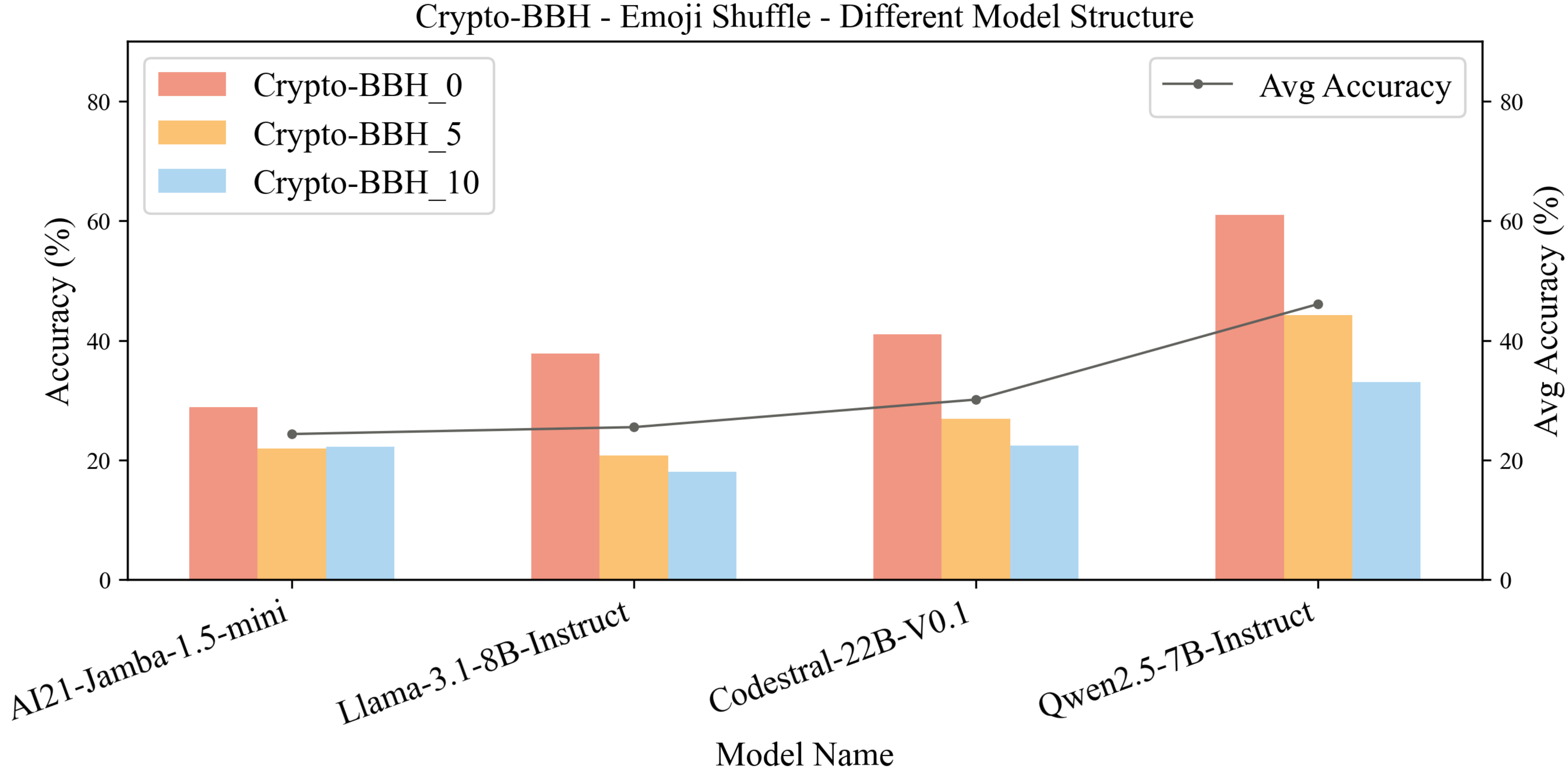}
    }
    \subfigure[The result of Crypto-Math]{
        \includegraphics[width=0.48\textwidth]{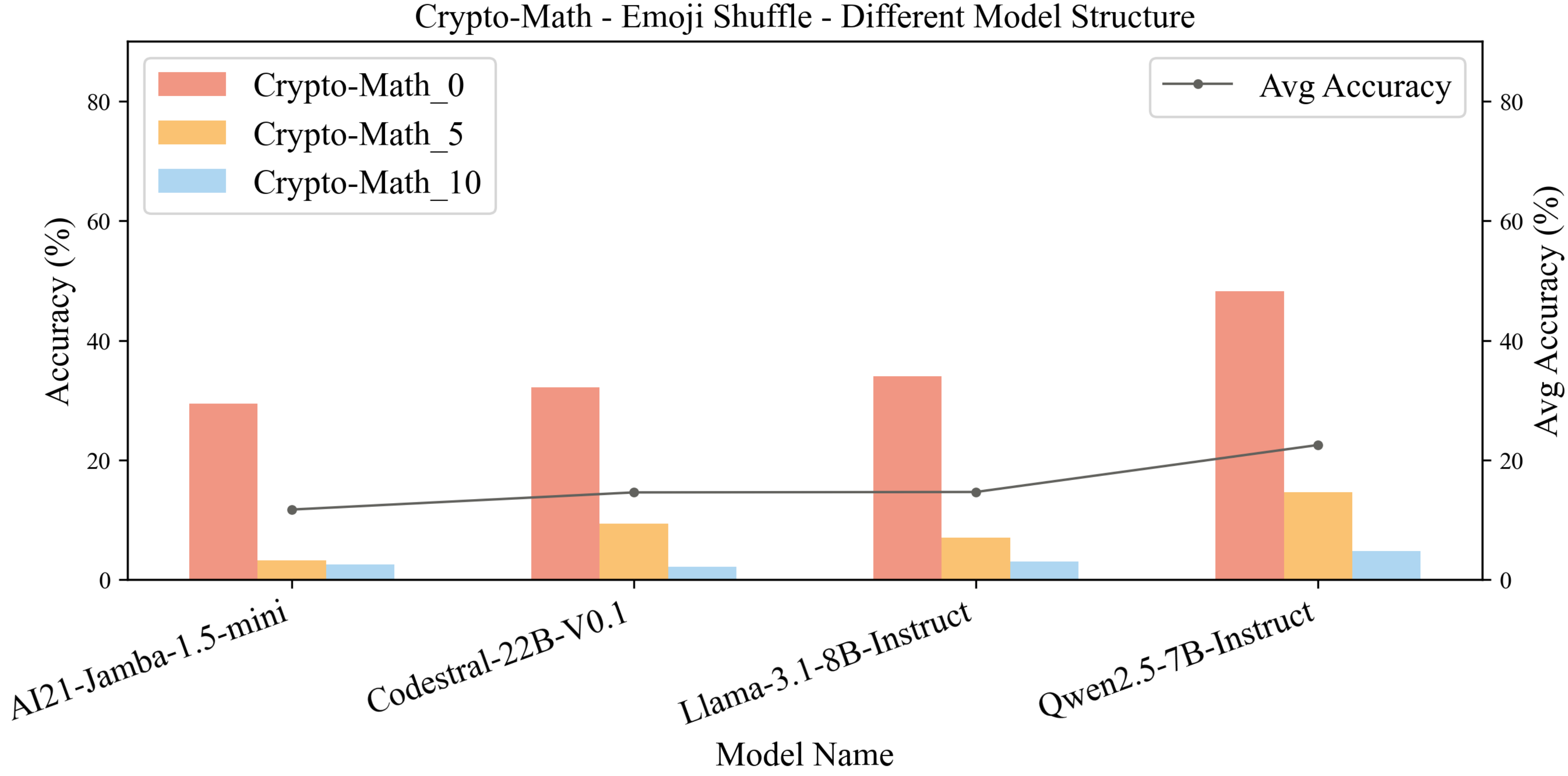}
    }
    \subfigure[The result of Crypto-MBPP]{
        \includegraphics[width=0.48\textwidth]{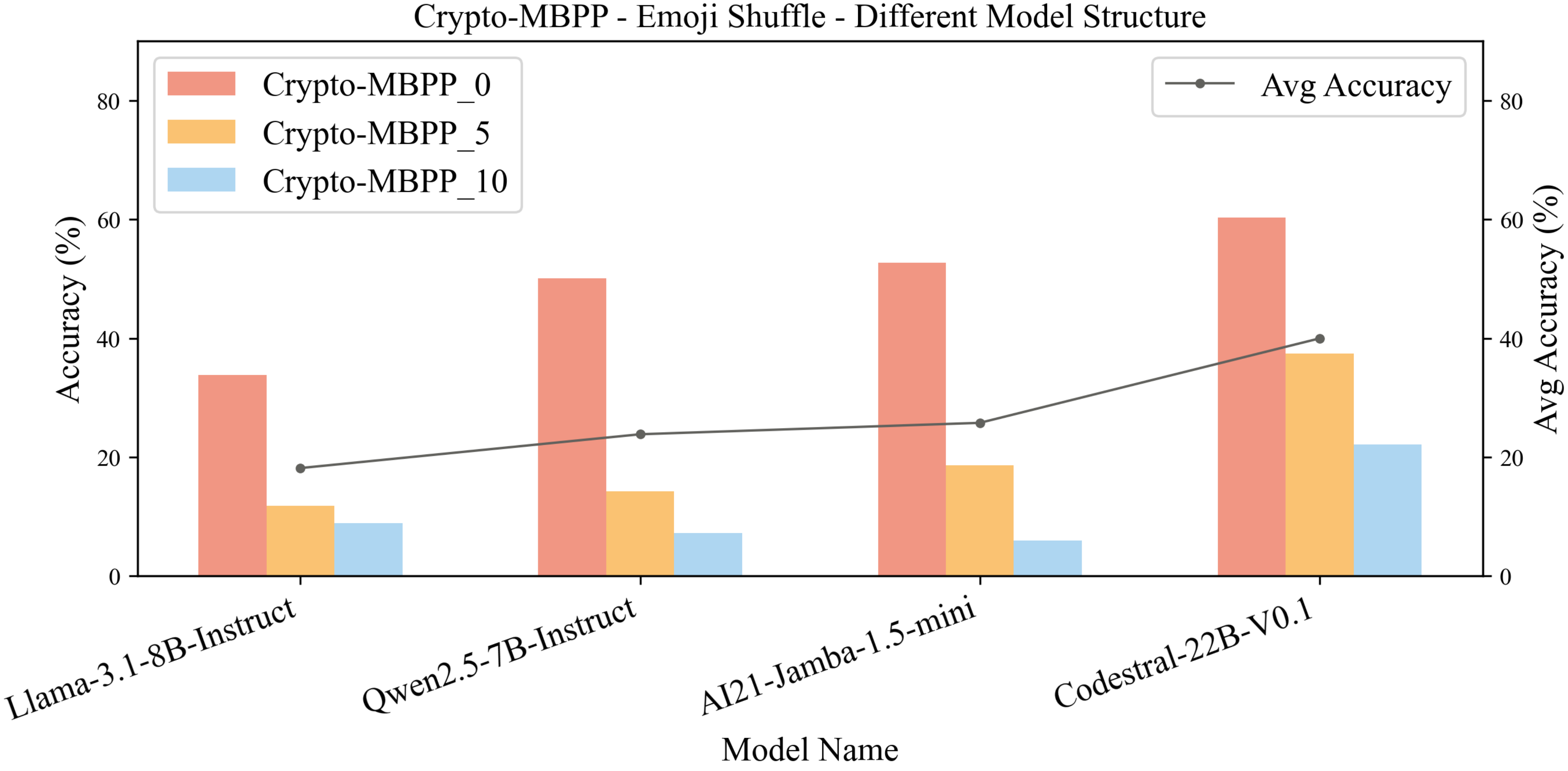}
    }
    \subfigure[The result of Crypto-MMLU]{
        \includegraphics[width=0.48\textwidth]{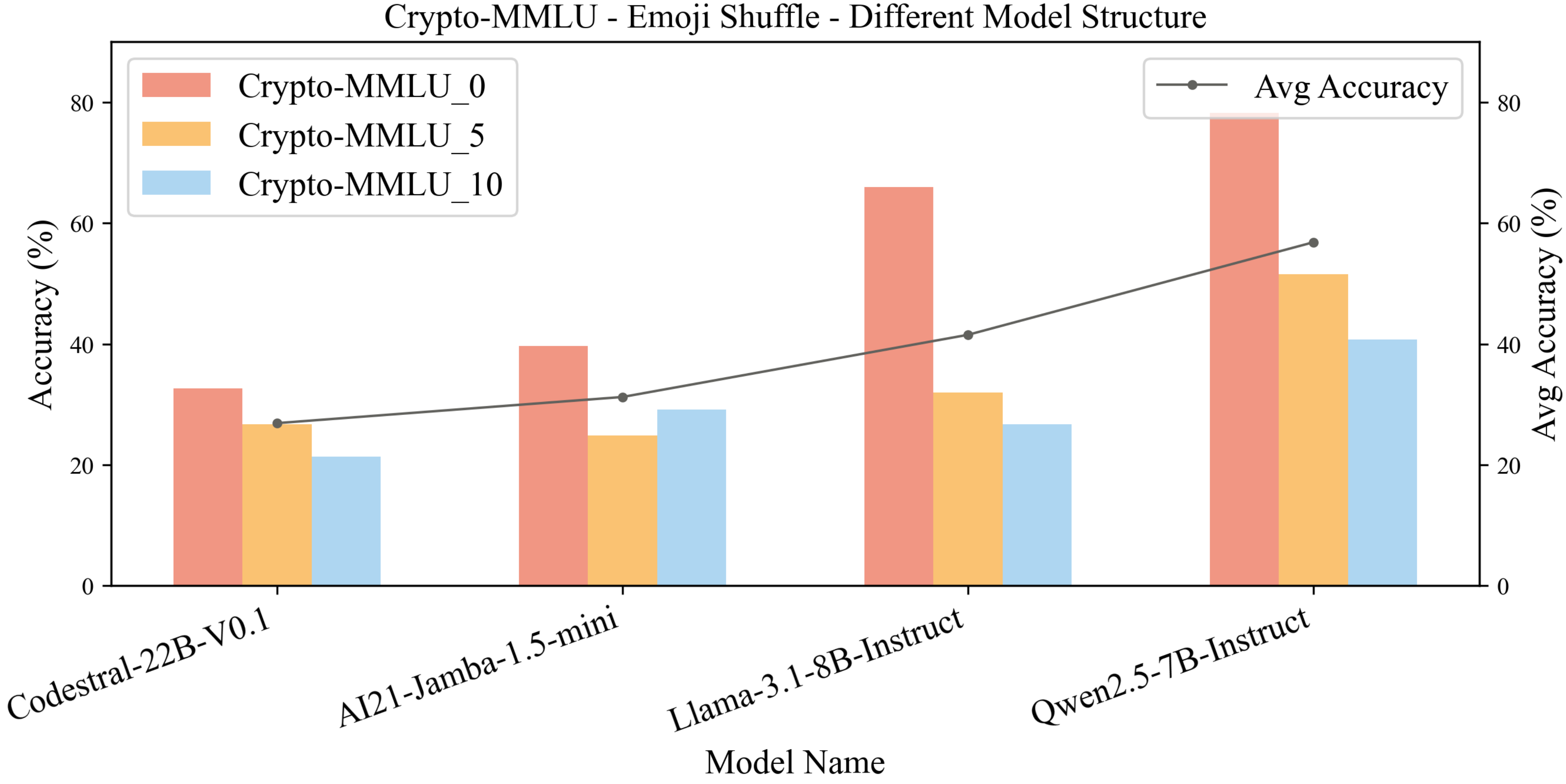}
    }
    \subfigure[The result of Crypto-MMLU-Num]{
        \includegraphics[width=0.48\textwidth]{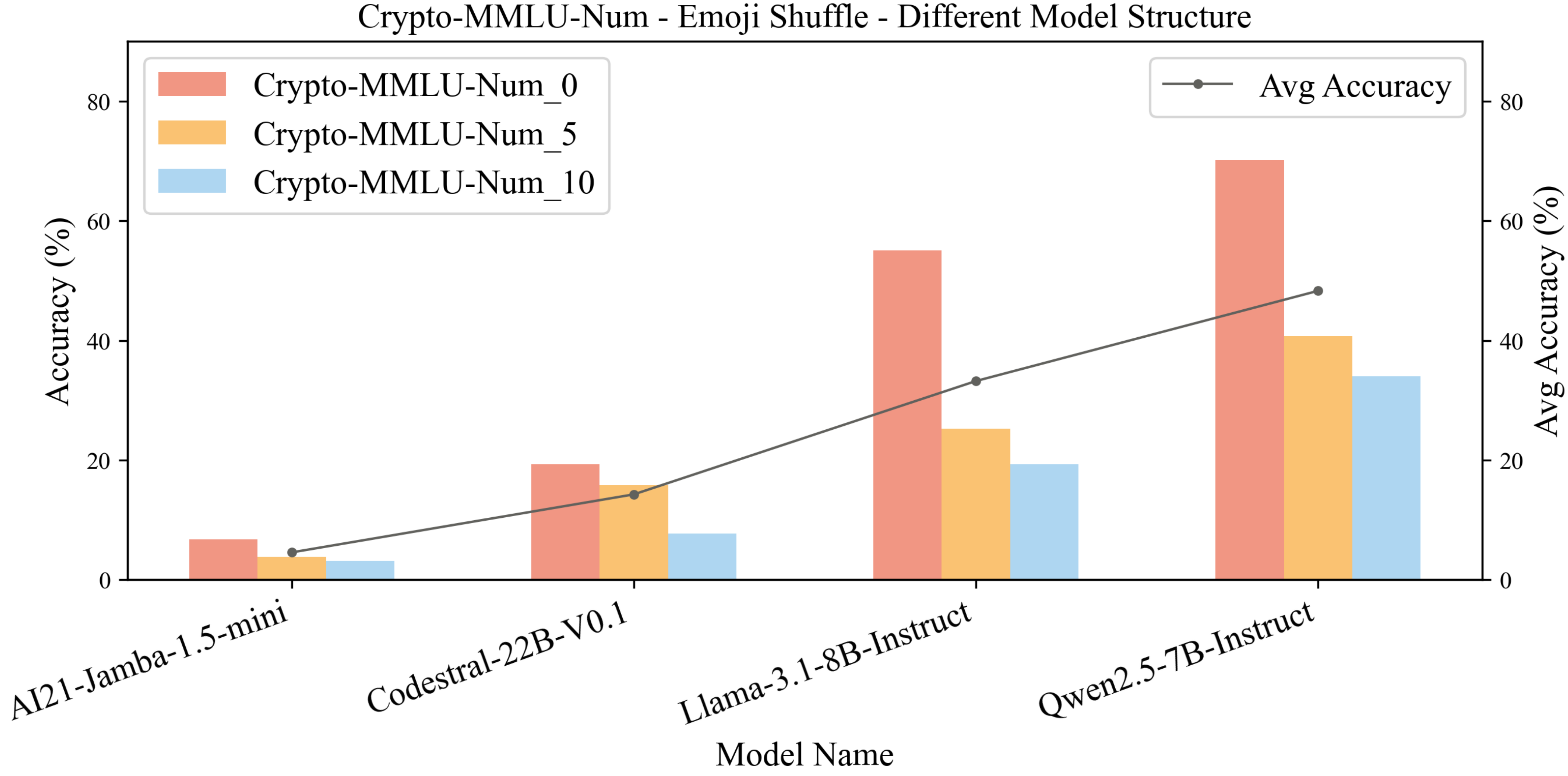}
    }
    \subfigure[The result of Crypto-MMLU-Alpha]{
        \includegraphics[width=0.48\textwidth]{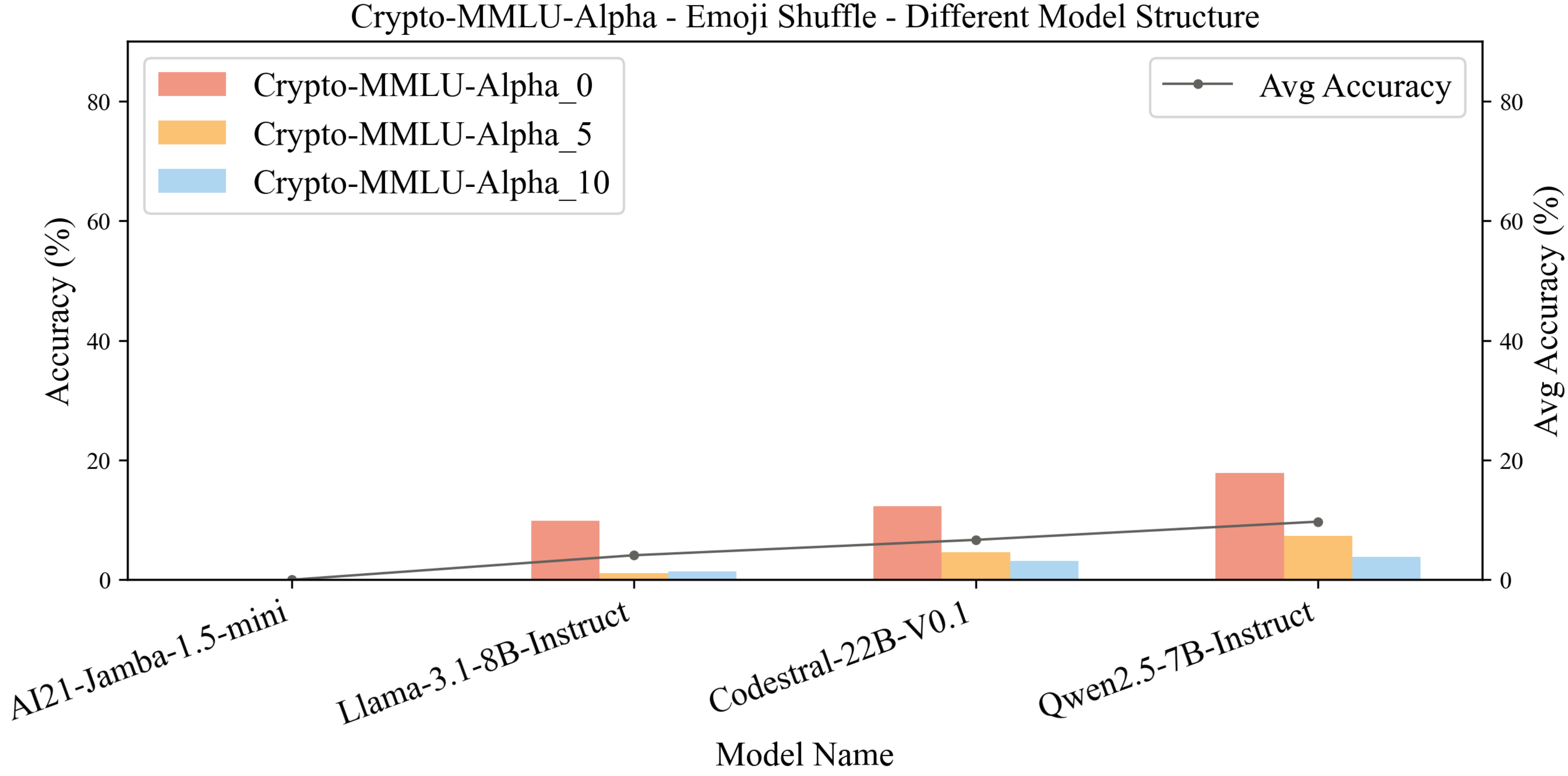}
    }
    \caption{The performance of different architectural models. In \textit{Domain\_Words}, \textit{Words} denotes the number of words encoded in the given question.}
    \label{fig:architectual-all}
    \vskip -0.2in
\end{figure}
\vfill

\newpage
\subsection{The Decoding Capacity of LLMs}
\label{appendix: decoding}
Content \ref{appendix: decoding} below shows the performance of different models in decoding.
\subsubsection{ROUGE-1 Score}
The statistical results of decoding accuracy using ROUGE-1\citep{lin2004rouge} are shown below.
\begin{figure}[H]
    \centering
    \vskip 0.2in
    \subfigure[The result of Crypto-BBH]{
        \includegraphics[width=0.48\textwidth]{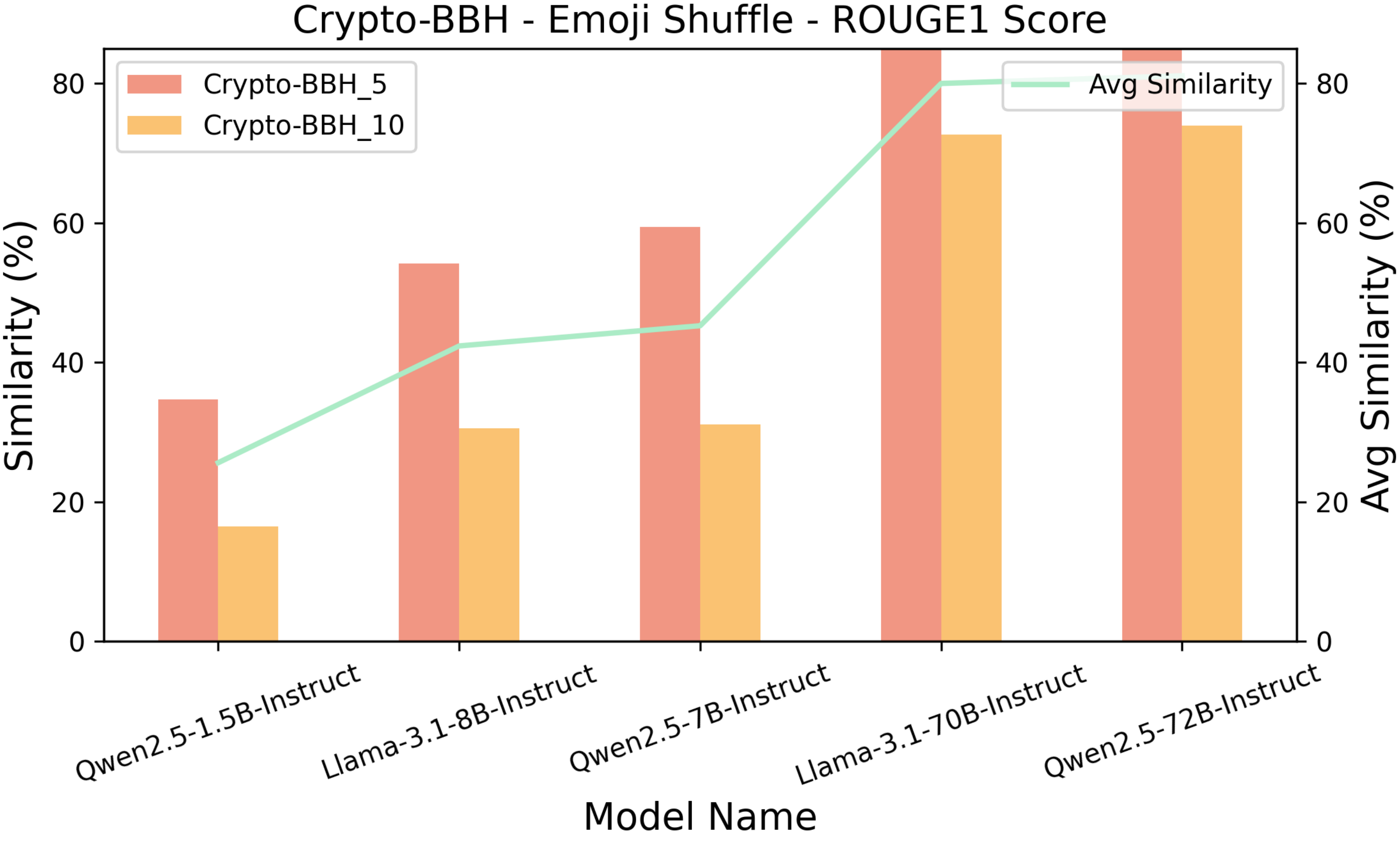}
    }
    \subfigure[The result of Crypto-Math]{
        \includegraphics[width=0.48\textwidth]{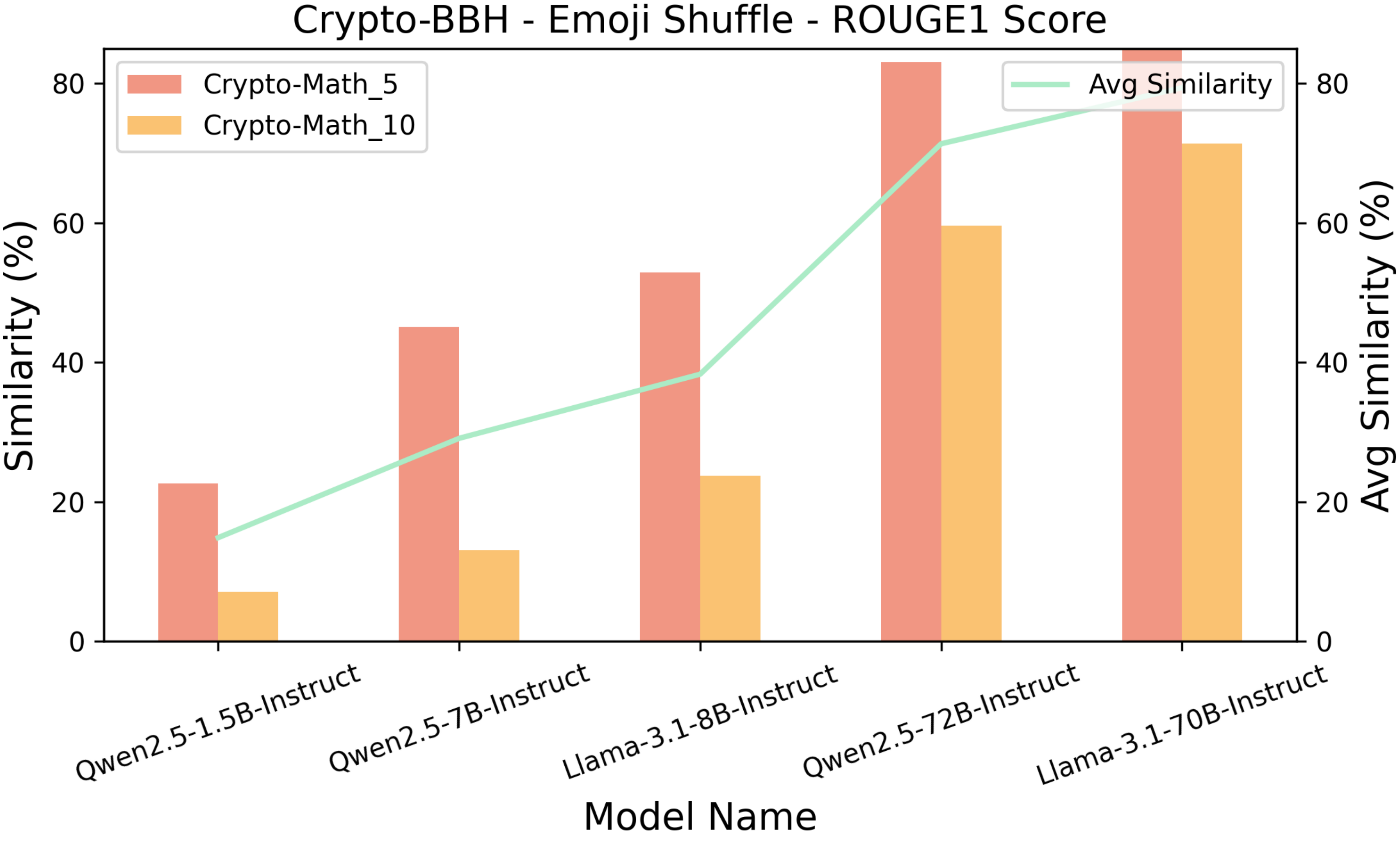}
    }
    \subfigure[The result of Crypto-MBPP]{
        \includegraphics[width=0.48\textwidth]{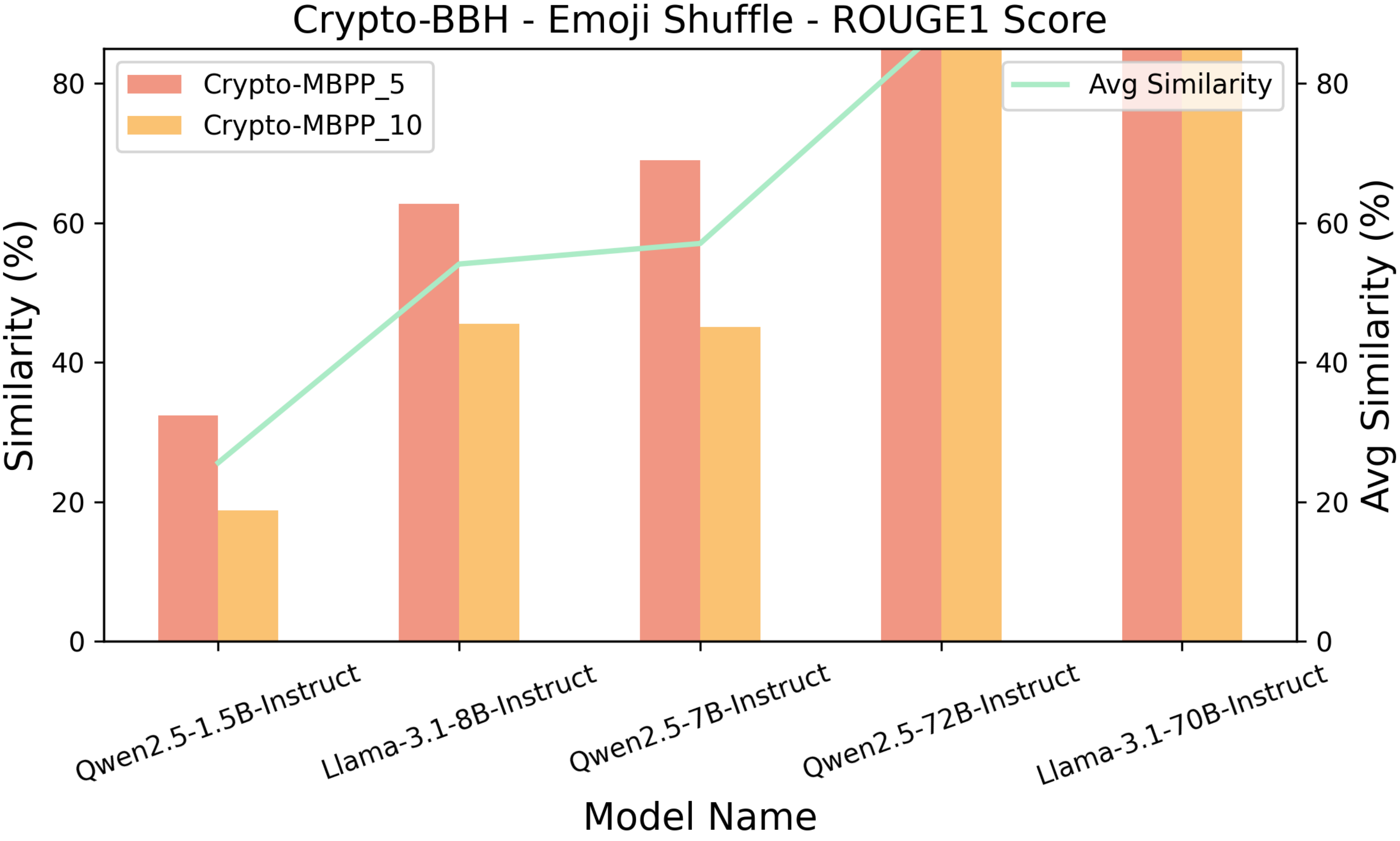}
    }
    \subfigure[The result of Crypto-MMLU]{
        \includegraphics[width=0.48\textwidth]{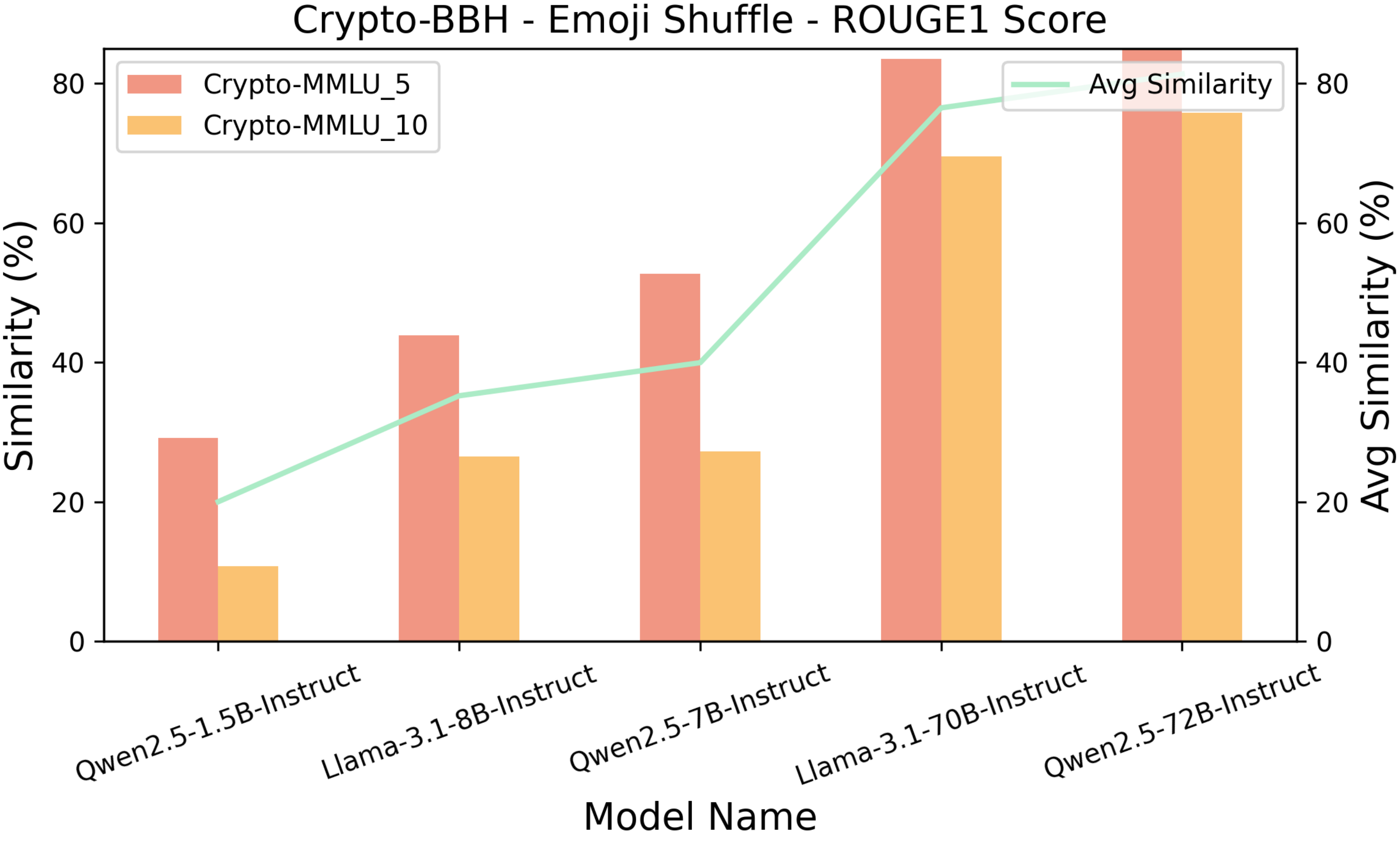}
    }
    \caption{The decoding performance of different models. In \textit{Domain\_Words}, \textit{Words} denotes the number of words encoded in the given question.}
    \label{fig:decoding-rouge-all}
    \vskip -0.2in
\end{figure}
\vfill

\newpage
\subsubsection{BLEU Score}
The statistical results of decoding accuracy using BLEU(1-gram)\citep{papineni2002bleu} are shown below.
\begin{figure}[H]
    \centering
    \vskip 0.2in
    \subfigure[The result of Crypto-BBH]{
        \includegraphics[width=0.48\textwidth]{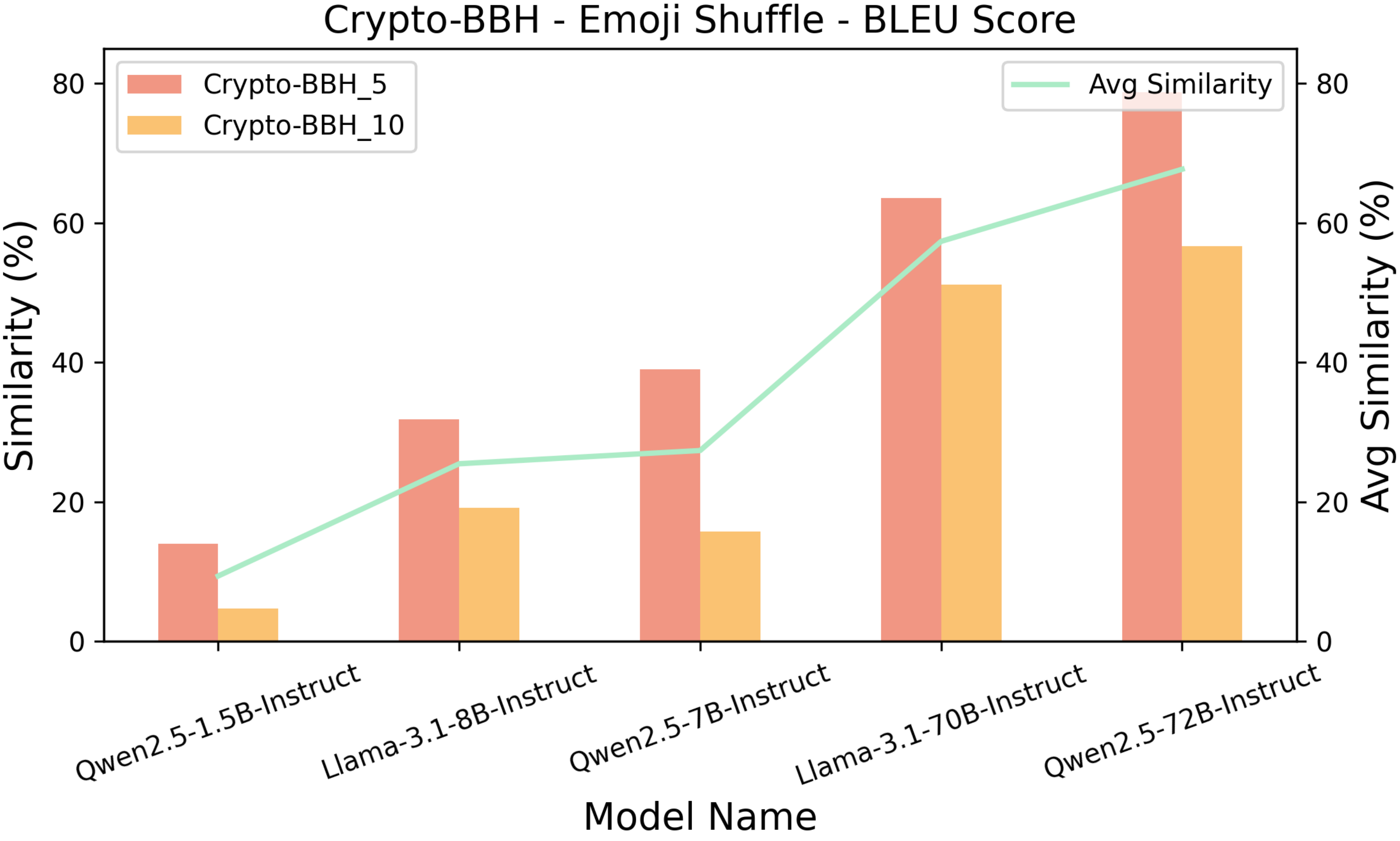}
    }
    \subfigure[The result of Crypto-Math]{
        \includegraphics[width=0.48\textwidth]{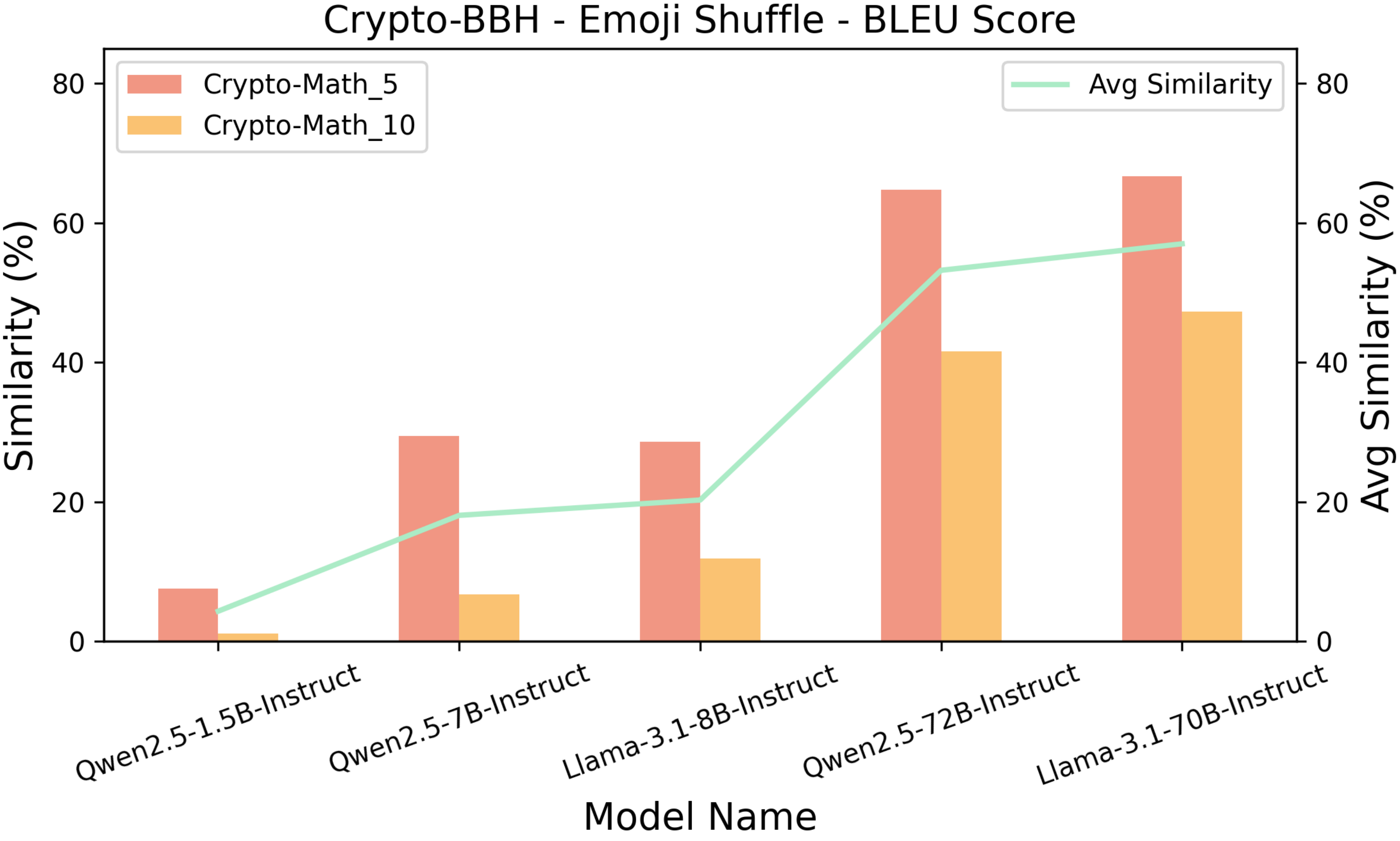}
    }
    \subfigure[The result of Crypto-MBPP]{
        \includegraphics[width=0.48\textwidth]{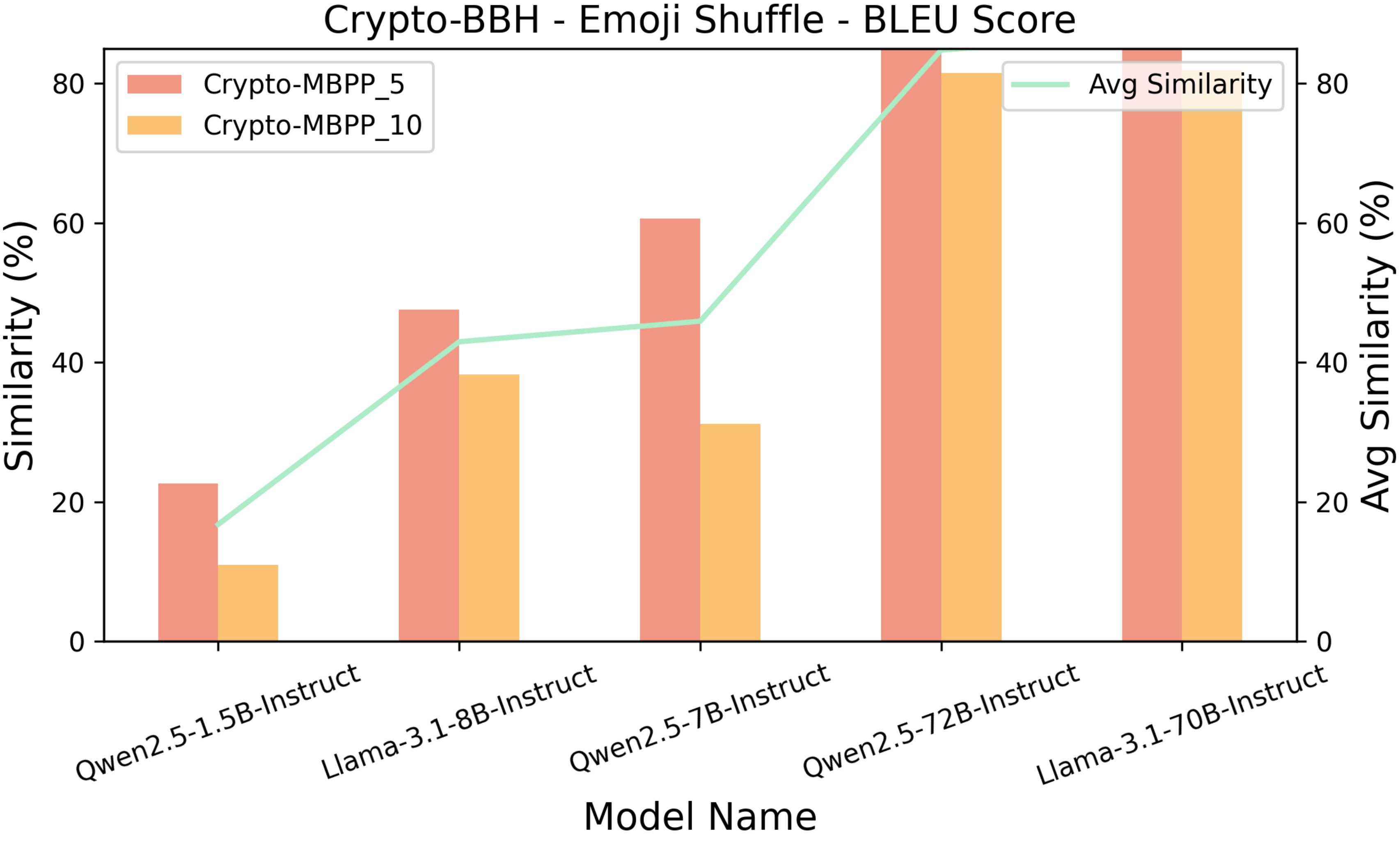}
    }
    \subfigure[The result of Crypto-MMLU]{
        \includegraphics[width=0.48\textwidth]{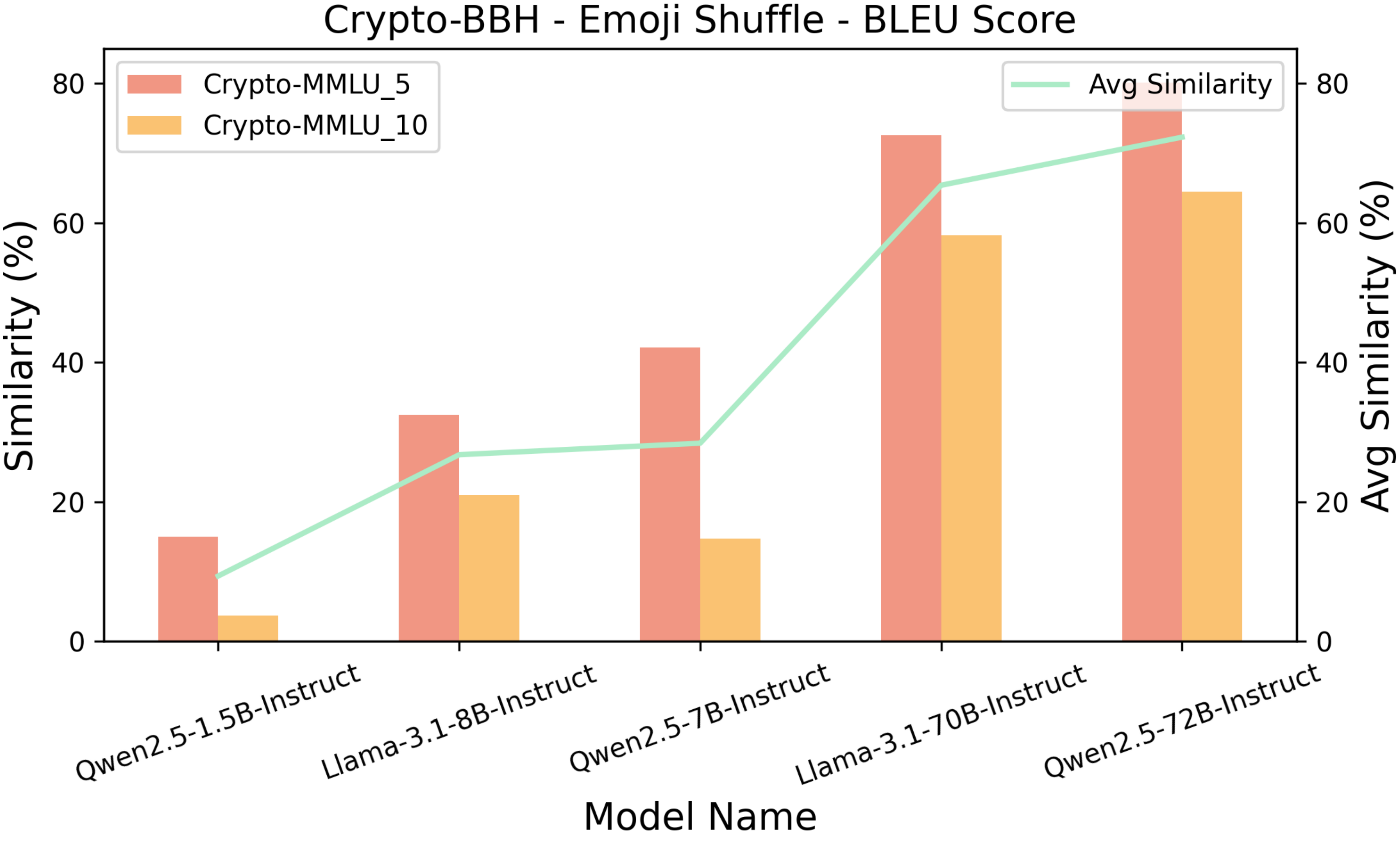}
    }
    \caption{The decoding performance of different models. In \textit{Domain\_Words}, \textit{Words} denotes the number of words encoded in the given question.}
    \label{fig:decoding-bleu-all}
    \vskip -0.2in
\end{figure}

\subsection{Variance of Accuracy and AUC for Different Models}
\label{Variance of AUC}
\begin{table}[H]
\centering
\caption{Variance of accuracy and AUC for closed-source models on Crypto-HighResolution. \textit{Num} stands for Accuracy, which corresponds to solving the question of \textit{Num} words being encoded.}
\vskip 0.15in
\resizebox{\linewidth}{!}{
\begin{tabular}{lcccccccccccc}
\toprule
 & \textbf{0} & \textbf{1} & \textbf{2} & \textbf{3} & \textbf{4} & \textbf{5} & \textbf{6} & \textbf{7} & \textbf{8} & \textbf{9} & \textbf{10} & \textbf{AUC}\\
\midrule
\textbf{Variance} & 0.0043 & 0.0055 & 0.0058 & 0.0058 & 0.0066 & 0.0069 & 0.0081 & 0.0094 & 0.0091 & 0.0116 & 0.0137 & \textbf{0.6221} \\
\bottomrule
\end{tabular}
}
\vskip -0.1in
\end{table}
\vfill

\newpage
\subsection{Correlation With Chatbot Arena}
Chatbot Arena~\citep{chiang2024chatbot}, which builds its evaluation system through human interaction, is regarded as the "gold standard" in the assessment of human-computer dialogue systems. Although this method has some biases (such as differences in user background and insufficient task diversity), its large scale and high ecological validity make it one of the most representative human evaluation frameworks. 

To investigate the correlation between our approach and human evaluation systems, this section employs the Spearman correlation to analyze the relationship between \benchmark{}, MMLU and Chatbot Arena rankings. The Spearman correlation is suitable for nonlinear data distributions and effectively reflects the monotonic relationship between variables, which is why it is widely used in analyzing the correlation between evaluation metrics and human preferences, providing data support for our findings.

\begin{figure}[H]
    \centering
    \vskip 0.2in
    \includegraphics[width=\linewidth]{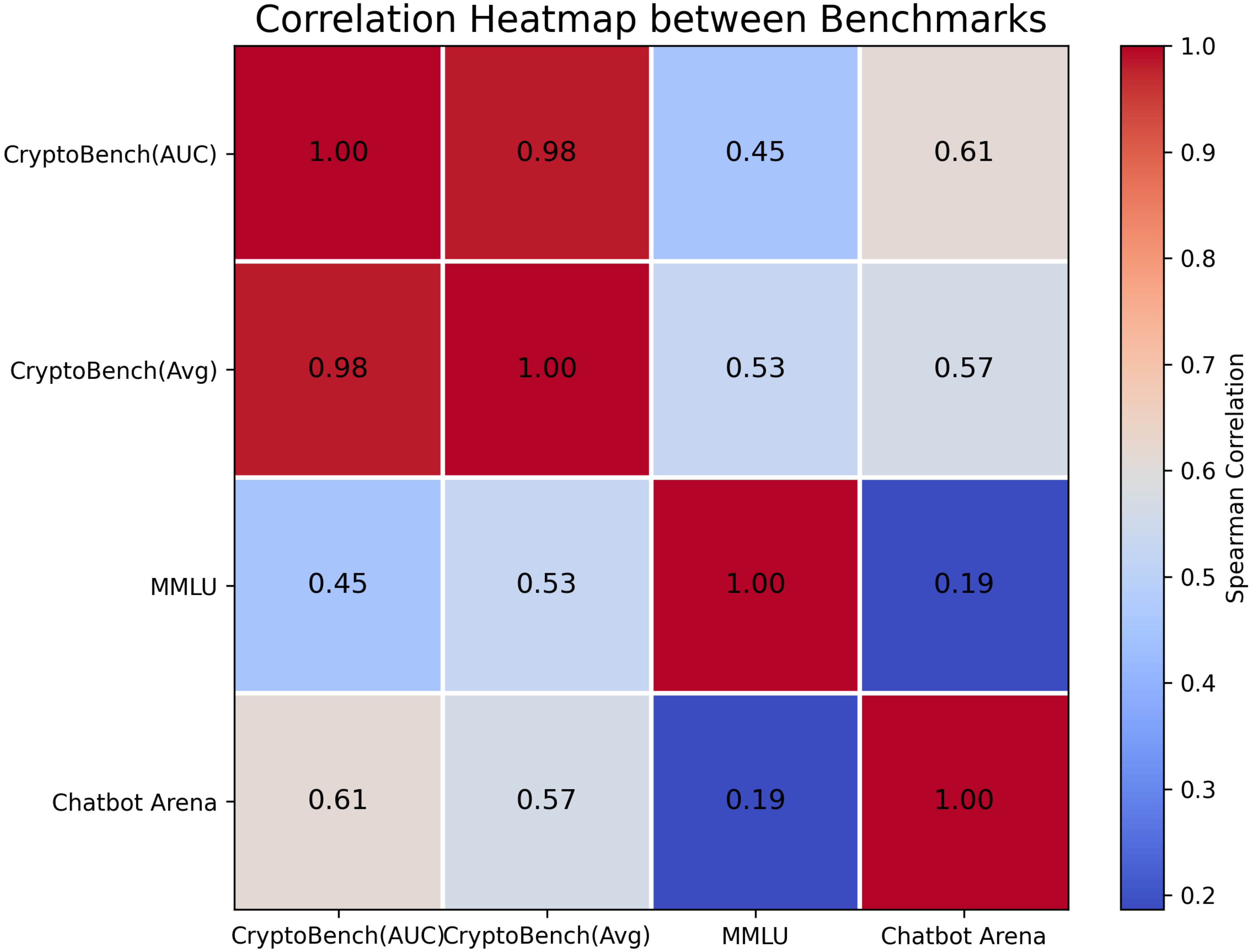}
    \caption{The correlation matrix for benchmarks. The closer the Spearman correlation is to 1, the more similar the rankings of the two benchmarks are.}
    \label{fig:correlation}
    \vskip -0.2in
\end{figure}

We calculate the Spearman correlation between \benchmark{}(AUC), \benchmark{}(Avg), MMLU, and Chatbot Arena. The AUC and Avg based on \benchmark{} yield a higher Spearman correlation(0.61, 0.57) compared to MMLU(0.19), as shown in Figure \ref{fig:correlation}.

\newpage
\section{Logit Lens Analysis(0\%/50\%/100\% encoding ratio)}
Since section \ref{sec:analysis} primarily utilizes the emoji shuffle encoding method with 0/3/5 words encoding, this section will present and supplement the experiments using 0\%/50\%/100\% encoding ratio, which provides more comprehensive evidence supporting the conclusions presented in the section  \ref{sec:analysis}.
\subsection{Base Morse Encoding Rule}
\begin{figure}[H]
    \centering
    \vskip 0.2in
    \subfigure[The result of Crypto-MMLU-BaseMorse on Llama-3.2-3B-Instruct]{
        \includegraphics[width=0.48\textwidth]{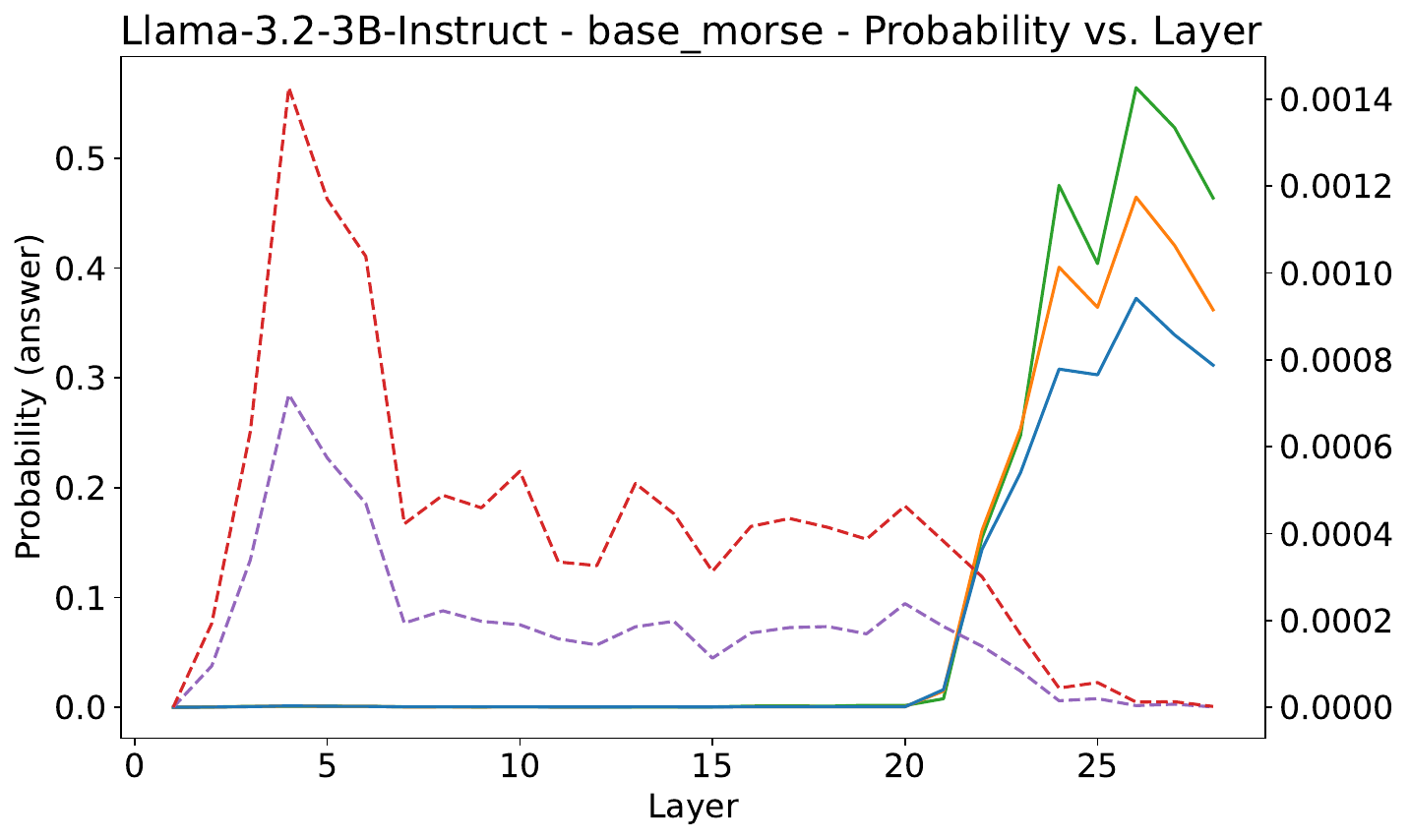}
    }
    \subfigure[The result of Crypto-MMLU-BaseMorse on Llama-3.1-8B]{
        \includegraphics[width=0.48\textwidth]{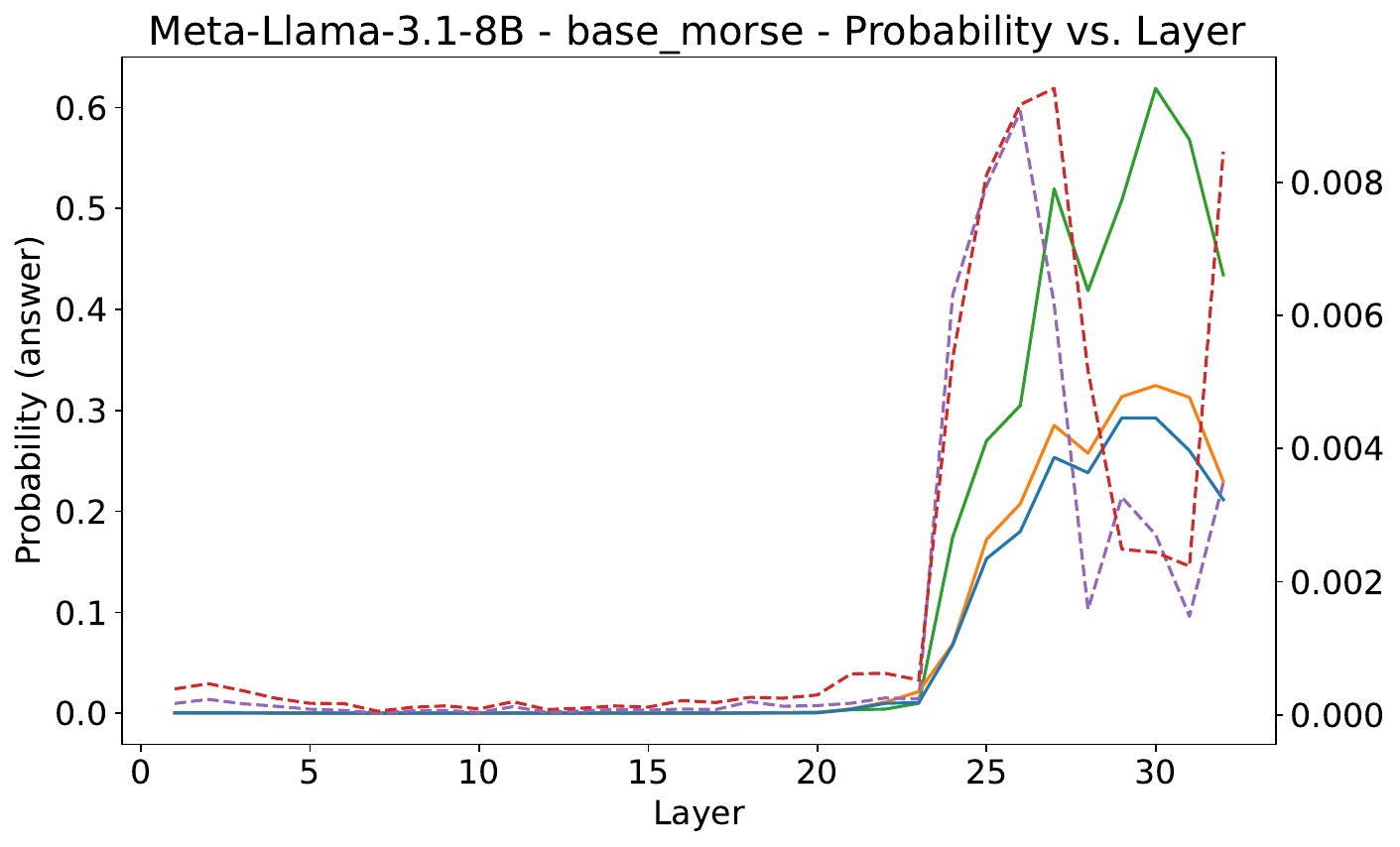}
    }
    \subfigure[The result of Crypto-MMLU-BaseMorse on Llama-3.1-8B-Instruct]{
        \includegraphics[width=0.48\textwidth]{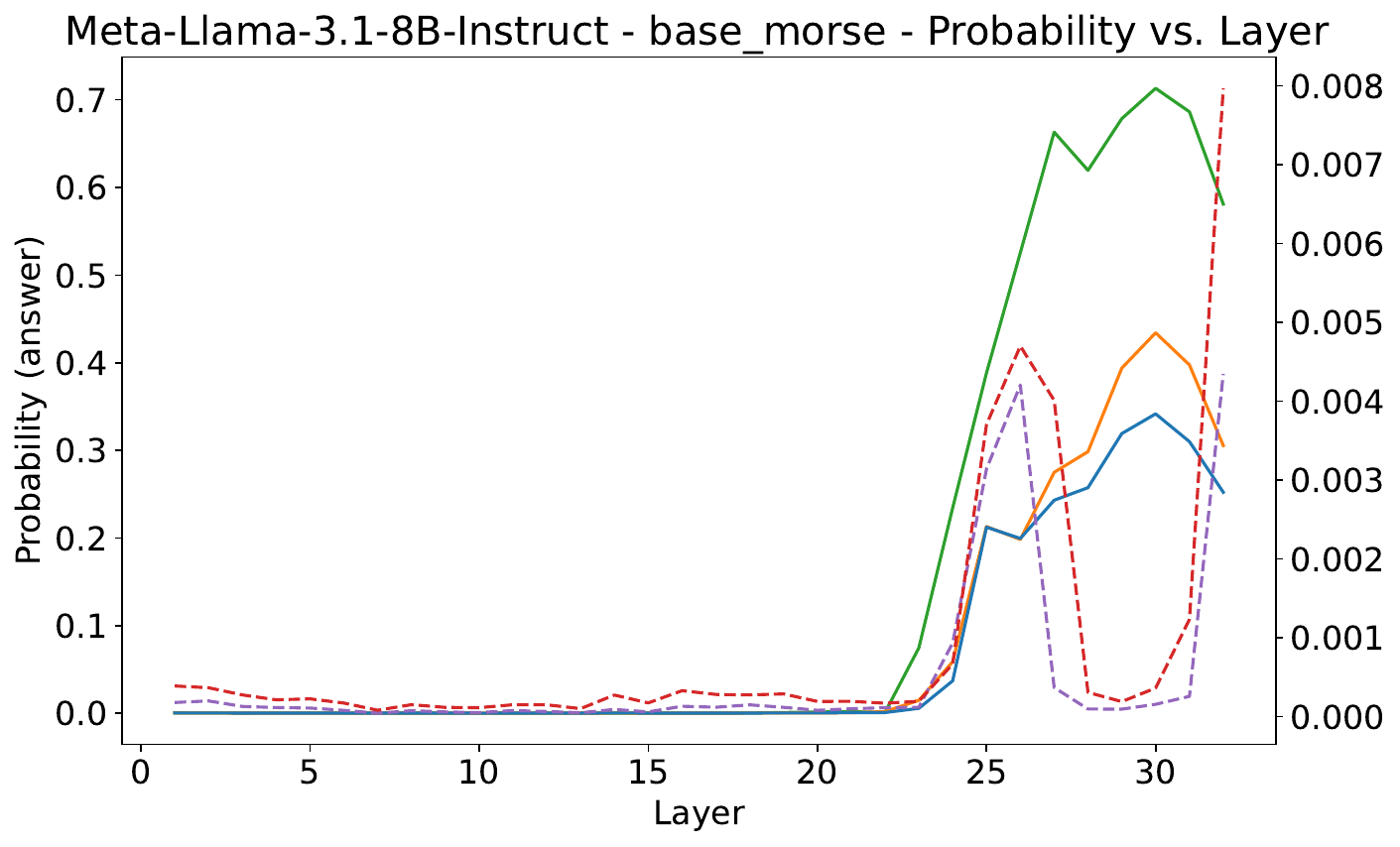}
    }
    \subfigure[The result of Crypto-MMLU-BaseMorse on Qwen2.5-0.5B-Instruct]{
        \includegraphics[width=0.48\textwidth]{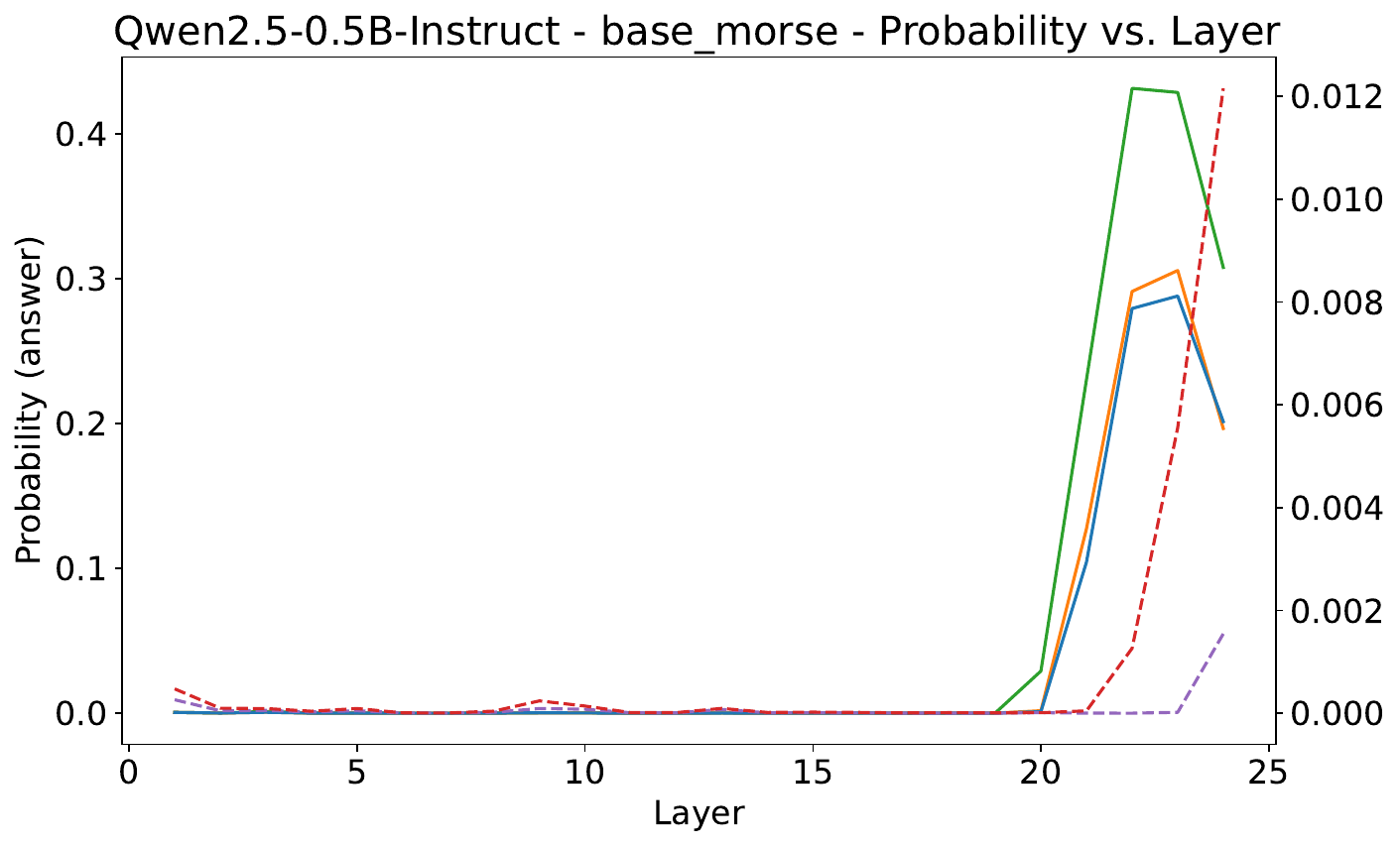}
    }
    \subfigure[The result of Crypto-MMLU-BaseMorse on Qwen2.5-1.5B-Instruct]{
        \includegraphics[width=0.48\textwidth]{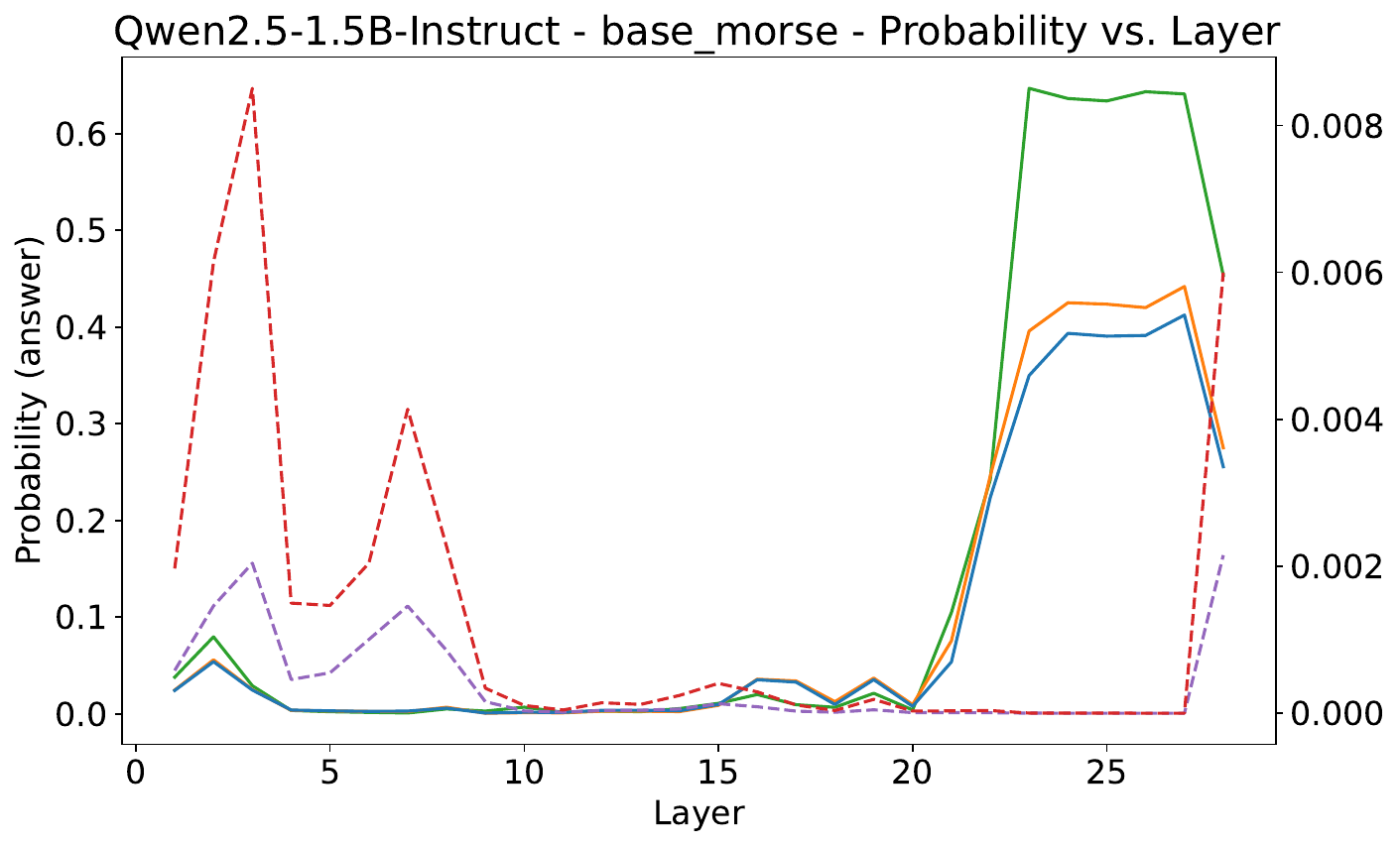}
    }
    \subfigure[The result of Crypto-MMLU-BaseMorse on Qwen2.5-3B]{
        \includegraphics[width=0.48\textwidth]{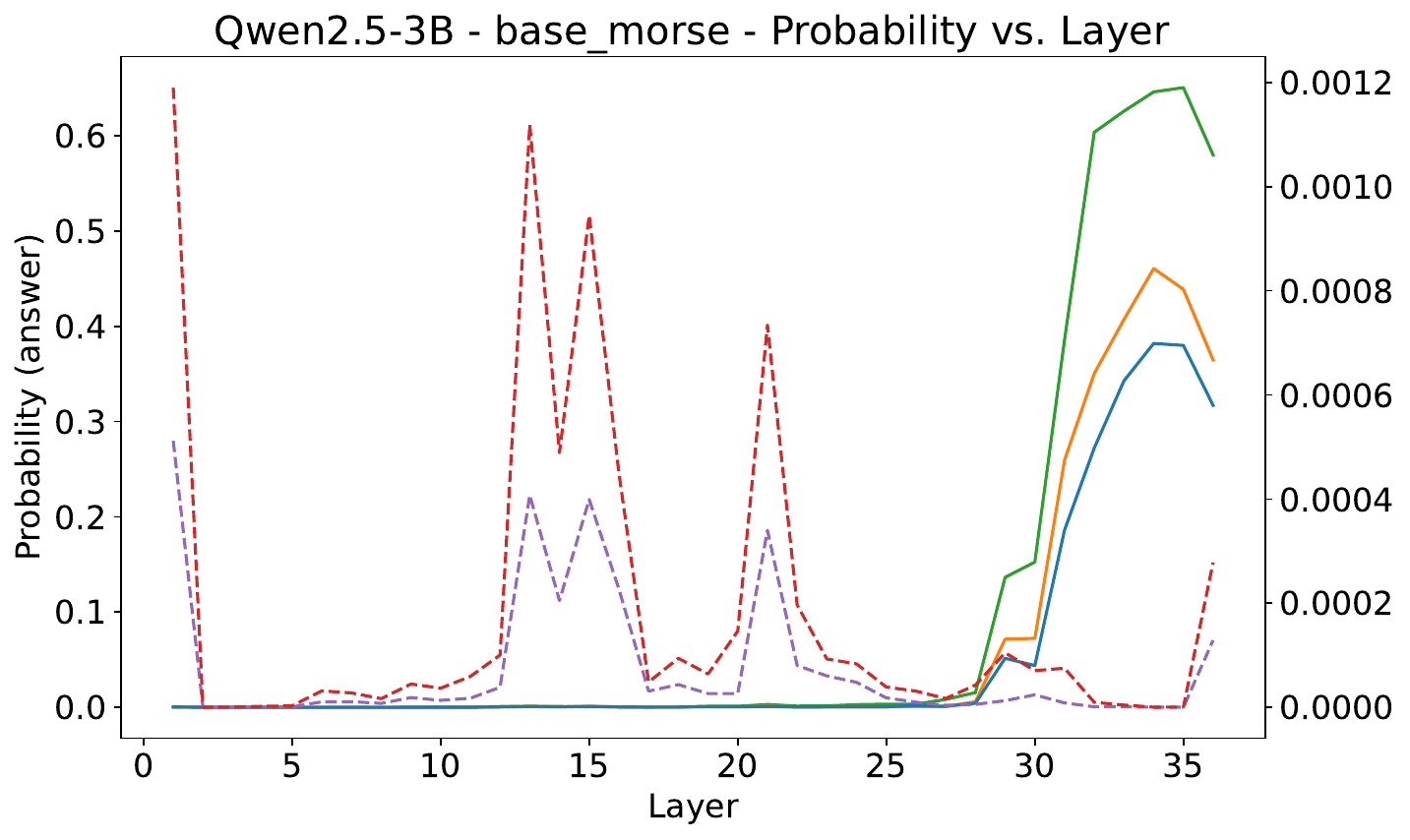}
    }

    \caption{The logit lens analysis on Crypto-MMLU-BaseMorse using $0\%/50\%/100\%$ encoding ratios.}
    \label{fig:appendix_logitlens0}
    \vskip -0.2in
\end{figure}
\vfill

\newpage
\begin{figure}[H]
    \centering
    \vskip 0.2in
    \subfigure[The result of Crypto-MMLU-BaseMorse on Qwen2.5-3B-Instruct]{
        \includegraphics[width=0.48\textwidth]{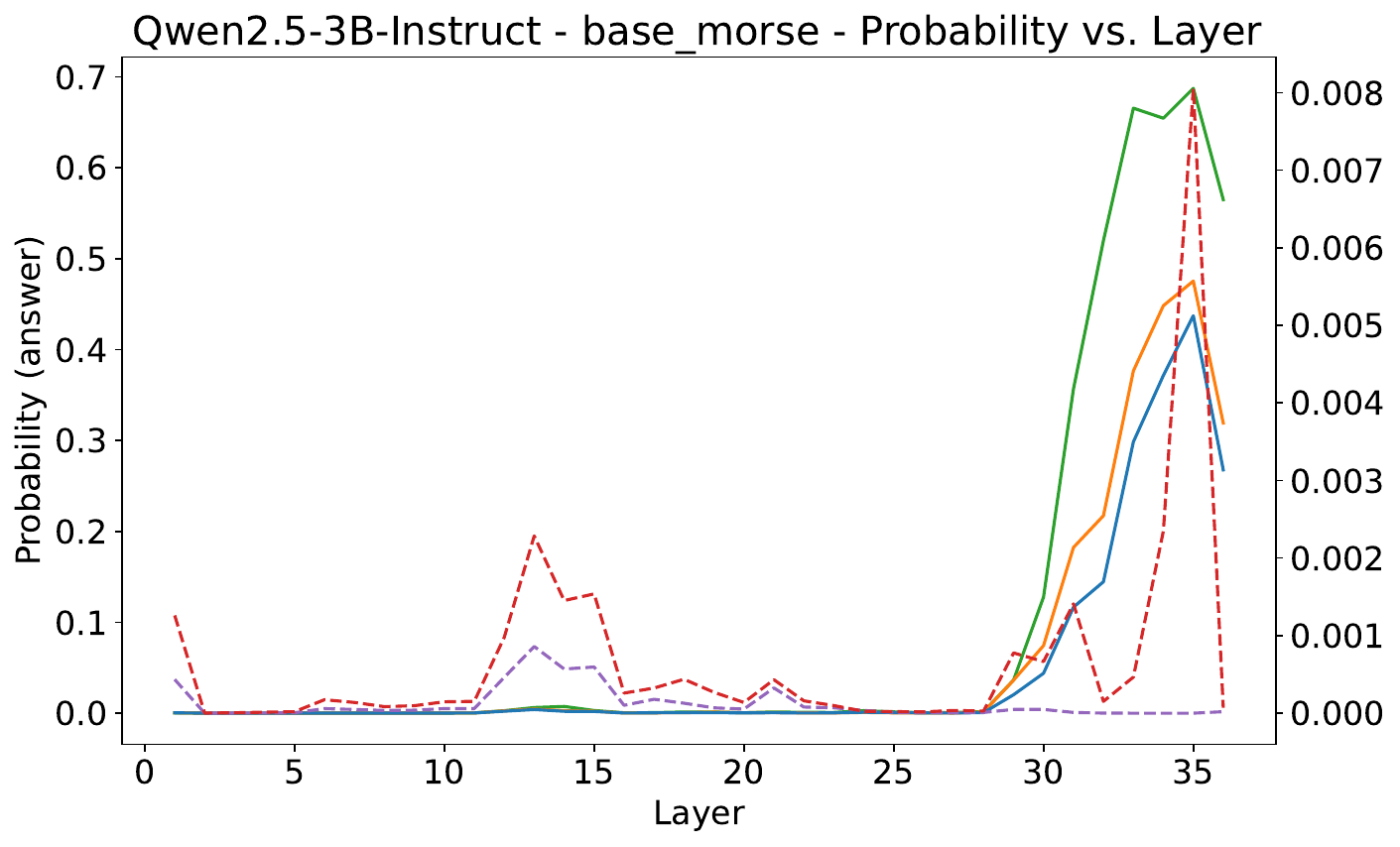}
    }
    \subfigure[The result of Crypto-MMLU-BaseMorse on Qwen2.5-7B]{
        \includegraphics[width=0.48\textwidth]{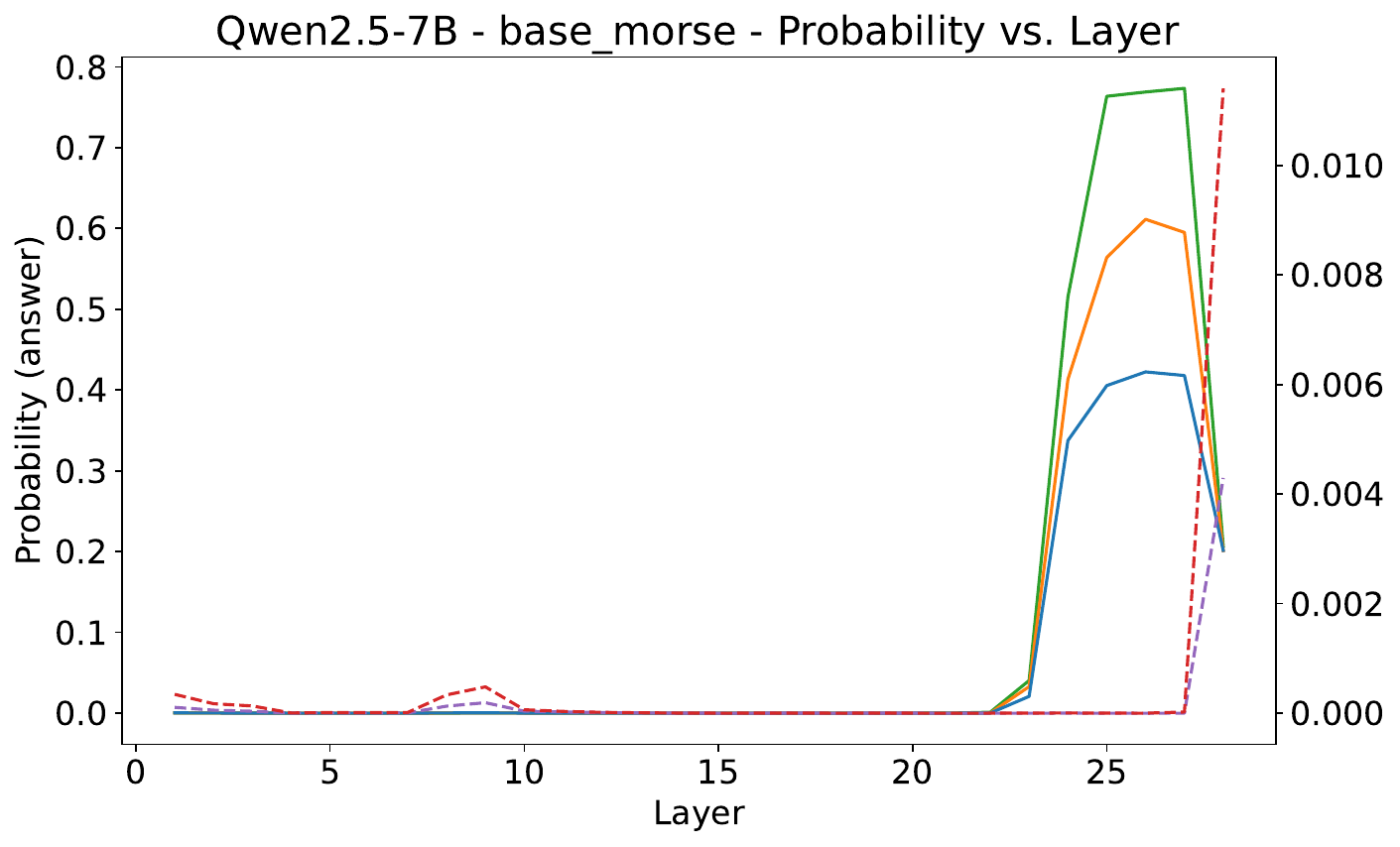}
    }
    \subfigure[The result of Crypto-MMLU-BaseMorse on Qwen2.5-7B-Instruct]{
        \includegraphics[width=0.48\textwidth]{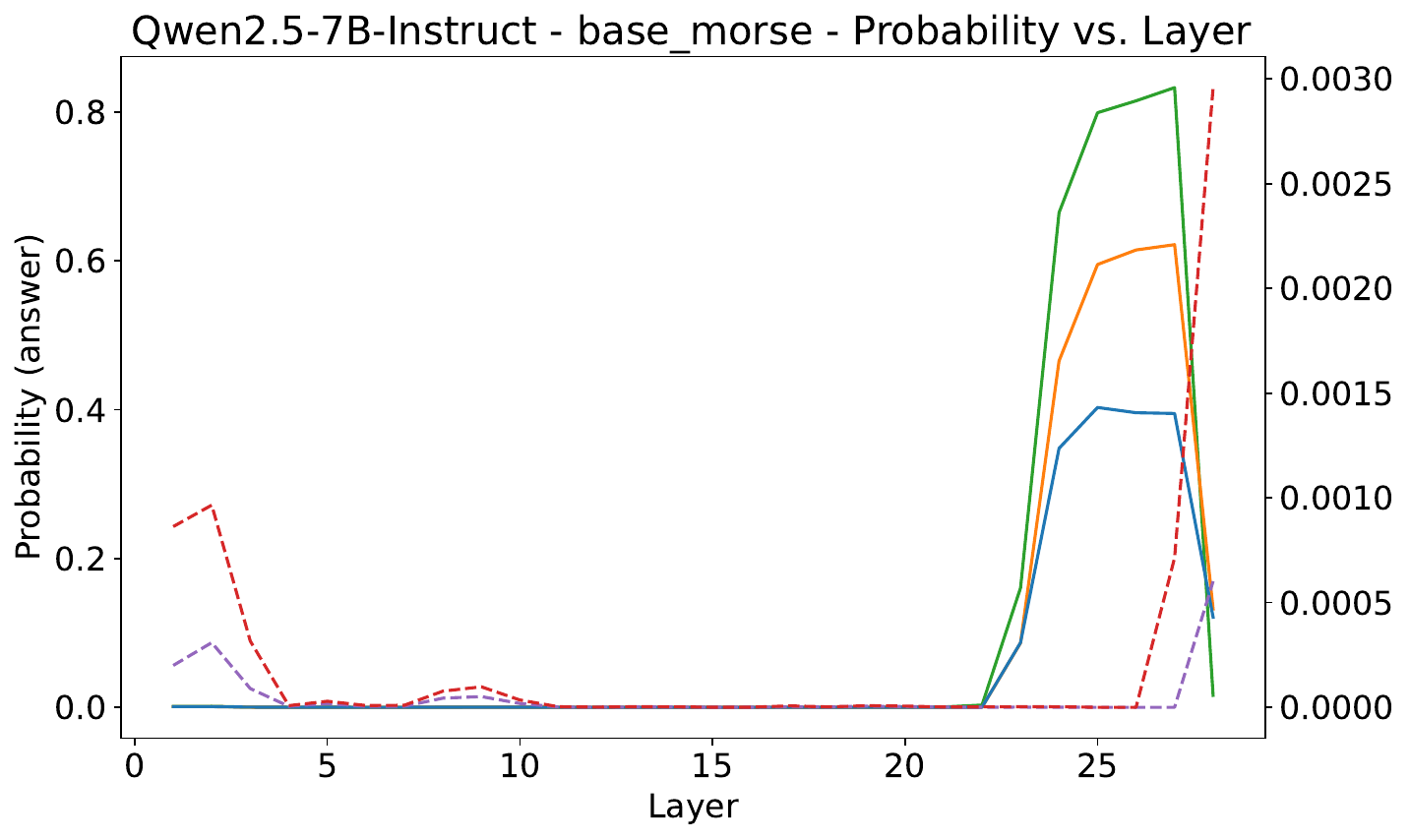}
    }
    \subfigure[The result of Crypto-MMLU-BaseMorse on Qwen2.5-14B-Instruct]{
        \includegraphics[width=0.48\textwidth]{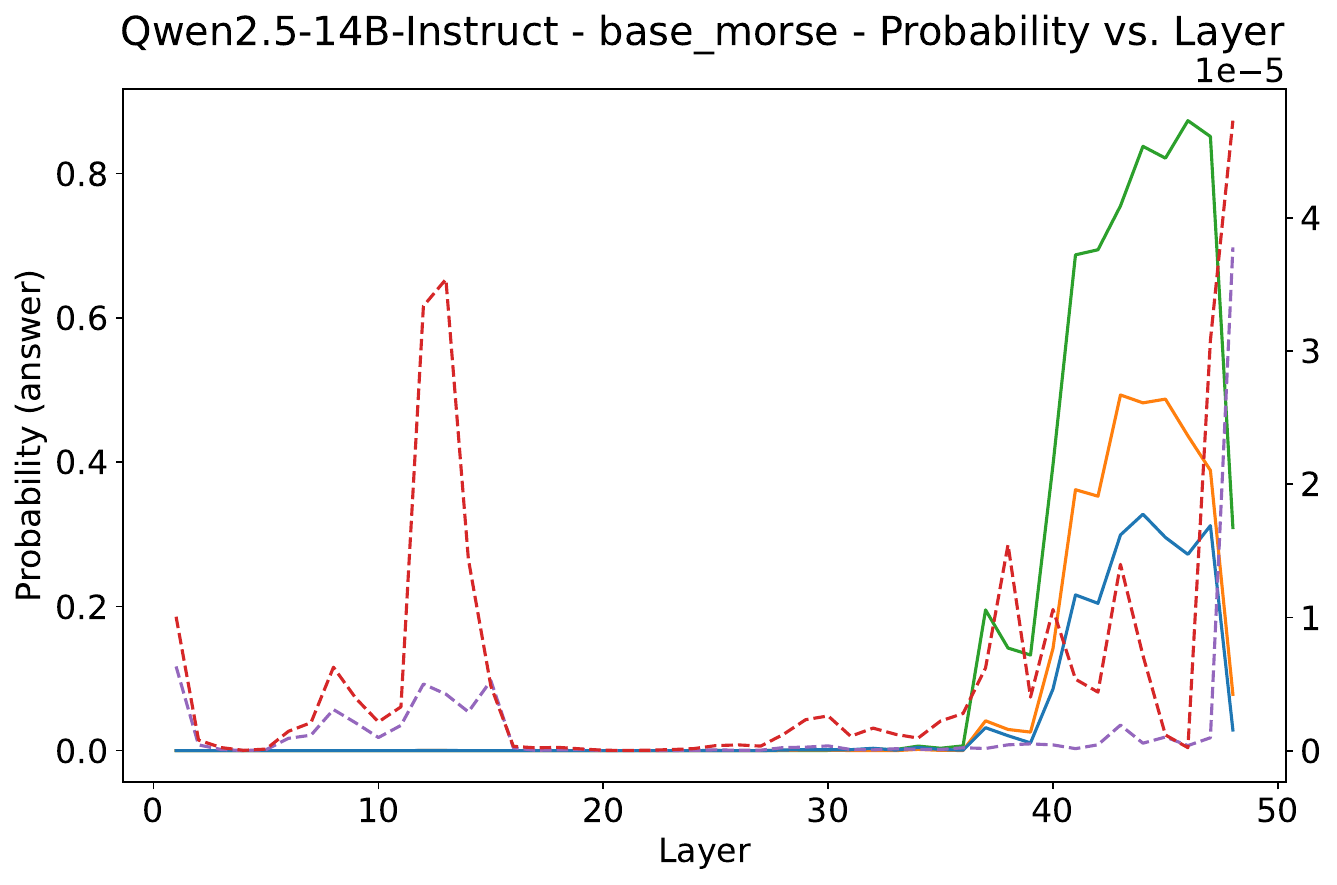}
    }
    \caption{The logit lens analysis on Crypto-MMLU-BaseMorse using 0\%/50\%/100\% encoding ratios.}
    \label{fig:appendix_logitlens1}
    \vskip -0.2in
\end{figure}

\subsection{Emoji Morse Encoding Rule}
\begin{figure}[H]
    \centering
    \vskip 0.2in
    \subfigure[The result of Crypto-MMLU-EmojiMorse on Llama-3.2-3B-Instruct]{
        \includegraphics[width=0.48\textwidth]{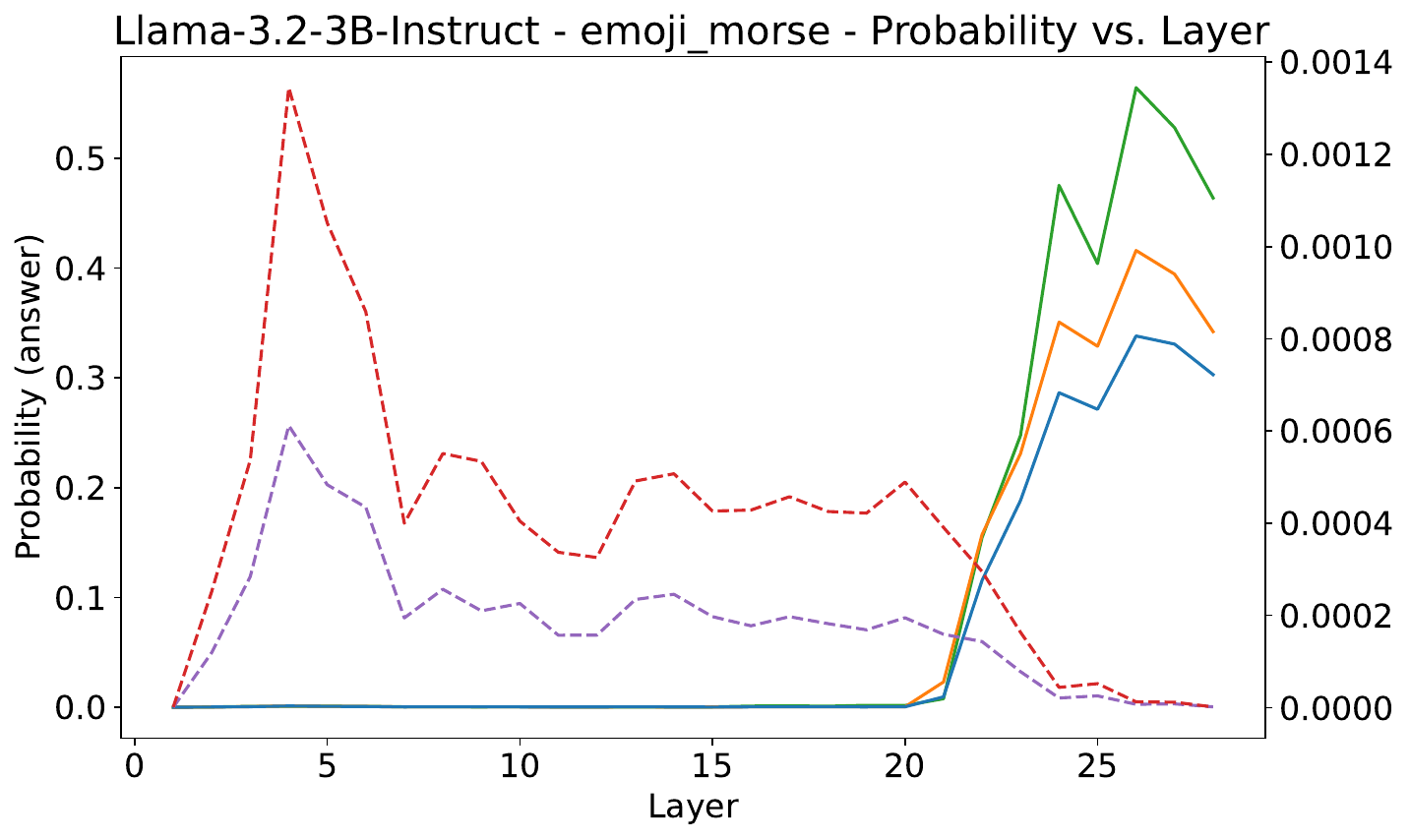}
    }
    \subfigure[The result of Crypto-MMLU-EmojiMorse on Llama-3.1-8B]{
        \includegraphics[width=0.48\textwidth]{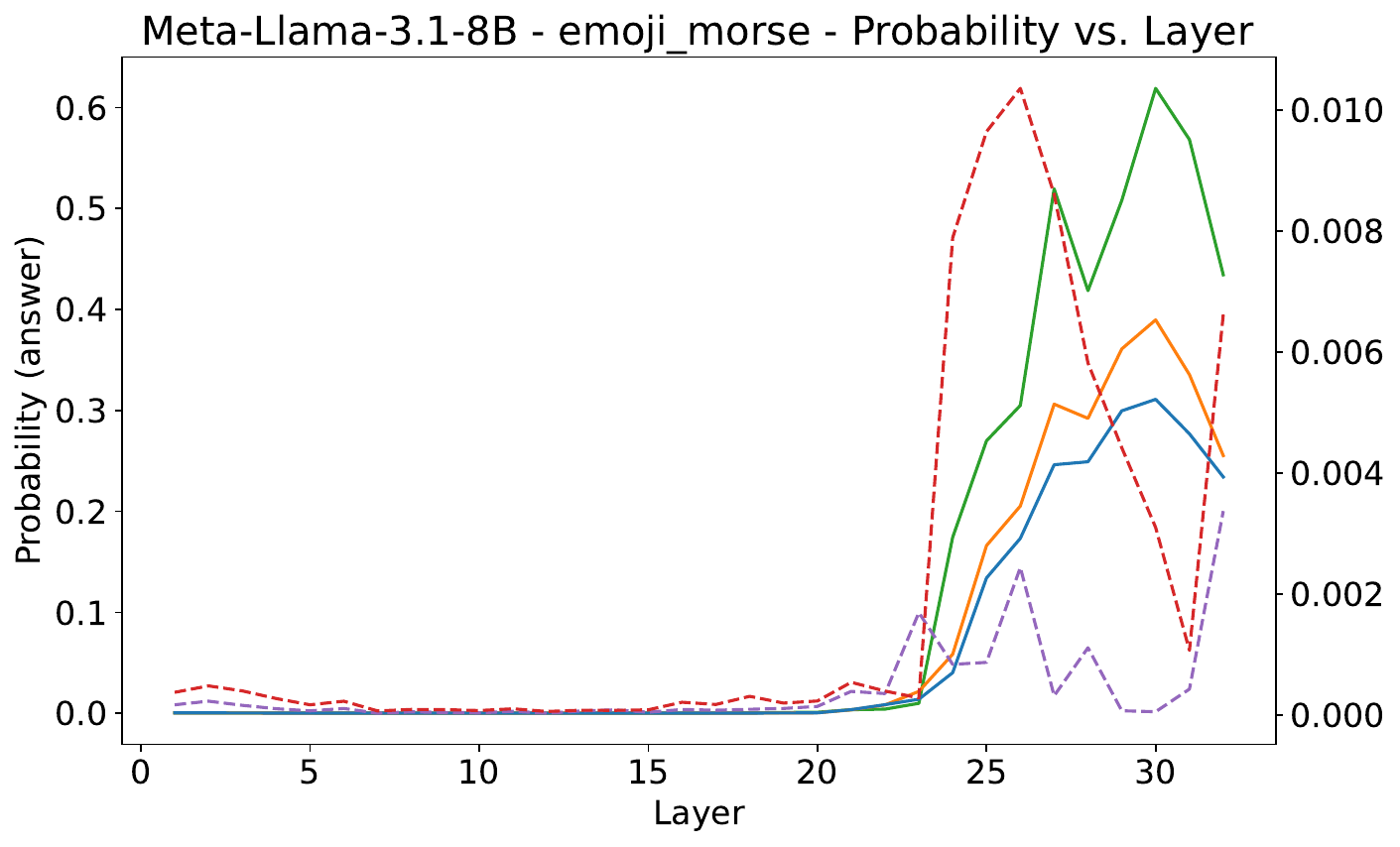}
    }
    \subfigure[The result of Crypto-MMLU-EmojiMorse on Llama-3.1-8B-Instruct]{
        \includegraphics[width=0.48\textwidth]{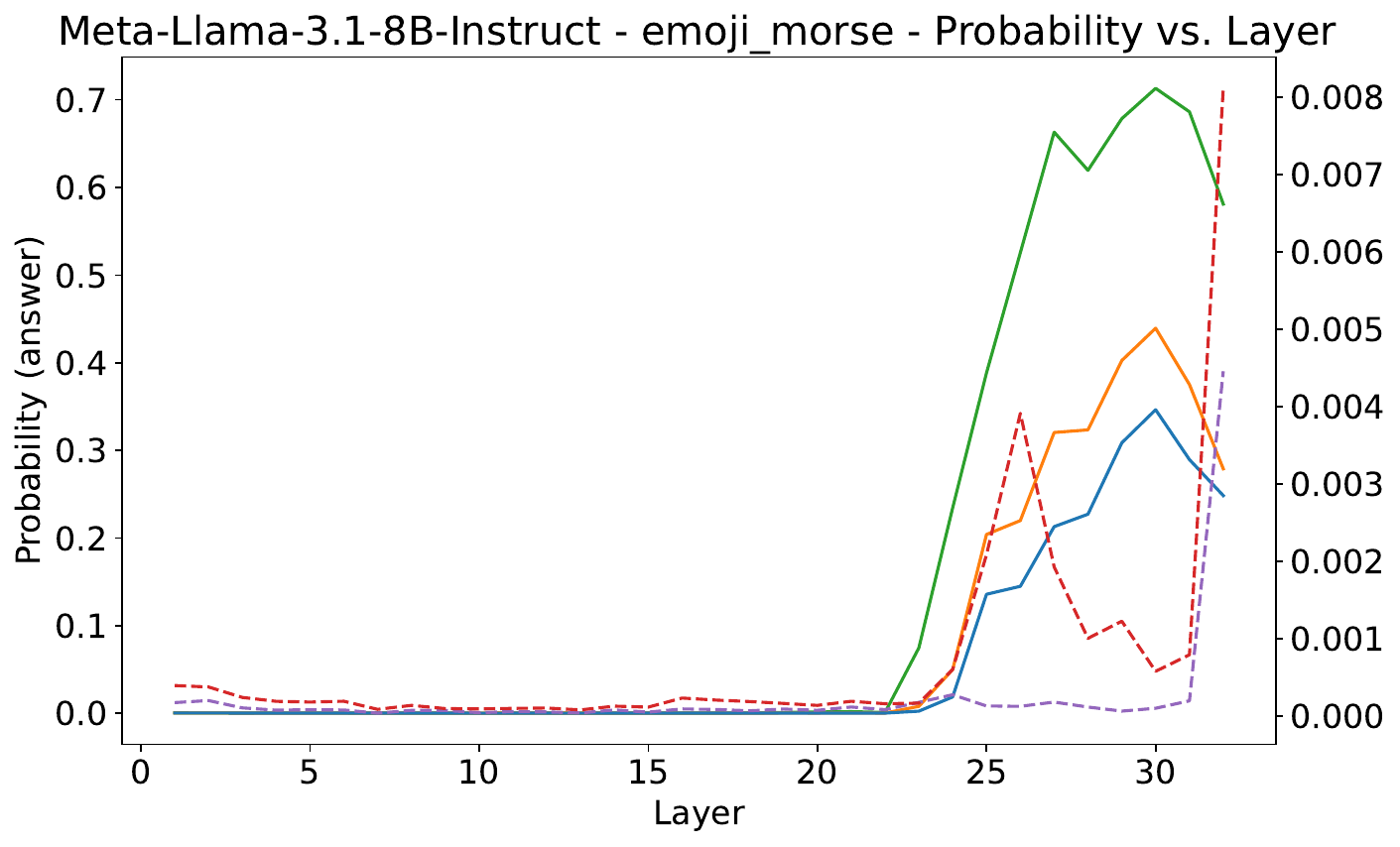}
    }
    \subfigure[The result of Crypto-MMLU-EmojiMorse on Qwen2.5-3B-Instruct]{
        \includegraphics[width=0.48\textwidth]{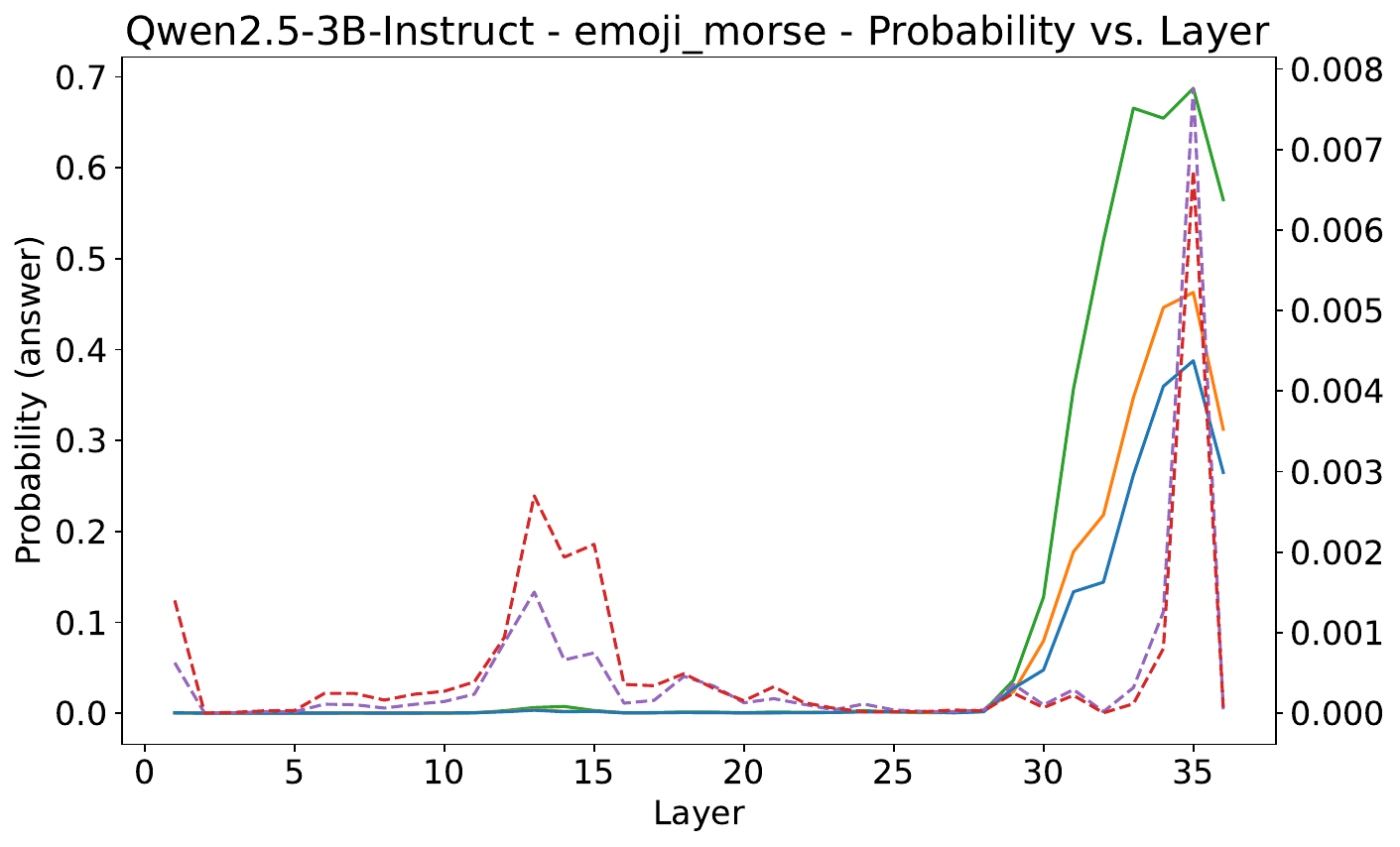}
    }
    \subfigure[The result of Crypto-MMLU-EmojiMorse on Qwen2.5-7B]{
        \includegraphics[width=0.48\textwidth]{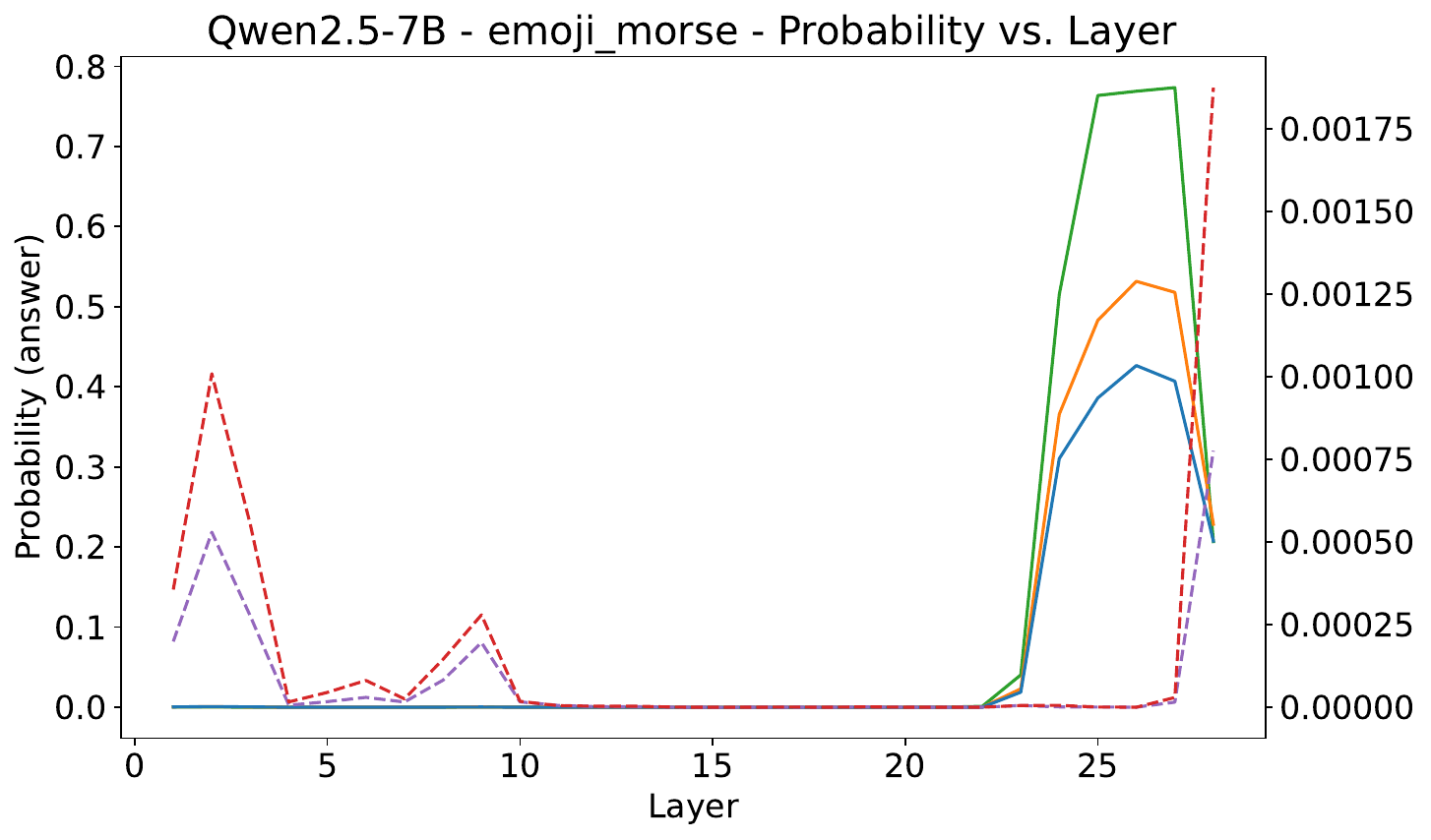}
    }
    \subfigure[The result of Crypto-MMLU-EmojiMorse on Qwen2.5-7B-Instruct]{
        \includegraphics[width=0.48\textwidth]{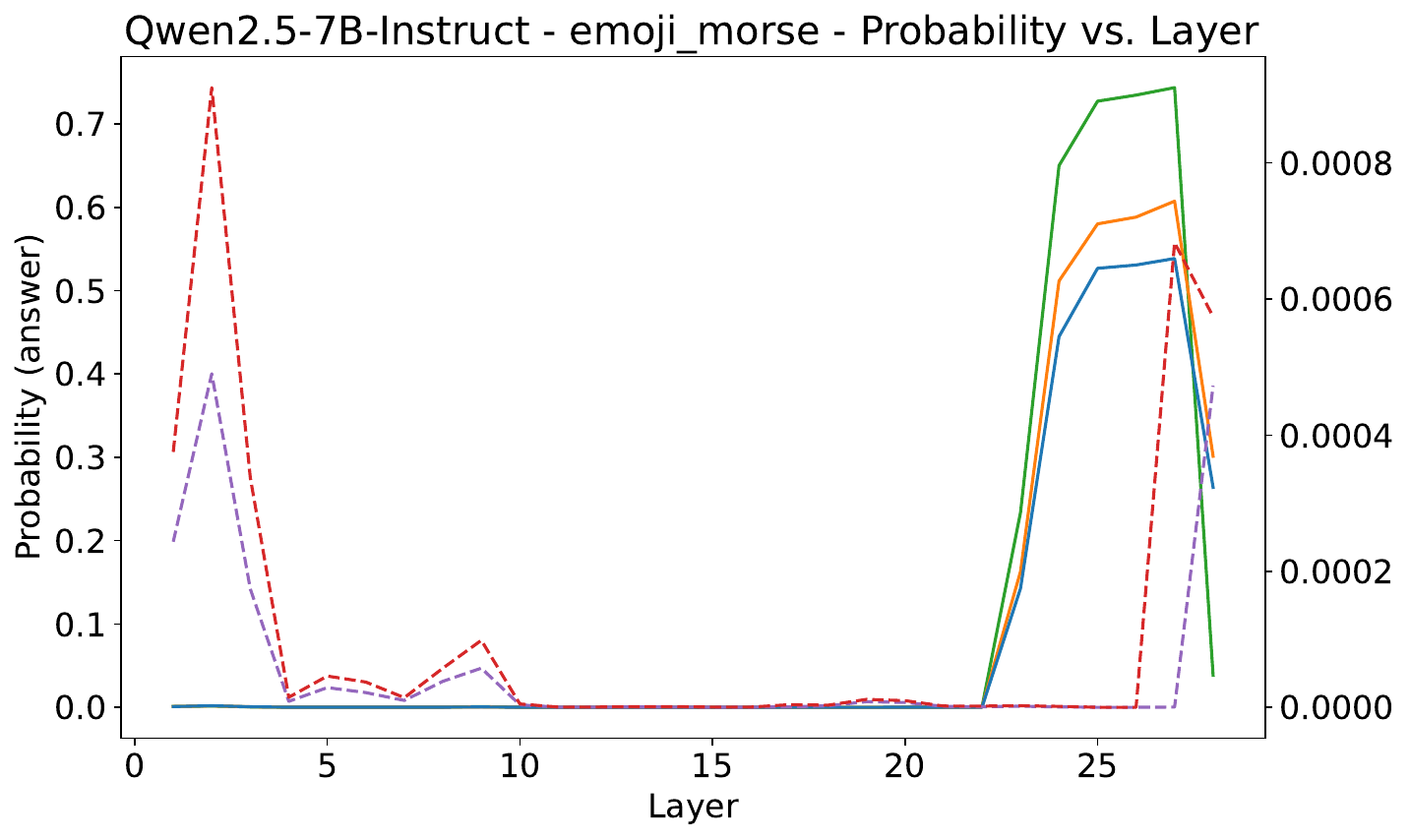}
    }
    \caption{The logit lens analysis on Crypto-MMLU-EmojiMorse using 0\%/50\%/100\% encoding ratios.}
    \label{fig:appendix_logitlens2}
    \vskip -0.2in
\end{figure}
\vfill
\newpage

\subsection{Emoji Shuffle Encoding Rule}
\begin{figure}[H]
    \centering
    \vskip 0.2in
    \subfigure[The result of Crypto-MMLU-EmojiShuffle on Llama-3.2-3B-Instruct]{
        \includegraphics[width=0.48\textwidth]{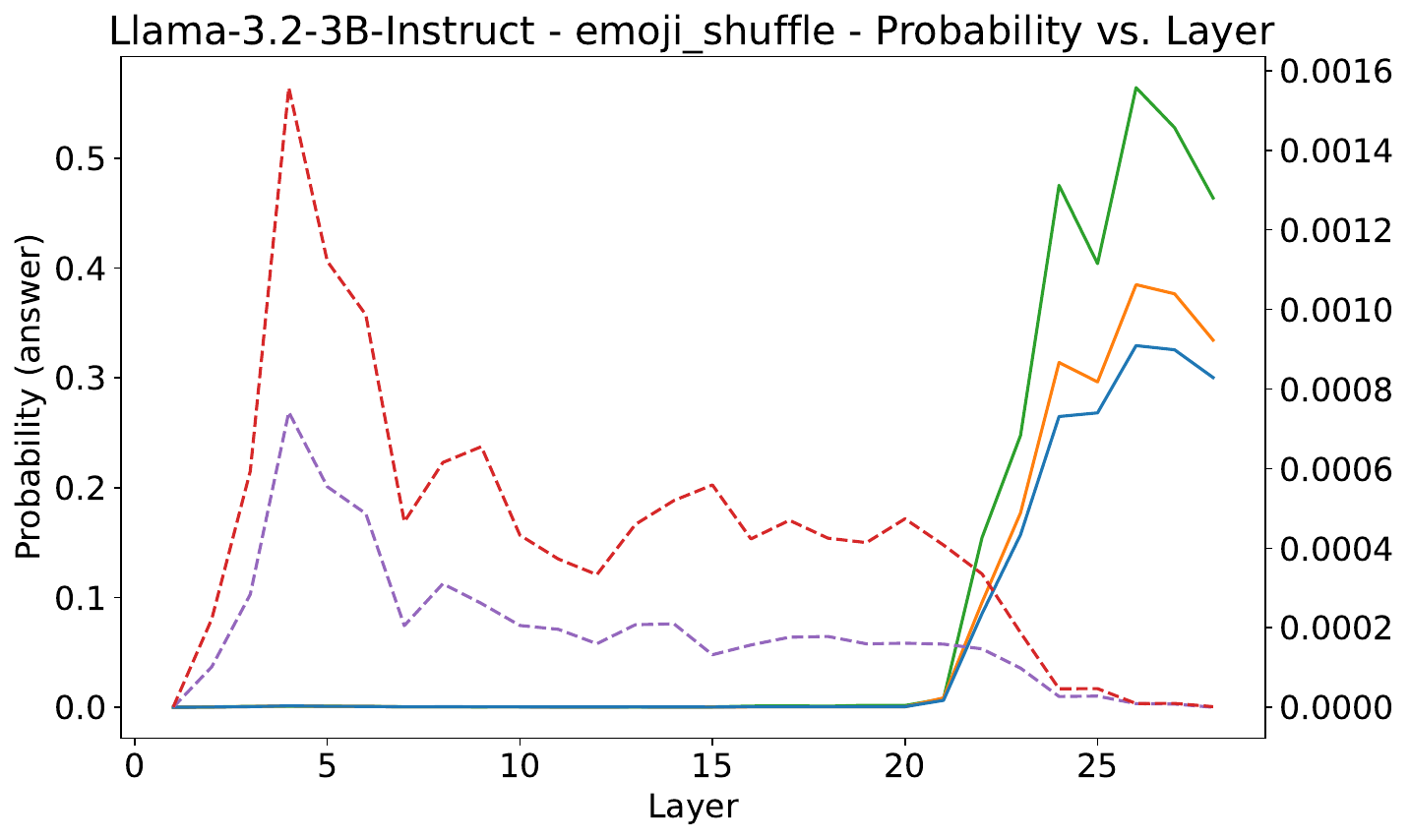}
    }
    \subfigure[The result of Crypto-MMLU-EmojiShuffle on Llama-3.1-8B]{
        \includegraphics[width=0.48\textwidth]{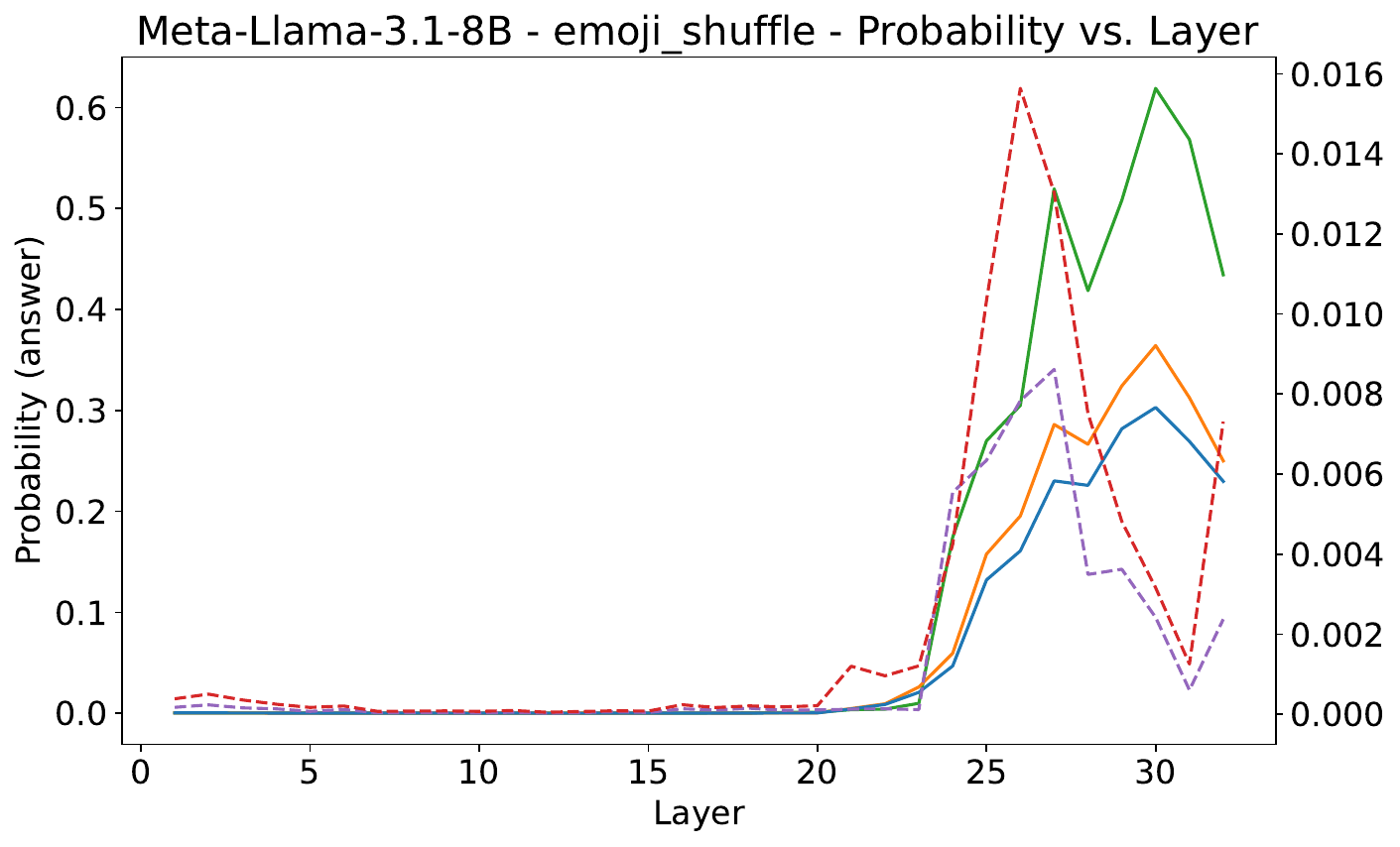}
    }
    \subfigure[The result of Crypto-MMLU-EmojiShuffle on Llama-3.1-8B-Instruct]{
        \includegraphics[width=0.48\textwidth]{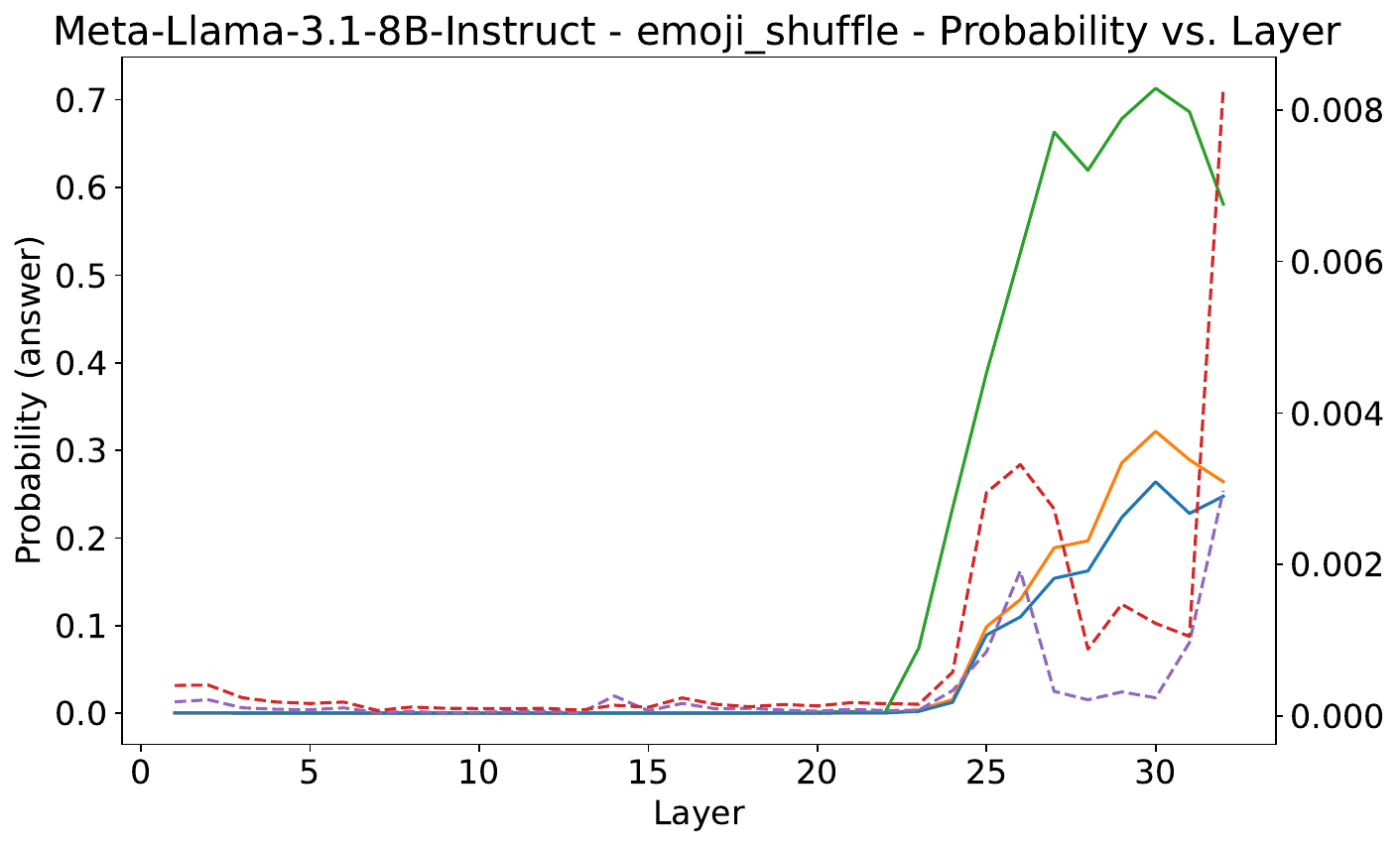}
    }
    \subfigure[The result of Crypto-MMLU-EmojiShuffle on Qwen2.5-3B-Instruct]{
        \includegraphics[width=0.48\textwidth]{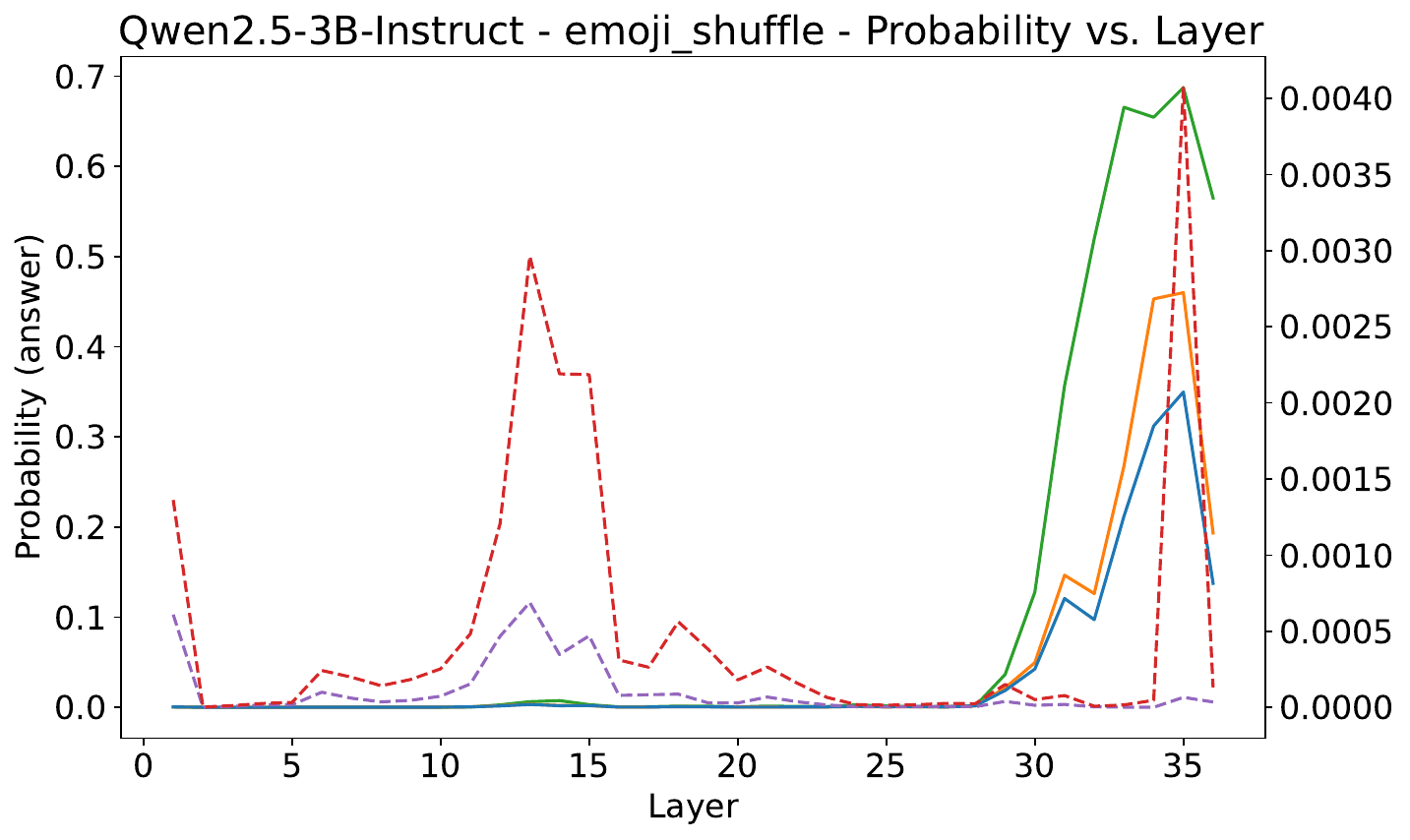}
    }
    \subfigure[The result of Crypto-MMLU-EmojiShuffle on Qwen2.5-7B]{
        \includegraphics[width=0.48\textwidth]{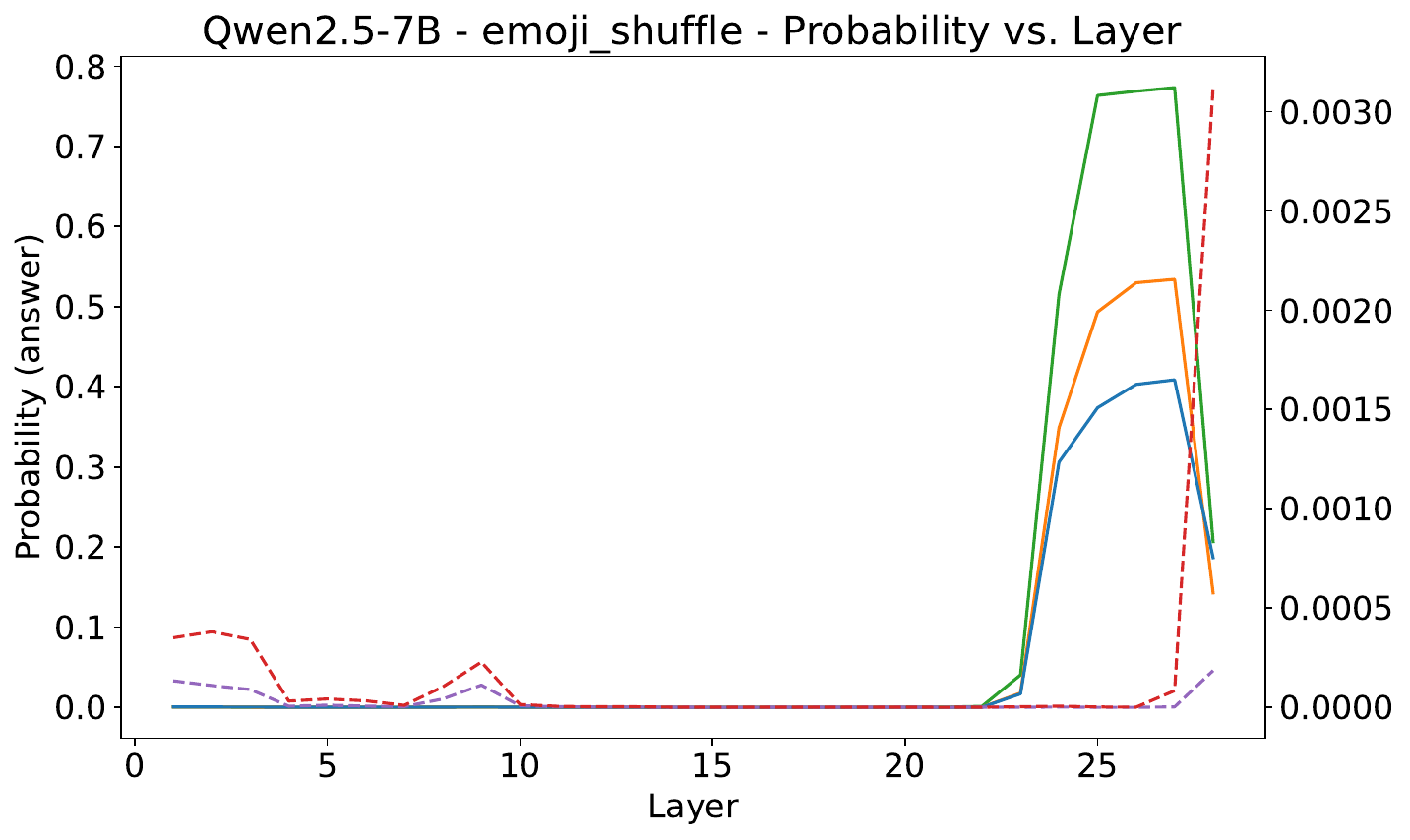}
    }
    \subfigure[The result of Crypto-MMLU-EmojiShuffle on Qwen2.5-7B-Instruct]{
        \includegraphics[width=0.48\textwidth]{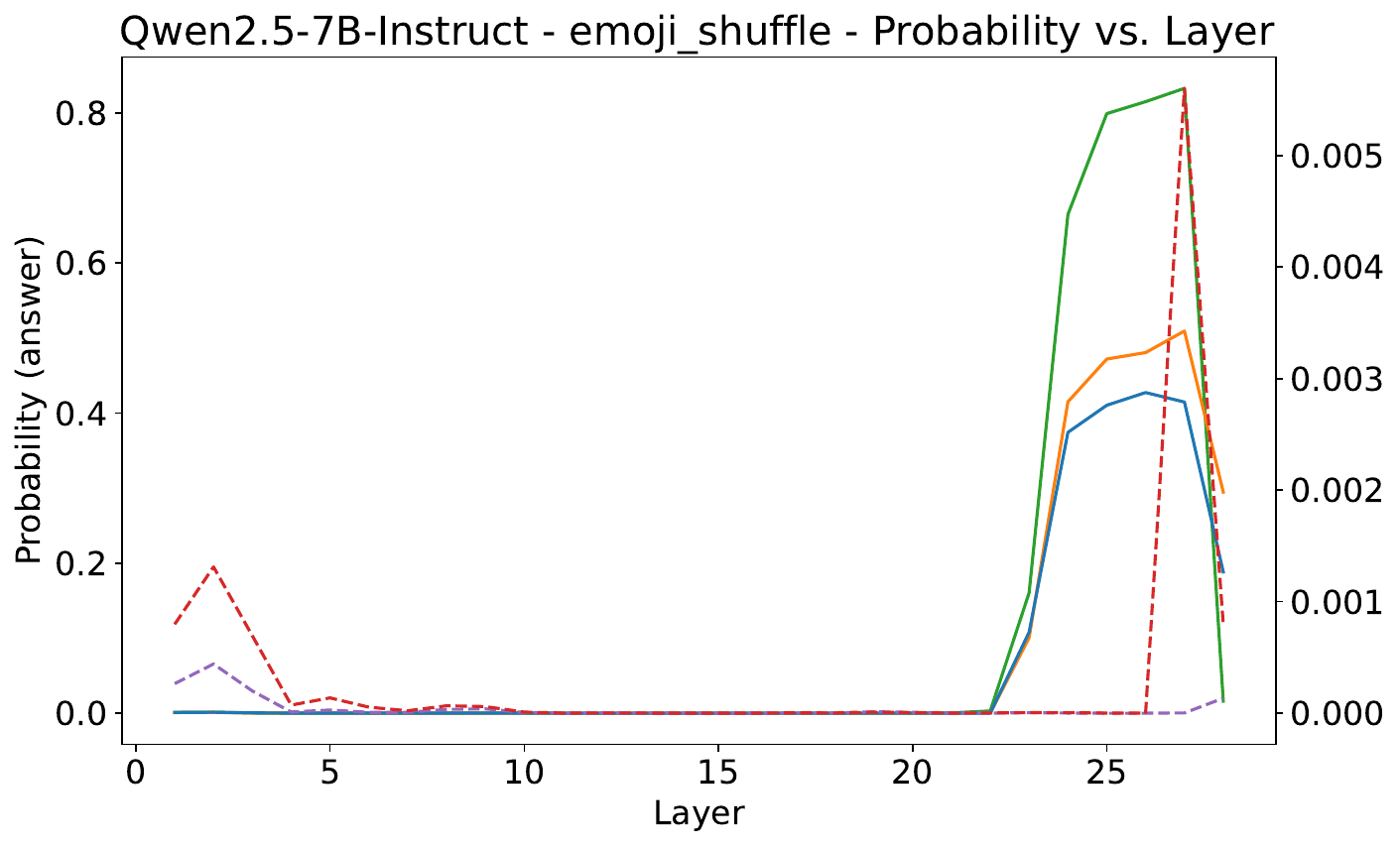}
    }
    \caption{The logit lens analysis on Crypto-MMLU-EmojiShuffle using 0\%/50\%/100\% encoding ratios.}
    \label{fig:appendix_logitlens3}
    \vskip -0.2in
\end{figure}
\vfill
\newpage

\section{Logit Lens Analysis(0/3/5 encoding words)}
Since sections \ref{sec:analysis} primarily utilizes the emoji shuffle encoding method with 0/3/5 words encoding, this section will present and supplement the entire experiments result using 0/3/5 encoding words, which provides more comprehensive evidence supporting the conclusions presented in the section  \ref{sec:analysis}.
\subsection{Base Morse Encoding Rule}

\begin{figure}[H]
    \centering
    \vskip 0.2in
    \subfigure[The result of Crypto-MMLU-BaseMorse on Llama-3.1-8B]{
        \includegraphics[width=0.48\textwidth]{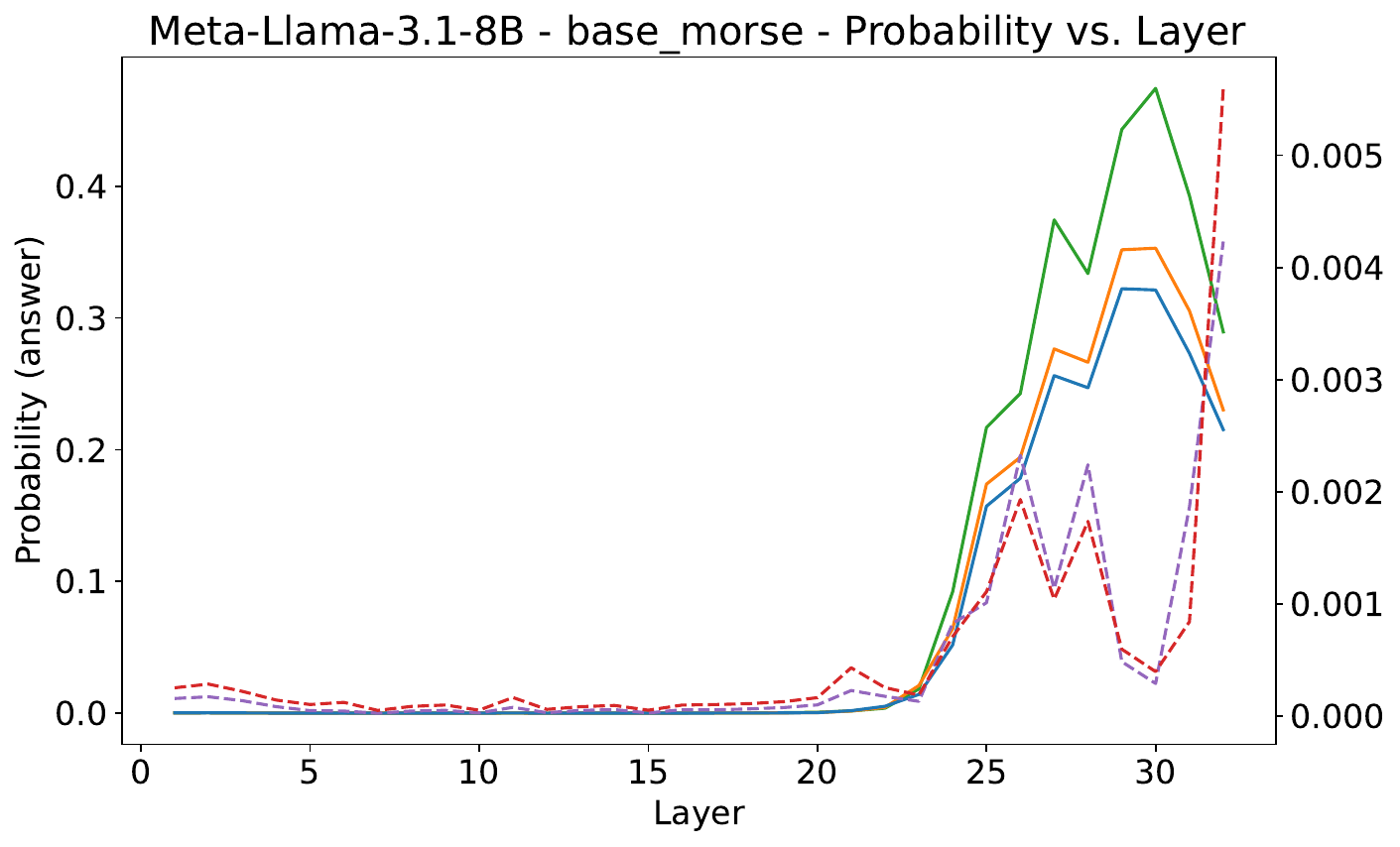}
    }
    \subfigure[The result of Crypto-MMLU-BaseMorse on Llama-3.1-8B-Instruct]{
        \includegraphics[width=0.48\textwidth]{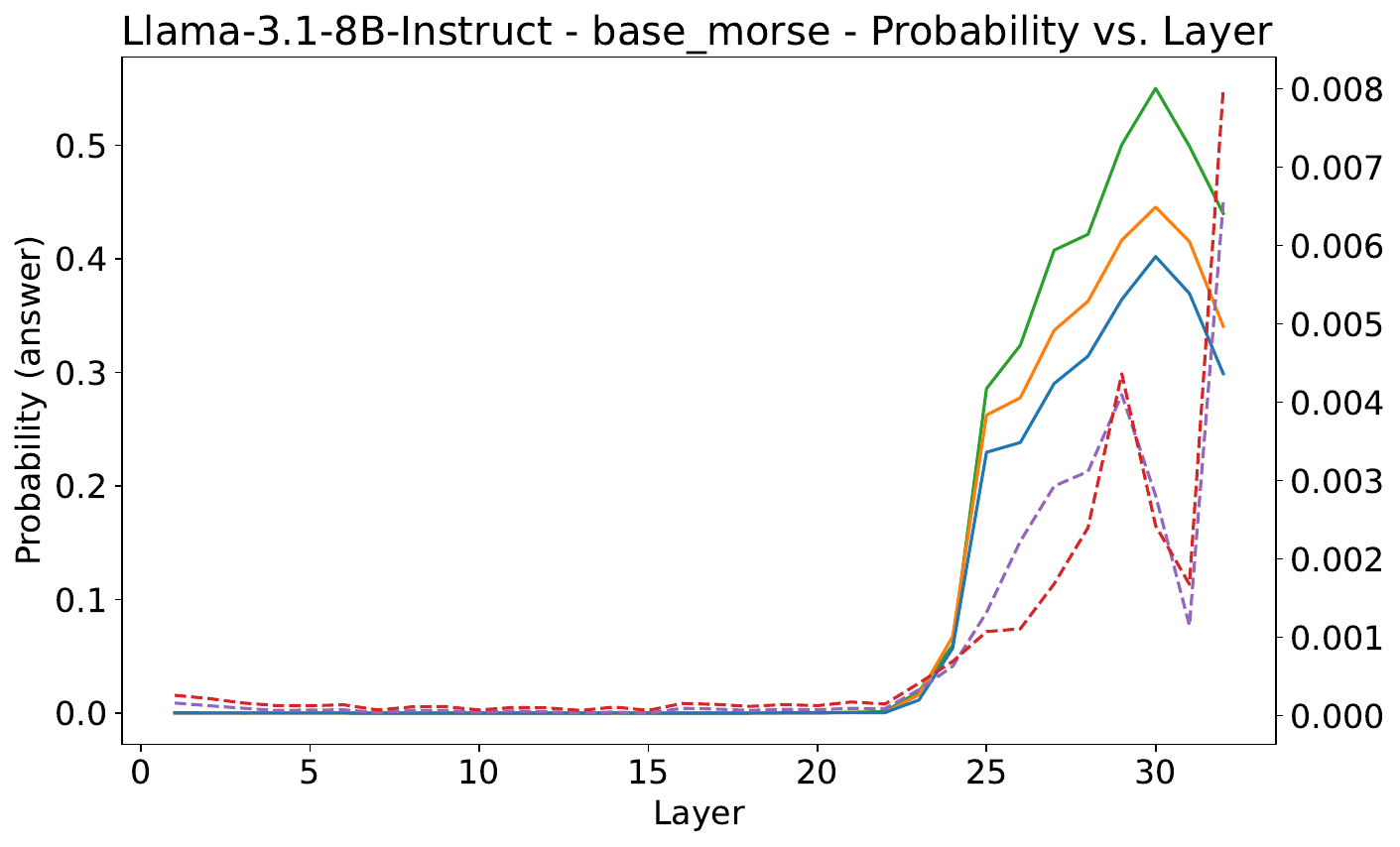}
    }
    \subfigure[The result of Crypto-MMLU-BaseMorse on Qwen2.5-0.5B-Instruct]{
        \includegraphics[width=0.48\textwidth]{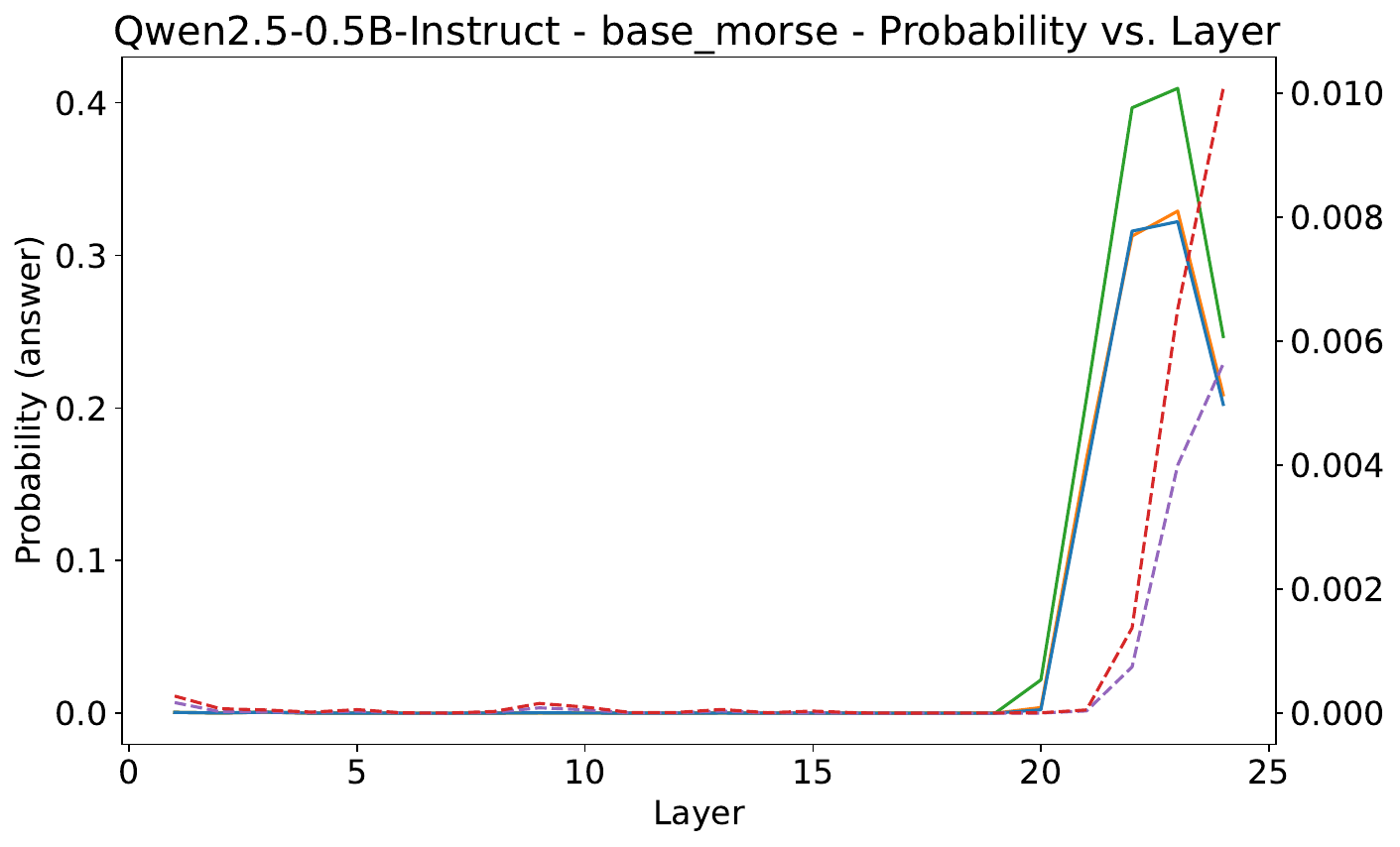}
    }
    \subfigure[The result of Crypto-MMLU-BaseMorse on Qwen2.5-1.5B-Instruct]{
        \includegraphics[width=0.48\textwidth]{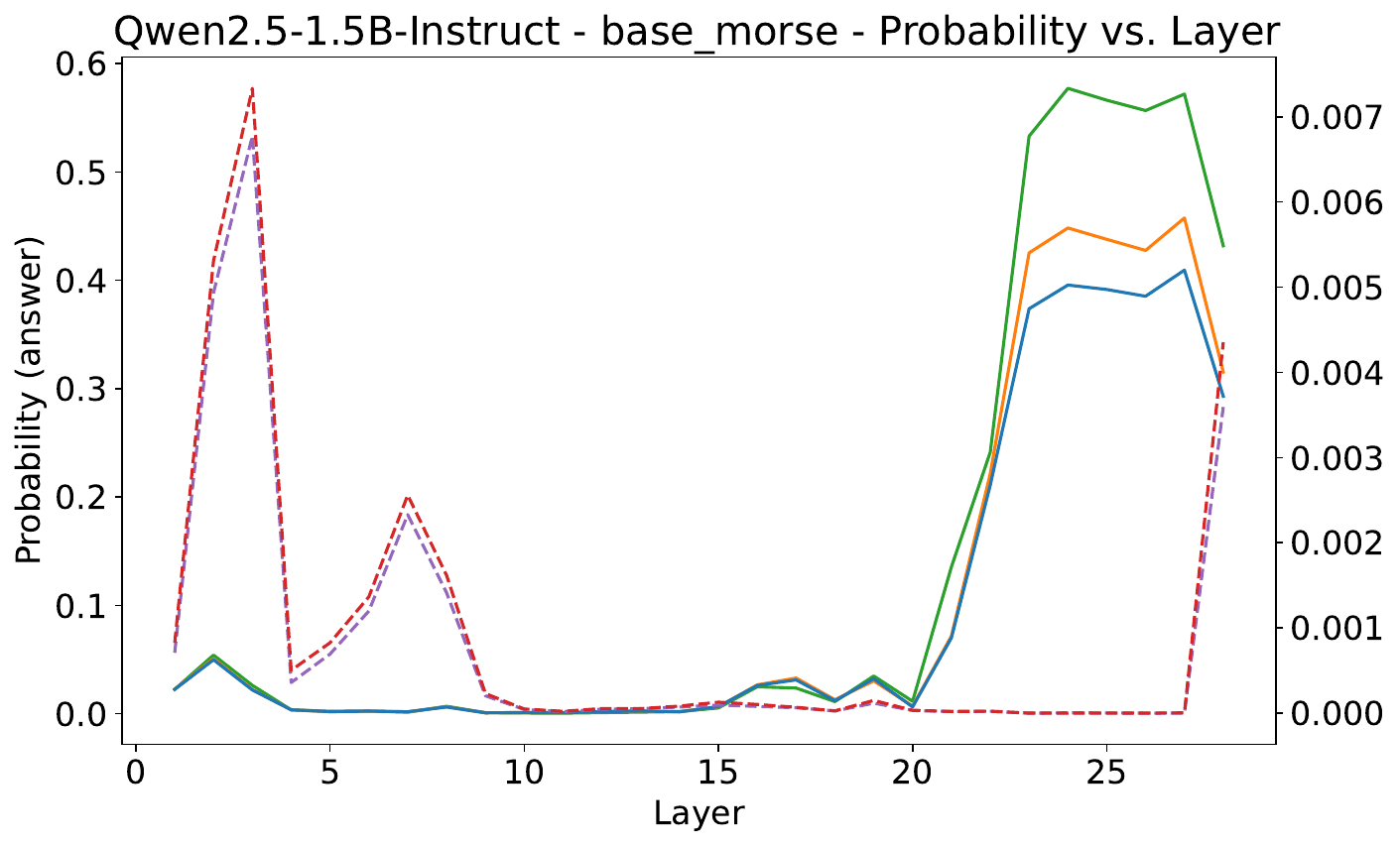}
    }
    \subfigure[The result of Crypto-MMLU-BaseMorse on Qwen2.5-3B]{
        \includegraphics[width=0.48\textwidth]{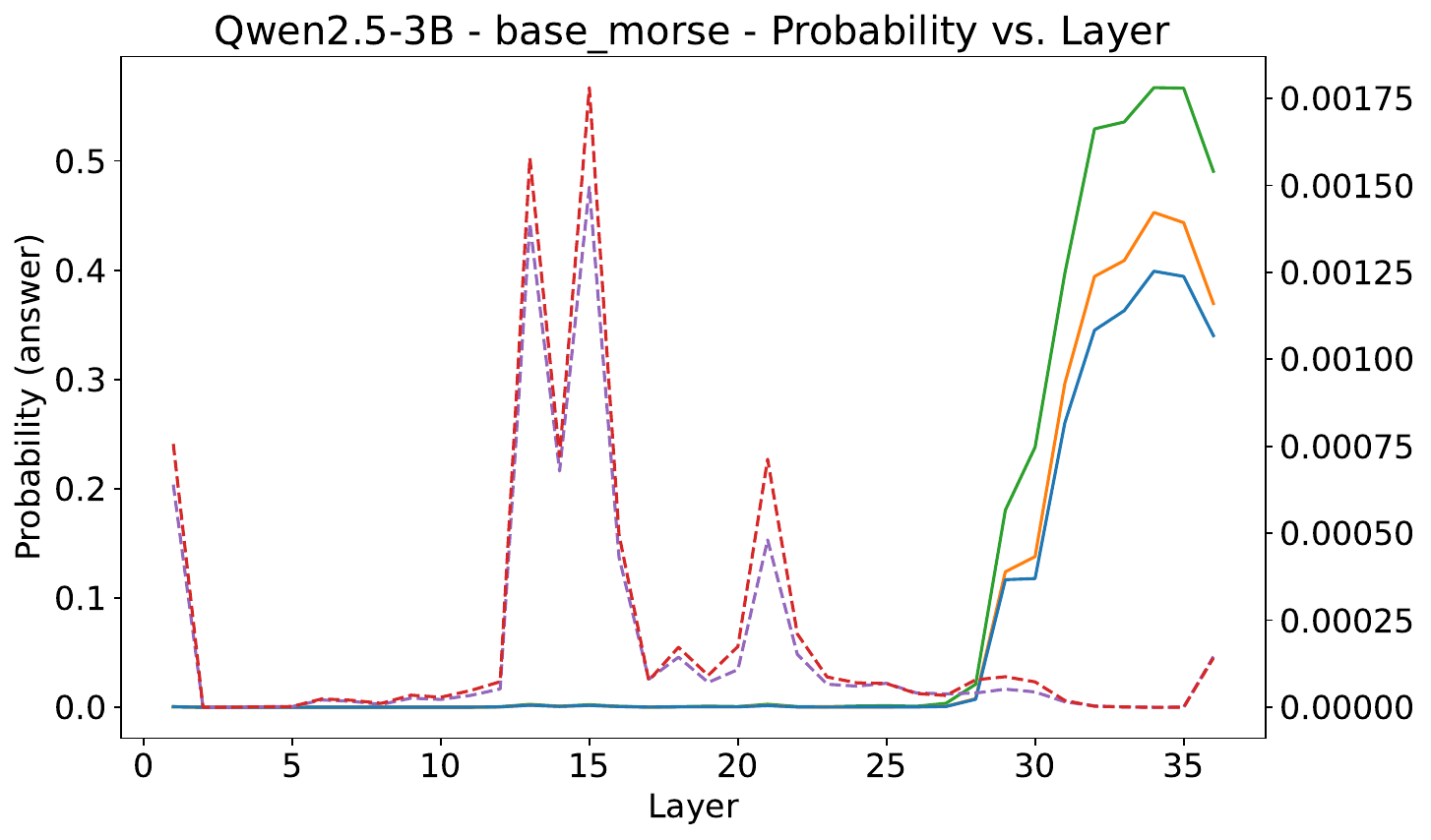}
    }
    \subfigure[The result of Crypto-MMLU-BaseMorse on Qwen2.5-3B-Instruct]{
        \includegraphics[width=0.48\textwidth]{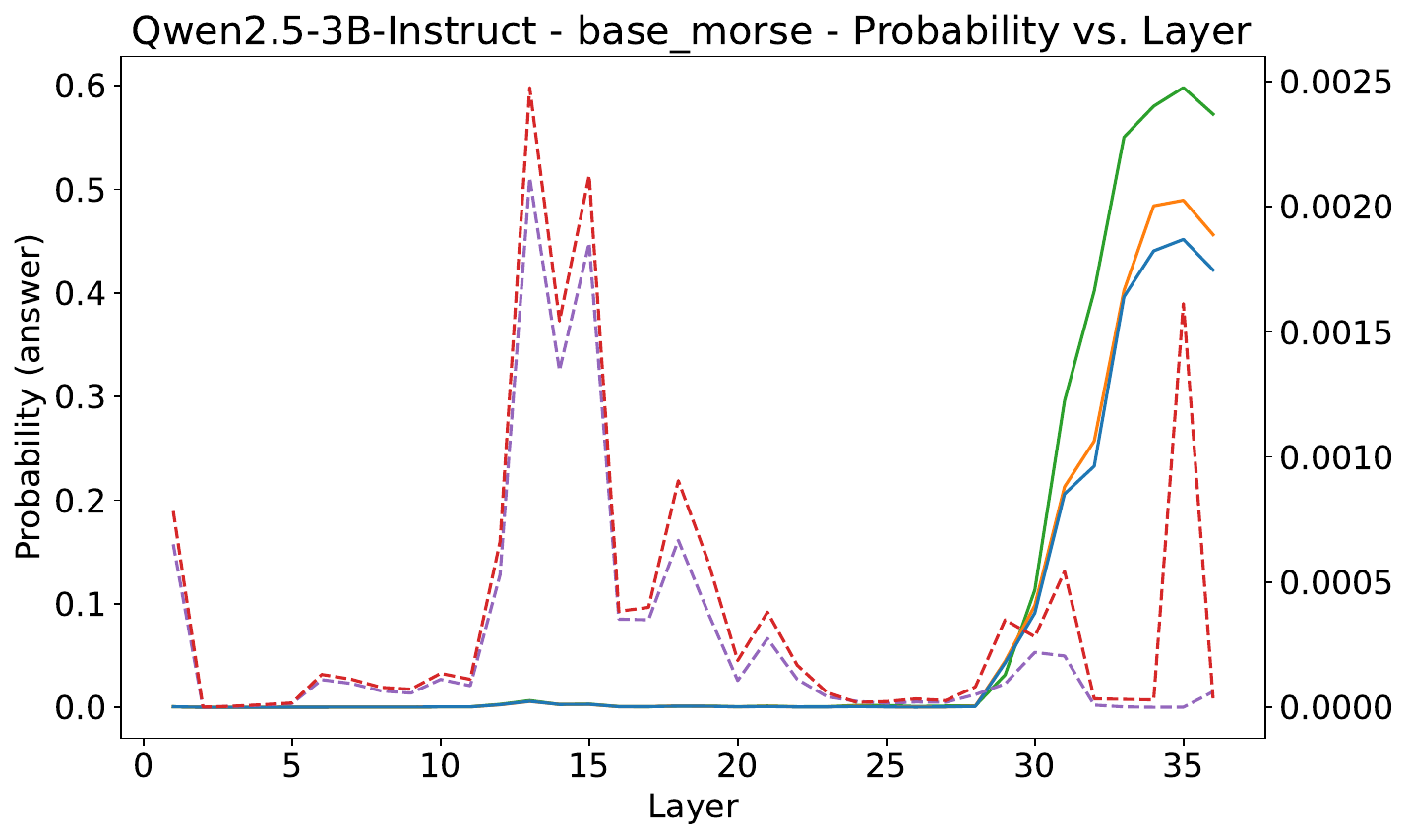}
    }
    \caption{The logit lens analysis on Crypto-MMLU-BaseMorse using 0/3/5 encoding words.}
    \label{fig:appendix_logitlens4}
    \vskip -0.2in
\end{figure}
\vfill

\newpage
\begin{figure}[H]
    \centering
    \vskip 0.2in
    \subfigure[The result of Crypto-MMLU-BaseMorse on Qwen2.5-7B]{
        \includegraphics[width=0.48\textwidth]{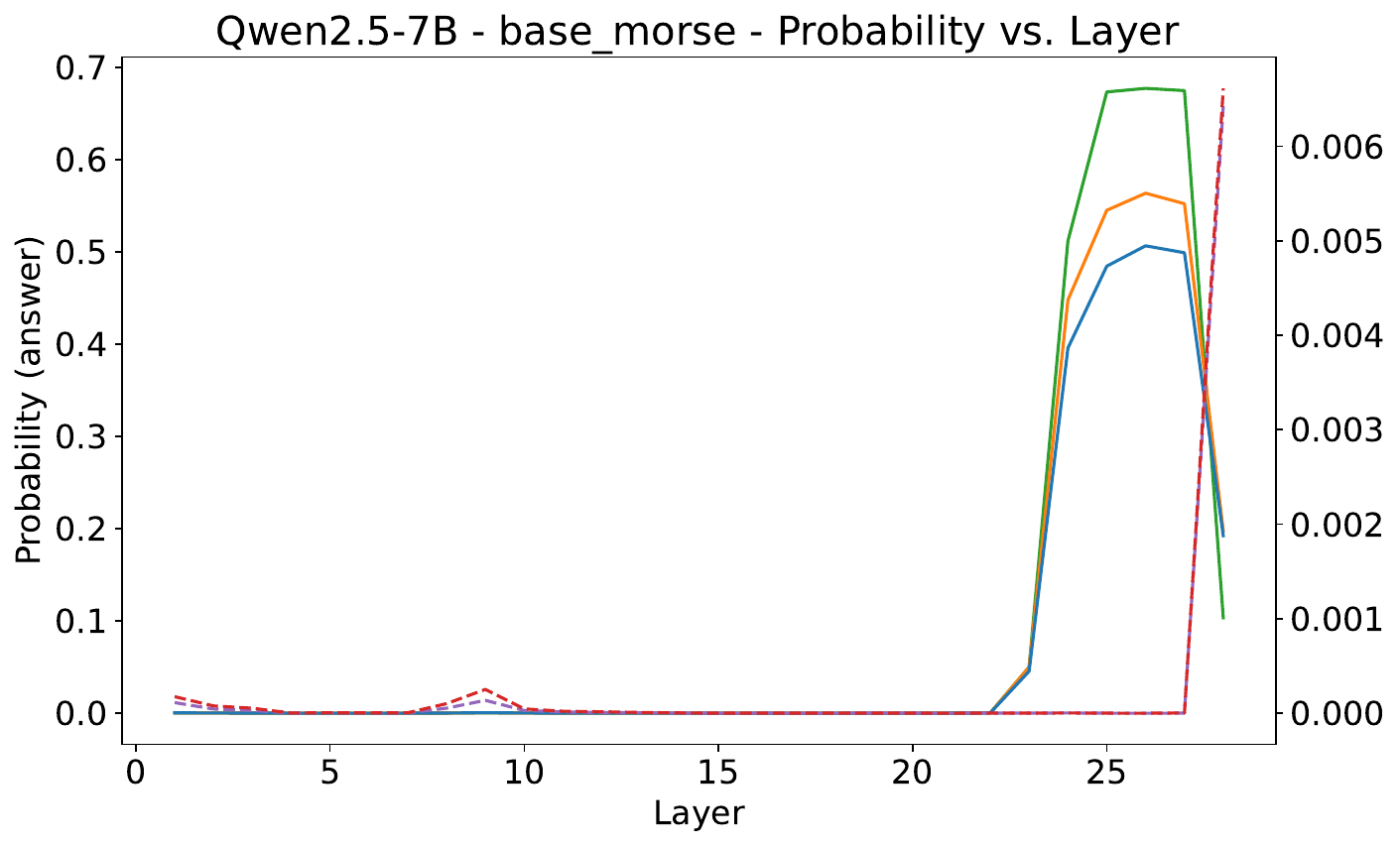}
    }
    \subfigure[The result of Crypto-MMLU-BaseMorse on Qwen2.5-7B-Instruct]{
        \includegraphics[width=0.48\textwidth]{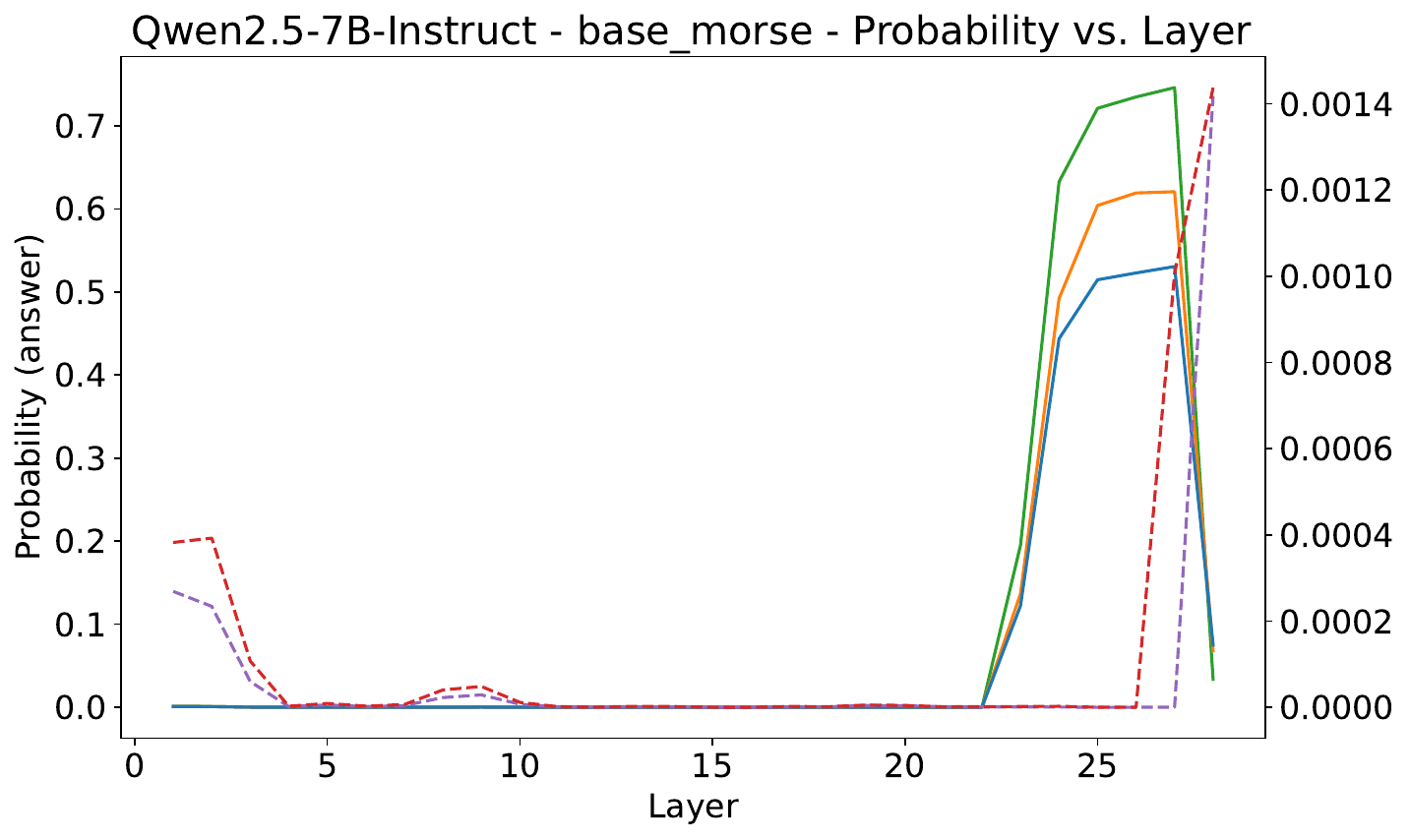}
    }
    \caption{The logit lens analysis on Crypto-MMLU-BaseMorse using 0/3/5 encoding words.}
    \label{fig:appendix_logitlens5}
    \vskip -0.2in
\end{figure}
\vfill

\subsection{Emoji Morse Encoding Rule}

\begin{figure}[H]
    \centering
    \vskip 0.2in
    \subfigure[The result of Crypto-MMLU-EmojiMorse on Llama-3.1-8B-Instruct]{
        \includegraphics[width=0.48\textwidth]{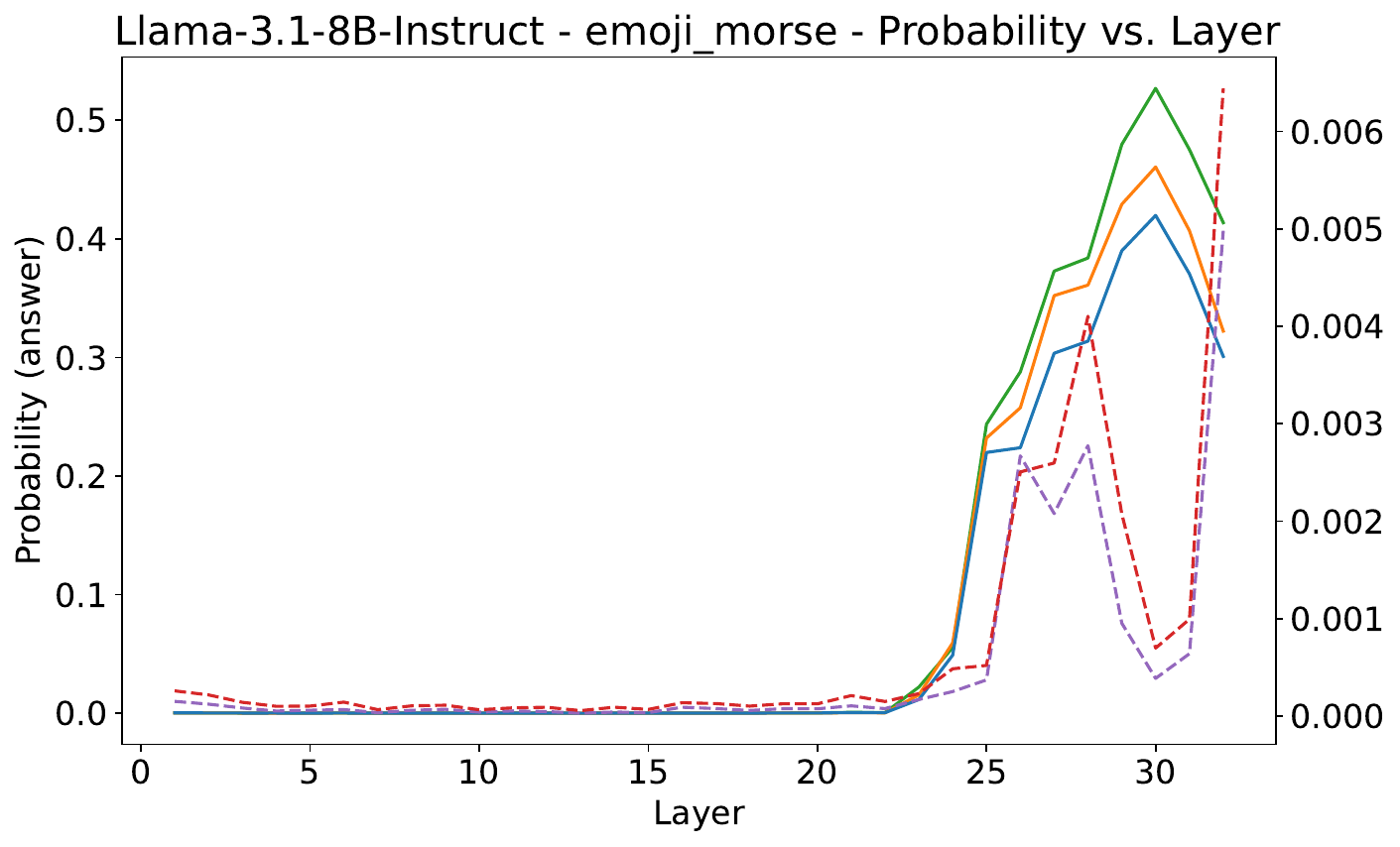}
    }
    \subfigure[The result of Crypto-MMLU-EmojiMorse on Qwen2.5-0.5B-Instruct]{
        \includegraphics[width=0.48\textwidth]{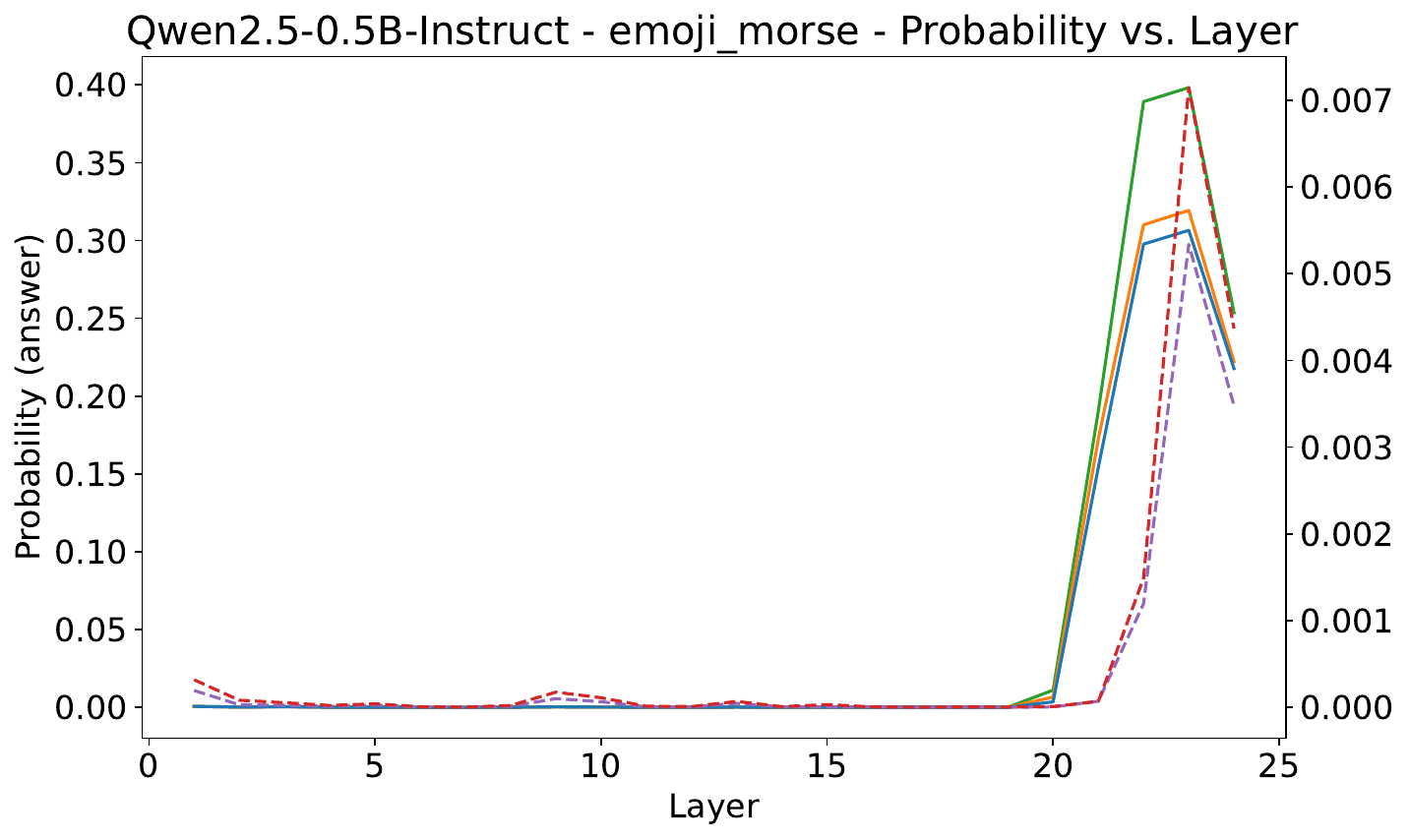}
    }
    \subfigure[The result of Crypto-MMLU-EmojiMorse on Qwen2.5-3B-Instruct]{
        \includegraphics[width=0.48\textwidth]{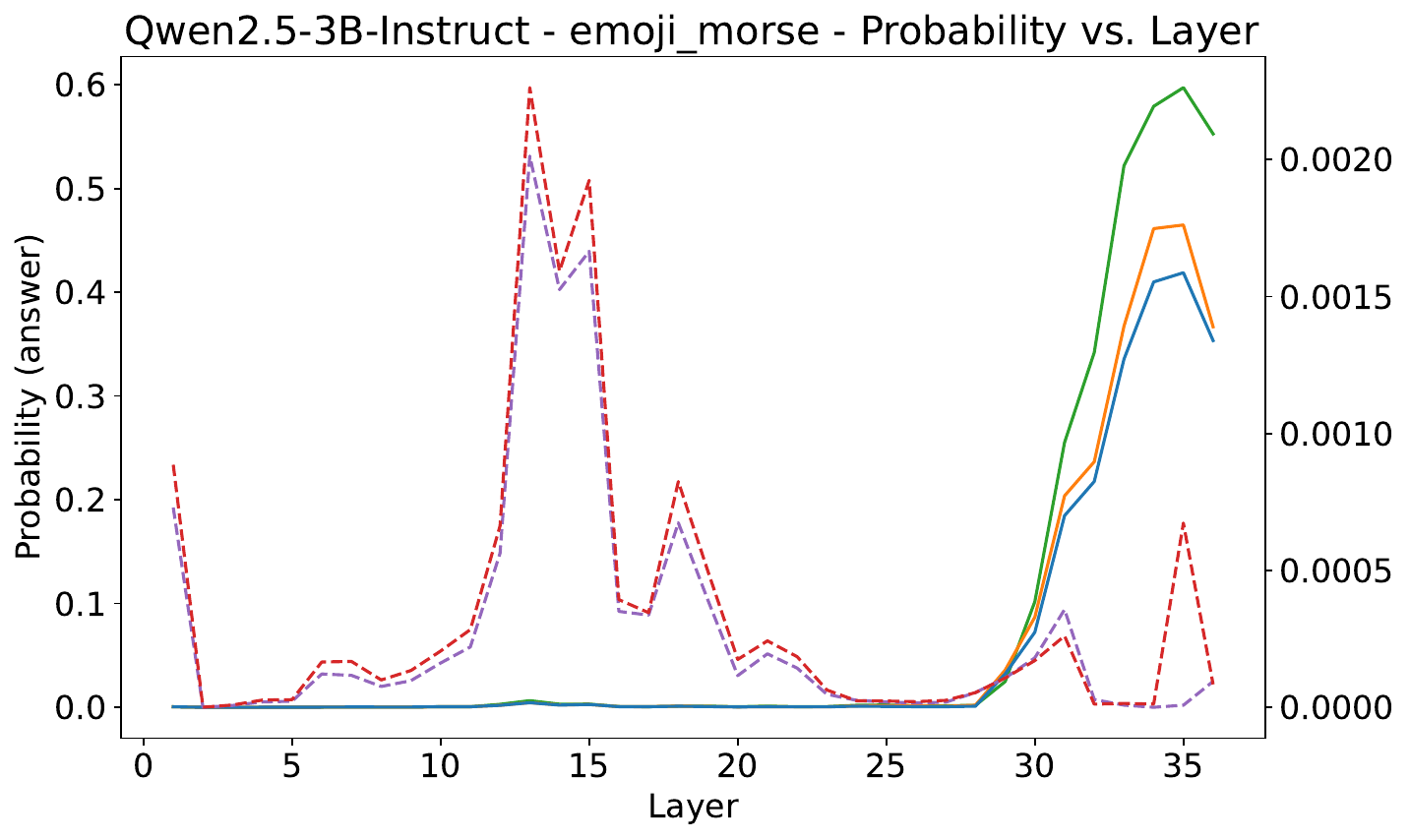}
    }
    \subfigure[The result of Crypto-MMLU-EmojiMorse on Qwen2.5-7B-Instruct]{
        \includegraphics[width=0.48\textwidth]{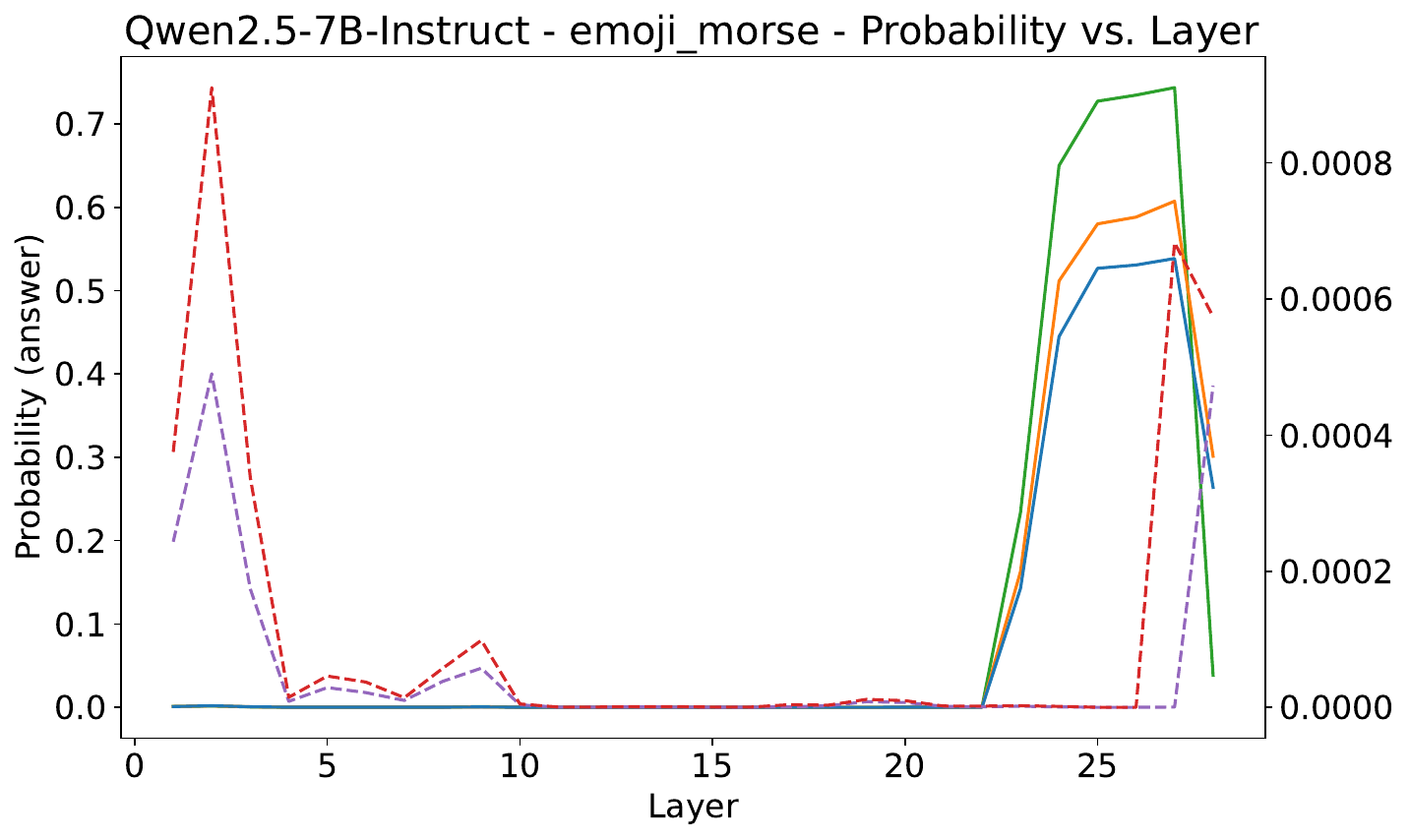}
    }
    \caption{The logit lens analysis on Crypto-MMLU-EmojiMorse using 0/3/5 encoding words.}
    \label{fig:appendix_logitlens6}
    \vskip -0.2in
\end{figure}
\vfill

\newpage
\subsection{Emoji Shuffle Encoding Rule}

\begin{figure}[H]
    \centering
    \vskip 0.2in
    \subfigure[The result of Crypto-MMLU-EmojiShuffle on Qwen2.5-0.5B-Instruct]{
        \includegraphics[width=0.48\textwidth]{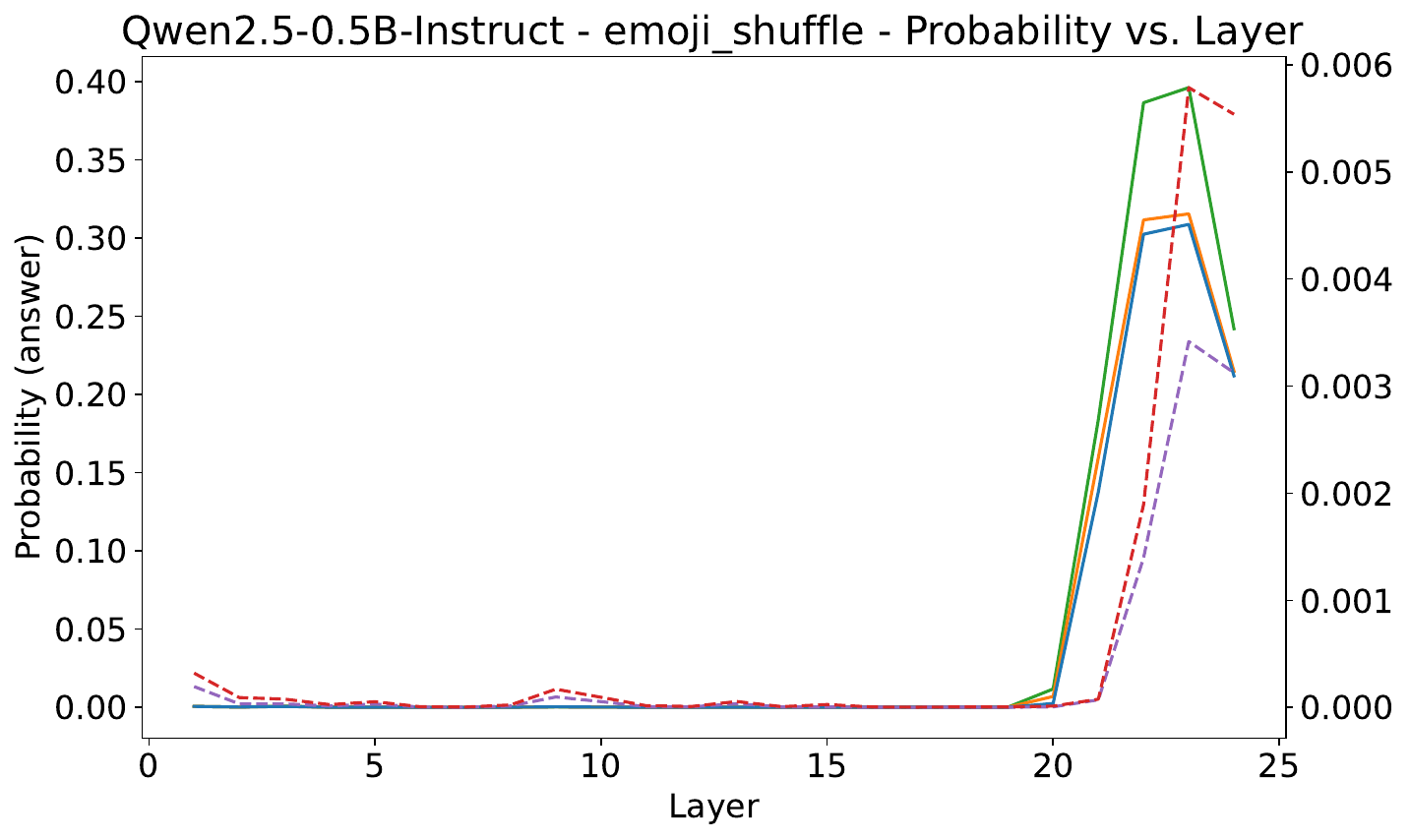}
    }
    \subfigure[The result of Crypto-MMLU-EmojiShuffle on Qwen2.5-1.5B-Instruct]{
        \includegraphics[width=0.48\textwidth]{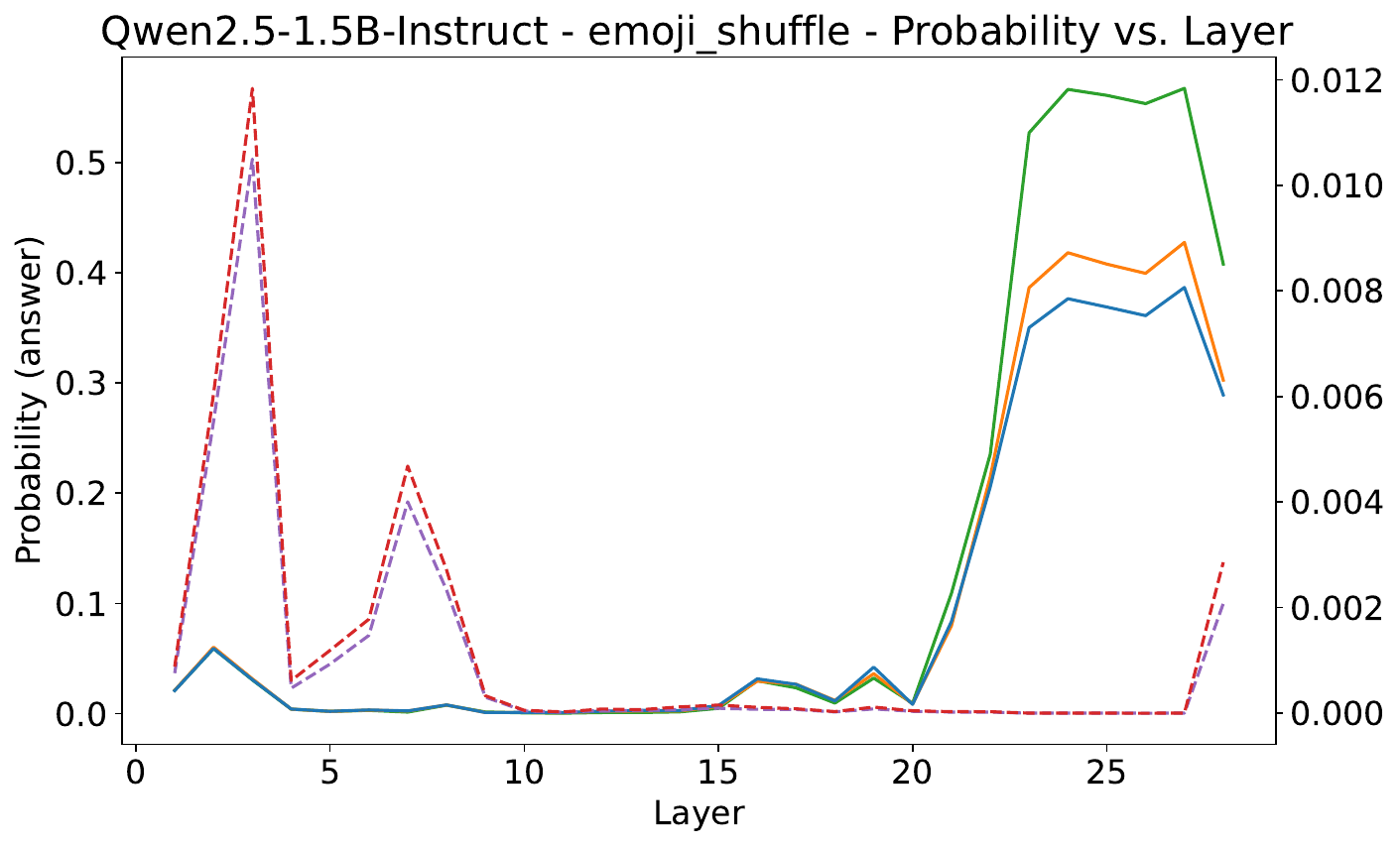}
    }

    \caption{The logit lens analysis on Crypto-MMLU-EmojiShuffle using 0/3/5 encoding words.}
    \label{fig:appendix_logitlens7}
    \vskip -0.2in
\end{figure}
\vfill

\section{Neuron Activation Analysis}
\label{Appendix_neuron}
The appendix \ref{Appendix_neuron} shows result of Neuron Activation Analysis not presented in section \ref{sec:analysis}.
\begin{figure}[H]
    \centering
    \vskip 0.2in
    \subfigure[The result of Neuron Activation on Llama-3.1-8B]{
        \includegraphics[width=0.48\textwidth]{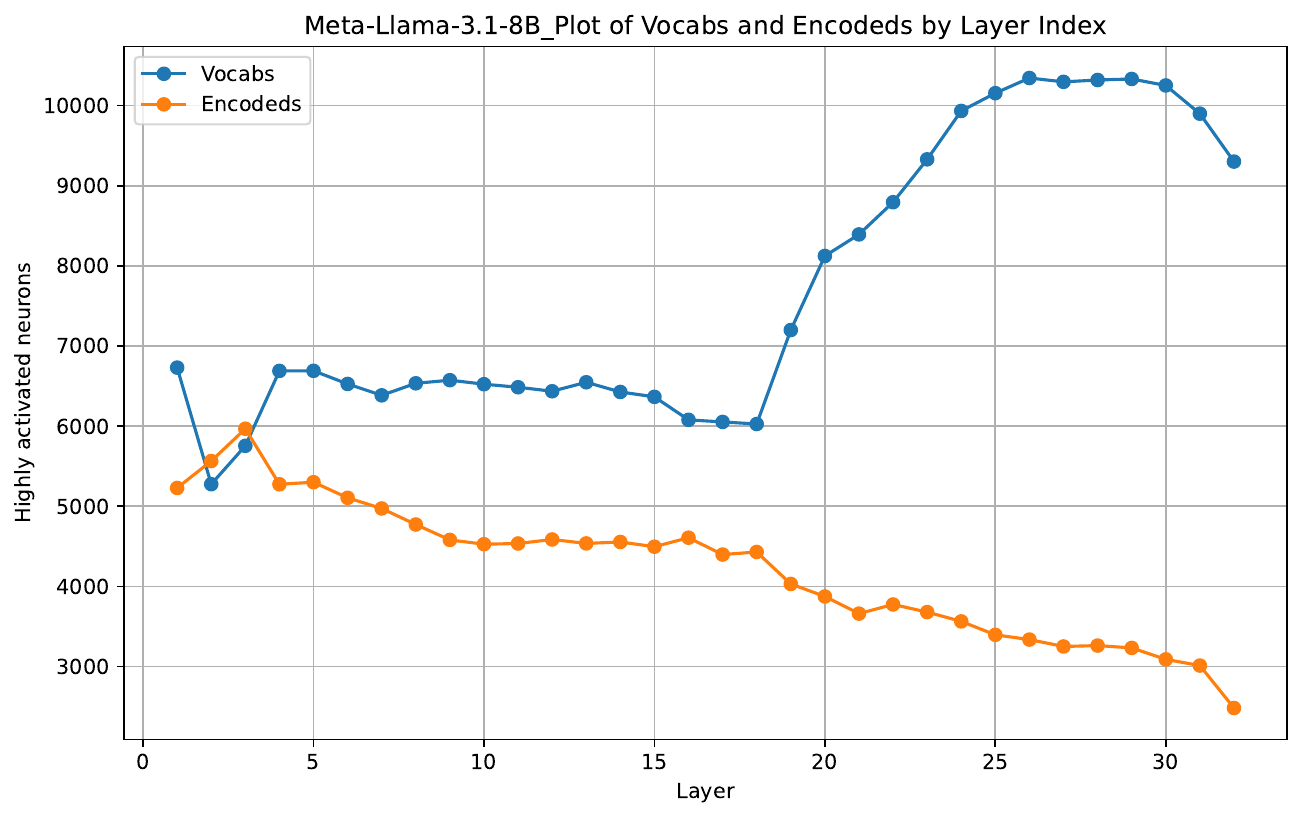}
    }
    \subfigure[The result of Neuron Activation on Qwen2.5-3B]{
        \includegraphics[width=0.48\textwidth]{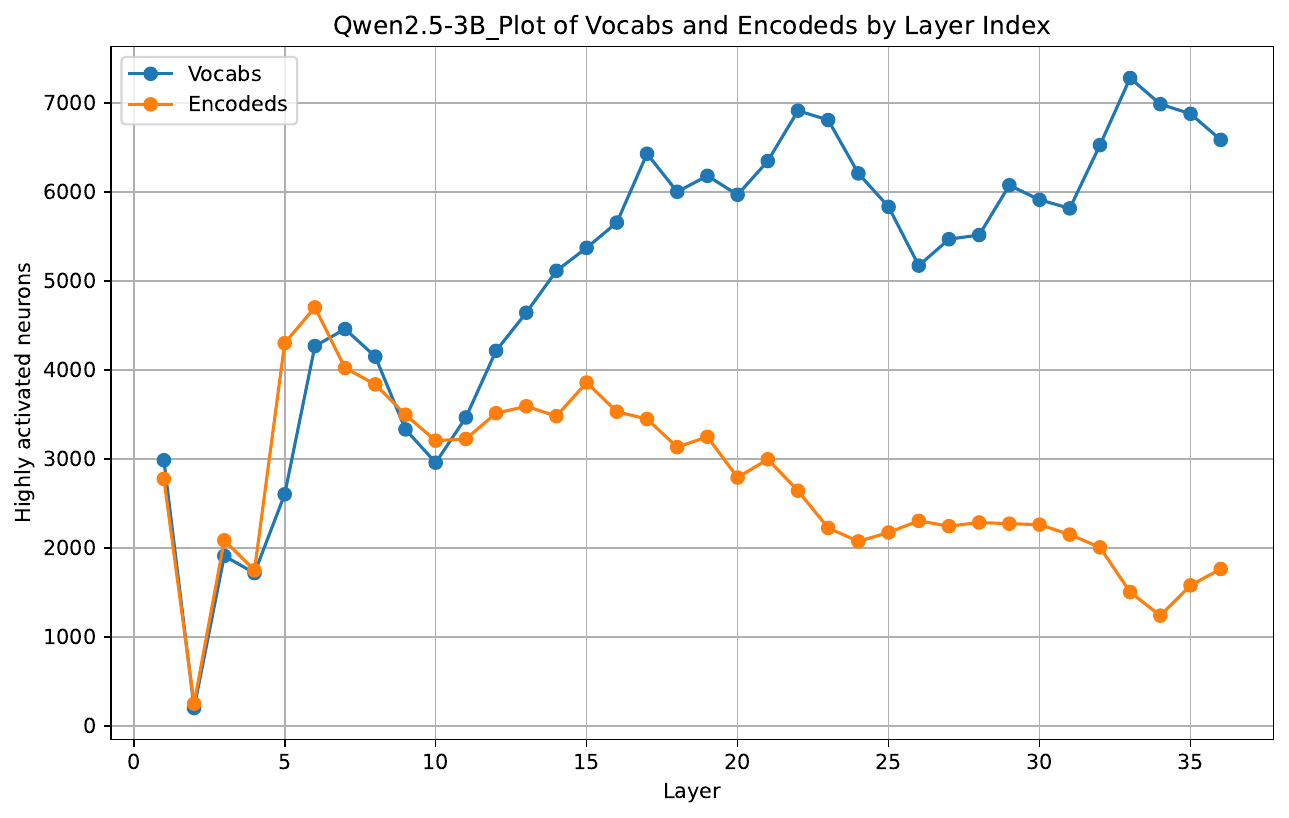}
    }
    \subfigure[The result of Neuron Activation on Qwen2.5-7B]{
        \includegraphics[width=0.48\textwidth]{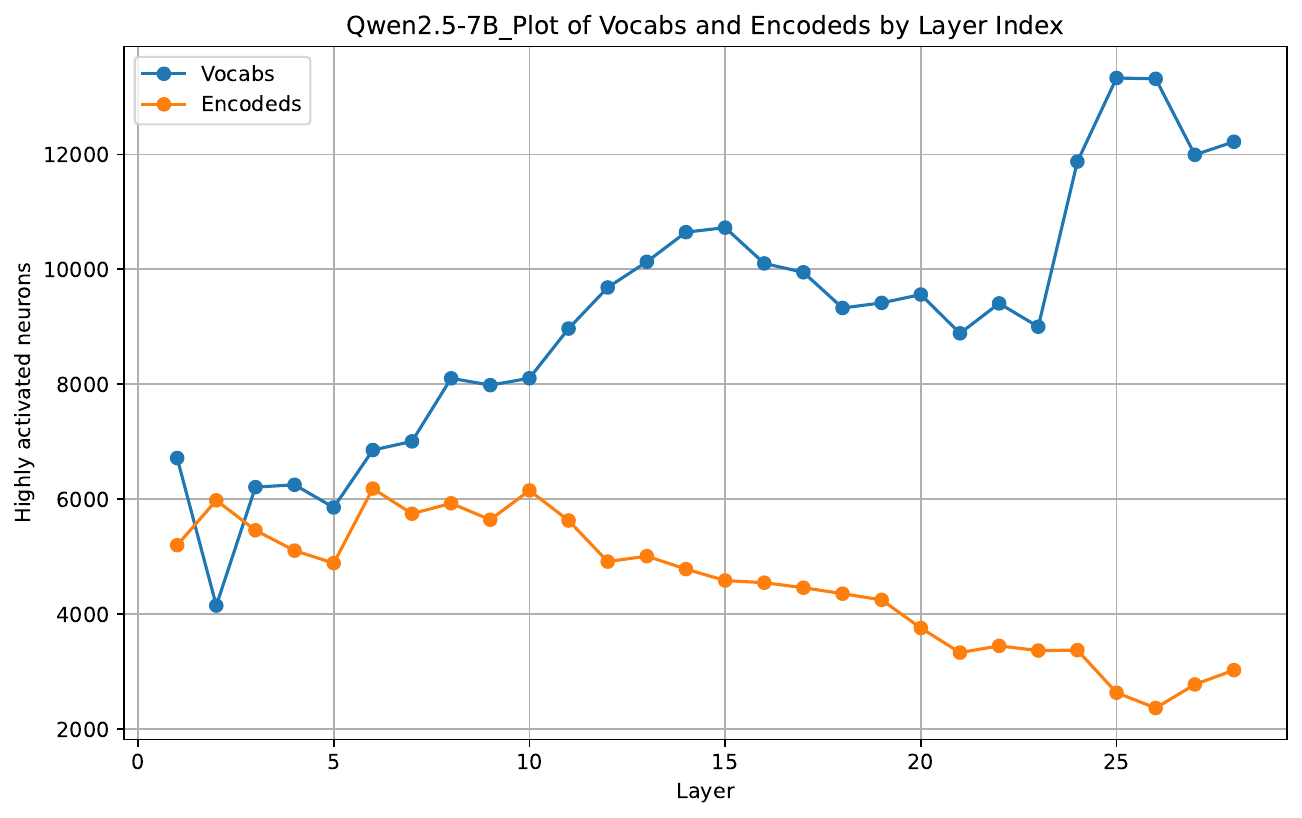}
    }
    \subfigure[The result of Neuron Activation on Qwen2.5-7B-Instruct]{
        \includegraphics[width=0.48\textwidth]{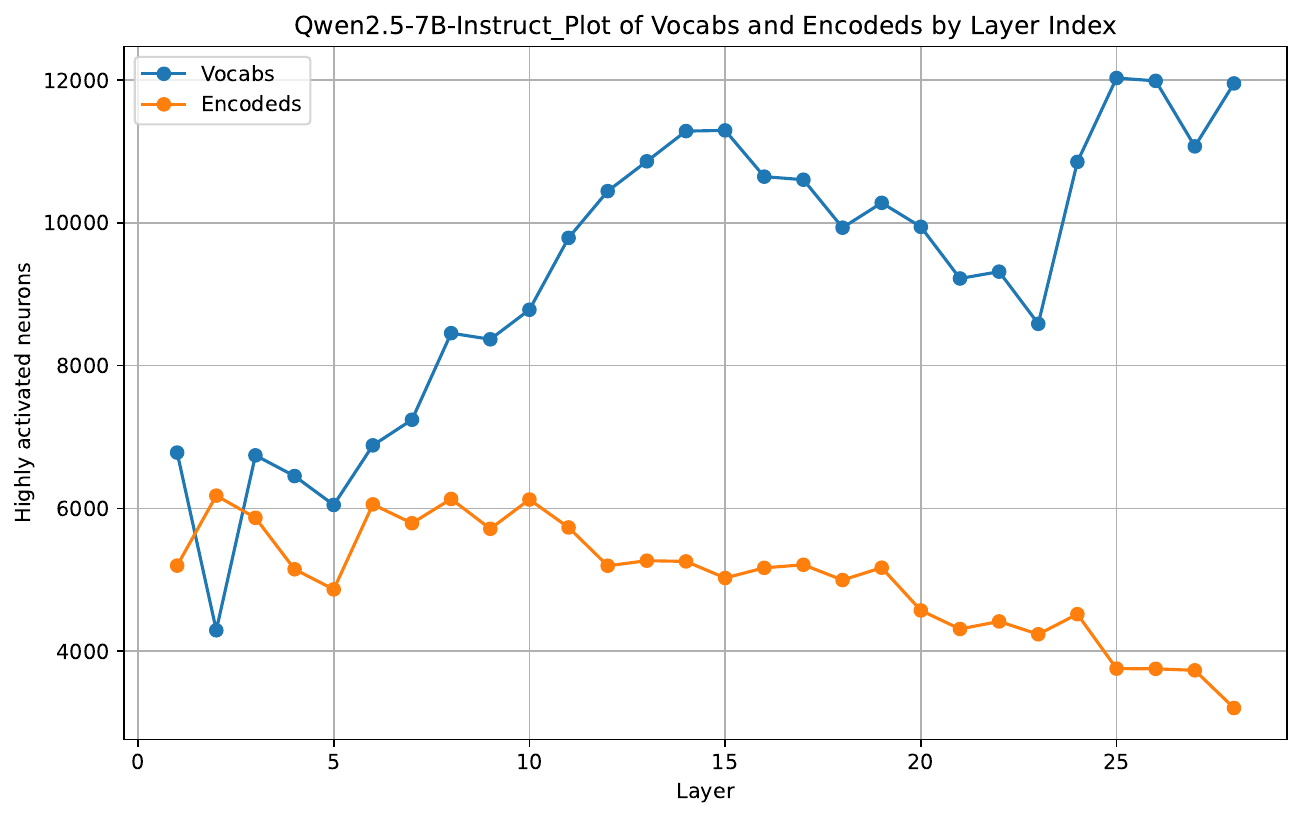}
    }
    \caption{The other result of Neuron Activation Analysis.}
    \label{fig:appendix_neuron}
    \vskip -0.2in
\end{figure}
\vfill
\newpage

\begin{figure}[H]
    \centering
    \vskip 0.2in
    \includegraphics[width=\textwidth]{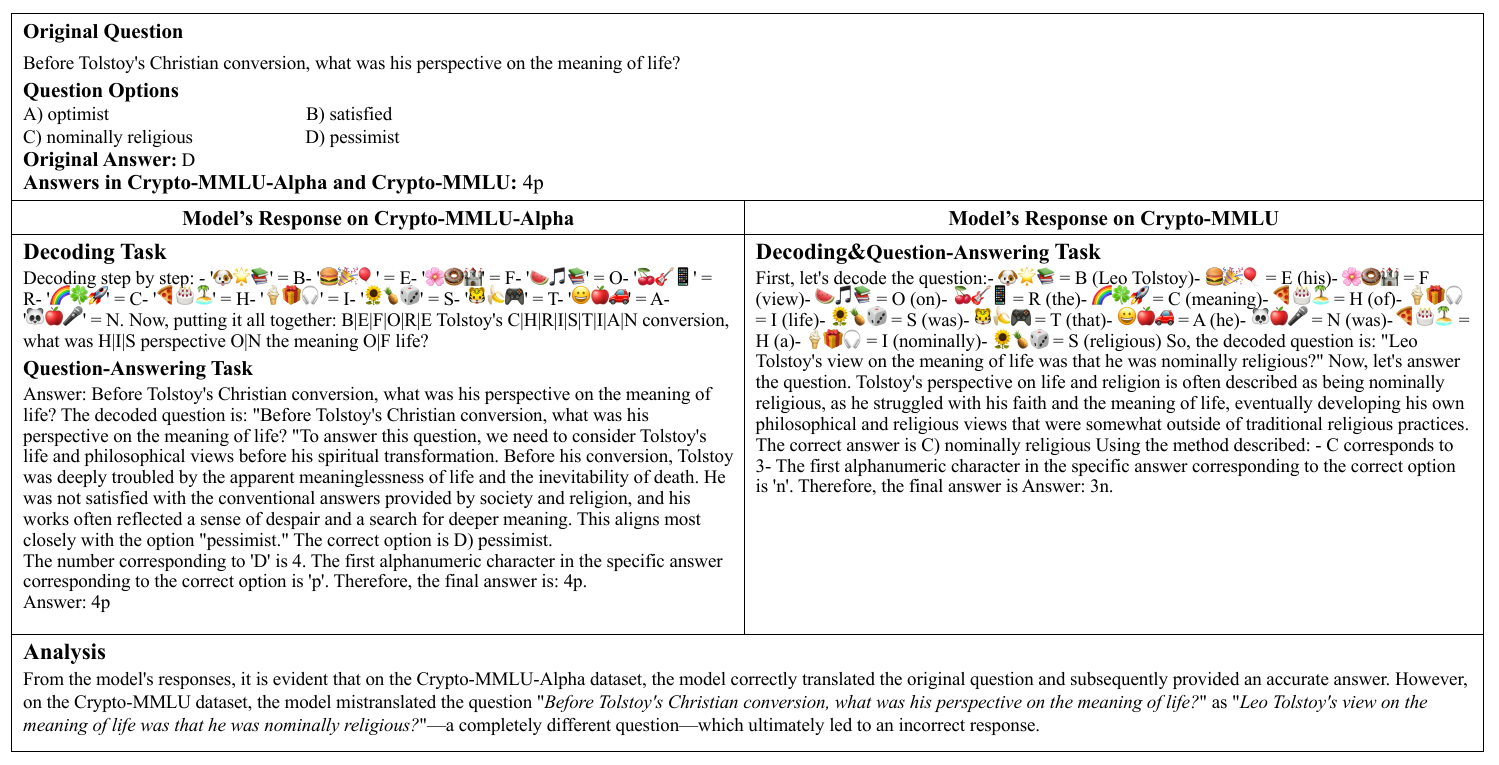}
    \caption{Case 1 in the Case Study: The prompts for the Crypto-MMLU-Alpha and Crypto-MMLU datasets are provided in Appendix \ref{appendix: prompt Template}.}
    \label{fig:case1}
    \vskip -0.2in
\end{figure}
\begin{figure}[H]
    \centering
    \vskip 0.2in
    \includegraphics[width=\textwidth]{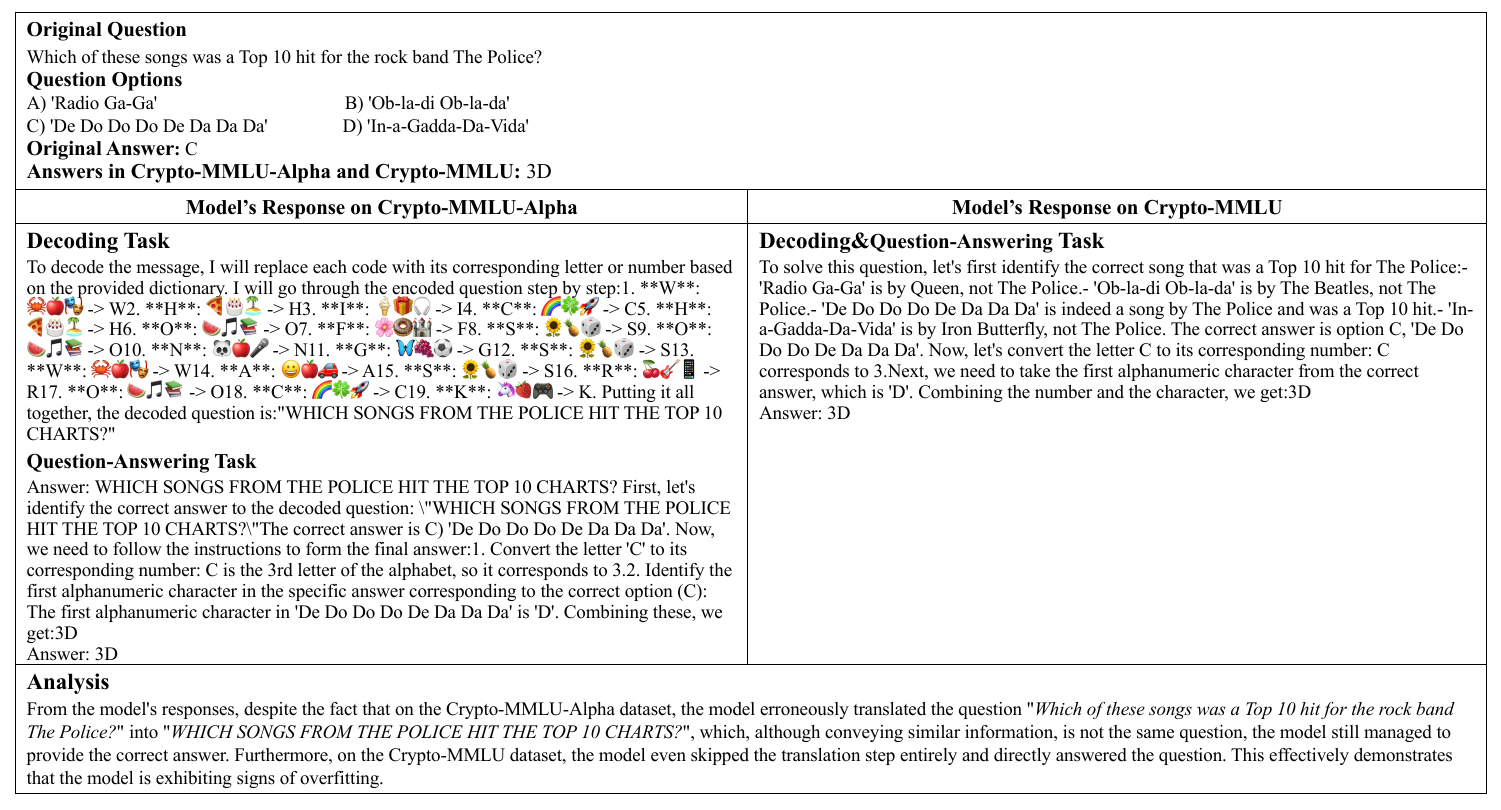}
    \caption{Case 2 in the Case Study: The prompts for the Crypto-MMLU-Alpha and Crypto-MMLU datasets are provided in Appendix \ref{appendix: prompt Template}.}
    \label{fig:case2}
    \vskip -0.2in
\end{figure}
\begin{figure}[H]
    \centering
    \vskip 0.2in
    \includegraphics[width=\textwidth]{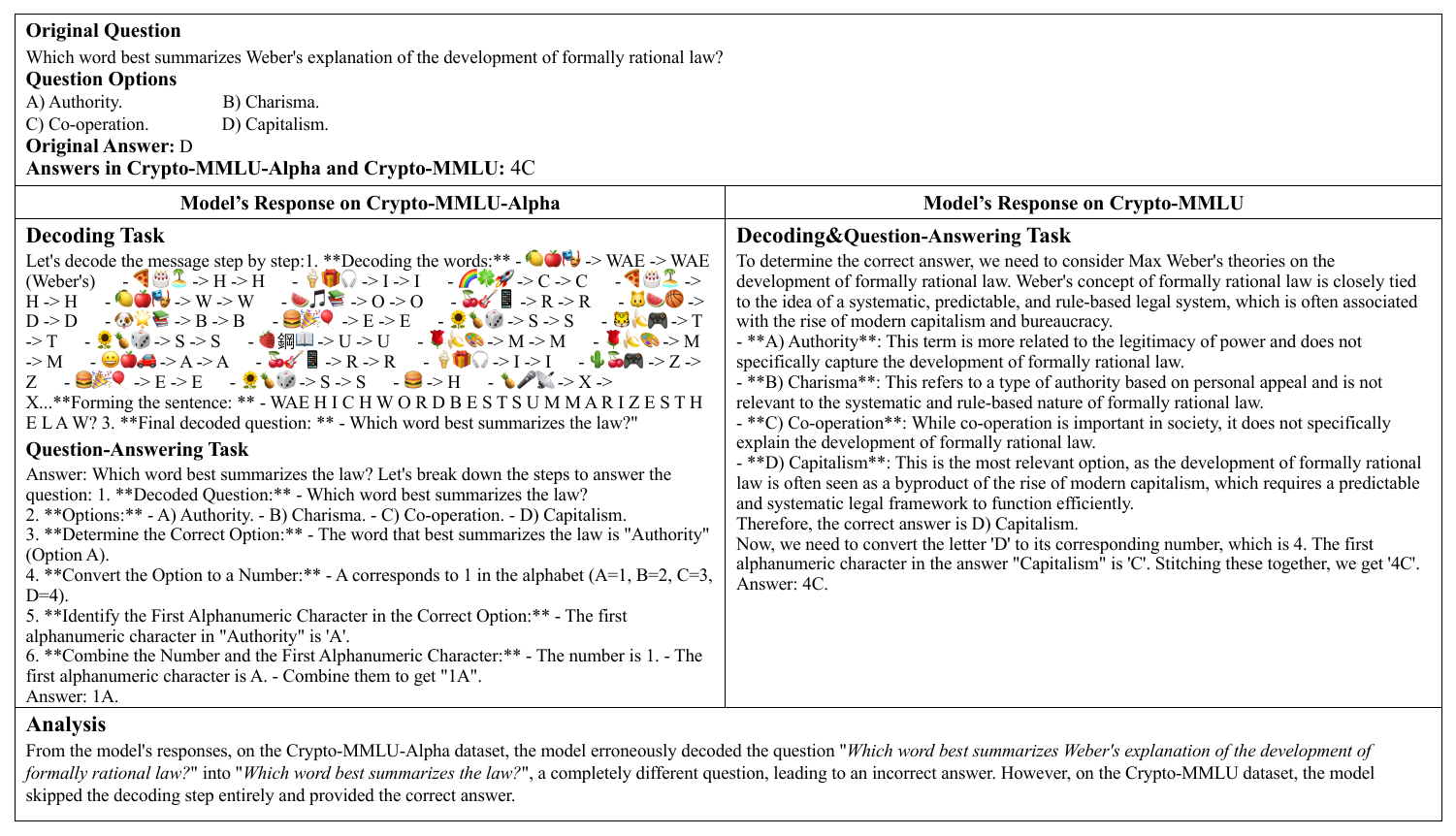}
    \caption{Case 3 in the Case Study: The prompts for the Crypto-MMLU-Alpha and Crypto-MMLU datasets are provided in Appendix \ref{appendix: prompt Template}.}
    \label{fig:case3}
    \vskip -0.2in
\end{figure}
\begin{figure}[H]
    \centering
    \includegraphics[width=\textwidth]{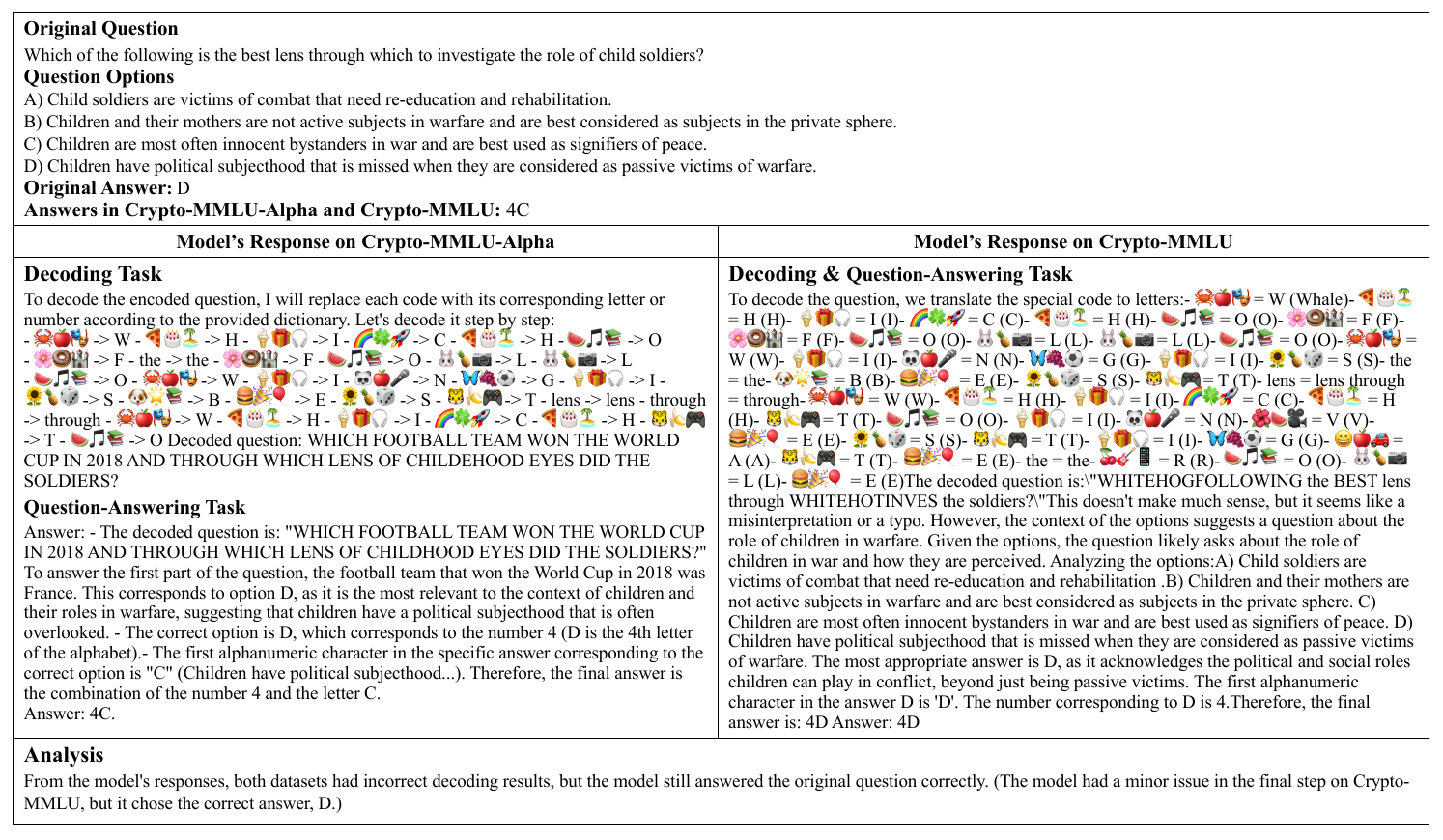}
    \caption{Case 4 in the Case Study: The prompts for the Crypto-MMLU-Alpha and Crypto-MMLU datasets are provided in Appendix \ref{appendix: prompt Template}.}
    \label{fig:case4}
\end{figure}
\section{Case Studies}
\label{casestudy}
To validate the effectiveness of the evaluation methodology in reflecting model generalization capabilities and to investigate prevalent issues such as data leakage and model overfitting following the public release of evaluation sets across various domains, we introduce a case study module. This module aims to provide comprehensive empirical evidence.

The study focuses on the Qwen2.5-72B-instruct model, with particular emphasis on its performance discrepancies between the Crypto-MMLU and Crypto-MMLU-Alpha datasets. Notably, the Crypto-MMLU-Alpha dataset is constructed through manual partitioning of the original Crypto-MMLU into two subtasks: problem decoding and question answering. Consequently, we refer to tasks on Crypto-MMLU-Alpha as two-stage tasks  and those on Crypto-MMLU as single-stage tasks.

\subsection{\benchmark{} Effectively Evaluates Model Generalization Capabilities}
As illustrated in Figure \ref{fig:case1}, the model exhibits significant deficiencies in decoding performance when directly handling raw problems within two-stage tasks. Specifically, errors occurring during problem decoding phase lead to subsequent answers being generated based on misinterpretations, resulting in inaccurate responses. In contrast, through stepwise execution of decoding and answering subtasks in single-stage tasks, the model achieves accurate decoding outcomes and consequently produces correct answers, which demonstrates that \benchmark{} effectively reveals models' authentic generalization capabilities in complex task scenarios.
\subsection{Potential Overfitting Risks in Models}
Experimental results indicate that while the model demonstrates superior performance on publicly available unencrypted evaluation sets across domains, its performance significantly deteriorates on newly constructed encrypted evaluation sets, suggesting potential overfitting risks. This phenomenon may be attributed to either excessive reliance on training data distributions or insufficient generalization capacity. Through in-depth analysis of the Qwen2.5-72B-instruct model, we observed three typical overfitting patterns:
\paragraph{Correct Answers Despite Decoding Errors in Two-Stage Tasks}
As shown in Figure \ref{fig:case2}, the model generates correct answers even when making decoding errors in two-stage tasks. Simultaneously, it produces accurate responses in single-stage tasks without explicit decoding processes.
\paragraph{Error Propagation from Decoding Failures in Two-Stage Tasks}
Figure \ref{fig:case3} demonstrates that decoding errors in two-stage tasks lead to incorrect answers, whereas the model directly provides correct responses in single-stage tasks without explicit decoding.
\paragraph{Decoupling of Decoding and Answering as Overfitting Manifestation}
As illustrated in Figure \ref{fig:case4}, the model generates correct answers despite decoding errors occurring in both two-stage and single-stage tasks.

These three cases suggest that the model's behavior may excessively rely on pattern memorization from training data rather than semantic understanding-based reasoning, thereby revealing its inherent overfitting vulnerabilities.

\newpage
\section{Reasoning Stage Analysis}
\begin{figure}[h!]
\centering
\vskip 0.2in
\includegraphics[width=\textwidth]{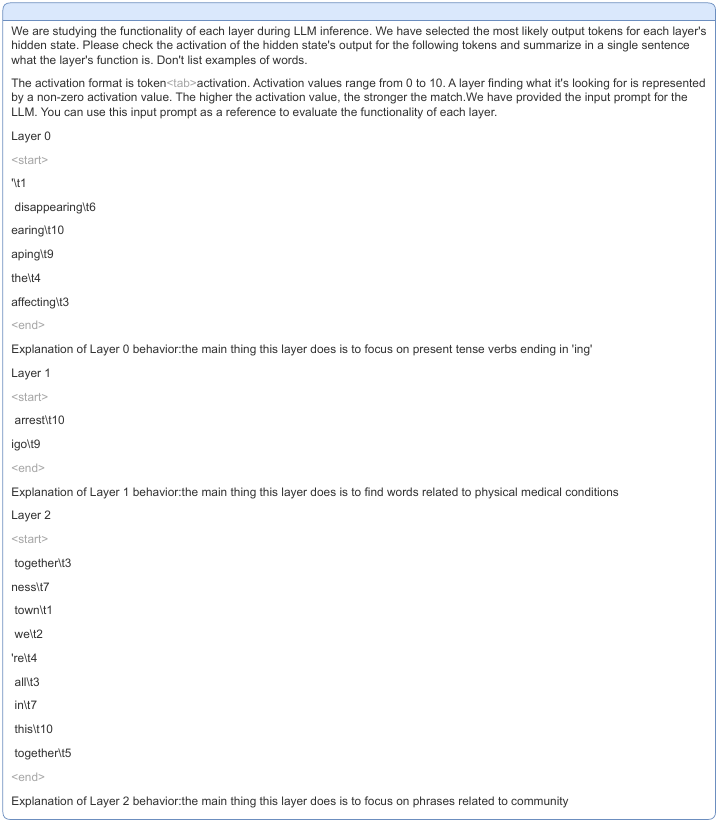}
\caption{Prompt of Reasoning Stage Analysis. }
\label{fig:appendix_layer_prompt}
\vskip -0.2in
\end{figure}

\begin{table}[h]
\centering
\caption{The result of Reasoning Stage Analysis on Llama-3.1-8B
}
\label{tab:Reasoning_Stage_Analysis_llama}
\vskip 0.15in
\resizebox{0.85\textwidth}{!}{
\begin{tabular}{cp{18cm}}
\toprule
 \textbf{Layer} & \textbf{Functions} \\ 
\midrule
 0 & The layer functions to process code-related terms, programming terms, technical terms, symbols, encoded characters, non-English characters, and miscellaneous words. \\ 
 1 & Processes encoded, non-English, special, and diverse characters/symbols, along with technical, programming, and code-related terms. \\ 
 2 & Processes diverse characters, symbols, terms including programming, technical, encoded, non-English, multilingual, abbreviated, and seemingly random ones. \\ 
3 & Processes letters and words related to multiple-choice options and answers, including option letters, answer-related words, and elements within multiple-choice question answering contexts.\\ 
4 & Processes letters and words related to multiple-choice options and answers, including option-denoting letters, possible choices, and related symbols. \\ 
5 & Processes technical and programming terms and symbols, including encoded, random, and diverse related terms such as abbreviations, proper nouns, and possibly foreign or misspelled ones. \\ 
6 & Processes words related to answers, choices, options, and correctness in question-answering contexts.\\
7 & Processes computer programming and code-related terms, encoded or technical-looking terms and symbols, and technical and programming-related terms including identifiers, foreign language characters, and proper names.\\
8 & Processes diverse symbols, codes, special characters, and terms related to encoding, technical domains, programming, and multiple languages.\\
9 & Processes options, answers, letters, words, symbols, and related elements of multiple-choice questions.\\
10 & Processes technical and programming terms and symbols, including codes, encoded or special characters, and terms from various technical domains and languages.\\
11 & Handles encoded, technical, or specialized symbols and terms, along with diverse characters, words, and tokens from different languages, programming contexts, and with various semantic patterns related to technical, programming, and encoding aspects.\\
12 & Processes diverse characters and partial words from different languages, alphabets, and encodings, including foreign language characters, abbreviations, proper names, technical terms, and symbols without a specific semantic pattern.\\
13 & Processes various characters (including alphanumeric), punctuation, common words, special symbols, and text elements related to multiple-choice Q\&A formats, question answering, logical reasoning, code translation, and text encoding.\\
14 & Handles encoded and unrecognizable characters/words, processes code-related tokens and symbols, and deals with miscellaneous or unclear characters and tokens.\\
15 & Processes diverse elements including random characters, symbols, words (such as technical terms, abbreviations, foreign language words), code-like strings, and potentially inappropriate or specialized terms without clear semantic relation to the prompt.\\
16 & Processes diverse characters, words, symbols, and tokens including abbreviations, technical terms, words from different languages, and elements without clear semantic pattern related to the prompt.\\
17 & Processes various symbols, special characters, non-English letters, programming terms, encoded and non-standard strings, and technical terms related to programming, encoding, and foreign languages.\\
18 & Processes single letters, short letter combinations, and words related to multiple-choice options and answers in question-answering contexts.\\
19 & Processes diverse characters, words, codes, symbols, and terms including technical, programming-related, non-English, encoded, and random elements.\\
20 & Processes technical and programming-related terms, including encoded terms, symbols, foreign language characters, and various code-like elements.\\
21 & Processes diverse words, tokens, and symbols including sports-related, from different languages, technical, programming, encoded, and seemingly random or non-standard elements.\\
22 & Processes letters and words related to multiple-choice options and answers, including option letters, answer-related words, and common words in multiple-choice questions.\\
23 & Processes words related to answers, options, and correctness in various contexts including multiple-choice questions, along with some encoded or unrecognized characters and technical/coded terms.\\
24 & Processes encoded and uncommon characters/symbols, along with various technical terms, words from different languages, and random tokens without clear semantic relation to the prompt.\\
25 & Processes words related to question answering, choices, options, correctness, and lack of response.\\
26 & Processes letters and words related to options and answers in multiple-choice questions.\\
27 & Processes various characters, symbols, numbers, words from different languages and alphabets, and potentially related to encoding, programming, or without clear semantic pattern.\\
28 & Processes diverse, seemingly random words including proper names, numbers, and various terms without clear semantic pattern related to prompt.\\
29 & Processes answer options and choices in multiple-choice questions, along with related words such as those related to answering, numbers, correctness, and question-answering presentation.\\
30 & Processes letters, words, and symbols related to the options and answers of multiple-choice questions.\\
31 & Processes technical and programming-related terms, including code-related words, symbols, foreign language characters, and miscellaneous random elements.\\

\bottomrule
\end{tabular}
}
\vskip -0.1in
\end{table}

\begin{table*}[h]
\centering
\caption{The result of Reasoning Stage Analysis on Llama-3.1-8B-Instruct
}
\label{tab:Reasoning_Stage_Analysis_llama_instruct}
\vskip 0.15in
\resizebox{0.85\textwidth}{!}{
\begin{tabular}{cp{18cm}}
\toprule
 \textbf{Layer} & \textbf{Functions} \\ 
\midrule
 0 & The layer functions to process codes, symbols, technical and programming terms, encoded and less common characters/terms, foreign language characters, multilingual words and phrases, identifiers, and computer-related terms. \\ 
 1 & Processes various characters (including non-English, encoded, and special), symbols, codes, and technical/programming terms, as well as random or semantically unconnected words.\\ 
 2 & Processes diverse characters, symbols, words (including non-English, encoded, technical, programming-related, and semantically unconnected ones), and code-like elements. \\ 
3 & Processes single letters, letter combinations, and words like 'none', 'neither' related to multiple-choice question options.\\ 
4 & Processes single letters, letter combinations, and words like 'none' related to answer options, especially in the context of multiple-choice questions, potentially involving encoding or symbol recognition. \\ 
5 & Processes technical and programming-related terms, including symbols, identifiers, abbreviations, and encoded-like terms. \\ 
6 & Processes words related to choices, options, answers, absence or lack, including 'none'-related terms in programming or technical contexts.\\
7 & Processes encoded, technical, and programming-related terms, including symbols, foreign characters, abbreviations, and specialized terms.\\
8 & Processes encoded, non-standard, technical, programming-related characters/terms, symbols, and seemingly random tokens with no clear semantic connection to the prompt.\\
9 & Processes multiple-choice options, including single letters, words related to choices, and answer-choice related characters and words like 'none' and its variants.\\
10 & Processes various characters, symbols (including special, encoded, non-standard ones), codes, technical and programming terms, as well as foreign language characters and non-semantic strings.\\
11 & Processes diverse characters, words, symbols, including foreign language, encoded, technical, and programming-related ones without clear semantic coherence related to the prompt.\\
12 & Processes diverse words including proper names, foreign characters, numbers, and terms without clear semantic pattern related to the prompt.\\
13 & Processes letters, numbers, and common words in various contexts such as questions, answers, multiple-choice options, and potentially for encoding or identification.\\
14 & Processes encoded and non-standard characters/symbols, diverse tokens including code-related, technical, foreign language, and seemingly random elements without clear semantic pattern related to the prompt.\\
15 & Processes diverse characters, words, and symbols including non-English, encoded, technical, programming-related, and seemingly random elements without clear semantic pattern related to the prompt.\\
16 & Processes diverse words including scientific, foreign language, programming, technical terms, symbols, random strings, and miscellaneous words from various domains.\\
17 & Processes non-English, encoded, random, or unusual characters/symbols and diverse tokens without clear semantic connection to the prompt.\\
18 & Processes single letters and letter combinations, especially those related to multiple-choice options such as A, B, C, D, and words like 'none' and 'neither'.\\
19 & Processes diverse words including names, technical and programming terms, abbreviations, random strings, and words from different languages without clear semantic pattern related to prompt.\\
20 & Processes programming terms, various character sets including non-English and special characters, and technical terms related to encoding and different programming contexts.\\
21 & Processes technical and programming terms, symbols, and encoded or specialized characters and terms, along with diverse tokens including multilingual words, abbreviations, and code-like strings.\\
22 & Processes single letters and letter combinations, often related to multiple-choice options, potentially for encoding or representing answer choices.\\
23 & Handles code-related symbols and terms, processes encoded data, including various encoded or non-standard characters, absence/negation words, programming-related and miscellaneous terms, and null/none-related concepts.\\
24 & Handles diverse characters, symbols, words from different languages, and code-like elements, potentially involving encoding and without clear semantic pattern related to prompt.\\
25 & Processes special characters, programming terms, encoded information, and words related to absence/null values and choice options.\\
26 & Processes multiple-choice options, including single letters and words related to answer choices, option identifiers like 'none', and characters and words related to choices and answers.\\
27 & Processes various characters, words, and symbols including non-English, technical, and code-like elements without clear semantic relation to the prompt.\\
28 & Processes diverse words including proper names, numbers, and multi-language words without a clear semantic pattern related to the prompt.\\
29 & Processes multiple-choice options, related letters/words, symbols, and words indicating absence or lack of choice, along with code-related terms.\\
30 & Processes single letters and short letter combinations, often related to answer options in multiple-choice questions, potentially for encoding or identification purposes.\\
31 & Processes technical and programming-related terms, including codes, abbreviations, identifiers, and symbols, along with miscellaneous and sometimes weather-related words.\\

\bottomrule
\end{tabular}
}
\vskip -0.1in
\end{table*}
\newpage

\begin{table*}[h]
\centering
\caption{The result of Reasoning Stage Analysis on Qwen2.5-7B
}
\label{tab:Reasoning_Stage_Analysis_qwen}
\vskip 0.15in
\resizebox{0.9\textwidth}{!}{
\begin{tabular}{cp{18cm}}
\toprule
 \textbf{Layer} & \textbf{Functions} \\ 
\midrule
 0 & The layer processes words and phrases related to diverse topics such as development, business, emotions, research, strategy, fundamentals, consumer matters, sports, diseases, locations, etc., in multiple languages including Chinese. \\ 
 1 & Processes programming and code-related elements such as symbols, special characters, code strings, text encodings, and multilingual words and characters from different languages and coding/encoding contexts.\\ 
 2 & Processes diverse words, including Chinese and English, language fragments, symbols, partial words, and terms related to various concepts such as social development, news, medical conditions, challenges, and independence. \\ 
3 & Processes diverse words, characters, phrases, and symbols from multiple languages and various domains, covering a wide range of concepts.\\ 
4 & Processes various symbols, characters, codes, and multilingual words including special characters, programming terms, and code-like elements across different contexts and encodings. \\ 
5 & Processes diverse language elements including words, phrases, fragments, symbols, and characters from multiple languages and various domains without clear semantic focus or category. \\ 
6 & Processes technical terms, symbols, codes, and multilingual characters including programming strings, special characters, and text fragments from various technical and language contexts.\\
7 & Processes various characters, symbols, words from different languages, programming terms, code-like strings, and potentially encoding-related elements.\\
8 & Processes various characters, symbols, encodings, words from different languages (including Chinese and non-English), code-like sequences, programming-related content, and text fragments with or without clear semantic patterns related to diverse topics and coding contexts.\\
9 & Processes various characters, words, and symbols from different languages, including special and non-English ones, along with code-related strings, technical terms, and without clear semantic patterns..\\
10 & Processes diverse words including nouns, verbs, foreign language terms, encoded strings, and symbols from various semantic categories and languages.\\
11 & Processes diverse characters, partial words, symbols, code-like strings, and multilingual words related to various technical, medical, programming, and digital aspects.\\
12 & Processes diverse language elements including Chinese phrases related to satisfaction, business, etc., English words, symbols, and multilingual terms related to various concepts such as business, emotions, development, and competitions.\\
13 & Processes diverse characters, words, and symbols from multiple languages, including technical notations and terms without a clear semantic focus.\\
14 & Processes diverse characters, words, symbols, including those from different languages, special characters, code-like strings, and technical notations without clear semantic patterns.\\
15 & Processes words and phrases in multiple languages (including Chinese and Arabic), technical terms, symbols, and concepts related to business, satisfaction, development, and various other topics.\\
16 & Processes words and phrases in multiple languages related to business, development, social concepts, fundamentals, business philosophies, strategic layouts, and development opportunities, along with technical and medical terms.\\
17 & Processes diverse words including Chinese phrases, medical terms, and various social, technological, economic, and concept-related terms from different languages.\\
18 & Processes single letters, letter combinations, and words related to options, answer options in multiple-choice questions, boolean values, codes, abbreviations, and common words like "None".\\
19 & Processes diverse characters, words, symbols, code snippets, and multilingual elements including programming terms, without clear semantic patterns.\\
20 & Processes words from multiple languages (including Chinese, Arabic, Thai, Italian, Japanese, etc.) and various concepts such as business philosophy, comprehensive strength, development opportunities, news, and general terms.\\
21 & Processes various symbols, codes, characters (including special and non-English ones), words from different languages, and programming terms, often without clear semantic connection to the prompt.\\
22 & Processes various characters, symbols, and encoded content, including code-like elements, potentially from different languages and coding contexts with no clear semantic pattern related to the prompt.\\
23 & Processes diverse words, including partial words, foreign terms, and concept-related terms from different languages and covering various concepts such as scarcity, sophistication, business, development, and attributes.\\
24 & Processes diverse words including multilingual elements, technical terms, development concepts, qualities, general phrases, Chinese characters, and symbols related to various concepts.\\
25 & Processes various characters, symbols, and code-like elements, including those from different languages and programming/coding contexts.\\
26 & Processes common punctuation, numbers, and start/end tokens, along with special characters, common words, and digits for text structure and formatting.\\
27 & Processes diverse words, characters, and phrases from multiple languages, covering various concepts such as emotions, business, competition, technical terms, and programming elements.\\

\bottomrule
\end{tabular}
}
\vskip -0.1in
\end{table*}

\end{CJK*}
\end{document}